\pdfoutput=1 
\documentclass[10pt]{article}
\renewcommand{\theequation}{\thesection.\arabic{equation}}
\renewcommand{\thefigure}{\thesection.\arabic{figure}}
\renewcommand{\thepage}{\roman{page}}
\makeindex

\usepackage{latexsym} 
\usepackage{makeidx}

\usepackage[twoside,width=28cc,height=38cc,a5paper]{geometry}
\usepackage{graphicx}
\usepackage[dutch]{babel} 

\begin{document}
\flushbottom
\newcommand{\BA}{\begin{eqnarray}}
\newcommand{\EA}{\end{eqnarray}}
\newcommand{\BE}{\begin{equation}}
\newcommand{\EE}{\end{equation}}
\newcommand{\BEN}{\begin{displaymath}}
\newcommand{\EEN}{\end{displaymath}}
\setcounter{tocdepth}{2}
\newcommand{\BO}{\\ \smallskip

\pagebreak[1]
{\bf Opgave} \small
\begin{quotation}
\noindent }

\newcommand{\EO}{\end{quotation} \normalsize \medskip}

\newcommand{\BEQ}{\begin{equation}}
\newcommand{\EEQ}{\end{equation}}
\newcommand{\BEA}{\begin{eqnarray}}
\newcommand{\EEA}{\end{eqnarray}}
\newcommand{\n}{{\bf n}}
\renewcommand{\r}{{\bf r}}
\newcommand{\R}{{\bf R}}
\newcommand{\Q}{{\bf Q}}
\newcommand{\q}{{\bf q}}
\newcommand{\ql}{{\bf q_\perp}}
\newcommand{\qt}{{\bf q_\perp}}
\newcommand{\kt}{{\bf k_\perp}}
\newcommand{\kl}{{\bf k_\perp}}
\newcommand{\p}{{\bf p}}
\newcommand{\I}{{\cal I}}
\renewcommand{\L}{{\cal L}}
\newcommand{\J}{{\cal J}}
\newcommand{\T}{{\cal T}}
\newcommand{\tOmega}{\tilde{\Omega}}
\newcommand{\Omegat}{\tilde{\Omega}}
\newcommand{\roo}{\mbox{\boldmath{$\rho$}}}
\newcommand{\rhoo}{\mbox{\boldmath{$\rho$}}}
\newcommand{\alfa}{\alpha}

\newcommand{\rmi}{{\rm i}}
\newcommand{\vrd}{\frac{1}{4}}
\newcommand{\rre}{\rm Re}
\newcommand{\rim}{\rm Im}

\newcommand{\vc}[1]{{\bf #1}}
\newcommand{\vcp}{\vc{p}}
\newcommand{\vcr}{\vc{r}}

\newcommand{\impur}{verontreiniging }
\newcommand{\impurs}{veront\-reinig\-ingen }

\newcommand{\eexp}[1]{{\rm e}^{\textstyle #1}}
\newcommand{\dpa}{\partial}
\newcommand{\dint}{{\rm d}}
\newcommand{\hlf}{\frac{1}{2}}
\newcommand{\drd}{\frac{1}{3}}
\newcommand{\twdrd}{\frac{2}{3}}
\newcommand{\zsd}{\frac{1}{6}}
\newcommand{\smv}{Schwarzschild-Milne vergelijking }
\newcommand{\tcrl}{{\cal T}}
\newcommand{\olandau}{{\cal O}}
\newcommand{\nolabel}{\nonumber}
\newcommand{\cof}{co\"effici\"ent }
\newcommand{\cofn}{co\"effici\"enten }
\newcommand{\transco}{trans\-missie\-co\"effi\-ci\"ent }
\newcommand{\bea}{\begin{eqnarray}}
\newcommand{\beatrix}{\begin{eqnarray}}
\newcommand{\claus}{\end{eqnarray}}

\newcommand{\idim}{\'e\'en\-di\-men\-si\-o\-na\-le }
\newcommand{\iiidim}{drie\-di\-men\-si\-o\-na\-le }

\newcommand{\GB}{\overline{G}}
\newcommand{\VB}{\overline{V}}

\newcommand{\figcap}[2] {\begin{center}
\raisebox{0.2mm}{\parbox[b]{2.5cm}{\caption{} \label{#1}} }
\parbox[t]{12cm}{{\sl #2}}
\end{center}}
\newcommand{\figcapw}[2]
{\begin{center}
\refstepcounter{figs}
\parbox[t]{6.125in}{\small \label{#1} {\bf Figure~(\ref{#1})}#2}
\end{center}}
\newcommand{\rightcap}[5]
{\parbox[b]{#2}{\hspace{#2} \vspace{#1}}
\parbox[b]{#3}{\parbox{2.5cm}{\caption{} \label{#4}} \\
[0.5cm] {\sl #5}}}
\newcommand{\beq}{\begin{equation}}
\newcommand{\eeq}{\end{equation}}
\newcommand{\beqa}{\begin{eqnarray}}
\newcommand{\eeqa}{\end{eqnarray}}
\newcommand{\bfig}{\begin{figure}}
\newcommand{\efig}{\end{figure}}
\renewcommand{\d}{{\rm d}}
\newcommand{\shc}{\mbox{sinhc}}
\newcommand{\etal}{{\it{et al. }}}
\def\to{\rightarrow}
\def\expect#1{\langle{#1}\rangle}

\setcounter{page}{-1}
\begin{center}
\Huge{Veelvoudige verstrooiing \\van golven}
\\
\bigskip
\bigskip
\Large{Theo M. Nieuwenhuizen}
\\
\Large{Instituut voor Theoretische Fysica}
\\
\Large{Universiteit van Amsterdam}
\\
\bigskip
\end{center}

 \vspace{2cm}

\newpage

\begin{center}
\Huge{Veelvoudige verstrooiing \\ van golven}
\bigskip \\
\Large{College voorjaar 1993}
\end{center}
\bigskip
\bigskip

 Abstract: 
  These are notes (in Dutch) of lectures on multiple scattering of waves, 
 given at the University of Amsterdam in 1993.
 At request of some of the students, now professors, they are made publicly available. 

The main part of the material is covered in the review paper:  
{\it Multiple scattering of classical waves: microscopy, mesoscopy, and diffusion,} 
by M. C. W. van Rossum and myself, Rev.  Mod. Phys. {\bf 71} (1999) 313 -- 371.

\section*{Voorwoord}
Dit dictaat behoort bij een college over veelvoudige
verstrooiing van golven, zoals licht. Basis voor dit college was het
artikel `Skin Layer of Diffusive Media' van Nieuwenhuizen en 
Luck \cite{NL11}, welk we in het vervolg met [NL] zullen aanduiden. 
Het college werd gegeven in de eerste helft
van 1993 aan de Universiteit van Amsterdam door Theo Nieuwenhuizen.
Gastcolleges werden verzorgd door Alexander L. Burin (Kurchatov Instituut,
Moskou) en Mark
van Rossum. De aantekeningen zijn uitgewerkt door Johannes de Boer,
Coen van Duin, Allard Mosk, Peter den Outer en Mark van Rossum.
We bedanken Jean-Marc Luck voor het verschaffen van enkele figuren.

Het dictaat is bedoeld als introductie tot het onderwerp en als naslagwerk, met referenties na ieder hoofdstuk.
Het beschrijft de behandelde stof en is aangevuld met zowel standaardtheorie als recente resultaten uit de 
werkgroep `Spectroscopie der Verdichte Materie' van het van der Waals-Zeeman Laboratorium van de Universiteit
van Amsterdam.

\bigskip

Redaktie: Theo Nieuwenhuizen en Mark  van Rossum.

\bigskip

Grafische vormgeving in LATEX: M.C.W. van Rossum.

Deze is er op gericht om 2 pagina's per A4 af te drukken.


\section*{Bij de tweede druk,  2014}
Twee decennia na dato blijkt dit dictaat nog steeds een werkzaam uitgangspunt te zijn voor
de introductie tot veelvoudige golfverstrooiing.
Op sommiger verzoek is het herzien en op arXiv geplaatst. 
Wederom met dank aan Mark van Rossum en ook aan Joop Hovenier.

In de tussentijd is een Engelse versie verschenen als deel van het overzichtsartikel [1].
Een aantal verdere onderwerpen is eveneens uitgewerkt.
De ruimtelijke verdeling van de totale transmissie wordt behandeld in [2]. 
Voor de toepassing op sterke voorwaartse verstrooiing, zie [3].
Voor de uitbreiding naar vectorgolven, zie [4].
Het geval van scalaire Rayleigh-verstrooiers wordt beschouwd in [5].


\newpage
\tableofcontents
\bigskip
\bigskip
\bigskip
\bigskip
\bigskip
\bigskip
\bigskip
\bigskip
\bigskip
\bigskip
\nopagebreak[3]

\noindent \copyright iets van dit diktaat mag zonder toestemming van de
uitgevers
gecopi\"eerd worden. Gehele of gedeeltelijke vertaling, in het bijzonder
mondelinge in \'e\'en of ander soort Engels, is ten strengste
verboden.

\renewcommand{\thepage}{\arabic{page}}
\setcounter{page}{1}
\renewcommand{\theequation}{\thesection.\arabic{equation}}
\renewcommand{\thefigure}{\thesection.\arabic{figure}}
\section{Inleiding}
\label{inleiding}
Het onderwerp van veelvuldige verstrooiing van golven heeft vele 
`klassieke' en `moderne' toepassinsgebieden. De eerste studies werden gedaan
door astrofysici voor de beschrijving van stralingstransport door 
steratmosferen en interstellaire wolken. Directe toepassingen van
veelvoudige lichtverstrooiing zijn onderzoek naar de strictuur en samenstelling van
planetaire atmosferen, wolken en mist. Ook kent iedereen witte verf 
en melkachtige vloeistoffen (karnemelk, luchthoudend water, 
suspensies van witte verf). Meer recente toepassingen 
zijn: verstrooiing van electronen door onzuiverheden bij lage temperaturen,
verstrooiing van neutronen, 
inverse verstrooiing van seismische golven en het gebruik van zichtbaar licht
voor het opsporen van afwijkingen in weefsels van levende wezens.

Om het probleem van veelvuldige lichtverstrooiing voor te stellen denke
men zich omgeven door een dichte mist. De eerste vraag is: hoever kan ik nog
scherp zien ? Deze afstand heet de {\it vrije weglengte} $\ell$\index{vrije
weglengte}\index{$l@$\ell$, vrije weglengte}, gedfini\"eerd als $\ell=1/n\sigma$, waarbij $n$ de verstrooiersdichtheid is en $\sigma$ hun werkzame doorsnede \footnote
{We zullen hier nog niet het onderscheid maken tussen de verstrooiings-vrije 
weglengte en de transport-vrije weglengte. We beperken ons in dit dictaat dus tot
isotrope verstrooiing.}. Voor wolken en mist kan $\ell$ vari\"eren 
van vele kilometers tot enkele meters.
Komt je in de mist een auto tegemoet, dan kun je zijn koplampen zien
als de afstand ongeveer \'e\'en vrije weglengte is. Een vergelijkbare situatie
treedt op wanneer er wolken voor de zon komen. Je kunt het directe
(niet verstrooide) zonlicht zien zolang de wolk minder dan \'e\'en
vrije weglengte dik is
 \footnote{Nogmaals: Stel, een auto komt je tegemoet in de mist. Wat zie je eerder: de koplampen of de diffuse gloed?
 Controleer het antwoord aan de hand van de vraag of het donker is wanneer de wolken te dik zijn om de zon te zien.}.

Dit dictaat behandelt ten eerste aspecten van de standaard
theorie voor stralingstransport ten gevolge van
veelvuldige verstrooiing van golven. De basis voor
dit werk werd in het midden van de vorige eeuw gelegd door astrofysici. 
Bekend zijn de boeken van Chandrasekhar \footnote{S. Chandrasekhar, geboren in 1910 in India als neefje van Raman,
heeft ondermeer de evolutie van witte dwergen verklaard. Boven de
`Chandrasekhar-limiet' van 1.2 zonnemassa's zal een ster exploderen,
daaronder wordt hij een witte dwerg. In 1983 kreeg Chandrasekhar de
Nobelprijs met W. A. Fowler voor zijn theoretische studies van de structuur
en ontwikkeling van sterren.}, van de
Hulst \footnote{Van de Hulst (Leiden) is de ontdekker van onder andere de 21
cm-lijn van
waterstof, die geleid heeft tot de bouw van de radiotelescoop in Dwingelo. Zijn boeken gaan vooral over enkelvoudige verstrooiing.}
~\cite{vdHulst}, alsmede die van Ishimaru~\cite{Ishim}.
Zij legden zich toe op de `stralingstransportvergelijking'
(`radiative transfer equation').

In dit college zijn we ge\"interesseerd in optisch dichte situaties.
De bron is dan vele vrije weglengtes verwijderd van het observatiepunt.
In de situatie met auto's in mist weten we dat er dan een {\it diffusieve
gloed}\index{gloed} is. Die kun je zien lang voordat de koplampen opduiken.
Ook voor wolken is het wel bekend: in plaats van het directe zonlicht
zien we diffusief licht dat zich min of meer homogeen door het hele wolkendek
voortplant \footnote{Interessant is overigens de situatie voor onweerswolken.
Dan kan er door de sterke electrische velden  absorptie van het 
invallende licht optreden, waardoor de wolken er dreigend zwart uitzien.}.

Na een algemene introductie (zie hier onder) bespreken we 
diffusief transport in hoofdstuk \ref{macroscopie}
en de stralings\-transport\-vergelijking in hoofdstuk \ref{mesoscopie}.
Vervolgens gaan we uit van het golfkarakter van het probleem. We 
voeren  de $t$-matrix in (hoofdstuk \ref{microscopie}) en de
 amplitude Greense functie c.q. de zelfenergie (hoofdstukken
\ref{greense} en \ref{intermezzo}).  Vervolgens wordt
de transportvergelijking in de ladder-benadering besproken en expliciet
opgelost voor half-oneindige media en voor plakken. 
De bekende oplossing wordt  uitgebreid
met effecten van interne reflecties, die optreden wanneer de
di\"electrische constante van het verstrooiend medium afwijkt van die van 
de omgeving, meestal lucht of glas. (hoofdstuk \ref{transport}-\ref{plak}).
In hoofdstuk \ref{terugstrooikegel} wordt de vorm van de terugstrooikegel
geanalyseerd. Vervolgens wordt besproken dat er vereenvoudigingen optreden
wanneer de verhouding van brekingsindices van het medium en het omringende
di\"electricum groot is (hoofdstuk \ref{groot}). Daarna wordt de situatie met
anisotrope verstrooiing onder de loep genomen (hoofdstuk~\ref{anisotrope}).
Tenslotte wordt aandacht geschonken aan verstrooiing door \'e\'en extra 
verstrooier (hoofdstuk \ref{invloed}) en aan spikkelcorrelaties (hoofdstuk
~\ref{correlatie}).

Voorlopig zullen we alle vectoraspecten van licht buiten beschouwing laten. 
We bestuderen dan eigenlijk akoestische golven. 
De resultaten kunnen als voorspelling voor licht gebruikt
worden, zolang de rol van de polarisatie ondergeschikt is.
Het probleem van verstrooiing van echt licht zal zeer analoog verlopen; 
verdere complicaties hebben voornamelijk met de vectorstructuur te maken.

\subsection{Lengteschalen}

De beschrijving van stralingstransport kan op drie afstandsschalen gebeuren:

-{\it Macroscopisch:} \index{macroscopie}Voor afstanden veel groter dan de
vrije weglengte  voldoet de lichtintensiteit aan een diffusievergelijking.
De diffusieconstante $D$ is een systeem parameter die uit de microscopie 
afgeleid moet worden. Enkele voorbeelden worden behandeld in hoofdstuk 2.

-{\it Mesocopisch:} \index{mesoscopie}Voor afstanden van de orde van
\'e\'en vrije weglengte
wordt de beschrijving gegeven door de stralingstransportvergelijking voor
de specifieke intensiteit. Dit
is de `Boltzmann' vergelijking voor het betreffende probleem. Op dit nivo
moet men de vrije weglengte $\ell$ en de transportsnelheid $v$
kennen. Voor afstanden veel groter dan \'e\'en vrije weglengte volgt dan
de diffusievergelijking met diffusieco\"efficient $D=\frac{1}{3}v\ell$.
Interferentie-effecten spelen geen rol hier.
Deze benadering wordt vanuit een mesoscopisch standpunt behandeld in
hoofdstuk 3. In verdere hoofdstukken komt hij telkens terug uit de microscopie.

-{\it Microscopisch:} \index{microscopie}Het probleem is
 vastgelegd door de bewegingsvergelijkingen
van golven in een medium met verstrooiers. Het golfkarakter en
interferenties komen naar voren. De typische afmeting van
verstrooiers is vaak \'e\'en micron, dus vergelijkbaar met de golflengte van 
zichtbaar licht. Resonantie-effecten zijn dan belangrijk. De mesoscopische 
Boltzmannbeschrijving volgt door de zogenaamde {\it ladderdiagrammen}
te beschouwen. 

De microscopische benadering zal veelal als uitgangspunt
dienen in dit dictaat.
Het voordeel van  deze beschrijving is dat hij de `uiteindelijke 
waarheid' bevat. Fundamentele aspecten, zoals de terugstrooikegel, 
de transportsnelheid, de doos van Hikami, kunnen alleen op dit nivo
bewezen worden.
Correcties kunnen systematisch berekend worden. De kleine
parameter in de theorie is de verhouding van golflengte
en vrije weglengte, $\lambda/\ell$.\index{$$l@$\lambda$, golflengte}
In typische gevallen is deze parameter van de orde van \'e\'en
tot tien procent.

Het nadeel van de microscopische beschrijving kan zijn dat hij erg 
gedetailleerd is. 
Men bedenke zich echter dat in de praktijk de precieze vorm van de verstrooiers
meestal niet eens bekend is. Voor diverse toepassingen kan men zelfs volstaan
met de diffusiebenadering.

\subsection{Zwakke localisatie, gesloten paden en de
terugstrooi\-kegel\index{localisatie!zwakke}} Zoals reeds opgemerkt, zijn er
correcties op de stralingstransportvergelijking.
Deze zijn afkomstig uit een onderliggende golfvergelijking.
Zolang de verhouding $\lambda/\ell$ klein is, zijn dit kleine correcties. Zij
worden daarom `zwakke localisatie' effecten genoemd. 

Voor electronen zijn de zwakke localisatie effecten voor het eerst geanalyseerd door
Altshuler, Aronov en Spivak~\cite{AAS} \footnote{ B.L. Altshuler,
A.G. Aronov, A.L. Larkin en B.Z.
Spivak hebben in 1992 de Hewlet-Packert Europhysics Prijs gewonnen voor hun
werk aan zwakke localisatie.}.  Zij stelden voor de
weerstand\index{$R@$R$, weerstand} $R$ van een mesoscopische geleider te
meten als functie van een extern magneet veld.
Hij vertoont dan kleine, reproduceerbare fluctuaties.~\cite{SharvinSharvin}
  Deze heten wel de
`magnetische vingerafdrukken' (`magneto fingerprints') van het systeem. 
Het blijkt dat de fluctuaties in de geleiding\index{geleiding}
\index{$G@$G$, geleiding}
$G=1/R$ een grillige functie van het veld zijn, maar gemiddeld de universele grootte
 $e^2/h\approx (25 k\Omega)^{-1}$ hebben.\index{$e@$e$, electronlading}
 \index{$h@$h$, constante van Planck}
 Dit gedrag valt eenvoudig te begrijpen
met behulp van de Landauer formule \index{Landauer formule}
\index{geleidbaarheid}voor de geleidbaarheid~\cite{FisLee}
\BEQ G=\frac{2e^2}{h}\sum_{a,b}T^{flux}_{ab} \label{Landauer}\EEQ
waarbij $T^{flux}_{ab}$\index{$Tab@$T_{ab}$ transmissieco\"efficient} de
(flux-)transmissie\-co\"efficient\index{transmissieco\"efficient} is van een
inkomende vlakke
electrongolf met golfvector $a$ (inkomend kanaal)\index{$a$, inkomend
kanaal}\index{inkomend kanaal}  naar uitgaande golfvector
(uitgaand kanaal) $b$.\index{$b$, uitgaand kanaal}\index{uitgaand kanaal}
 De factor $2$ in vgl. (\ref{Landauer}) komt van de
spinontaarding van de electronen. $T$ is 0 (1) als het kanaal dicht (open)
is. De waarden tussen 0 en 1 komen nauwelijks voor, daarom zullen
fluctuaties in
de som optreden als een kanaal open of dicht gaat; dit leidt tot een effect 
van orde $e^2/h$.

De universele geleidbaarheidsfluctuaties (`uni\-versal conduc\-tance
fluctua\-tions', `UCF' ) worden beschreven door de
$C_3$\index{correlatiefuncties}\index{$C3@$C_3$, correlatie} correlatie
functie, die verderop besproken wordt. In principe geven deze effecten een
verlaging van het geleidingsvermogen. Dit wordt veroorzaakt doordat 
de kans om terug te keren op de uitgangspositie een waarschijnlijkheid
heeft die groter is dan men in de `klassieke' of `Boltzmann'-theorie aanneemt.
Laten we kijken naar de amplitude $E$ voor een effect dat opgebouwd is uit
 bijdragen $C_p$ van vele interferentiepaden\index{interferentie} $p$.
Bijdragen waarbij
het zelfde pad in de omgekeerde richting doorlopen wordt noemen we $D_p$.
(Indien er niet zo'n pad is, nemen we natuurlijk $D_p=0$).
\BEQ E=\sum_p (C_p+D_p) \EEQ
De bijbehorende intensiteit is
\BEQ |E^2|=\sum_p ( C^*_pC_p+D^*_pD_p)+\sum_p(C^*_pD_p+D^*_pC_p)
+\sum_{p\neq p'}(C^*_p+D^*_p)(C_{p'}+D_{p'}) \EEQ
In de Boltzmanntheorie wordt alleen de {\it waarschijnlijkheden} voor
individuele processen meegenomen, dus alleen de termen $C^*_pC_p
+D^*_pD_p$. In een golftheorie met tijdsomkeerinvariantie
zijn de bijdragen voor gesloten paden hetzelfde, ongeacht 
hun omloopszin, d.w.z. $C_p=D_p$.
 Hieruit volgt dat de termen $C^*_pD_p+D^*_pC_p$ even groot zijn:
 {\it er is een extra factor 2 voor gesloten paden}. Dit beschrijft een
vergrote kans om terug te keren naar het uitgangspunt.

In de optica is er een `voorloper' van dit verschijnsel, namelijk de 
terugstrooi-kegel. (Het is niet echt een voorloper maar eerder het equivalent).
 Het `gesloten pad' hier betekent dat je naar het
gereflecteerde signaal kijkt precies in de richting waarmee je instraalt
(`terugstrooirichting'). Zoals voorspeld in door Barbaranenkov in
1973~\cite{Barbara} en waargenomen door onafhankelijke groepen 
(in 1984 door Kuga en Ishimaru \cite{KugaIsh} en in 1985 door zowel
van Albada en Lagendijk~\footnote{Ad Lagendijk heeft in 1991 de Physica-prijs
gekregen en de Physica-lezing `Licht is alles' uitgesproken.} als door Wolf
en Maret \cite{vALag},
\cite{WMM}), is er {\it gemiddeld} in de terugstrooirichting een
intensiteitspiek die maximaal twee keer zo groot is als de achtergrond. Hoewel
de hoogte ervan een effect van orde \'e\'en is,
is het een zwakke-localisatie-effect
omdat het een smalle piek is met breedte $\lambda/\ell$ rond de 
terugstrooirichting. Kijkt men niet precies in de terugstrooirichting
dan neemt het effect af omdat de interferentie destructief wordt;
de propagatoren worden dan `massief'\footnote{Vergelijk met de vier
fundamentele krachten: bij zwaartekracht en electromagnetische kracht is het
interactiedeeltje (graviton, photon) massaloos, de dracht is dan oneindig;
bij kernkrachten zijn de interactiedeeltjes massief, de kracht valt dan
exponentieel af op grote afstanden.}.

Voor electronen
is het analogon de onderdrukking van lusdiagrammen (`loops') door een extern
magneetveld. Let wel: de door interferentie\"effecten veroorzaakte
vergroting van de terugkeerkans wordt teniet gedaan door een veld. Dus het
aanbrengen van een veld {\it vergroot} het geleidingsvermogen
\footnote{Hier laten we spin-baan-verstrooiing buiten beschouwing; die leidt
namelijk tot tegenovergestelde effecten.}.

De universele geleidbaarheidsfluctuaties\index{universele geleidbaarheids
fluctuaties} en de versterkte terugstrooikegel
hebben aanleiding gegeven tot intensieve studie van mesoscopische fluctuaties.
Recente bijdragen zijn te vinden in de boeken  uitgegeven door van Haeringen
en Lenstra ~\cite{vHL} en van  Altshuler, Lee en Webb.~{\cite{AWL}}

\subsection{De transportsnelheid\index{transportsnelheid}}
Een effect waarvan het
belang pas recentelijk ingezien is, is het volgende~\cite{vATLT}.
Stel je voor dat straling door resonantie effecten een tijdje
$\tau_{dw}$\index{$$t@$\tau_{dw}$, verblijftijd} binnen of in de buurt van
iedere verstrooier verblijft (verblijftijd, oplaadtijd, `dwell time',
`charging time')\index{verblijftijd}\index{oplaadtijd}.
De reistijd naar andere verstrooiers is natuurlijk gelijk aan de gemiddelde
afstand tussen twee verstrooiingen, $\ell_{sc}$,\index{$lsc@$\ell_{sc}$,
verstrooiingsvrije weglengte}  gedeeld door de lichtsnelheid in het medium,
$\tau_{sc}=\ell_{sc}/c$.\index{$$tsc@$\tau_{sc}$, reistijd}
De gemiddelde tijd tussen per verstrooiing is daarom
\BEQ \tau=\tau_{sc}+\tau_{dw}.\label{tau=}\EEQ
\index{$$t@$\tau$, totale tijd per verstrooiing}
{\it De transportsnelheid\index{transportsnelheid} is dus  kleiner dan de
lichtsnelheid}, namelijk 
\BEQ v=\frac{\ell_{sc}}{\tau}=
c\frac{\tau_{sc}}{\tau_{sc}+\tau_{dw}}.\index{$v@$v$, transportsnelheid}\EEQ
Dit effect is van belang als de grootte van de verstrooiers
van de orde van de golflengte is, omdat er dan resonanties op kunnen treden.
De reductie in de snelheid is typisch een factor $10-100$
 voor een volumefractie van $25\%$ TiO$_2$
 ~\footnote{Spreek uit: tie-joo-twee.}
 deeltjes in lucht
~\cite{vATLT}. Hetzelfde effect treedt op bij vastgeprikte atomen als
de ingestraalde frequentie dichtbij de resonantiefrequentie is.
In
zo'n geval kan de reductiefactor typisch $10^5-10^6$ zijn in het centrum van
 de lijn \footnote{Dit effect wordt vaak uitgesmeerd door de thermische
beweging van de resonante atomen, zowel in een gas als in een vaste stof.
Deze beweging zorgt voor Doppler verbreding van de lijn. Is deze verbreding
groter dan de natuurlijke lijnbreedte, dan wordt het resonantieeffect 
uitgesmeerd.}.
De lage waarde van de transportsnelheid verklaart dat de experimenteel
bepaalde diffusieconstante veel kleiner is dan men naief zou verwachten
\cite{vATLT}.

\subsection{Sterke localisatie: de Anderson overgang en het 
Ioffe-Regel criterium\index{localisatie!sterke}\index{Anderson localisatie}}

Men kan zich de vraag stellen: als je de hoeveelheid wanorde in het
systeem vergroot, kan dan de diffusieconstante (of geleidbaarheid) van het
systeem naar nul gaan? Deze vraag werd beantwoord in het beroemde
artikel van Anderson in 1958
\footnote{Tezamen met N.F. Mott en J.H. van Vleck heeft P.W.
(`Phil') Anderson in 1977 de Nobelprijs gekregen voor hun
fundamentele onderzoek aan de electronische structuur van
magnetische en wanordelijke systemen.}~\cite{PWA}.
Het is experimenteel bekend dat een amorfe geleider een overgang kan maken
naar een isolator (metaal-isolator overgang, `metal-insulator transition', 
`MIT'). Dit gebeurt wanneer alle relevante eigenfuncties van het systeem
exponentieel gelocaliseerd worden, de zogenaamde Anderson overgang. 
Ioffe en Regel hebben opgemerkt dat de vrije weglengte in een systeem 
nooit kleiner kan zijn dan de golflengte. Met andere woorden: voorbij dat
punt is de theorie van zwakke localisatie niet meer geldig. In drie dimensies leidt dit
tot het Ioffe-Regel criterium\index{Ioffe-Regel criterium} voor het optreden
van Anderson localisatie indien \BEQ k\ell=\frac{2\pi\ell}{\lambda}=1. \EEQ
\index{$k$, golfgetal}In \'e\'en dimensie blijken alle
toestanden exponentieel gelocaliseerd in
de thermodynamische limit. Quasi-metallisch gedrag treedt wel op wanneer de
systeemlengte kleiner is dan de localisatielengte. In twee dimensies is de 
situatie hetzelfde, maar de localisatielengte kan astronomisch groot zijn,
in welk geval een mesoscopisch systeem meer op een niet-gelocaliseerd systeem (metaal) lijkt.

De vraag of er in optische systemen een Anderson overgang kan optreden is nog
steeds niet volledig beantwoord. Experimenteel zou een gelocaliseerd systeem
moeten werken als een perfecte, maar niet vlakke spiegel. Inmiddels is het
wel duidelijk geworden dat het niet eenvoudig zal zijn zo'n systeem te maken.

Een theoretische beschrijving voor de Anderson overgang is de zelf-consistente
theorie van Volhardt en W\"olfle. Dit is een min of meer recht-toe-recht-aan
methode. Een recent overzicht is te vinden in
\cite{VW}. 

 Een meer fundamentele aanpak gaat uit van een veldentheorie
voor het systeem. De metallische fase wordt beschreven door een 
intensiteitsachtig veld (samengesteld veld, `composite field'). De belangrijke
fluctuaties treden alleen op in de `orientatie' van het veld en vari\"eren
langzaam; de grootte van het veld is in essentie vast.~\cite{Wegner}
Dit is analoog aan de situatie met Heisenbergspins; ook zij hebben
vaste lengte maar een veranderlijke orientatie. Deze aanpak leidt daarom tot
een {\it niet-lineair sigma-model} voor de Anderson overgang. Hiermee heeft
men de renormalisatiegroep vergelijkingen in $2+\epsilon$ dimensies bepaald
tot op orde $\epsilon^5$~\cite{Hikami2}. Dit legt de kritieke exponenten
vrij nauwkeurig vast in $d=3$.

Wij zullen in dit dictaat sterke localisatieaspecten verder 
buiten beschouwing laten.

\newpage

\renewcommand{\thesection}{\arabic{section}}
\section{Macroscopie: de diffusiebenadering}
\setcounter{equation}{0}\setcounter{figure}{0}

\label{macroscopie}\index{macroscopie}

\subsection{De transmissieco\"efficient van een plak en de wet van Ohm}

Beschouw transmissie van licht door een {\it plak\index{plak}}\index{$L@$L$,
dikte} met dikte $L$ (`plane parallel
geometry' of `slab'). Zoals aangegeven in figuur \ref{skin.eps} valt een
vlakke golf
onder een hoek $\theta_a$\index{$$h@$\theta_a$, inkomende hoek} in op een
laag met verstrooiers. De brekingsindex
van het medium met verstrooiers is $n_0$; daarbuiten is hij $n_1$. De
verhouding van brekingsindices,
\BEQ m=\frac{n_0}{n_1}, \index{$m$,
brekingsindexverhouding}\index{$n_0$, brekingsindex in het medium}
\index{$n_1$, brekingsindex buiten}   \EEQ
is groter dan \'e\'en indien een droge
stof in lucht geplaatst
is ($n_1\approx 1$). Het verschil in brekingsindices leidt er toe dat
de binnenkomende bundel een hoek $\theta_a'$ heeft in het medium (refractie).
De verstrooiers zijn lukraak geplaatst met homogene dichtheid. Dit
leidt tot een diffusief gedrag van de lichtintensiteit. Men kan
zich dat eenvoudig voorstellen omdat iedere verstrooiing het licht weer in een
andere richting stuurt. We beschouwen voorlopig het geval van isotrope 
verstrooiing, waar deze verstrooiing in iedere richting even sterk is.

\begin{figure}
\centerline{\includegraphics[width=12cm]{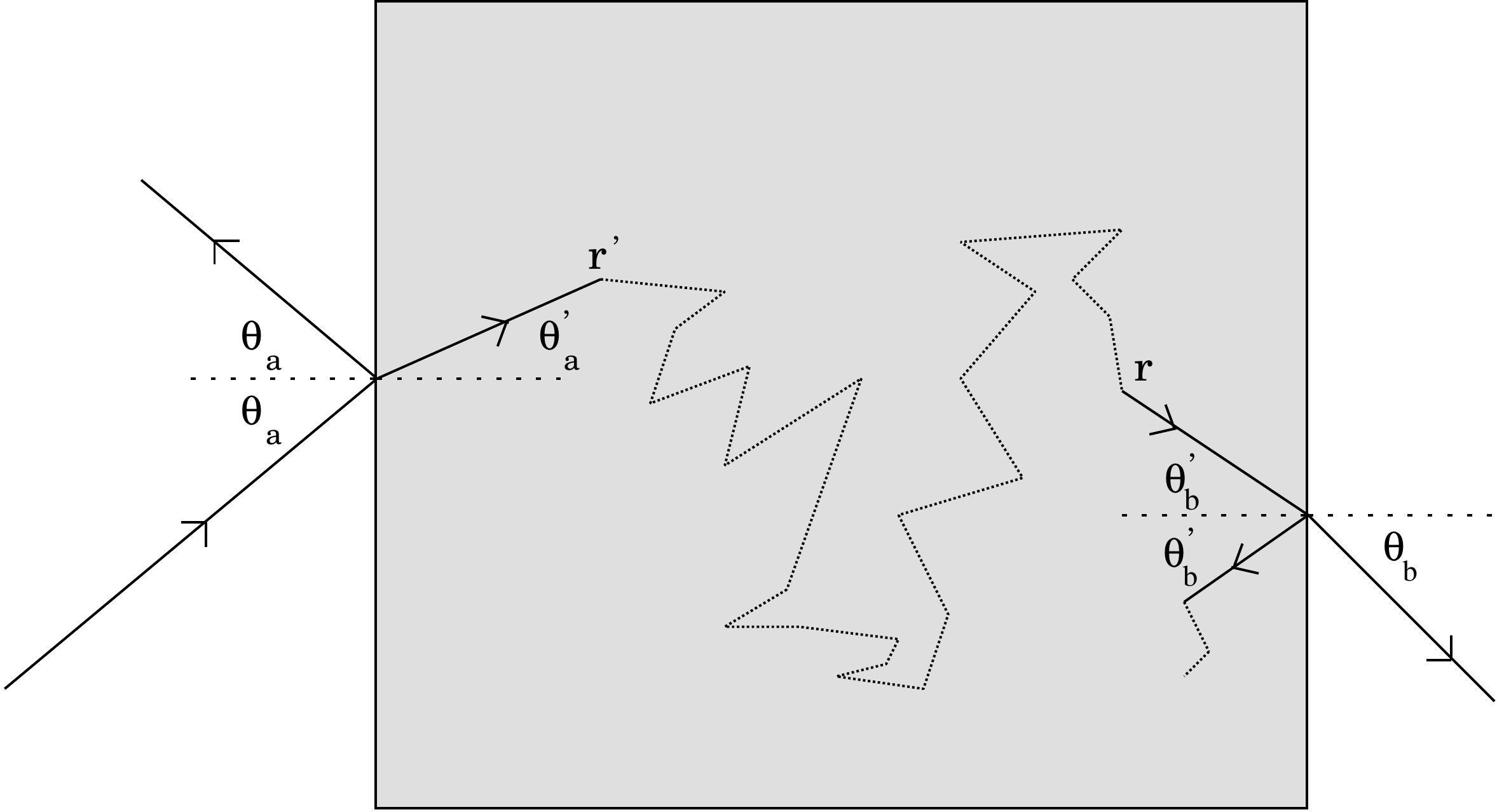}}
\caption{Schematische voorstelling van veelvoudige verstrooiing in een
plak. Een bundel die binnen komt onder hoek $\theta_a$ wordt
afgebogen naar hoek $\theta_a'$ in het medium en genereert ook een speculaire reflectie.
Nadat diffusief tranport plaatsgevonden heeft, veroorzaakt een golf, die
onder hoek $\theta_b'$ aankomt op \'e\'en van de wanden, een speculaire
reflectie en een uitgaande bundel met hoek $\theta_b$.}
\label{skin.eps}
\end{figure}

In een stationaire situatie voldoet de diffuse lichtintensiteit
$I$\index{$Ic@$I$, diffuse intensiteit} aan de stationaire diffusie
vergelijking
\BEQ \frac{\partial}{\partial t}I({\bf r},t)=D\nabla^2 I({\bf r},t)=0 \EEQ
Hier is $D$ de diffusieconstante.\index{$D@$D$, diffusieconstante} In een
stationaire situatie speelt hij geen rol (we kunnen hem eruit delen).

De inkomende intensiteit in $z=0$ is $I_0$. \index{$I0@$I_0$, sterkte
inkomende  bundel} De niet-verstrooide
intensiteit zal in de plak afvallen als
\BEQ\label{LB} I_{in}(z)=I_0\eexp{-z/\ell} \index{$Iin@$I_{in}$,
onverstrooide
inkomende intensiteit} \EEQ Deze exponenti\"ele afval is bekend onder de
naam `Lambert-Beer wet'\index{Lambert-Beer wet}.
Hij beschrijft het lot van het directe zonlicht of de koplampen van de
auto zoals besproken in de inleiding.

De afval (\ref{LB}) van de niet-verstrooide intensiteit gaat hand in hand
met het opbouwen van de diffuse intensiteit. Dit gebeurt in een randlaag
met een dikte van essentieel \'e\'en vrije weglengte. Later zal dit
uitgebreid besproken worden bij het oplossen van de Schwarzschild-Milne 
vergelijking.  In de diffusiebenadering neemt men eenvoudig aan dat
de inkomende intensiteit na een kleine diepte geheel omgezet is
in diffuse intensiteit. Het probleem van de vlakke golf invallend op de laag
wordt vervangen door de laag dikker te `denken' en aan te nemen dat
een bundel diffuus licht op een `injectiediepte'\index{injectiediepte}
$-z_0$\index{$z_0$, injectiediepte}
in de verstrooiende laag geinjecteerd wordt. De waarde van $z_0$ moet uit
de mesoscopie berekend worden; het is duidelijk dat $z_0$ van de orde
van de vrije weglengte is. Voor isotrope verstrooiing in de afwezigheid
van interne reflecties geldt $z_0=0.71044609\, \ell$. We zullen  nu echter
$z_0$ gelijk nemen aan $\ell$.

We kiezen de $z$-as z\'o dat de invallende bundel vanuit $z=-\infty$
in $z=0$ op het medium valt. In de huidige beschrijving heeft de 
diffuse intensiteit dus de randvoorwaarde
\BEQ I(z=-z_0)=I_0 \EEQ
Voor een plak met dikte $L$ geldt aan de uittreezijde op analoge wijze
dat licht binnen een randlaagje ter dikte van $z_0$ zonder verdere 
verstrooiing uit het medium kan propageren. De diffuse intensiteit voldoet
dan ook aan de randvoorwaarde
\BEQ I(z=L+z_0)=0 \EEQ
Voor onze plak-geometrie hangt $I$ alleen van $z$ af, zodat vgl. (1)
reduceert tot $I''(z)=0$, met oplossing $I(z)=a+bz$.
Voor ons probleem vinden we
\BEQ I(z)=I_0 \frac{L+z_0-z}{L+2z_0} \label{mcw14} \EEQ
De doorgelaten intensiteit zal in essentie gelijk zijn aan de intensiteit
 in $z=L$. Voor kleine $z_0$ is dit equivalent met $z_0$ maal de
afgeleide van de intensiteit aan de uittreerand. De transmissieco\"efficient,
$T$,\index{transmissieco\"efficient!diffusiebenadering} is dus gegeven door
\index{$T@$T$, transmissieco\"efficient}\BEA \label{Tequiv}
T&=&\frac{I(z=L)}{I_0}=-\frac{z_0}{I_0} \frac{dI(z)}{dz}
\left|_{z=L+z_0}\right. \nonumber\\ 
&=&\frac{z_0}{L+2z_0}\approx \frac{\ell}{L} \EEA
Zonder ons te bekommeren om voorfactoren zien we hier de afhankelijkheid
$T\sim \ell/L$, die we kennen voor het geleidingsvermogen van een weerstand met
lengte $L$ (wet van Ohm)\index{Ohm, wet van}. Let wel dat voor
electrontransport bij niet te lage
temperaturen de diffusie veroorzaakt wordt door botsing met fononen, terwijl
deze voor zeer lage temperaturen een gevolg is van verstrooiing door
onzuiverheden. 
 
\subsection{De diffusiepropagator\index{diffusiepropagator}}
Laten we nu kijken naar de oplossing van de diffusievergelijking.
De meest algemene vorm is
\BEQ\label{D1} \partial_t I(\r,t)=D\nabla^2 I(\r,t)-D\kappa^2I(\r,t) \EEQ
De tweede term in het rechterlid beschrijft absorptie. De absorptielengte
is\index{absorptielengte}\index{$Lezabs@$L_{abs}$,
absorptielengte}\index{$$k@$\kappa$,
inverse absorptielengte} \BEQ L_{abs}=\frac{1}{\kappa}.
\EEQ De oplossing van (\ref{D1}) met beginvoorwaarde $I(\r,0)=\delta(\r)$ is
\BEQ\label{D2}  I(\r,t)=(4\pi Dt)^{-3/2}\eexp{-\frac{r^2}{4Dt}-D\kappa^2 t}
 \EEQ
\BO Toon aan dat de algemene oplossing is
 \BEQ \label{D3} I(\r,t)=\int d^3\r' I(\r',0)
 (4\pi Dt)^{-3/2}\eexp{-\frac{(\r-\r')^2}{4Dt}-D\kappa^2 t} \EEQ
\EO
Vaak werkt men met de Fourier-Laplace getransformeerde
\BEQ \label{D4}
I(\q,\Omega)= \int d^3\r \eexp{-i\q\cdot\r} \int_0^\infty dt \eexp{i\Omega t}
 I(\r,t) \index{$q@$\q$, macroscopische impuls}\index{$$X@$\Omega$,
macroscopische frequentie}\index{macroscopische impuls}\index{macroscopische
frequentie}\EEQ \BO Toon aan dat de  oplossingen (\ref{D2}) en (\ref{D3})
equivalent zijn aan
\BEQ \label{D5} I(\q,\Omega)=\frac{1}{ D\q^2+D\kappa^2-i\Omega } ; \qquad
I(\q,\Omega)=\frac{ I(\q ,{\bf t} =0) } { D\q^2+D\kappa^2-i\Omega }
\EEQ \EO
Omdat deze uitdrukkingen divergeren in de limiet $\q,\Omega\to 0$,
noemt men dit wel de `diffusie-pool'\index{diffusie-pool} of het
`diffuson'\index{diffuson}.
We zullen veelal ge\"interesseerd zijn in plak geometri\"en. Er is dan
translatie invariantie in het $\rhoo=(x,y)$-vlak. De twee-dimensionale
impulsvector noteren  we als $\ql=(q_x,q_y)$. De oplossing
 met beginvoorwaarde\index{$$r@$\rhoo$, twee dimensionale plaatsvector} \BE
I(\r,t=0)=
S(\qt) \exp(i\ql\cdot\rhoo) \delta(z-z') \EE moet dus voldoen aan de
vergelijking\index{$S@$S$, bron}
\BEQ \label{D6} DI''(z)-D\qt^2I(z)+i\Omega
I(z)-D\kappa^2 I(z)=-S(\qt)\delta(z-z') \EEQ Dit kunnen we schrijven als
\BEQ \label{D7} I''(z)=M^2I(z)-\frac{S(\qt)}{D}\delta(z-z') \EEQ
met `massa' $M$ gedefinieerd door
\BEQ \label{D8} M^2=\qt^2+\kappa^2-i\frac{\Omega}{D}. \EEQ
We noemen dit massa omdat het de exponenti\"ele afval van de
intensiteit beschrijft, zie voetnoot
in hoofdstuk \ref{inleiding}. Voor een plak met randvoorwaarden $I(0)=I(L)=0$
is de oplossing van vgl. (\ref{D6})
\BEA \label{D9} I(z,\qt,\Omega)=\frac{S(\qt)}{D}\,\,
\frac{\sinh(Mz_<)\sinh(ML-Mz_>)}
{M\sinh(ML)}\nonumber\\
{\rm waarbij}\qquad z_<={\rm minimum}(z,z');\qquad z_>={\rm maximum}(z,z')\EEA
\BO Neem $z'$ vast en geef de uitdrukking voor $z<z'$ en voor $z>z'$.
Toon  aan dat aan vgl. (\ref{D7}) voldaan is. \EO
In de stationaire, vlakke en absorptieloze limiet ($\Omega=\kappa=0$,
$\ql={\bf 0}$ zodat $M=0$) reduceert (\ref{D9}) tot de `tentfunctie'
\BEQ \label{D10} I(z;z';\qt={\bf 0},\Omega=0)=\frac{S(\qt)}{D}\,\,
\frac{z_<(L-z_>)}{L} \EEQ

\subsection{Plaatsbepaling van een object in een diffusief medium}
\index{localisatie!van objecten}
Diffusietheorie is nuttig gebleken bij
het bepalen van de positie van een object in een melkachtige vloeistof.
Stel je voor dat een object (bijv. een haar) in een glas melk zit
(of een bromvlieg in een glas karnemelk\index{karnemelk}). Neem aan dat je
het  object niet met het blote oog kan waarnemen. Lukt dat wel met een
intense, welgedefinieerde lichtbron, zoals een laser?
Dit probleem werd experimenteel en theoretisch beschouwd door
den Outer et al.~\cite{dOuter}. Het is een
inversie probleem: wanneer we de
intensiteit vlak aan de rand van het medium op alle plaatsen meten,
kunnen we dan de vorm en positie van het object reconstrueren? Zulke 
vragen zijn van interesse in de medische fysica.
De gebruikte methode vereist namelijk geen operatie (hij is
niet-invasief).
Mogelijke toepassingen kunnen zijn het opsporen van borstkanker,
 een hersenbloeding
(veel bloed veroorzaakt veel absorptie) en de ziekte van Altzheimer
(weinig doorbloeding van een deel van de hersenen).

Omdat de stationaire diffusie vergelijking dezelfde vorm heeft als de Laplace
vergelijking\index{Laplace vergelijking} in de electrostatica, kunnen we het
effect van het
object in een multipool-expansie\index{multipoolexpansie} uitdrukken. Daar
de multipoolmomenten als
machten in de straal gaan, zullen alleen de eerste twee (lading en 
dipoolmoment) voor kleine objecten van belang zijn.
De intensiteits-verstoring aan de rand van het medium veroorzaakt door
het object berekenen we met behulp van
beeldladingsmethoden\index{beeldladingen}.
Indien het object licht absorbeert, wordt het beschreven worden door een
lading; een tegenovergestelde lading beschrijft dan het genereren
van extra
straling door de invallende straling. Indien er geen absorptie is maar wel
een verschil is tussen de diffusieconstante
van het object en het omringende medium, zal het object alleen voor 
extra of juist voor minder verstrooiing zorgen dan de omgeving. In zo'n geval
wordt het object beschreven door een dipool (Rayleighverstrooier).
De precieze vorm van het object komt tot uitdrukking in hogere multipolen,
die voor kleine objecten nauwelijks waarneembaar zijn.

\subsubsection{Het transmissie- en reflectieprofiel
in twee dimensies}
We zullen het transmissie- en reflectieprofiel berekenen voor een plak
van een verstrooiend medium met daarin een cylinder met
straal $a$, klein vergeleken met $L$, de dikte van de plak.
\index{$a$, straal van verstrooier}
In de praktijk kan de cylinder een dun potlood zijn, wat straling absorbeert,
of een glasfiber, die niet absorbeert maar minder verstrooit dan zijn
omgeving
\footnote{We nemen aan dat er geen intensiteit door de einden van de
glasfiber weglekt.}.
Experimenteel kun je cylinders of draden makkelijk 
 manipuleren en zij geven sterkere signalen dan bolletjes.
Het probleem van verstrooiing aan een ondergedompeld bolletje wordt
behandeld in hoofdstuk \ref{invloed}.
 
Als lichtbron nemen we een vlakke golf. De coordinaten van het
hart van de draad zijn $(x,0,L-Z_0)$ met $-\infty<x<\infty$, dus 
$Z_0$ meet de afstand van het hart van de draad tot het
uittreevlak.\index{$Z0@$Z_0$, afstand tot uittreevlak}
De stationaire diffusievergelijking voor  de lichtintensiteit
$I(\r)$ is gegeven door:
\BEQ \nabla^2 I(\r) = 0. \label{difvgl} \EEQ
Met randvoorwaarden aan de voor- en achterkant van de plak:                 
\BEQ
I(x,y,0)=I_0, \mbox{\ \ } I(x,y,L)=0,  \label{rnds}
\EEQ
De diffusievergelijking binnen het object is gegeven door
\BEQ
\nabla^2 I(\r) = \kappa^2 I(\r). \label{difvglab}
\EEQ
We veronderstellen dus uniforme absorptie in het object, zodat we een
absorptie lengte $L_{abs}\equiv 1/\kappa$ kunnen introduceren.  Het volledige
probleem is nu om vergelijking (\ref{difvgl}) en (\ref{difvglab}) op te
lossen
onder de randvoorwaarden (\ref{rnds}) en met extra randvoorwaarden aan het
oppervlak van het object
\BEA
I_{uit}(a^{+}) & = & I_{in}(a^{-}), \label{conrab}\\ 
D_{1}\frac{\partial I_{uit}}{\partial n}\mid_{a^{+}}
                 & = & D_{2}\frac{\partial I_{in}}{\partial n}\mid_{a^{-}}.
\label{conarab}
\EEA
De afgeleide langs de normaal, $\partial/\partial n={\bf n}\cdot\nabla$, 
wordt genomen op de rand van\index{$n@${\bf n}$, normaalvector}
het object. Het omringende medium heeft diffusieconstante $D_{1}$ en het
object heeft diffusieconstante $D_{2}$. Voor een glasfiber kunnen we
$D_2=\infty$ stellen. Vergelijkingen (\ref{conrab}) en (\ref{conarab})
beschrijven
continuiteit en continu-differentieerbaarheid van de intensiteit op de rand,
 wat neerkomt op fluxbehoud aan het oppervlak van het object.


We mogen nu de $x$-coordinaat buiten beschouwing laten en we voeren de
complexe coordinaat $\zeta=L-z+iy$ in.
De probeer-oplossing voor de diffusievergelijking buiten het object is nu
\BEQ
\frac{I_{uit}(\zeta)}{I_{0}}=\frac{L-z}{L}+ q\mbox{Re}\{
 \ln\frac{\sin\frac{\pi}{2L}(\zeta -Z_0)} {\sin \frac{\pi}{2L}(\zeta+
Z_0)}\}
+p\mbox{Re}\{ \cot \frac{\pi}{2L}(\zeta-Z_0)+\cot 
\frac{\pi}{2L}(\zeta+Z_0)\}
\label{int}
\EEQ
 Binnen het object geldt
\BEQ
\frac{I_{in}(\zeta)}{I_0}=
A\mbox{I}_0(\kappa\rho)+ B 
\mbox{I}_1(\kappa\rho)\cos\phi.
\label{inbol}
\EEQ
I$_0$ en I$_1$ zijn gemodificeerde Besselfuncties van orde
0 en 1, $\rho=\sqrt{(z-L+Z_0)^2+y^2}$ en $\phi=\arctan{y/(L-Z_0-z)}$.

De eerste term in vgl.~(\ref{int}) beschrijft de ongestoorde intensiteit
(\ref{mcw14}),
de tweede term geeft het effect van absorptie aan (analoog aan een lading
in electrostatica) en de derde is de dipool-term als gevolg van extra
verstrooiing. De constanten $q$, $p$, $A$ en $B$ worden vastgelegd 
door de randvoorwaarden op het oppervlak van het object.\index{$p$,
dipoolmoment} \index{dipool} \index{$q$, lading}
Voordat we de probeer-oplossing invullen maken we eerst gebruik van het feit
dat op het oppervlak geldt: $\mid
\zeta-Z_0\mid= a\ll L$, zodat daar
\BEA
\frac{I_{uit}(\zeta )}{I_0}  & = & 
\frac{Z_0+(L-Z_0-z)}{L}+q[\ln\frac{\pi\rho}{2L
\sin\frac{\pi}{L}Z_{0}}+
O(\frac{a^2}{L^2})] \nonumber\\
                &   &  \mbox{\ \ \ \ \ \ \ \ \ \ \ \ }
+p[\frac{2L(L-Z_0-z)}{\pi\rho}
+\cot\frac{\pi}{L}Z_0+O(\frac{a}{L})]. \label{uitbol}
\EEA
\BO
Leid af dat invullen van (\ref{inbol}) en
 (\ref{uitbol}) in (\ref{conarab}) en
(\ref{conrab}) geeft:
\BEA
q & = & \left[ \frac{Z_{0}}{L}\ + p\cot (\frac{\pi Z_0}{L})\right]
\frac{D_2\kappa a\mbox{I}_1 (\kappa a)}
{D_{1}\mbox{I}_0(\kappa a)+D_2\kappa a\mbox{I}_1 (\kappa a)
\ln(L/a^*)},
\nonumber \\
p & = & \frac{\pi a^{2}}{2L^{2}}
\frac{D_1\mbox{I}_1(\kappa a)-D_2\kappa a\mbox{I}_0(\kappa a)
+D_2\mbox{I}_1 (\kappa a)} 
{D_1\mbox{I}_1(\kappa a)+D_2\kappa a\mbox{I}_0(\kappa a)-
D_2\mbox{I}_1(\kappa a)}, 
\nonumber\\
A & = &\frac{\mbox{I}_0(\kappa a)D_1q}
{\kappa a \mbox{I}_1 (\kappa a) D_2},\nonumber \\ 
B & = & \frac{ 
a}{ 
\mbox{I}_1 (\kappa a)}(\frac{1}{L}
 + p\frac{2L}{\pi a^{2}}), \label{cofqpAB2}
\EEA
met de herschaalde straal 
\BEA
 a^{*}  = \frac{\pi a}{2 \sin (\pi Z_0 /L) }. \nonumber
\EEA \EO

Binnen de diffusiebenadering zijn de transmissie- en de 
reflectieco\"efficient
gedefinieerd door vergelijking (\ref{Tequiv}).
Voor de transmissieco\"efficient vinden
we\index{transmissieco\"efficient}\index{reflectieco\"efficient} \BEA
T(y) & \equiv & \left.
-\frac{\ell }{ I_{0} } \frac{\partial I(\r)}{\partial z}\right|_{z=L}
 , \\
 &= & \frac{\ell}{L} \left\{ 1 \ -\ q
\frac{\pi\sin\frac{\pi}{L} Z_{0}}{\cosh\frac{\pi y}{L}-\cos\frac{\pi}
{L}Z_0}
\ - \ 2\pi p
\frac{ 1-\cosh \frac{\pi y}{L} \cos \frac{\pi}{L}Z_0}
{(\cosh \frac{\pi y}{L} -\cos \frac{\pi}{L}Z_0 )^{2}}
\right\} ,   \nonumber
\label{TY}
\EEA
en voor de reflectieco\"efficient
\BEA
R(y) & \equiv & \left. 1\
+ \frac{\ell}{I_{0} }\frac{\partial I(\r)} {\partial z} \right|_{z=0}
,\\
     & =      &1-\frac{\ell}{L}\left\{1 \ +\ q
\frac{\pi\sin\frac{\pi}{L}Z_0}{\cosh\frac{\pi y}{L}+\cos\frac{\pi}{L}Z_0}
\ - \ 2\pi p
\frac{ 1+\cosh \frac{\pi y}{L}\cos\frac{\pi}{L}Z_0}
       {(\cosh\frac{\pi y}{L}+\cos\frac{\pi}{L}Z_0)^{2}} \right\}.  \nonumber
\label{BY}
\EEA

Het effect op de transmissie is veel duidelijker omdat daar de achtergrond
een stuk kleiner is. In figuren \ref{peter1} en \ref{peter2}  zijn de
transmissiemetingen aan een plak verstrooiend medium met daarin een
potloodstaafje of een glasfiber te zien. Het potlood absorbeert licht,
vandaar zijn zwarte kleur. De glasfiber zorgt alleen voor verminderde
verstrooiing. In de figuren  is ook de beste aanpassing van de theoretisch
uitgerekende lijnvorm (\ref{TY}) getekend. Voor het potlood vindt men $q\neq
0$, $p=0$ en voor de glasfiber $q=0$, $p\neq 0$, zoals verwacht. Men ziet dat
de metingen zeer goed door de diffusie theorie beschreven worden. De reden is
dat de effecten van een klein object groot zijn: een object leidt tot een
{\it diffusief verbrede verstoring}\index{diffusief verbrede verstoring}, die
qua vorm alleen van de systeemgrootte $L$ afhangt; slechts de sterkte van de
verstoring, $p$ of $q$, hangt van de straal van het object af. Voor meer
details van dit onderzoek, zie referentie~\cite{dOuter}.
\BO Binnen de gemaakte benadering is de uitgaande flux op de rand van de 
cylinder gegeven door $-\ell \partial I_{uit}/\partial n $. 

$\bullet$ Toon aan dat dit gelijk is aan
$I_0\cos{\phi}\,\ell/L-I_0 q\ell/\rho$
De netto intensiteit die uitgaat van de cylinder is dus 
$-2\pi q\ell I_0$; met andere woorden, er wordt een hoeveelheid 
$+2\pi q\ell I_0$ geabsorbeerd. Kun je dit ook direct aantonen?

$\bullet$ Toon aan dat dit precies gelijk is aan het verlies in
totaal uitgestraalde intensiteit. 
(Je krijgt dat door $T(y)$ en $R(y)$ over $y$ te integreren).

$\bullet$ Toon op dezelfde manier aan dat de termen met $p$ tezamen geen
verandering van de totaal uitgestraalde intensiteit veroorzaken.\EO

\begin{figure}
\centerline{\includegraphics[width=10cm]{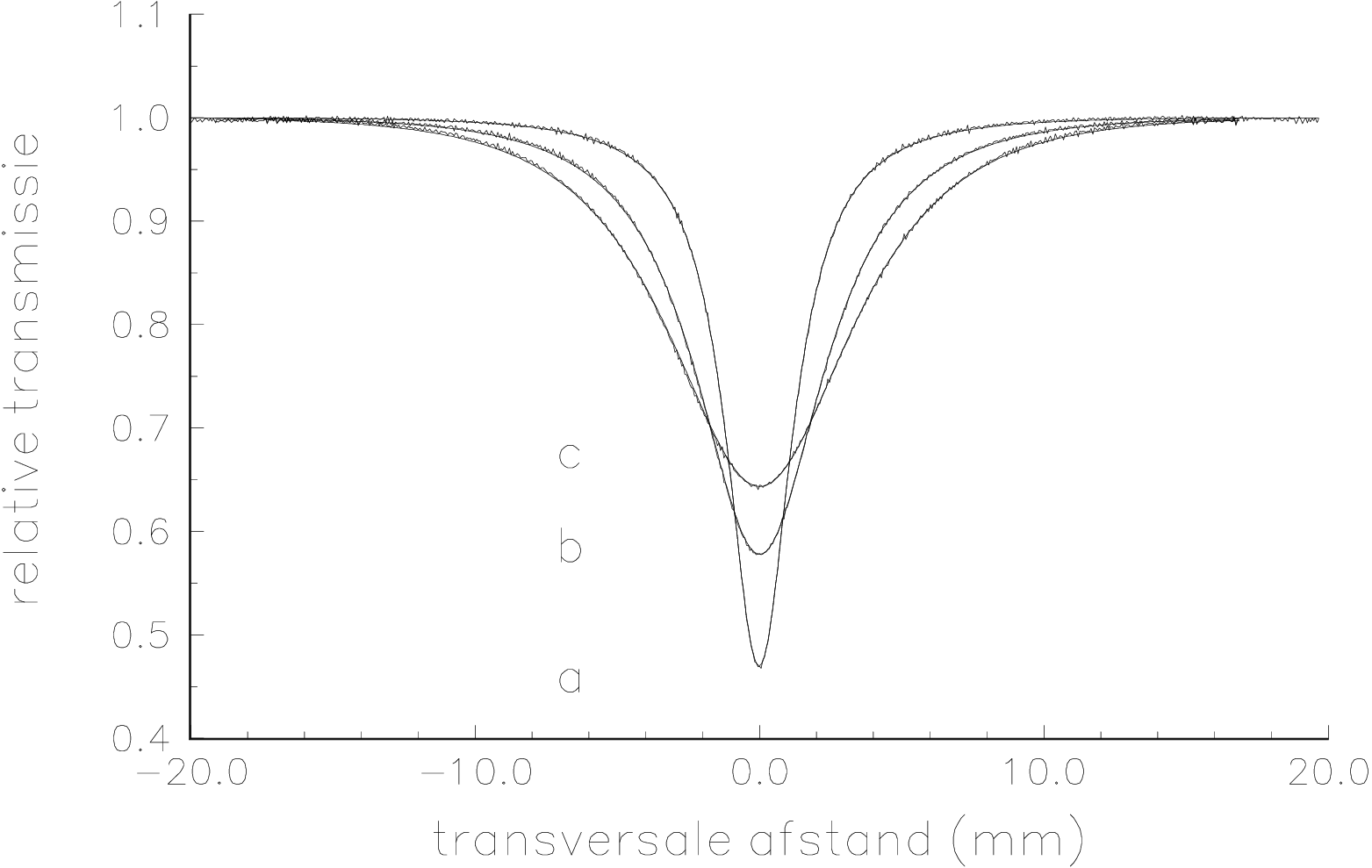}}

\label{peter1}
\caption{De doorgelaten intensiteit als functie van de
transversale afstand voor een  cuvet gevuld met een diffuus verstrooiende
suspensie, en verschillende posities van een potloodstaafje met doorsnede
\o~=350~$\mu m$:  a) $Z_0$ = 1.00~$mm$,  b) $Z_0$~= 2.60~$mm$, c) $Z_0$
=4.10~$mm$. De aanpassingen (`fits') van de berekende curves geven: a) $Z_0
= 1.41
\pm 0.05$~$mm$, b) $Z_0 = 2.86 \pm 0.05$~$mm$, c) $Z_0 =4.33 \pm 0.05$~$mm$.
Voor alle gevallen is gebruikt:
 $p=0$, $\ell = 20\pm 3$~$\mu m$, cuvetdikte $L$~= 10.0~$mm$.}
\end{figure}

\begin{figure}
\centerline{\includegraphics[width=10cm]{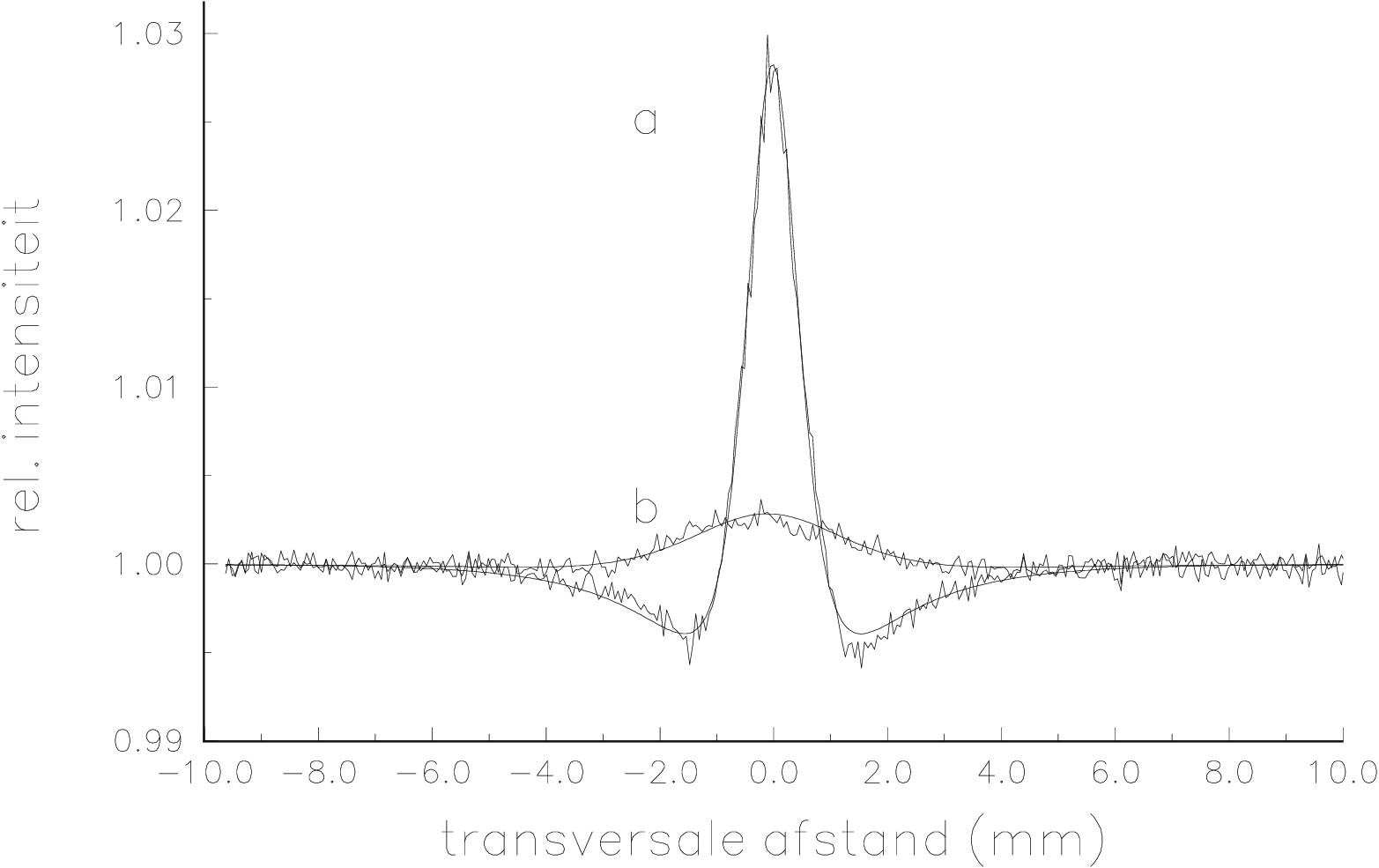}}
\label{peter2}
\caption{De doorgelaten intensiteit als functie van de tranversale
afstand voor een cuvet gevuld met een diffuus verstrooiende suspensie,
met daarin een
transparante glasfiber met doorsnee \o = 350 $\mu m$ op de posities: a)
$Z_0=0.74\pm 0.05$, en b)~$Z_0$=2.5~$mm$. Er geldt $\ell=20\pm 3$~$\mu m$, 
cuvetdikte $L=5.0$~$mm.$ De aanpassingen aan  de berekende curves geven: a)
$Z_0$~=~0.60~$\pm$~0.02~$mm$, $D_2 =\infty$, b) $Z_0 =2.4\pm 0.1 mm$,
$D_2 =\infty$ en in beide gevallen $q=0$.}
\end{figure}

\subsection{Mesoscopische verstrooiers}

De monopoolsterkte $q$ en dipoolsterkte $p$ van een gegeven mesoscopische verstrooier kan 
bepaald worden, zie de theorie en voorbeelden in \cite{LanN1998,LN1999}.

\newpage

\renewcommand{\thesection}{\arabic{section}}
\section{Mesoscopie: de stralingstransportvergelijking}
\setcounter{equation}{0}\setcounter{figure}{0}

\label{mesoscopie}\index{mesoscopie}

Voor veel toepassingen van stralingstransport hoeft men slechts
het gedrag op afstanden van \'e\'en vrije weglengte te kennen. De standaard
theorie wordt nu kort besproken en aangevuld met de behandeling
 van interne reflecties binnen dit kader.

\subsection{De specifieke intensiteit\index{specifieke intensiteit}}

Bij de bestudering van een stralingsveld, is het zinvol
de hoeveelheid\index{stralingsenergie}\index{$$x@$\omega$, interne
frequentie} \index{$Uo@$U$, stralingsenergie}
stralingsenergie, $dU$, in een gegeven frequentie interval
 $(\omega-\frac{1}{2}\Omega,\omega+\frac{1}{2}\Omega)$
te beschouwen. Deze is  getransporteerd door een oppervlakteelement
$d\sigma$ in richtingen vallend binnen de ruimtehoek $d\n$
 gecentreerd rond de richting $\n$, 
gedurende een tijd $dt$
\footnote{Hier volgen we de introductie van Chandrasekhar~\cite{Chandra2}
 in een enigzins
andere notatie en soms andere normering voor wat betreft factoren $4\pi$.}.
\begin{figure}[htbp]
\caption{}
\centerline{\includegraphics[width=3cm]{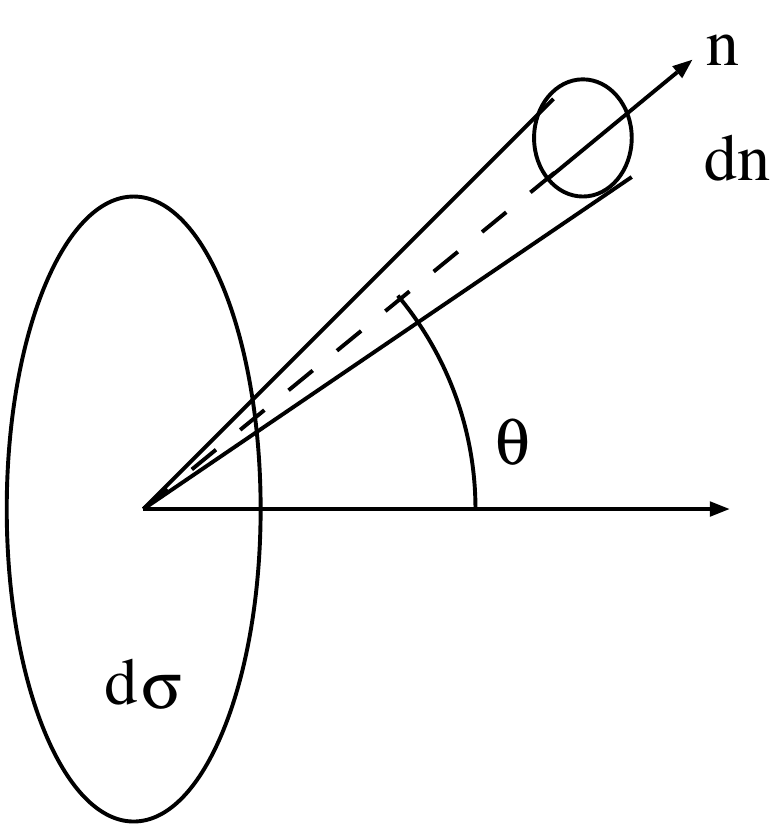}}

\label{chandra}
\end{figure}
Deze energie wordt als volgt
uitgedrukt in de `specifieke intensiteit' $\I$
\BEQ dU=\I\cos{\theta}\,\Omega d\sigma d\n dt \EEQ
waarbij $\theta$ de hoek is tussen de richting $\n$ van de uitgaande straling
en de normaal van $d\sigma$. Deze constructie noemt Chandrasekhar een
`pencil of radiation'; we zullen de term `stralingskegel' gebruiken.
De specifieke intensiteit\index{specifieke intensiteit}\index{$Ia@$\I$,
specifieke intensiteit} wordt ook
wel `diffuse intensiteit' of kortweg `intensiteit' genoemd. Hij hangt in het
algemeen af van de positie ${\bf r}=(x,y,z)$, de richting $\n$, de
frequentie $\omega$ en de tijd $t$.
Wij zullen veelal ge\"interesseerd zijn in stationaire, monochromatische
situaties voor een plak met axiale symmetrie. 
Dan hangt $\I$ alleen af van $z$ en $\mu=\cos\theta$,\index{$$m@$\mu$,
cosinus van hoek} waarbij $\theta$ de hoek is tussen de $z$-as en de
stralingsrichting. 
Propageert de straling een afstand $ds$ in de gegeven richting, dan is er
een stralingsverlies door verstrooiing in andere richtingen
en, eventueel, door absorptie. Dit wordt beschreven door
\BEQ d\I=-n\sigma_{ex} \I ds \EEQ
Hier is $n$ de dichtheid\index{$n$, verstrooiersdichtheid} van verstrooiers
en $\sigma_{ex}$\index{$$sex@$\sigma_{ex}$, extinctie werkzame doorsnede}
\index{extinctie werkzame doorsnede}
de {\it extinctie-werkzame-doorsnede}, zie hoofdstuk \ref{microscopie}.
De oplossing van deze
vergelijking beschrijft de afval van de niet-verstrooide intensiteit, 
$\I(z)\sim\exp(-z/\ell_{sc})$ met verstrooiings-vrije
 weglengte (`scattering mean free path') gegeven door
\BEQ \ell_{sc}=\frac{1}{n{\sigma_{ex}}}\index{verstrooiings-vrije
weglengte}\EEQ

In de afwezigheid van 
absorptie wordt de stralingsenergie verstrooid met een snelheid 
\BEQ {\sigma_{ex}} n ds\times \I cos\theta \Omega d\sigma d\n \EEQ
ofwel, met $dN=n\cos\theta d\sigma ds$
\BEQ {\sigma_{ex}} \I dN\Omega\,d\n \EEQ
De {\it fasefunctie}\index{$p$, fasefunctie}\index{fasefunctie}
$p(\cos\Theta)=p(\n\cdot\n')$
geeft aan dat de fractie \BEQ {\sigma_{ex}} \I dN\Omega\,d\n
\times p(\cos\Theta)\frac{d\n'}{4\pi} \EEQ
verstrooid wordt binnen een element $d\n'$ rond de richting 
$\n'$, die zelf een hoek $\Theta$ maakt met de inkomende richting $\n$.
~\footnote{We zullen  verderop de uitgaande richting met $\n$ en de
binnenkomende richting met $\n'$ aangeven.}
Het albedo\index{albedo}\index{$a$, albedo} van een verstrooier is
gedefinieerd als (zie ook hoofdstuk \ref{microscopie})
\BEQ a=\int \frac{d\n}{4\pi}p(cos\Theta). \EEQ
Indien er geen absorptie optreedt geldt er $a=1$. Van bijzondere 
interesse is {\it isotrope verstrooiing}\index{isotrope verstrooiing}
\BEQ p(\cos\Theta)=1 \EEQ
Deze beschrijft $s$-golf verstrooiing, in het bijzonder
verstrooiing van electronen aan kleine onzuiverheden.
De fasefunctie voor {\it Rayleigh verstrooiing}\index{Rayleigh verstrooiing}
\BEQ p(\cos\Theta)=\frac{3}{4}(1+\cos^2\Theta)\EEQ
beschrijft dipool-verstrooiing of
$p$-golf verstrooiing, zoals optreedt bij verstrooiing van licht
 aan objecten veel kleiner dan de golflengte.

De {\it emissieco\"efficient}\index{emissieco\"efficient} $j$ is
gedefinieerd\index{$j$, emissieco\"efficient}
zodanig dat een aantal deeltjes $dN$ een hoeveelheid stralingsenergie
\BEQ j dN\,d\n\,\Omega\,dt \EEQ
uitzendt in een frequentie interval $(\omega-\frac{1}{2}\Omega,
\omega+\frac{1}{2}\Omega)$ binnen een tijdje
$dt$ in een ruimtehoekelementje $d\n$. Uit een stralingskegel
in richting $\n'(\theta',\phi')$$=$$(\sin\theta'\cos\phi',$
$\sin\theta'\sin\phi',$ $\cos\theta')$
 wordt dus een hoeveelheid
\BEQ {\sigma_{ex}} dN\,\Omega\,dn\,p(\theta\phi;\theta'\phi')
\I(\theta',\phi')
\frac{sin\theta'\, d\theta' d\phi'}{4\pi} \EEQ
verstrooid in de richting $\n(\theta,\phi)$. De emissieco\"efficient
is dus
\BEQ j(\theta,\phi)=\frac{{\sigma_{ex}}}{4\pi}\int_0^\pi\int_{-\pi}^\pi
p(\theta,\phi;\theta'\phi')\I(\theta',\phi')sin\theta'\,d\theta' d\phi' 
\EEQ

Met deze ingredienten kunnen we nu de stralingstransportvergelijking
afleiden. Beschouw straling in een frequentieinterval $(\omega-\frac{1}{2}
\Omega,\omega+\frac{1}{2}\Omega)$
in een tijdje $dt$ dat door een klein cylindertje (`pilledoosje')
met oppervlak $d\sigma$ en hoogte $ds$ gaat. Wanneer de normaal van $d\sigma$
evenwijdig is aan $\n$, is het verschil in stralingsenergie
tussen beide oppervlakken per definitie 
\BEQ \frac{d\I}{ds}ds \Omega d\sigma d\n dt=d\I \Omega d\sigma
d\n dt\EEQ Dit verschil ontstaat uit de onbalans tussen ge\"emitteerde
straling \BEQ j n d\sigma ds \Omega d\n dt \EEQ
en het stralingsverlies
\BEQ {\sigma_{ex}} nds\times \I \Omega d\sigma d\n dt \EEQ
Er volgt
\BEQ \frac{d\I}{ds}=n j-{\sigma_{ex}} n \I\EEQ
Defini\"eren we de `bronfunctie'\index{bronfunctie}
\index{$J@$\J$, bronfunctie}
$\J=j/{\sigma_{ex}}$ op plaats ${\bf r}$,
\BEQ  \J({\bf r};\theta,\phi)=
\int_0^\pi\int_{-\pi}^\pi p(\theta,\phi;
\theta'\phi')\frac{\sin\theta'd\theta' d\phi'}{4\pi}
\I({\bf r};\theta',\phi').\label{bronfie} \EEQ
Zoals $\I$ de straling aangeeft die {\it aankomt} in $\r$, zo beschrijft
 $\J$ de {\it verstrooide weggaande} straling.
Nu volgt hieruit de {\it
stralings\-transport\-vergelijking}\index{stralingstranportvergelijking}
\index{transportvergelijking}
\BEQ \frac{1}{n\sigma_{ex}}\, \frac{d\I}{ ds}=\J-\I \EEQ

\subsection{De plak-geometrie}

Voor een plak-geometrie $0\le z\le L$ hangen fysische grootheden alleen 
van $z$ af. Het is handzaam de `optische diepte'\index{optische
diepte}\index{optische dikte} \BEQ \tau=\frac{z}{\ell_{sc}}=n{\sigma_{ex}}
z\EEQ in te voeren. De optische dikte van de plak is dan
\BEQ b=\frac{L}{\ell_{sc}}.\EEQ

Laat $\theta$ de hoek tussen de positieve $z$-as en de
stralingsrichting aangeven, en $\phi$ de hoek met de positieve
$x$-as. Voor het gemak bekijken we een ingaande 
bundel die rotatie-symmetrisch is
rond de $z$-as, dwz. {\it de inkomende bundel is een kegel} met openingshoek
$\theta_a$ en intensiteit $I_0/2\pi$, onafhankelijk van $\phi$.

We vinden  de dimensieloze vorm van de stralingtransportvergelijking
\footnote{Omdat we
$\tau$ anders defini\"eren dan Chandrasekhar, leidt dit
tot een extra min-teken in vgl. (\ref{RTE})}.
\BEQ\label{RTE}
\mu \frac{d\I(\tau,\mu,\phi)}{d\tau}=\J(\tau,\mu,\phi)-\I(\tau,\mu,\phi)
\EEQ
waarbij \BE \mu=\cos \theta. \EE
Voor een vlakke golf die met intensiteit  $I_0$ in richting
$\n_a(\theta_a,\phi_a)$ invalt op het grensvlak $z=0$, is 
de randvoorwaarde $\I(0,\mu,\phi)$$=
I_0\delta(\mu-\mu_a)\delta(\phi-\phi_a)$, waarbij $\mu_a=\cos\theta_a>0$. 
\subsubsection{Isotrope verstrooiing}\index{isotrope verstrooiing}

Laten we bovenstaande vergelijking omschrijven tot een 
integraalvergelijking voor het geval van een half-oneindig medium.
Voor $\mu<0$ volgt er uit vgl. (\ref{RTE}) voor de straling in de $-z$ richting
(teruggaande richting)
\BEQ \I(\tau,\mu,\phi)=\int_\tau^\infty {\cal J}(\tau',\mu,\phi)
\eexp{-(\tau'-\tau)/|\mu|}\frac {d\tau'}{|\mu|} \EEQ
terwijl voor $\mu>0$ voor de intensiteit in de $+z$ (doorgaande) 
richting geldt 
\BEQ \I(\tau,\mu,\phi)=\I(0,\mu,\phi)\eexp{-\tau/\mu}+\int_0^\tau 
{\cal J}(\tau',\mu,\phi)\eexp{-(\tau-\tau')/\mu}\frac{d\tau'}{\mu} \EEQ
Voor {\it isotrope}
 verstrooiing hangt ${\cal J}$ niet van $\mu$ en $\phi$ af. We
vinden dan een gesloten vergelijking voor 
\BEQ \Gamma(\tau)=\frac{1}{I_0}\int d\mu' d\phi'\I(\tau,\mu'\phi'),
\index{$$C@$\Gamma$, genormeerde diffuse intensiteit}\EEQ
die hier gelijk is aan $4\pi\J/I_0$ \footnote{Chandrasekhar gebruikt de
notatie $4\pi J$ voor $\Gamma$}.
Combineren van de vorige twee vergelijkingen levert voor een bundel die
onder richting $\mu_a,\phi_a$ invalt
\BEQ  \Gamma(\tau)= \eexp{-\tau/\mu_a}+\int_0^\infty d\tau'
\int_0^1 \frac{d\mu}{2\mu} \eexp{-|\tau'-\tau|/\mu}\Gamma(\tau') \EEQ
Deze `Boltzmann' vergelijking\index{Boltzmann vergelijking} zal later ook
uit de ladder-diagram\-men te
voorschijn komen. Deze geven dan de microscopische onderbouwing voor de
hier gegeven beschrijving. Correcties op de ladderdiagrammen geven 
de beperkingen van de stralingstransportvergelijking.

\subsubsection{Anisotrope verstrooiing\index{anisotrope verstrooiing} en
Rayleigh verstrooiing\index{Rayleigh verstrooiing}}
Wanneer de verstrooiing anisotroop is, is er vaak nog cylinder symmetrie
ten aanzien van de binnenkomende richting.
Dit treedt op wanneer de verstrooiers bolvormig zijn. Voor willekeurig 
gevormde verstrooiers met lukrake orientaties is er gemiddeld gesproken ook
cylinder symmetrie. In zulke gevallen hangt de fasefunctie
 $p(\cos\Theta)$ alleen van
$\phi-\phi'$ af, en kan de middeling over $\phi$ uitgevoerd worden. Dit
definieert, met $\mu=\cos\theta$,
\BEQ p_0(\mu,\mu')=\int\frac {d\phi}{2\pi}p(\theta\phi;\theta'\phi')
=\int\frac{d\phi}{2\pi}\frac{d\phi'}{2\pi}p(\theta\phi;\theta'\phi') \EEQ
Voor Rayleigh verstrooiing heeft men $p=3(1+\cos^2\Theta)/4$.
Gebruikend dat 
\BEQ \cos\Theta=\n\cdot\n'=\sin{\theta}\sin{\theta'}
\cos(\phi-\phi')+\cos{\theta}\cos{\theta'}\EEQ
 vind je\index{$p_0$, gemiddelde fasefunctie}
\BEQ p_0(\mu,\mu')=\frac{3}{8}(3-\mu^2-\mu'^2+3\mu^2\mu'^2)\EEQ

Vanwege de vorm van de $\J$ integraal, is het handig in
te voeren\footnote{Chandrasekhar gebruikt de notatie 
$4\pi K$ voor 
$\Delta$}:\index{$$D@$\Delta$, tweede moment
van de specifieke intensiteit}
\BEQ \Gamma(\tau)=\frac{2\pi}{I_0}\int_{-1}^1 d\mu \I(\tau,\mu)\qquad
     \Delta(\tau)=\frac{2\pi}{I_0}\int_{-1}^1 d\mu \mu^2\I(\tau,\mu)
\label{GammaDelta} \EEQ \BO
Toon aan dat deze functies aan de volgende gekoppelde 
integraalvergelijkingen voldoen
\BEA \Gamma(\tau)&=&\eexp{-\tau/\mu_a}     
+(\frac{9}{16}E_1-\frac{3}{16}E_3)*\Gamma+
(\frac{9}{16}E_3-\frac{3}{16}E_1)*\Delta            \label{GammaDeltavgl}
\\
\Delta(\tau)&=&\mu_a^2\eexp{-\tau/\mu_a}
+(\frac{9}{16}E_3-\frac{3}{16}E_5)*\Gamma+
(\frac{9}{16}E_5-\frac{3}{16}E_3)*\Delta \nonumber
\EEA
Hier hebben we ingevoerd de exponenti\"ele integralen
$E_k$\index{exponenti\"ele integraal}\index{$Ek@$E_k$ exponenti\"ele
integraal} \BEQ E_k(x)=\int_0^1 \frac{d\mu}{\mu}\mu^{k-1} \eexp{-x/\mu}=
\int_1^\infty \frac{dy}{y^k}\eexp{-xy} \EEQ en het product
\BEQ (E*f)(\tau)=\int_0^\infty d\tau' E(|\tau-\tau'|)f(\tau').\EEQ
\EO

\subsection{De transport-vrije weglengte en de absorptielengte
\index{transport-vrije weglengte}\index{vrije weglengte}\index{absoptielengte}}
We hebben reeds gezien dat een uitgezonden lichtsignaal exponentieel 
uitdooft als functie van de afstand tot de bron. Dit komt 
doordat er meer en meer uit verstrooid wordt. De karakteristieke afstand is
de verstrooiingsvrije weglengte, $\ell_{sc}$.
 Stel je nu voor dat de verstrooiing
ineffectief is, doordat hij sterk in de voorwaartse richting plaats vindt.
Dan moet de diffusieconstante erg groot zijn. Met andere woorden, in de 
identiteit $D=\frac{1}{3}v\ell$ moet $\ell$ een andere lengteschaal  
voorstellen. Intuitief verwacht men dat dit de afstand is waarover de 
{\it richting} van de straling verloren gaat. Deze lengteschaal heet de
transport-vrije-weglengte (`transport mean free path'), $\ell_{tr}$. 
\index{$ltr@$\ell_{tr}$, transport-vrije-weglengte}
 We zullen nu laten
zien onder welke aannamen deze uit de stralingstransportvergelijking volgt.

De tijdsafhankelijke stralingstransportvergelijking heeft de vorm
\BEQ\label{tSTV}
\tau\frac{\partial}{\partial t}\I({\bf r},{\bf n},t)+
\ell_{sc}\n\cdot\nabla \I({\bf r},{\bf n},t)=
\int\frac{d\n'}{4\pi} p(\n,\n')\I({\bf r},{\bf n}',t)-\I({\bf r},{\bf n},t)
\EEQ
Hierin is $\tau=\tau_{sc}+\tau_{dw}$ uiteraard de gemiddelde tijd 
per verstrooiing uit vergelijking (\ref{tau=}). 
We kunnen nu de locale stralingsdichtheid
$I$\index{stralingsdichtheid}\index{$I@$I$, diffuse intensiteit}
en de locale stroomdichtheid ${\bf J}$\index{stroomdichtheid}\index{$J@${\bf
J}$, stroomdichtheid} defini\"eren als \BEQ I(\r,t)=\int d\n \I(\r,\n,t)\qquad
{\bf J}(\r,t)=\frac{\ell_{sc}}{\tau}\int d\n \I(\r,\n,t)\n.\EEQ
Integreren van vgl. (\ref{tSTV}) geeft de
continuiteitsvergelijking\index{continuiteitsvergelijking}
\BEQ \partial_t I(\r,t)+\nabla\cdot{\bf J}(\r,t)=-\frac{1-a}{\tau}I(\r,t).\EEQ
Hierbij is het albedo\index{albedo} ingevoerd: 
\BEQ a=\int\frac{d\n}{4\pi}p(\n,\n').\EEQ
Indien $a=1$ is er geen absorptie.
Vermenigvuldigen we (\ref{tSTV}) met $\n$ en integreren we over
 $\n$ dan krijgen
we
\BEQ \label{JJ}
\frac{\tau^2}{\ell_{sc}}\partial_t {\bf J}+
\frac{\tau}{\ell_{sc}}{\bf J}+\ell_{sc}\nabla\cdot \int d\n \I(\r,\n,t)\n\n=
\int d\n d\n' p(\n,\n') \I(\r,\n',t)\n\EEQ
Voor verstrooiing die gemiddeld genomen sferisch symmetrisch is, geldt dat
de integraal $\int d\n p(\n,\n')\n$ 
alleen maar evenredig met $\n'$ kan zijn. Neem je het inproduct met $\n'$, 
dan vind je het gemiddelde van de cosinus van de verstrooiingshoek,
\BEQ\label{avcos}
\langle \cos\Theta\rangle= \langle\n\cdot\n'\rangle=\int \frac{d\n}{4\pi}
p(\n,\n')\n\cdot\n'=\int \frac{d\n}{4\pi}p(\cos\Theta)\cos\Theta
=\int_{-1}^1\frac{d\mu}{2}p(\mu)\mu.\EEQ
Dus kan (\ref{JJ}) geschreven worden als
\BEQ
\frac{\tau^2}{\ell_{sc}}\partial_t {\bf J}+
\frac{\tau}{\ell_{sc}}(1-\langle\cos\Theta\rangle){\bf J}
=-\ell_{sc}\nabla\cdot \int d\n \I(\r,\n,t)\n\n
\EEQ
Het rechterlid hangt niet alleen van $I$ en ${\bf J}$ af.
Neemt men echter aan dat 
intensiteitsverdeling bijna isotroop is, dan is de
stroom veel kleiner dan de dichtheid. Men mag dan de benadering maken
~\cite{Ishimaru}
\BEQ\label{IIJ} \I(\r,\n,t)\approx I(\r,t)+\frac{3\tau}{\ell_{sc}}
\n\cdot{\bf J}(\r,t)+\cdots,\EEQ
omdat hogere orde asymmetrieen \footnote{meestal uitgedrukt in Legendre 
polynomen; tweede en hogere ordes worden dus verwaarloosd.}
 dan klein zijn. In deze benadering volgt
\BEQ \frac{\tau^2}{\ell_{sc}}\partial_t {\bf J}+
(1-\langle \cos\Theta\rangle)\frac{\tau}{\ell_{sc}}{\bf J}=-
\frac{\ell_{sc}}{3}\nabla I.\EEQ
Voor processen langzaam in de tijd vinden we dus
\BEQ\label{hulp} {\bf J}(\r,t)=-D\nabla I(\r,t).\EEQ
Ingevuld in de continuiteitsvergelijking leidt dit tot de gewenste
diffusievergelijking voor de dichtheid
\BEA \partial_t I(\r,t)&=&D\nabla^2 I(\r,t)-\frac{1-a}{\tau}I(\r,t)\nonumber\\
&\equiv &D\nabla^2 I(\r,t)-D\kappa^2 I(\r,t).\EEA
De diffusieconstante $D$ is gegeven door
\BEQ D=\frac{\ell_{sc}^2}{3\tau(1-\langle \cos\Theta\rangle)}\equiv
\frac{1}{3}\,v\,\ell_{tr}. \EEQ
Hier is $v=\ell_{sc}/\tau=c\tau_{sc}/(\tau_{sc}+\tau_{dw})$ de eerder 
besproken transportsnelheid\index{transportsnelheid}.
Verder treedt op de {\it transport-vrije-weglengte}
\BEQ \ell_{tr}=\frac{1}{1-\langle \cos\Theta\rangle } \ell_{sc}\EEQ
die de gemiddelde afstand, waarover de richting van de straling
verloren gaat, aangeeft. Inderdaad zal bij sterke voorwaartse verstrooiing 
$\langle \cos\Theta\rangle $ dichtbij \'e\'en liggen, waardoor
de transport-vrije-weglengte groot wordt.
Ook vonden we een expliciete uitdrukking voor de inverse absorptielengte
 $\kappa$ en de absorptielengte $L_{abs}$:
\BEQ L_{abs}\equiv\frac{1}{\kappa}\equiv \sqrt{\frac{\ell_{sc}\ell_{tr}}
{3(1-a)}}.\EEQ 
\BO Laat zien dat je met behulp van (\ref{hulp}) de aanname
 $\mid \int d\n \I(\r,\n,t)\n\mid \ll \int d\n \I(\r,\n,t)$ kunt 
 uitdrukken als de voorwaarde dat de dichtheid
$I(\r,t)$ langzaam varieert over \'e\'en vrije weglengte. \EO

\subsubsection{De  verstrooiings-vrije weglengte     
 versus de transport-vrije-weglengte}
We herhalen nog even kort \index{kort, maar krachtig} de belangrijkste
verschillen tussen de twee vrije weglengtes die we tegen gekomen zijn
voor het algemene geval van niet-isotrope verstrooiing:

\begin{itemize}

\item{De  verstrooiings-vrije weglengte is in essentie de afstand tussen 
nieuwe verstrooiingen. Hij treedt in het bijzonder op in het uitdoven van een 
uitgezonden golf (de Lambert-Beer wet), in de transportsnelheid
 $v=\ell_{sc}/\tau$ en in de vleugels van de terugstrooikegel.} 
 
\item{De transport-vrije-weglengte is de afstand waarover de richting van
de golf verloren gaat. Hij treedt op in diffusieve aspecten zoals 
de diffusieconstante $D=v\ell_{tr}/3$, 
de transmissiecoefficient en de piek van de terugstrooikegel.}
\end{itemize}

\subsection{De injectiediepte\index{injectiediepte}, interne
reflecties\index{interne reflectie} en de  verbeterde
diffusiebenadering\index{diffusiebenadering!verbeterde}}
\label{vbdb}

In het vorige hoofdstuk is de injectiediepte $z_0$ slordig behandeld. Wat we
eigenlijk willen zeggen is dat voor een lineaire oplossing van de
diffusievergelijking voor een medium dat in $z=0$ begint, de precieze
vorm is $I(z)=z+z_0$; die is dus nul op een afstandje $z_0$ buiten het
medium \footnote{ De verdeling $I(z)=z+z_0$ zullen we later tegenkomen
als de grote-$z$-uitdrukking van de homogene oplossing van de 
Schwarzschild-Milne vergelijking. $z_0$ is dan door de precieze vorm
van deze vergelijking vastgelegd. Deze eigenschap proberen we nu op een
ad hoc manier in de diffusiebenadering in te bouwen.}.
Dit kan ook uitgedrukt worden als de voorwaarde
\BEQ I(0)=z_0 I'(0) \label{II'0}\EEQ

Indien de wanden van het systeem gedeeltelijk reflecteren, zal licht
langer opgesloten zijn in het medium. Hierdoor zal de transmissieco\"efficient 
kleiner worden met een factor van orde \'e\'en.
 In de praktijk is dit een relevant effect daar het verstrooiiende medium
vrijwel altijd een andere brekingsindex heeft dan het omringende  medium,
meestal lucht bij een droge stof of glas bij een vloeistof.
Lagendijk, Vreeker en de Vries hebben als eersten op het belang
van deze effecten
gewezen\cite{LVdV}. Later zag men in dat de waarde van $z_0$ erdoor
verandert. Voor een \'e\'endimensionaal probleem geven zij de uitdrukking
\BEQ z_0=\frac{1+{\overline R}}{1-{\overline R}}\ell_{sc} \EEQ
met daarin ${\overline R}$ de gemiddelde
reflectieco\"efficient\index{$R@${\overline
R}$, gemiddelde reflectieco\"efficient }
\index{reflectieco\"efficient!gemiddelde}. 

Voor drie dimensies
is de analoge relatie door Zhu, Pine en Weitz~\cite{ZPW} uitgewerkt.
We geven nu een vereenvoudigde afleiding van hun resultaat. 
Voor een probleem waarbij alleen de 
$z$-afhankelijkheid relevant is, wordt de relatie (\ref{IIJ})
tussen de specifieke intensiteit $\I$, de stralingsdichtheid $I$
 en stroomdichtheid ${\bf J}$
\BEQ \I(z,\mu)=I(z)+\frac{3}{v}\mu J_z(z) \EEQ
met $\mu=\cos\theta$ gerelateerd aan de richting van de straling t.o.v.
de $z$-as en $v=\ell_{sc}/\tau$. Er volgt
\BEQ J_z(z)=\frac{v}{4\pi}\int_{-1}^{1}\mu d\mu\int_0^{2\pi}d\phi
\, \I(z,\mu) \EEQ
Hieruit lees je af dat de stralingsstroom per eenheid van ruimtehoek
$d\Omega=d\mu d\phi$ gelijk is aan
~\footnote{Let op de subtiele $4\pi$-behandeling.}
\BEQ \frac{dJ_z}{d\Omega}=v\cdot\mu\cdot\frac{\I}{4\pi}
=\frac{1}{4\pi}\left\{ \mu vI(z)+3\mu^2J(z)\right\} \EEQ
De totale straling op diepte $z=0$ in de positieve $z$-richting is dus
\BEA \label{J+z}
J_{+z}(0)&=&\frac{v}{4\pi}\int_0^1\mu d\mu \,2\pi\,
\I(0,\mu)\nonumber\\
&=&\frac{v}{4}I(0)+\frac{1}{2}J_z(0)\nonumber\\
&=&\frac{v}{4}I(0)-\frac{v\ell_{tr}}{6}I'(0) \EEA
waarbij de diffusieve stroom (\ref{hulp}) ingevuld is.

Indien er geen interne reflecties aan de wand zijn, moet 
$dJ_z/d\Omega$ nul zijn voor alle $\mu>0$. Eisen we slechts dat $J_{+z}$
nul is en vergelijken we met vgl. (\ref{II'0}), dan vinden we
\BEQ z_0=\frac{2}{3}\ell_{tr} \qquad{\rm diffusie\, benadering;\,\,geen\,
 interne\, reflecties}\EEQ
Voor isotrope verstrooiing ligt deze uitdrukking niet ver van de exacte waarde
$z_0=0.71044\ell$ uit de stralingstransportvegelijking. Hij is niet
exact omdat we alleen de {\it hoek-gemiddelde stroom} nul gezet hebben.

Laten we nu aannemen dat de brekingsindex van het verstrooiende medium,
$n_0$, verschilt t.o.v. die van zijn omgeving $n_1$. De verhouding
\BEQ m=\frac{n_0}{n_1} \EEQ is groter dan \'e\'en voor een droog medium 
in lucht maar kan kleiner dan \'e\'en worden indien het medium geklemd
wordt tussen glasplaten met een hoge brekingsindex.
In beide situaties zullen interne reflecties optreden aan het oppervlak.
De reflectieco\"efficient\index{reflectieco\"efficient} is gegeven door
\BEQ R(\mu)=\left| \frac{\mid\mu\mid-\sqrt{m^{-2}-1-\mu^2}}
		  {\mid\mu\mid+\sqrt{m^{-2}-1-\mu^2}}\right|^2. \EEQ
\index{$R@$R(\mu)$, reflectieco\"efficient} 
Hij is gelijk aan \'e\'en in
geval van totale reflectie, namelijk
wanneer het argument van de wortels negatief wordt. We kunnen nu de inwendig
aan het oppervlak gereflecteerde stroom berekenen  
\BEA J_{+z}^{refl}
&=&-\int_{-1}^0 d\mu\int_0^{2\pi}d\phi R(\mu)\frac{dJ_z}{d\Omega}
=-\int_{-1}^0 \frac{d\mu}{2} R(\mu)\left(v\mu I(0)+
3\mu^2 J_z(0)\right)\nonumber\\
&=&\frac{v}{2}C_1 I(0)-\frac{3}{2}C_2J_z(0)
=\frac{v}{2}C_1I(0)+\frac{v\ell_{tr}}{2}C_2I'(0)\EEA
waarbij
\BEQ \label{C1=C2=}
C_1=\int_0^1 d\mu \mu R(\mu)\qquad C_2=\int_0^1 d\mu \mu^2 R(\mu). \EEQ
\index{$C1@$C_1$, moment van $R(\mu)$}\index{$C2@$C_2$, moment van
$R(\mu)$} Door dit hoekgemiddelde gelijk te stellen
aan het hoekgemiddelde (\ref{J+z}) vindt men de relatie (\ref{II'0})
met de voorspelling
\BEQ \label{z0===}
 z_0=\frac{2}{3}\,\frac{1+3C_2}{1-2C_1}\,\ell_{tr} \EEQ
Deze relatie was afgeleid door Zhu, Pine en Weitz.
Voor het geval van isotrope verstrooiing  werd later in [NL] opgemerkt
dat zij exact wordt in de limiet van groot brekings\-index\-verschil 
($m\to 0$ of $m\to\infty$). De reden is eenvoudig dat de
wanden dan als goede spiegels werken, waardoor de intensiteit ook dichtbij
de wanden bijna constant wordt.
 Anders gezegd, de oppervlaktelaag heeft een
grote dikte $z_0$. Op een stukje ter grootte van enkele vrije weglengtes
na, is de diffusiebenadering van toepassing op de hele oppervlaktelaag.
  
In de afleiding van vgl. (\ref{z0===}) is alleen gebruik gemaakt van 
de fasefunctie bij het invullen van de diffusieconstante. Dit leidde tot
het optreden van de transport-vrije-weglengte. 
Je zou daarom verwachten dat (\ref{z0===})
 ook geldig is voor willekeurige anisotrope
verstrooiing in de limiet van een groot brekingsindexverschil. 
Dit is fysisch te verklaren daar de wanden dan werken als goede spiegels.
Omdat er dan aan de randen veel meer verstrooiingen voorkomen,
leidt dit ook dichtbij de rand tot een vrijwel isotrope verdeling van 
de straling. En voor een isotrope verdeling maakt
het niet meer uit of de fasefunctie al dan niet isotroop is.
Dit punt wordt in hoofdstuk \ref{anisotrope} verder aan de orde gesteld.

\renewcommand{\thesection}{\arabic{section}}
\section{Microscopie: de $t$--matrix en werkzame doorsnedes}
\setcounter{equation}{0}\setcounter{figure}{0}

\label{microscopie}\index{microscopie}

Verstrooiing van een golf wordt beschreven door de $t$-matrix van het
object, die op zijn beurt de werkzame doorsnede vastlegt. Voor het geval
van \'e\'en enkele verstrooier bespreken we nu
deze grootheden door  oplossing van de golfvergelijking.

\subsection{De Schr\"odinger- en de scalaire golfvergelijking}
In dit college worden verstrooiingsverschijnselen beschouwd van twee
op het eerste gezicht heel verschillende soorten golven:
waarschijnlijkheidsgolven
(de electron golffunctie) en klassieke (licht-) golven.
Waarschijnlijkheidsgolven worden beschreven door de Schr\"odinger
vergelijking\index{Schr\"odinger vergelijking} : \begin{eqnarray}
\label{schrd1}
\frac{-\hbar^2}{2m}\nabla^2\Psi + V\Psi &=& E\Psi
 \index{$V$, potentiaal}
\index{$hst@$\hbar$, constante van Planck'} 
\index{$$Y$@$\Psi$, golffunctie; golfamplitude} \end{eqnarray}
Om het rekenen eenvoudiger
te maken, kiezen we eenheden waarin de constante 
 ${\hbar^2}/{2m}$  gelijk is aan $1$. Daarmee
wordt de Schr\"o\-dinger vergelijking
\begin{eqnarray}
-\nabla^2\Psi + V\Psi &=& E\Psi
\end{eqnarray}
We kunnen bijvoorbeeld puntverstrooiers invoeren door de potentiaal te
kiezen als
$V({\bf r}) = -\sum_i u\delta({\bf r}-{\bf R}_i)$. De constante $-u$ is
hierin de `kale' verstrooiingssterkte\index{$u$, kale verstrooiingssterkte}
van de puntdeeltjes.\index{$Ri@ ${\bf R}_i$, verstrooierspositie}

Akoestische golven\index{akoestische golven} worden beschreven door de
klassieke golfvergelijking: \begin{eqnarray}
\label{klassiek1}
\nabla^2\Psi - \frac{\varepsilon({\bf r})}{c^2}\frac{\dpa^2}{\dpa t^2}\Psi
&=&0 \end{eqnarray}
Waarbij $\varepsilon$ 
\index{$$e@$\varepsilon$, genormeerde massadichtheid; dielectrische constante}
de genormeerde locale massadichtheid is.
We zullen deze vergelijking ook voor licht toepassen. We `vergeten' daarbij
dat licht een vectorgolf is.
$\varepsilon$ is dan de di\"electrische constante, $c$ is de
voortplantingssnelheid in het geval dat $\varepsilon=1$.
\index{$c$, lichtsnelheid; geluidssnelheid}
De klassieke golfvergelijking kunnen we tijdsonafhankelijk maken door 
monochromatische golven te beschouwen,
$ \Psi({\bf r},t) = \Psi({\bf r})\eexp{i \omega t}$. Hierbij is
$\omega$ de frequentie. Dit komt overeen
met het doen van een tijdsonafhankelijk experiment met monochromatisch licht.
Ook in deze vergelijking kunnen we puntverstrooiers\index{puntverstrooier}
invoeren, ditmaal door te stellen $ \varepsilon({\bf r}) = 1+\sum_i \alfa
\delta({\bf r}-{\bf R}_i)$. $\alpha$ is de
`polariseerbaarheid' van de verstrooier.\index{$$a@$\alpha$,
polariseerbaarheid}\index{polariseerbaarheid}
Hierdoor zijn de Schr\"odinger en de klassieke golfvergelijking 
formeel in dezelfde vorm te schrijven:
\begin{eqnarray}
\label{beiden}
&-\nabla^2\Psi - u\sum_i \delta({\bf r}-{\bf R}_i) \Psi = k^2\Psi
\\ \nonumber
&\mbox{waarin  }  \left\{
  \begin{array}{ll}
  u=\mbox{constant} & \mbox{in de Schr\"odinger
vergelijking; $k=\sqrt{E}$} \\
  u=\alfa k^2   & \mbox{in de klassieke golfvergelijking; $
k=\omega/c $}  \end{array}
  \right.
\end{eqnarray}
Hierdoor kunnen veel van de resultaten hieronder afgeleid worden zonder te
specificeren welke soort golven verstrooid wordt. In sommige gevallen zal de
energie-afhankelijkheid van $u$ voor `licht'-golven een essentieel verschil
blijken te maken.

\subsection{De $t$--matrix en resonante puntverstrooiers} Voor de
microscopische beschrijving beginnen we met de enkelvoudige verstrooiing van
een vlakke golf. Dit proces wordt beschreven door de $t-$matrix van de
verstrooier.
\subsubsection{De puntverstrooier in  \'e\'en
dimensie}\index{puntverstrooier}
Om de betekenis en eigenschappen van de $t-$matrix  uit te
leggen is het handig om met een eenvoudig voorbeeld te beginnen: een golf in
\'e\'en dimensie die aan een puntvormige verontreiniging (met
verstrooiingssterkte $u$) verstrooid wordt. In de quantummechanica is de
golfvergelijking die dit probleem beschrijft: \begin{eqnarray} \label{1dim}
-\Psi''(x)-u\delta(x-x_0)\Psi(x) &=& E\Psi(x)\\ E&=&k^2 \end{eqnarray}

Neem aan dat de inkomende golf, $\eexp{ikx}$, aankomt uit $-\infty$, dan kunnen
we de oplossing van het probleem vinden door te stellen:
\begin{eqnarray}
\Psi = \left\{
  \begin{array}{ll}
    \eexp{ikx}+A\eexp{-ik(x-x_0)} &\mbox{voor $x \leq x_0$} \\
    B\eexp{ik(x-x_0)}    &\mbox{voor $x\geq x_0$}
  \end{array}
\right.
\end{eqnarray}
We kunnen $A$ en $B$ oplossen door dit in te
vullen in de golfvergelijking (\ref{1dim}). De oplossing luidt:
\begin{eqnarray}
\Psi &=& \eexp{ikx}-\frac{t}{2ik}\eexp{ik|x-x_0|}\eexp{ikx_0} \label{opl} \\
\label{t-1d}
t &=& \frac{u}{1-iu/(2k)}
\end{eqnarray}
De factor $t$ noemen we de `$t-$matrix'\index{$t$-matrix}. Deze naam komt
voort uit het feit dat $t$ in het algemeen een matrix is.

\subsubsection{De Greense functie in de plaats-ruimte}\index{Greense
functie} De Greense functie van de (ongestoorde)
golfvergelijking is gedefinieerd door :
\begin{eqnarray}
\label{defgreen}
(-\frac{\dpa^2}{\dpa x^2}-k^2)G(x,x')=\delta(x-x')
\end{eqnarray}
Door middel van invullen is aan te tonen dat Greense functie
gelijk is aan :
\begin{eqnarray}
\label{1dgreenx}
G(x,x')=\frac{\eexp{ik|x-x'|}}{-2ik}
\end{eqnarray}
Hierbij gebruik je dat \BE\frac{\partial^2}{\partial x^2} |x-x'| =
\frac{\partial}{\partial x} \frac{x-x'}{|x-x'|}=2\delta(x-x') \EE
Fysisch gezien geeft de Greense functie de respons op plaats $x$
als gevolg van een verstoring op plaats $x'$. Voor alle $x$ tussen $-\infty$
en $+\infty$ kan de oplossing (\ref{opl}) geschreven worden als
\BE \label{effect}
\Psi=\eexp{ikx} + G(x,x_0)t \eexp{ikx_0}
\EE
De eerste term is natuurlijk de inkomende golf.
De tweede term beschrijft de amplitude van de inkomende golf in $x_0$, de
verstrooiingssterkte $t$ en het effect in $x$ van de verstrooiing in $x_0$.

\subsubsection{De Greense functie in de impuls-ruimte}
Een meer systematische manier om de Greense functie te bepalen is het
Fourier-trans\-for\-me\-ren van vergelijking (\ref{defgreen})
\begin{eqnarray}
\nonumber
\int_{-\infty}^\infty \dint x\,\eexp{-ip(x-x')} \{(-\frac{\dpa^2}{\dpa
x^2}-k^2)G(x,x') - \delta(x-x')\}&=&0\\
\nonumber
(p^2-k^2)\int_{-\infty}^\infty \dint x\,\eexp{-ip(x-x')}G(x,x') &=&
\eexp{ip0}
\\
(p^2-k^2)G(p) &=& 1 \label{pgreen1d} \\  \nonumber
G(p) &=& \frac{1}{p^2-k^2}
\end{eqnarray}
Merk op dat de Fourier-getransformeerde niet afhangt van het referentiepunt
$x'$ dat we gebruikten bij het uitvoeren van de transformatie. Dit is een gevolg
van de translatie-invariantie van het systeem zonder verstrooier.
De Fourier getransformeerde Greense functie is eenvoudig van vorm maar heeft
in dit geval weinig praktisch nut, we willen zijn vorm in `de echte ruimte'.
 Die krijgen we door terug te transformeren:
\begin{eqnarray}
G(x,x') = \int_{-\infty}^\infty \frac{\dint p}{2 \pi}\eexp{ip(x-x')} G(p)
\end{eqnarray}
Deze integraal is echter slecht gedefinieerd omdat hij
in twee punten divergent is. De `truc' om deze divergentie op te heffen is
het optellen van een (eindig maar willekeurig klein) imaginair deel bij de
noemer
$p^2-k^2$, zodat de polen niet meer op de re\"ele as liggen. De integraal
kan dan berekend worden door middel van Cauchy-integratie (ook wel
contourintegratie geheten) (zie ref. \cite{Paepe}, hoofdstuk 2 voor een
goede introductie in Cauchy-integratie).
De gebruikelijke keuze is te stellen
\begin{eqnarray}
G(p)=\frac{1}{p^2-k^2-i0}
\end{eqnarray}
waarbij $i0$ staat voor $i\epsilon$, met $\epsilon$ re\"eel, positief en
willekeurig klein. Deze keuze van het teken noemen we de {\em geretardeerde}
Greense functie,\index{Greense functie!geretardeerd} $G_R$,
de andere keuze van het teken van de imaginaire bijdrage heet de {\em
geavanceerde}\index{Greense functie!geavanceerd} Greense functie,
$G_A=G_R^{\ast}$, en correspondeert met een onfysische `terug
in de tijd lopende' oplossing van de vergelijkingen.\footnote{We zouden ook
het teken van de imaginaire bijdrage ook andersom kunnen kiezen voor de
geretardeerde Greense functie. Van belang is dat we
dit overal consequent blijven doen. Voor de intensiteit heb je altijd het
product $G_A G_R$ nodig.} We geven $G_A$ ookwel aan met
$\overline{G_R}=\overline{G}$.\index{$G_R$, geretardeerde Greense
functie}\index{$G_A=\overline{G_R}=\overline{G}$, geavanceerde Greense
functie} 
\begin{figure} 
 \centerline{\includegraphics[width=8cm]{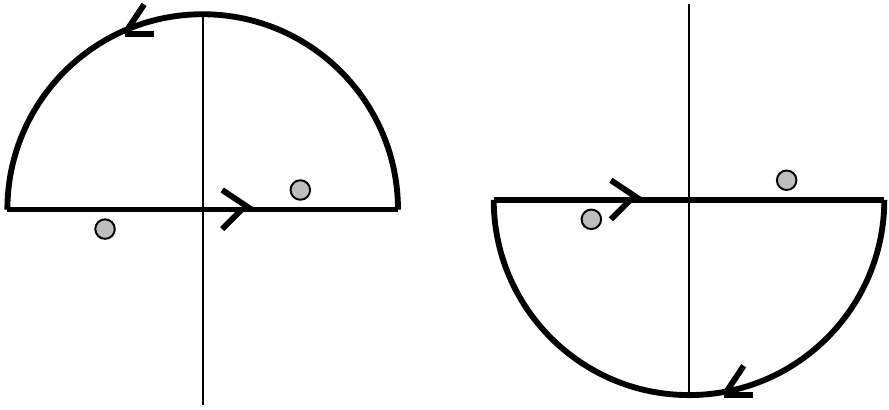}
}
\caption{Integratiekrommen in het complexe vlak; voor $x>x'$ en $x<x'$}
\label{fig:krommen}
\end{figure}

Nu de polen niet meer op de re\"ele as liggen, kiezen  we
het integratiecontour. Voor $x-x'>0$ wordt dat een halve
cirkel
in het bovenhalfvlak, voor $x-x'<0$ een halve cirkel in het onderhalfvlak,
zodat de integrand voldoende snel naar 0 gaat voor ${\rm Im}\, p \to
\infty$.
De integratiekrommen en de posities van de polen zijn gegeven in figuur
 \ref{fig:krommen}. Steeds ligt \'e\'en pool binnen het integratiecontour.
Gebruik van  de Cauchy stelling levert precies (\ref{1dgreenx}).

\subsection{Expansie van de $t$--matrix}
Uit het voorbeeld in \'e\'en dimensie zien we dat de $t$--matrix gegeven wordt
door
\begin{eqnarray}
\label{t-g}
t(x_0)=\frac{u}{1-uG(x_0,x_0)}
\end{eqnarray}
Een alternatieve definitie
van de $t$-matrix
die voor alle dimensies gebruikt kan worden
is de som van de amplitudes
van de 1 maal verstrooide golf + de 2 maal aan hetzelfde deeltje verstrooide
golf + ... . Deze reeks heet de Born-reeks:
\begin{eqnarray}
\label{t-expansie}
t &=& u+uGu+uGuGu+uGuGuGu+... \\
{\bf x}&=& \input{4.pic}
\end{eqnarray}
De kromme lijntjes in de diagrammen verbinden
verstrooiingen aan \'e\'en en dezelfde verstrooier. De $t-$matrix geven we
in het vervolg met {\bf x} aan. In het algemeen moet deze vergelijking
gelezen worden als: \BA\label{TVT}
t(\r,\r')&=&-V(\r)\delta(\r-\r')+V(\r)G(\r,\r')V(\r') 
\nonumber \\ &&-\int d^3\r'' V(\r)
G(\r,\r'') V(\r'') G(\r'',\r') V(\r') + \ldots \EA
Dit is een iteratieve  wijze om het verstrooiingsprobleem `op te lossen'.
Voor Mie-verstrooiing heb je er niet veel aan.
In ons \'e\'endimensionale geval met $V(x)=-u\delta(x-x_0)$ wordt dit:
$t(x,x')=\delta(x-x_0)\delta(x'-x_0) t(x_0)$, met $t(x_0)$ uit (\ref{t-g}).
Hier dient opgemerkt te worden dat in de $t$-matrix expliciet de Greense
functie voorkomt. Omdat de Greense functie niet noodzakelijk plaats-
en tijdsonafhankelijk is, is de $t$-matrix dat ook niet. De `standaard'
$t$-matrix van \'e\'en verstrooier in een oneindig uitgebreid vacuum is
dan ook alleen in d\`at geval van toepassing. Zijn er andere objecten
(zoals andere verstrooiers of wanden) aanwezig, dan heeft dat gevolgen voor
de waarde van de Greense functie en dus ook voor de $t$-matrix:
{\em De $t$-matrix beschrijft het effect van een verstrooier in zijn
locale omgeving.}

\subsection{De puntverstrooier in drie dimensies}\index{puntverstrooier}
\subsubsection{De Greense functie in drie dimensies}
In drie dimensies luidt de vergelijking voor verstrooiing aan een
puntverstrooier met verstrooiingssterkte $u$ :
\begin{eqnarray}
\label{3dgolfvgl}
-\nabla^2 \Psi({\bf r}) - u \delta^3({\bf r}-{\bf r}_0)\Psi({\bf r}) =
k^2\Psi({\bf r})
\end{eqnarray}
De bijbehorende vergelijking voor de ongestoorde Greense functie luidt
\begin{eqnarray}
\label{3dgreenvgl}
(-\nabla^2-k^2)G({\bf r},{\bf r'})=\delta^3({\bf r}-{\bf r'})
\EA
Door Fourier transformatie van (\ref{3dgreenvgl})
kunnen we, analoog aan het \idim geval,
de Greense functie in de ${\bf p}$-ruimte bepalen. Deze blijkt precies
dezelfde te zijn als (\ref{pgreen1d}), het \idim geval.
De  Greense functie in de plaatsruimte is echter
verschillend omdat we nu een \iiidim Fourier transformatie moeten
uitvoeren:
\BEA
\label{greenr3d}
G({\bf r},{\bf r'})&=&\int \frac{d^3 {\bf p}}{(2\pi)^3} \frac{\eexp{i {\bf
p} \cdot({\bf r-r'} )}}{p^2-k^2-i0}\nonumber\\
&=&\int_0^\infty\frac{pdp}{2\pi^2}\frac{\sin p|\r-\r'|}{|\r-\r'|}
\frac{1}{p^2-k^2-i0}\nonumber\\
&=&\int_{-\infty}^\infty\frac{pdp}{2\pi^2}\frac{\eexp{ip|\r-\r'|}}{2i|\r-\r'|}
\frac{1}{p^2-k^2-i0}
\EA
Je vindt dus
\BE G({\bf r},{\bf r'})=  \frac{\eexp{ik|{\bf r}-{\bf r'}|}}{4 \pi |{\bf
r}-{\bf r'}|} \EE
Dit is nauw verwant met de Yukawa potentiaal\index{Yukawa potentiaal} voor
mesonen.

\subsubsection{De tweede-orde Born benadering\index{Born benadering!tweede
orde}}
Merk op dat de Greense functie in drie dimensies niet eindig is voor ${\bf r}
 \to {\bf r'}$ en
dat daarmee de Born reeks (\ref{t-expansie}) zinloos is. Een veel gebruikte
benadering om dit probleem `op te lossen' is het weglaten van de derde en
hogere
orden uit de Born reeks, en van de overgebleven termen ook nog het (divergente)
re\"ele deel weg te laten. Het resultaat is de `tweede-orde Born benadering' 
\begin{eqnarray}
\label{2eordeborn}
t = u+i\, u^2 {\rm Im}\, G({\bf r},{\bf r})
\end{eqnarray}
Voor het beschouwde bulk systeem wordt dit
\BEQ t=u+iu^2\frac{k}{4\pi}.\EEQ
Omdat ${\rm Im}\,t >0$ behoudt men de eigenschap van verstrooiing. Meer
precieze details, zoals resonant gedrag, worden op deze manier over het
hoofd gezien.
Voor electronen wordt de tweede orde Born benadering vaak toegepast. Men
beschouwt meestal Gaussische\index{wanorde!Gaussische} wanorde met $\langle
V(\r) \rangle=0$ en $\langle
V(\r) V(\r') \rangle =u^2 \delta (\r-\r')$. Dit leidt tot $t = iu^2\,{\rm
Im}\,G({\bf r},{\bf r})$$=$$iu^2k/4\pi$.
 Voor licht is de tweede-orde Born echter minder
geschikt indien er resonantie effecten zijn.

\subsubsection{Expansie rond de oorsprong}
Een manier om de divergentie te regulariseren is de Greense
functie te ontwikkelen rond ${\bf r}={\bf r'}$ en de oneindige bijdrage
te vervangen door een eindige:
\begin{eqnarray}
\nonumber
G({\bf r}\to{\bf r'}) &=&\frac{1}{4\pi|{\bf r}-{\bf r'}|} + \frac{ik}{4\pi}
+ O(|{\bf r}-{\bf r'}|) \\
\label{greg1}
G^{\mbox{reg}}({\bf r},{\bf r})&=& \Lambda + \lim_{{\bf r}'\rightarrow {\bf
r}}
\left[ G({\bf r},{\bf r'})-\frac{1}{4\pi |{\bf r}-{\bf r'}|} \right]
=\Lambda + \frac{ik}{4\pi} \index{$Greg@$G^{reg}$,
geregulariseerde Greense functie}\index{$$L@$\Lambda$, interne parameter}\EA
Deze subtractie is te rechtvaardigen daar een verstrooier altijd een eindige
afmeting heeft, bijvoorbeeld een bol met straal $a$. Dan zal de `cut-off'
$\Lambda$ van de orde $1/a$ zijn. $\Lambda$ is te beschouwen als een interne
parameter van de verstrooier, net als $u$. We kunnen nu de $t$- matrix
uitrekenen met behulp van $G^{\mbox{reg}}$. Deze methode is ook toepasbaar is
als de Greense functie van het systeem anders is dan de `vacuum' functie
(\ref{greenr3d}) want de divergentie van $G$ is altijd
$\frac{1}{4\pi|{\bf r}-{\bf r'}|}$. Het is duidelijk dat $\Lambda$ een
positieve constante is die niet van $k$ afhangt.

\subsubsection{Regularisering\index{regularisatie} van de impulsintegraal}
Een andere methode, ook toegepast in de quantumveldentheorie, is het
regulariseren van de Fourier-integraal die we gebruiken om Greense functie
terug te transformeren naar de $\r$-ruimte. Dat gebeurt door de divergente
bijdrage af te splitsen en te vervangen door een eindige constante:
\begin{eqnarray}
\nonumber
G({\bf r},{\bf r})&=& \int \frac{\dint^3{\bf p}}{(2\pi)^3} G({\bf p})
\\ \nonumber
&=& \int \frac{\dint^3{\bf p}}{(2\pi)^3}
 \{ \frac{1}{{ p}^2}+(G({\bf p}) - \frac{1}{{p}^2}) \}
\\ \nonumber
&=& \int \frac{\dint^3{\bf p}}{(2\pi)^3}\frac{1}{{ p}^2} +
 \int \frac{\dint^3{\bf p}}{(2\pi)^3}
 \{ G({\bf p}) - \frac{1}{{ p}^2} \}
\\
G^{\mbox{reg}}({\bf r},{\bf r})&=& \Lambda +
\int \frac{\dint^3{\bf p}}{(2\pi)^3}
\{ G({\bf p}) - \frac{1}{p^2} \}
\\ \nonumber
&=& \Lambda + 4 \pi \int_0^\infty \frac{\dint p}{(2\pi)^3}
 \frac{-k^2p^2}{p^2(p^2-k^2-i0)}
\\
&=& \Lambda + \frac{ik}{4\pi}
\end{eqnarray}
Hierbij is alleen in de laatste twee stappen de expliciete uitdrukking voor
$G({\bf p})$ gebruikt. In de laatste stap is gebruik gemaakt van dezelfde
Cauchy-integraal als eerder bij de \'e\'endimensionale 
Greense functie en van het feit dat de integrand even is.
Deze subtractie is ook toepasbaar indien de Greense functie afwijkt
van de vorm (\ref{greenr3d}), bijvoorbeeld bij de aanwezigheid van wanden of
andere verstrooiers.
\BO
De divergentie voor grote $\p$ kan men verwijderen door het systeem op een
rooster te zetten. De minimale afstand en dus de maximale impuls worden
begrensd. Op een
simpel kubisch rooster met roosterafstand $a$ volgt de Greense functie uit:
\index{$a$, roosterafstand}
\BE-\frac{1}{a^2} \Delta G_{{\bf
r},{\bf r'}}
- k^2 G_{{\bf r},{\bf r'}}= \frac{1}{a^3} \delta_{{\bf r},{\bf r'}}
\EE
De rooster-Laplaciaan is gedefinieerd door: \BE (\Delta f)_\r= \sum_{i=1}^6
\left( f_{ {\bf r}+\roo_i,{\bf r'} }-f_{{\bf r},{\bf r'}} \right).
\index{$$D@$\Delta$, rooster-Laplaciaan}\EE
De verbindingsvectoren tussen naburige roosterpunten zijn
$\roo_i=-\roo_{i+d}=a\hat{{\bf e}}_i$, $i=1, \ldots, d$.
De discrete Fouriertransformatie is gedefinieerd door
\BEQ G(\p)=a^3\sum_{\r} G_{\r\r'}\eexp{-i\p\cdot(\r-\r')} \EEQ
Dit geeft  (zie ook volgende hoofdstuk) 
\BE
G(\p)=\left( \frac{2}{a^2} (3-\cos p_x a- \cos p_y a-\cos p_z
a)-k^2-i0 \right)^{-1} \EE
De terugkeer Greense functie  is \BE
 G_{\r,\r}=\int_{-\pi/a}^{\pi/a}  \frac{d^3 \p}{(2\pi)^3} G(\p) \EE
Toon aan dat voor $k=0$ deze divergeert als $W/a$ voor $
a\rightarrow 0$. De voor\-factor $W$ is een zogenaamde Watson-integraal.
Geef de expliciete uitdrukking voor $W$ en zijn numerieke waarde.
\EO

\subsubsection{Resonanties}\index{resonantie}
Als we in de uitdrukking voor de $t$-matrix (\ref{t-g}) de
geregulariseerde vorm van Greense functie invullen
wordt de vorm anders dan in het \'e\'endimensionale geval (\ref{t-1d}).
We kunnen de `matrix' in dezelfde vorm terugbrengen door een effectieve
verstrooiingssterkte\index{effectieve
verstrooiingssterkte}\index{$U@$U_{\rm eff}$, effectieve
verstrooiingssterkte} $U_{\rm eff}$ in te voeren : \begin{eqnarray}
\nonumber
t &=& \frac{u}{1-uG^{\mbox{reg}}({\bf r},{\bf r})}
\nonumber \\ &=& \frac{u}{1-u(\Lambda+\frac{ik}{4\pi})}
\nonumber \\&\equiv& \frac{U_{\rm eff}}{1-U_{\rm eff}\frac{ik}{4\pi}}
\label{ueff}
\EA
Dit definieert: \BA
U_{\rm eff} &\equiv& \frac{u}{1-u\Lambda}
\end{eqnarray}
Voor Schr\"odinger golven veroorzaakt dit geen essentieel verschil: aangezien
$u$ en $\Lambda$ constanten zijn is $U_{\rm eff}$ dat ook.
Voor scalaire golven is er w\`el een belangrijk verschil. Dan is $u$
frequentieafhankelijk, (zie vergelijking (\ref{beiden})) : $u = \alfa k^2$.
Noem $k_*$ het resonante golfgetal.
Identificeren we nu\, 
\footnote{Dat kan mits $\alfa >0$, d.w.z. $\varepsilon_{\rm verstrooier} >
 \varepsilon_{\rm medium} $.}
$\Lambda \equiv 1/(\alfa k_*^2)$ dan krijgen we
\BA
t&=&\frac{\alfa k^2k_*^2}{k_*^2-k^2-ivk_*^2k^3/4\pi} \\
U_{\rm eff} &=& \frac{\alfa k^2}{1-k^2/k_*^2}
\end{eqnarray}
Je ziet dat $U_{\rm eff} \to \infty$ als $k \to k_*$.\index{$k_*$,
resonantie golfgetal} $t$ wordt dan gelijk aan
$t_* = 4\pi i/k=2i\lambda$. Dit duidt op een resonantie met
sterke verstrooiing want $\lambda \gg 1/\Lambda$, m.a.w. de effectieve
verstrooiingslengte $t$ is veel groter dan de diameter
 $\approx 1/\Lambda$
van de verstrooier (zie onder vergelijking
(\ref{greg1})).

De resonantie is een {\em interne}
resonantie van de verstrooier, vergelijkbaar met de s-resonantie van een
Mie-bol. De resonantie wordt (net als alle resonanties, ook die van de
Mie-bol) sterk
be\"{\i}nvloed door de {\em omgeving} waarin de verstrooier zich bevindt.
Dit wordt hier beschreven door de Greense functie.
Merk op dat $t\approx \alfa k^2$ voor $k\rightarrow 0$. Dit leidt tot de
Rayleigh wet: $\sigma \sim \omega^4$ voor $\omega \rightarrow 0$. Hoewel dit
kwalitatief correct is, zal in het algemeen de voorfactor $\alfa $ bij
$\omega=0$ anders zijn dan bij de resonantie.

\subsection{Vergelijking met de scalaire Mie verstrooier}
In 1908 loste Mie de verstrooiing van vectorgolven aan een bolletje
exact op.
Wij beperken ons hier weer tot scalaire golven.
We vergelijken de puntverstrooier uit het vorig deel met een
Mie verstrooier\index{Mie verstrooier}.
De $t-$matrix voor scalaire s-golf
verstrooiing\index{$t$-matrix!Mieverstrooier}
is (zie bijvoorbeeld Merzbacher blz. 238 \cite{merz}).
\BE
t=\frac{4\pi \eexp{-2ika}}{mk \cot (mka) -ik} -\frac{4\pi \eexp{-ika} \sin ka}{k}
\EE
waarbij $a$ de straal van de bol is en $m$ de verhouding tussen de
golfgetallen binnen en buiten de verstrooier. De eerste resonantie is bij het
golfgetal $k_*=\pi/(2ma)$, als de cotangens nul wordt. Het idee is nu
om de straal van de bol naar nul te laten gaan, terwijl het golfgetal dicht
bij resonantie blijft.
Neem de limiet $a \rightarrow 0, m\rightarrow \infty$,  zodanig dat $
k_*$ vast blijft. Er geldt voor $k$ dichtbij $k_*$
\BA
mk \cot mka &\approx& \frac{m^2k^2 a \pi}{2ma}
(\frac{1}{k}-\frac{1}{k_*}) \nonumber \\
&\approx& \frac{\pi^4}{32m^2a^3 } (\frac{1}{k^2}-\frac{1}{k_*^2})
\EA
Ofwel
\BE
t=4\pi \left( \frac{\pi^4}{32m^2 a^3}(\frac{1}{k^2}-\frac{1}{k_*^2})-ik
\right)^{-1} \label{tmie} \EE
Dit moet vergeleken worden met de uitdrukking (\ref{ueff}) , \BE
t=\frac{U_{\rm eff}}{1-iU_{\rm eff}k/4\pi}=4\pi \left( \frac{4\pi}{\alfa}
(\frac{1}{k^2}-\frac{1}{k_*^2}) -ik \right)^{-1} \label{tmie2} . \EE
De $\delta-$potentiaalsterkte is 
\BE \alfa=\int d^3 r (\epsilon(r)-1)=\int_{r<a}d^3 r [m^2-1]
=\frac{4\pi a^3}{3} (m^2-1). \EE Ofwel als we de limiet nemen: 
\BE \frac{4\pi}{\alfa}=\frac{3}{a^3 m^2}. \EE
 Schrijven we (\ref{tmie}) in de vorm van
(\ref{tmie2}) dan vinden we $4\pi/\alfa \approx \pi^4 /32 m^2 a^3$.
 Omdat het verschil een factor $\pi^4/96$ is, zien we dat de
 resultaten binnen 2\% overeenstemmen.
In formule (\ref{greg1}) werd $1/\Lambda$ een maat voor de straal van de
verstrooier genoemd.
We vinden hier als $m \gg 1$ \BE \frac{1}{\Lambda}=\frac{k_*^2
(m^2-1)a^3}{3}\approx \frac{\pi^2}{12} a \EE
Inderdaad is $1/\Lambda$ een maat voor de straal van de verstrooier.
We hebben aldus een eenvoudige uitdrukking voor de $t-$matrix gevonden die
de essentiele fysica ook kwantitatief beschrijft dichtbij de s-golf
resonantie van de verstrooier.

\subsection{Werkzame doorsnedes\index{werkzame doorsnede} en het albedo}

In de lite\-ratuur worden dikwijls verschil\-lende `werkzame door\-snedes'
gedefinieerd \cite{vdHSP}, \cite{Ishi}.
Wij zullen hier de drie belangrijkste behandelen.

 Laat een vlakke golf onder richting $\n'$ invallen.
Zoals besproken in vgl. (\ref{effect}) kan bij verstrooiing aan \'e\'en
object de totale golf geschreven worden als
\BEQ \Psi(\r)=\eexp{ik\n'\cdot\r}+\int d^3\r'd^3\r'' G(\r,\r')t(\r',\r'')
\eexp{ik\n'\cdot\r''}  \label{mcw20}
\EEQ
De $t$-matrix in de plaatsruimte is gedefinieerd in vgl. (\ref{TVT}).
Zijn Fourier getransformeerde is\index{$pv@$\p$, interne impuls}
\BEQ t(\p,\p')=\int d^3\r
d^3\r'\eexp{-i\p\cdot\r+i\p'\cdot\r'}t(\r,\r')\label{mcw21} \EEQ
Laten we aannemen dat het zwaartepunt van de verstrooier zich in de oorsprong 
bevindt. Ver van de oorsprong geldt
\BEQ G(\r,\r')\approx \frac{\eexp{ikr-ik\n\cdot\r'}}{4\pi r}
\qquad{\rm met}\quad \n=\frac{\r}{r}.\EEQ
Vul dit in in formule (\ref{mcw20}), dan vind je met (\ref{mcw21}) dat de
verstrooide golf de vorm heeft: \BEQ \Psi_{sc}(\r)\approx
\frac{\eexp{ikr}}{4\pi
r}t(k\n,k\n') \EEQ 
De {\it verstrooiings-werkzame doorsnede}\index{werkzame
doorsnede!verstrooiings} (`scattering cross section')
is per definitie de over een boloppervlak ge\"integreerde intensiteit,
genormeerd op de inkomende intensiteit.
Dit is gelijk aan het oppervlak waarover je de inkomende bundel, bij 
loodrechte inval, zou moeten integreren om even veel intensiteit op te vangen.
\BEA
\sigma_{sc}&=&\int_{4\pi} r^2d\n\frac{\vert t(k\n,k\n')\vert^2}{(4\pi r)^2}
\nonumber\\ &=&\frac{1}{(4\pi)^2} \int_{4\pi}d\n \vert t(k\n,k\n')\vert^2
\EEA  \index{$$ssc@$\sigma_{sc}$, verstrooiingswerkzame doorsnede}
Je ziet dat de $t$-matrix alleen voor impulsen met $|\p|=k$ nodig is
(`verre veld', `op de massaschil', `on shell'). Men  schrijft
daarom vaak $t(\n,\n')$ in deze uitdrukking.
Een {\it isotrope puntverstrooier} heeft dus de 
verstrooiings-werkzame-doorsnede
\BEQ \sigma_{sc}=\frac{\bar t t}{4\pi} \label{sigmasc}\EEQ
De tweede belangrijke grootheid is de 
extinctie-werkzame doorsnede\index{werkzame doorsnede!extinctie}. Hij geeft
aan hoeveel intensiteit
uit de doorgaande bundel is verdwenen. Laten we aannemen dat de vlakke
golf vanaf $z=-\infty$ langs de $z$-as invalt. We kijken naar de 
intensiteit in een kleine ruimtehoek rond de $z$-as voor grote $z$.
Omdat dan $r\approx z+(x^2+y^2)/2z$, is de totale golf gegeven door
\BEQ \Psi(\r)=\eexp{ikz}+t(\n,\n)\frac{\eexp{ikz}}{4\pi z}
\eexp{ik(x^2+y^2)/2z} \EEQ
De intensiteit is dan voor grote $z$
\BEQ \Psi^*\Psi=1+{\rm Re}\frac{t(\n,\n)}{2\pi z}\eexp{ik(x^2+y^2)/2z} \EEQ
Dit integreren we over $x$ en $y$ in een oppervlakje $A$\index{$A$,
oppervlak} loodrecht op de $z$-as. De voorwaarde dat $x$ en $y$ veel kleiner
zijn dan $z$ zorgt ervoor dat in de tweede term Gaussische integralen
optreden. Daarom valt de $z$-afhankelijkheid er uit. We schrijven het
resultaat als \BEQ \int_A dxdy \Psi^*\Psi=A-\sigma_{ex}\EEQ
De {\it extinctie-werkzame doorsnede} (`extinction cross
section')\index{extinctie werkzame doorsnede}\index{$$sex@$\sigma_{ex}$,
extinctie werkzame doorsnede}
is het oppervlak waarover je de binnenkomende bundel moet integreren om
er evenveel intensiteit uit weg te halen. Hij is gelijk aan
\footnote{Zowel $\sigma_{sc}$ als $\sigma_{ex}$ zullen van de
inkomende richting afhangen indien de verstrooier een onregelmatige vorm
heeft.}
\BEQ\label{sigmaex} \sigma_{ex}(\n)=\frac{{\rm Im}\,t(\n,\n)}{k}. \EEQ
Voor een puntverstrooier heb je natuurlijk
\BEQ\label{sigmaexpunt} \sigma_{ex}=\frac{{\rm Im}\,t}{k} \EEQ

Het {\it albedo}\index{albedo!definitie}\index{$a$, albedo} van de
verstrooier is
gedefinieerd als de verhouding van de verstrooide en de weggenomen intensiteit:
\BE \label{albedo} a=\frac{\sigma_{sc}}{\sigma_{ex}} \EE
Voor pure verstrooiing verwacht je $a=1$; dit heet het {\it optisch
theorema}\index{optisch theorema}. Het is echter ook mogelijk dat er
absorptie
optreedt. Dan kun je de {\it absorptie-werkzame doorsnede} defini\"eren als
\BEQ \sigma_{abs}\equiv \sigma_{ex}-\sigma_{sc}
\quad\Rightarrow\quad  \sigma_{ex}=\sigma_{sc}+\sigma_{abs} \EEQ
Het albedo is dan gelijk aan
\BEQ a=\frac{\sigma_{sc}} {\sigma_{sc}+\sigma_{abs}} \EE

Vanwege de verre-veld-constructie zal het optisch theorema niet
zonder meer van toepassing zijn in een systeem met veel verstrooiers.
Dat zie je eenvoudig omdat je dan niet direct kunt zeggen wat je
met $k$ bedoelt. Niettemin moet een systeem bestaande uit verstrooiers
die ieder voor zich niet absorberen, voldoen aan behoud van energie.
Dit wordt opgelegd door de {\it Ward identiteit}\index{Ward identiteit}, een
generalisatie van het optisch theorema.

Voor een sferische verstrooier is de extinctie-werkzame doorsnede
hoekonafhankelijk. Het is dan zinvol de {\it fasefunctie} te defini\"eren
\BEQ p(\cos\Theta)=p(\n\cdot\n')=\frac{\mid t(\n,\n')\mid^2}
{4\pi\sigma_{ex}}
=\frac{\mid t(\n,\n')\mid^2}
{4\pi}\cdot\frac{k}{{\rm Im}\,t}.\EEQ 
Er geldt dan \BEQ \int\frac{d\n}{4\pi}p(\n\cdot\n')=\frac{\sigma_{sc}}
{\sigma_{ex}}=a \EEQ

Voor een isotrope puntverstrooier heb je de hoekonafhankelijke uitdrukking
$p=a$, in afwezigheid van absorptie: $p=1$. \BO\begin{itemize}
\item{ Toon aan dat je voor dipool- of $p$-verstrooiing de Rayleigh-fasefunctie
$p(\cos\Theta)=3(1+\cos^2\Theta)/4$ krijgt.}
\item{Neem aan dat de brekingsindex een klein imaginair gedeelte
heeft, $m=m_r+im_i$. Bepaal het albedo van de scalaire Mie-verstrooier tot in
eerste orde in $m_i$\index{albedo!Mie-verstrooier}.} 
\item{Neem aan dat de parameter $u=u_r-iu_i$ een klein complex gedeelte
heeft. Bepaal het albedo van de puntverstrooier tot op eerste orde in
$u_i$.}\end{itemize} \EO

\renewcommand{\thesection}{\arabic{section}}
\section{Greense funkties van wanordelijke systemen}\index{Greense functie}
\setcounter{equation}{0}\setcounter{figure}{0}

\label{greense}

In tegenstelling tot het vorige deel wordt nu verstrooiing aan {\it
vele} verstrooiers beschouwd. De gemiddelde Greense functie is
gerelateerd aan twee belangrijke begrippen: de toestandsdichtheid en de
zelfenergie. We kijken weer naar waarschijnlijkheidsgolven met vaste 
energie $E=k_0^2$ of akoestische golven met
vaste frequentie $\omega=c k_0$. Door de aanwezigheid van vele verstrooiers
zal $k_0$ veranderen in het effectieve golfgetal\index{$K@$K$, effectief
golfgetal} \BEQ K=k+\frac{i}{2\ell_{sc}}. \EEQ
Voor akoestische golven beschrijft dit 
een fasesnelheid $v_{ph}=\omega/k=ck_0/k$ 
\index{$v_{ph}$,fasesnelheid}\index{fasesnelheid}
en een verstrooiings-vrije weglengte $\ell_{sc}.$

\subsection{De toestandsdichtheid\index{toestandsdichtheid}}
Beschouw een Schr\"odinger eigenwaarde probleem
$H\Psi_\lambda = E_\lambda \Psi_\lambda $.
Uit de Greense functie is de toestandsdichtheid $\rho(E)$ te berekenen.
De toestandsdichtheid telt alle toestanden met een zekere energie-eigenwaarde:
\BE \rho(E)=\frac{1}{N} \sum_{\lambda} \delta(E-E_{\lambda}),
\EE waarbij $N$ het aantal roosterplaatsen is.\index{$N$, aantal 
roosterplaatsen}
Anderzijds geldt ook (het spoor is ongevoelig voor de basis-keuze):
\BE \mbox{Tr} \, g(E)=\mbox{Tr}  \frac{1}{H-E} =\sum_\r \left(
\frac{1}{H-E} \right)_{\r,\r} = \sum_{\lambda} \frac{1}{E_{\lambda} - E} \EE
\index{$g$, niet-gemiddelde Greense functie}
Waarbij de definitie $g(E)\equiv (H-E)^{-1}$ gebruikt is.
Gebruik nu een stelling uit de complexe analyse\BE
\frac{1}{x-i 0} = \mbox{PV} \frac{1}{x} +i\pi \delta(x) \EE
Hierbij geeft PV de `principal value' of hoofdwaarde aan, gedefinieerd door:
\BE
\mbox{PV} \frac{1}{x} f(x) =\lim_{\epsilon \rightarrow 0}
\left[ \int_{-\infty}^{-\epsilon} \frac{f(x)}{x} +\int^{\infty}_{\epsilon}
\frac{f(x)}{x} \right] .\EE We vinden de vaak gebruikte\index{$$r@$\rho$,
toestandsdichtheid} formule \BE \rho(E)=\frac{1}{\pi N} \mbox{Im Tr }g(E+i0)
=\frac{1}{\pi N} \mbox{Im Tr }G(E+i0)
\; . \EE
Hier is $G$ de {\it gemiddelde} Greense functie of {\it amplitude} Greense
functie. 
De gelijkheid
geldt omdat de fluctuaties voor grote $N$ klein zijn. Dit heet
`zelfmiddeling'.\index{zelfmiddeling}

\subsection{Het Lloyd model\index{Lloyd model}}
We proberen inzicht te krijgen in enkele eigenschappen van Greense funkties
in wanordelijke systemen.
Er zijn vele soorten wanorde. De  twee belangrijkste zijn
willekeurige geplaatste verstrooiers (bijv. een TiO$_2$-suspensie),
 ook wel topologische wanorde\index{wanorde!topologische} genoemd, en
willekeurige potentialen op een rooster (bijv. electronen in een
verontreinigd kristal). Op een rooster wordt willekeurige plaatsing
van identieke
verstrooiers ook wel binaire wanorde genoemd. We zijn ge\"interesseerd
in gemiddelde grootheden,
b.v. gemiddelde Greense funktie of de gemiddelde toestandsdichtheid
\footnote{Ook de verdeling van grootheden over verschillende
wanorde configuraties kan interessant zijn. De breedte van de verdeling
kan groter
zijn dan men in eerste instantie verwacht, zoals bijvoorbeeld bij
`universele geleidings fluctuaties'.}. Vooralsnog worden alleen zog.
amplitude eigenschappen bestudeerd. We bekijken het
Lloyd model. In dit model is door
een speciale distributie van random potentialen de gemiddelde Greense
functie exact oplosbaar in willekeurige dimensie. We bekijken hier
het \'e\'endimensionale geval. Beschouw de Schr\"odinger
vergelijking\index{Schr\"odinger vergelijking! op rooster} op een
rooster met roosterafstand 1. De Hamiltoniaan is:
\BE
H=-\Delta+V_\r
\EE
Op iedere plaats $\r$ is de waarde van de potentiaal getrokken uit de
verdeling:\BE
p(V_\r)=\frac{\Gamma}{\pi
[(V_\r-\VB)^2 +\Gamma^2]}\; . \EE Dit is een Lorentz-verdeling,
gepiekt
rond $\VB$\index{$V2@$\VB$, gemiddelde potentiaal} en met een breedte
$\Gamma$.\index{$$C@$\Gamma$, breedte verdeling}
De potentiaalsterktes op verschillende plaatsen zijn ongecorreleerd.
Op het rooster is de Schr\"odinger vergelijking
een matrix vergelijking:
$\sum_ j H_{ij} \Psi_j=E \Psi_i$. De rooster-Laplaciaan in \'e\'en
dimensie is gedefinieerd als:
\BA
\Delta_{i,i}=-2 &&  \nonumber \\
\Delta_{i,i \pm 1 }=1 &&   \nonumber \\
\Delta_{i,j}=0 && \mbox{ in overige gevallen}
\EA
Er is hier wat keuzevrijheid, want we kunnen ook koppeling met tweede
buren, etc., meenemen. We willen in ieder geval dat we de
continu\"um Laplaciaan terugkrijgen als de
roosterafstand naar nul gaat ( de continu\"um-limiet).
De Fourier getransformeerde van de rooster Laplaciaan werkend op een
willekeurige functie $f$ is: \BA
{\cal F} [\Delta f]&=&
\sum_{m=1}^N \Delta f(m) \eexp{imp}  =\sum_{n=1}^N f(n)
[-2\eexp{ipn}+\eexp{ip}\eexp{ipn}+ \eexp{-ip}\eexp{ipn} ] \nonumber \\
&=&\sum_{n=1}^N f(n)
[2\cos p-2] \eexp{ipn} \nonumber \\ &=&[2\cos p-2] {\cal F} [f]
\EA
Waarbij $N$ het aantal roosterpunten is.
Men vindt dus: $ \Delta(p)= 2\cos p -2$.
Inderdaad voor kleine $p$, ofwel bij langzame variatie van $f$ t.o.v. de roosterafstand,
geldt de bekende continuum relatie: ${\cal F}[-\Delta f] = p^2
f(p)$. (In $d$ dimensies verandert men $\Delta_{\r,\r}$ in $-2d$; de Fourier
getransformeerde is dan: ${\cal F}[\Delta]=2 \cos p_1 +\ldots+ 2\cos p_d
-2d$.)
De operator $H-E$ ziet er in \'e\'en dimensie op het rooster als volgt
uit: \BA
\left( \begin{array}{cccccc}
     2-E+V_1 &-1& 0 & 0 & \cdots & -1 \\
       -1& 2-E+V_2& -1& 0 & \cdots & 0 \\
       0&-1&2-E+V_3&-1& \cdots &0 \\
       \vdots & \vdots & \vdots & \vdots & \ddots & \vdots \\
       -1 &0&0&0&\cdots & 2-E+V_N
       \end{array}
\right)
\EA
(De getallen in de rechter bovenhoek en linker onderhoek zijn bepaald door
de keuze van randvoorwaarden, hier periodiek).
In willekeurige dimensie kan de bijbehorende Greense funktie van een
dergelijk systeem gevonden worden
door: \BE g_{\r,\r'}=\left( \frac{1} {-\Delta-E+V}
\right)_{\r,\r'}= \frac{\mbox{codet}(-\Delta-E+V)_{\r,\r'}}
{\mbox{det}(-\Delta-E+V)} \EE Het is voor niet al te grote systemen
mogelijk de matrix $g$ numeriek op te bepalen. Echter in het Lloyd model
is een exacte oplossing voor de gemiddelde Greense functie mogelijk.

De {\it gemiddelde} Greense funktie\index{gemiddelde Greense
functie}\index{$G@$G$, gemiddelde of amplitude Greense functie}
of {\it amplitude} Greense functie\index{amplitude Greense functie} $G$, wordt
verkregen door integratie over
alle mogelijke wanorde configuraties. Dit betekent een integratie over alle
waarden van de potentiaal op elke plaats.
\BA
G_{\r,\r'}&=&\prod_i \int dV_i p(V_i) g_{\r,\r'}
\EA
Middelen we eerst over de potentiaal op positie 1, dit geven we aan met een
accent (de overige potentialen houden we in eerste instantie constant) \BA
g'_{\r,\r'} &=& \int dV_1
p(V_1) g_{\r,\r'} \nonumber \\ &=&\int dV_1 \frac{\Gamma}{\pi
((V_1-\VB)^2 +\Gamma^2)} \frac{\mbox{codet}(-\Delta-E+V)_{\r,\r'}}
{\mbox{det}(-\Delta-E+V)} \EA
De eerste noemer heeft polen in $V_1=\VB \pm i \Gamma$.
De tweede noemer heeft polen bij ree\"ele energie eigenwaarden.
Als we kiezen $\mbox{Im} E >0$, liggen alle polen van de tweede noemer boven
de ree\"ele as van $V_1$. We sluiten de integratiecontour naar beneden en
vinden:
\BE g'_{\r,\r'}= \left( \frac{1} {-\Delta-E+V'} \right)_{\r,\r'} \EE
waarbij $V'=(\VB-i\Gamma, V_2, \ldots V_N)$.
Bij de overige integraties gaan de zelfde argumenten op en alle potentialen
worden uiteindelijk vervangen door $V_\r \Rightarrow \VB-i\Gamma$
\BE G_{\r,\r'}=
\left( \frac{1}
{-\Delta-E+\VB-i\Gamma} \right)_{\r,\r'} \label{GGG}\EE
De matrix (\ref{GGG}) dient nog gediagonaliseerd te worden.
We beperken ons nu weer tot het \'e\'endimensionale geval.
De diagonalisatie gaat gemakkelijk in de impuls-ruimte.
\BA
G_{x,x'}&=& \int_{-\pi}^{\pi} \frac{dp}{2\pi} \frac{\eexp{ip(x-x')}}{2-2\cos p-E+\VB-i\Gamma}
\nonumber \\
&=&\oint_{|z|=1} \frac{dz}{2\pi i z}
\frac{z^{(x-x')}}{2-z-z^{-1}-E+\VB-i\Gamma} \nonumber \\
&=&-\oint_{|z|=1}  \frac{dz}{2\pi i}  \frac{z^{(x-x')}}
{(z-\eexp{iK})(z-\eexp{-iK})}
\nonumber
\EA
Waarbij $\eexp{-iK} = \frac{1}{2}(2-E+\VB-i\Gamma)+
\sqrt{\frac{1}{4}(2-E+\VB-i\Gamma)^2-1}$.
De noemer heeft polen in $z=\exp(\pm iK)$ en afhankelijk van het teken van
$(x-x') $ heeft de teller polen in nul of oneindig.
We nemen $x>x'$ (de tellerpool ligt dan in oneindig) en gebruiken de
residu\"en stelling. Vanwege symmetrie
van $G$ hoeft het geval $x<x'$ niet apart te worden uitgewerkt.
\BA G_{x,x'}&=& -\frac{2 \pi i}{2\pi i} \frac{z^{|x-x'|}}{z-\eexp{-iK}}
\left|_{z=\eexp{iK}} \right. \nonumber \\
&=&  \frac{\eexp{iK|x-x'|}}{-2i\sin K}  \label{sinh}
\EA

We beschouwen eerst de situatie waarbij de breedte verdeling van de
potentiaal nul is. Als $\Gamma=0$ wordt het spectrum van
eigenwaarden
beperkt tot $0<E-\VB<4$. De parameter $K$ is nu re\"eel positief.
Stel je $K=k$, dan geeft formule (\ref{sinh}) \BE
G_{x,x'}=\frac{\eexp{ik|x-x'|}}{-2i\sin k} \EE met 
$k=\frac{1}{2} \arccos( 2-E-\VB)$. Daar $\Gamma=0$ hebben hebben
alle potentialen dezelfde grootte. Van wanorde is hier geen sprake. We
vinden de Greense functie op het rooster voor de ongestoorde ($H=-\Delta$)
situatie met een verschoven energie.


Als $\Gamma>0$ strekt het spectrum zich uit, er zijn oplossingen
voor $-\infty <E < \infty$.
Het imaginaire deel van $K$ is nu ongelijk aan nul. Met
$K\equiv k+i/(2\ell)$ vind je
\BE
G_{x,x'}=\frac{\eexp{ik|x-x'|} \,\eexp{-|x-x'|/2\ell}}{-2 i\sin K},
\EE Dus $G$ valt af als \BE |G_{r,r'}|^2\sim
\eexp{-|r-r'|/\ell} \EE
De onverstrooide intensiteit valt exponentieel af over \'e\'en vrije
weglengte, dit geldt ook voor andere dimensies. Dit wordt wel de wet van
Lambert-Beer genoemd.
Voor kleine $\Gamma$ kan men aantonen dat $\ell=\sin (k) /\Gamma$, de
vrije weglengte wordt hier dus kleiner aan de bandrand.
De toestandsdichtheid wordt \BE \rho(E)=\frac{1}{2\pi} \mbox{Re}
\frac{1}{\sin K} \EE In figuur \ref{dos} is de
toestandsdichtheid voor verschillende wanordesterktes afgebeeld.


\begin{figure}[h]
\centerline{\includegraphics[width=9cm]{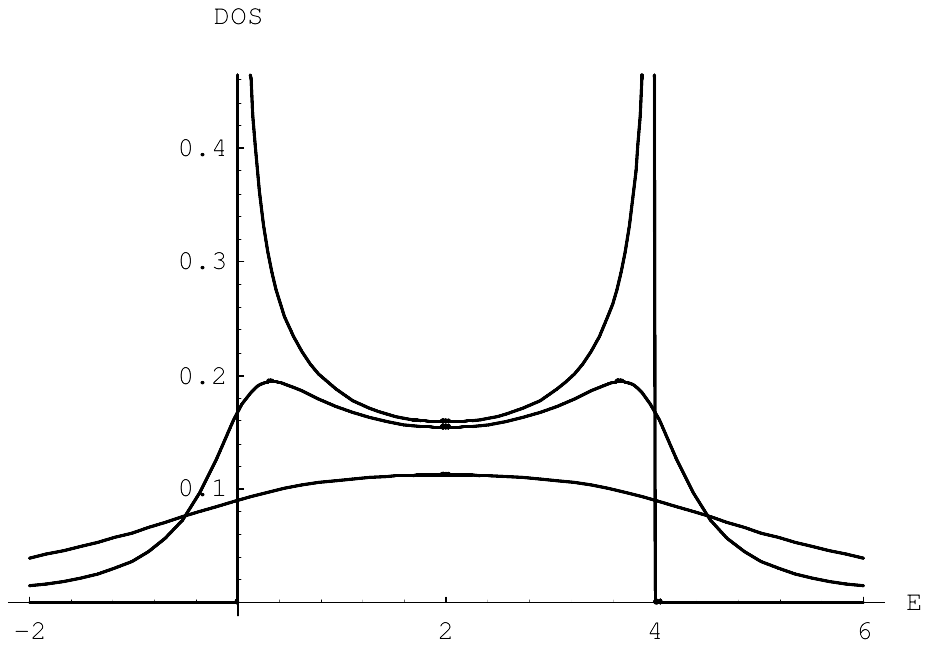}}

\vspace{-7cm}

\caption{De toestandsdichtheid van het \'e\'endimensionale Lloyd model voor
$\Gamma=0, \frac{1}{2}$ en $2$ (van boven naar beneden bij $E=2$). Ten gevolge van de wanorde verdwijnen de van
Hove singulariteiten en krijgt de toestandsdichtheid een langzame 
staart $\rho(E)\sim 1/E^2$. $\VB$ is nul gesteld.} 
\label{dos} \end{figure}

\newpage 


\subsection{Diagrammatische ontwikkeling van de
zelfenergie}\index{zelfenergie} \index{$$Sig@$\Sigma$, zelfenergie}
Het model uit de vorige paragraaf is een uitzondering omdat exacte oplossing
van $G$ mogelijk is\footnote{Exact oplosbare modellen bestaan in \'e\'en
dimensie. B.I. Halperin beschouwt een Gaussische witte ruis potentiaal op
een lijn \cite{halp}, Th.M. Nieuwenhuizen beschouwt exponenti\"ele
verdelingen
op een 1D rooster \cite{thmn6},\cite{thmn7}. Alleen het Lloyd model is
oplosbaar in alle
dimensies.}. In praktijk zal men vrijwel altijd een
benaderingsstrategie moeten gebruiken om relevante grootheden uit te
rekenen.
Diagrammen helpen dan om een systematische storingsreeks op te schrijven.
Stel dat we ditmaal een systeem hebben met willekeurig geplaatste gelijk
sterke verstrooiers (mistdeeltjes, `melkdeeltjes', bomen,..). De potentiaal
 is dan in de puntlimiet $V(\r)=-\sum_{i=1}^N u \delta(\r-{\bf R}_i)$. De
Greense  funktie voor dit probleem is in figuur \ref{gvg} weergegeven.
\begin{figure}[htbp]
\centerline{\includegraphics[width=12cm]{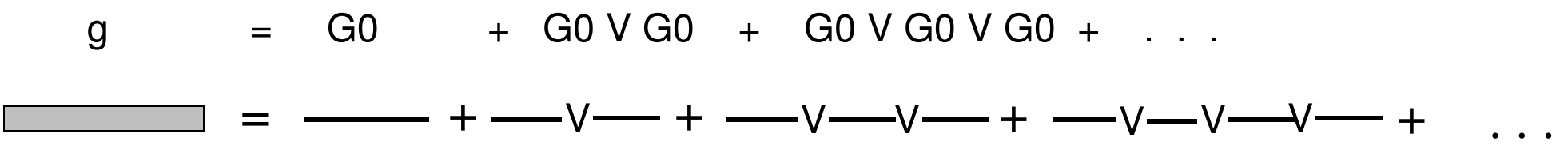}}

\caption{De aangeklede Greense funktie.} \label{gvg}
\end{figure}
Met $G_0$ wordt de zog. kale\index{propagator!kale} of naakte propagator
bedoeld, de propagator in het medium zonder verstrooiers. De gezochte grootheid is $g$, de Greense functie van
het wanordelijke medium. $g$ wordt ook wel de
aangeklede\index{propagator!aangeklede} of volle propagator genoemd.
Om de Greense funktie precies te weten, zou men alle diagrammen uit
moeten rekenen (en de som moet dan ook nog convergeren !). Dit is
meestal onmogelijk. We maken daarom de benadering dat de verstrooiersdichtheid $n$
klein is. Door
deze aanname zullen collectieve effecten van meerdere deeltjes klein zijn.

We middelen over alle wanorde configuraties, d.w.z. over 
alle posities van de $N$
verstrooiers in het volume $V$, en sorteren de diagrammen (zie 
figuur \ref{sigma}).

\begin{figure}[htbp]
\centerline{\includegraphics[width=10cm]{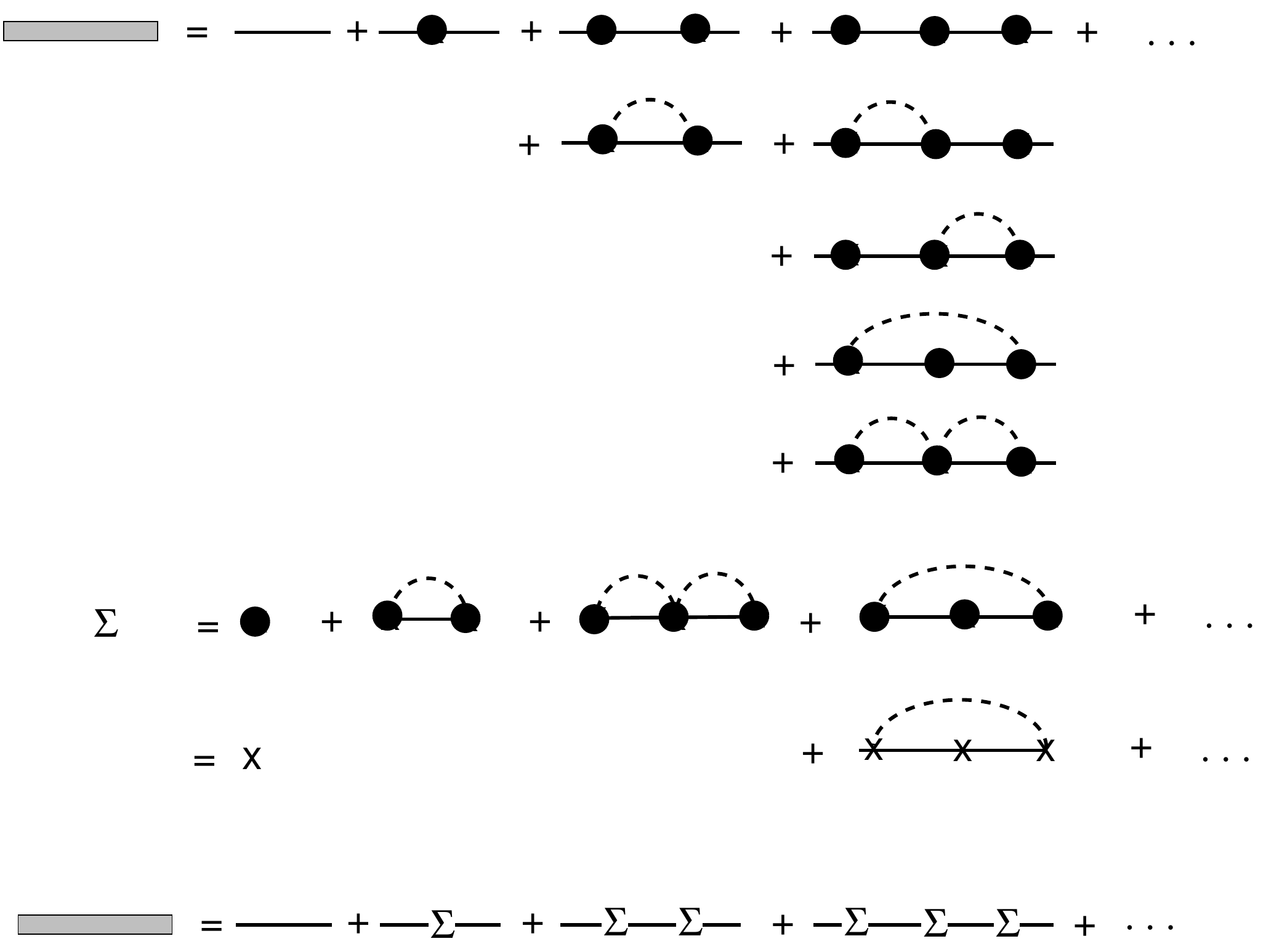}}
\caption{De gemiddelde Greense funktie (de dikke lijn). De dunne lijnen 
stellen de kale Greense functies voor; de gestreepte lijnen
geven aan dat de verstrooiing aan dezelfde potentiaal plaatsvindt. De
zelfenergie $\Sigma$ bevat alleen
irreducibele\index{diagrammen!irreducibele}
diagrammen en genereert alle diagrammen via de onderste relatie.} \label{sigma}
\end{figure}

\BA
\label{Ggen}
G=\langle g \rangle &=&G_0+G_0 \Sigma G_0+G_0 \Sigma G_0 \Sigma G_0 +\ldots
\nonumber \\          &=&\frac{1}{G_0^{-1}-\Sigma}
\nonumber \\          &=&\frac{1}{-\nabla^2-k_0^2-\Sigma}
\EA
In de laagste orde van de verstrooiersdichtheid hebben we $\Sigma$ al
uitgerekend in het vorige hoofdstuk. Hij is namelijk evenredig met de 
$t-$matrix!
Ten gevolge van het middelen is er ook nog een dichtheidsafhankelijkheid,
die al uit de eerste orde Born benadering gevonden kan worden \BE \VB=-u\int
\frac{d^3\R_1}{V}
\cdots \frac{d^3\R_N}{V} \sum_{i=1}^N  \delta(\r-\R_i)=-u\sum_{i=1}^N \int
\frac{d^3{\bf R}_i}{V} \delta(\r-\R_i) = -\frac{uN}{V} =-nu.\EE 
\index{$N$, aantal verstrooiers}\index{$V$, volume}
Hetzelfde argument gaat op voor elke orde in de Bornreeks, dus

\BE \label{Gmetnt}
\Sigma=nt+O(n^2)
\EE
Dit wordt ook wel de {\it
onafhankelijke-verstrooiers-benadering}\index{onafhankelijke verstrooier
benadering} (`independent scatterer approximation') genoemd,
m.a.w. we werken in eerste orde van de dichtheid. In figuur \ref{sigma}
correspondeert deze benadering met $\Sigma=x$. Het effectieve
golfgetal is \BE K=\sqrt{k_0^2+\Sigma}\equiv k+\frac{i}{2\ell}. \EE
Dit leidt tot een complexe brekingsindex\index{brekingsindex},
\BE m=\frac{K}{k_0}\approx \sqrt{1+\frac{nt}{k_0^2}} \EE

De Greense functie (\ref{Ggen}) neemt tenslotte een vorm aan die we vaak gaan gebruiken

\BE 
\label{Gpknt}
G({\bf p},k_0)= \frac{1}{p^2-k_0^2-nt} , \qquad k_0=\frac{\omega}{c}.
\EE

\subsection{Zelfconsistentie}
Zoals gezegd kunnen we nooit alle diagrammen berekenen. Het is meestal
mogelijk zonder veel werk, hogere orde diagrammen van een bepaald type
mee te nemen. In geval van de zelfenergie kunnen we in plaats van $t=u/[1-u
G_0 (r,r)]$, namelijk $t=u/[1-u G(r,r)]$ nemen. Dit noemen we een {\it
zelfconsistente}\index{zelfconsistentie} benadering. Je kunt nagaan dat nu
ook het laatste
diagram in de uitdrukking voor $\Sigma$ in figuur \ref{sigma} in de
zelfenergie wordt
meegenomen. Dit betekent dat ook sommige bijdragen van willekeurige orde in
$n$ zijn berekend. In het algemeen moet men bij deze techniek goed opletten geen
bijdragen dubbel mee te nemen. \BO Hoewel we nu al veel meer bijdragen
meenemen missen we er ook nog. Bij vier keer strooien is er \'e\'en diagram
dat
wel in de zelfenergie zit, maar niet de zelfconsistente $t-$matrix. Welk ?
\EO
Deze methode leidt tot een rijker
(= fysischer) gedrag van het systeem; de zelfconsistentie beschrijft
verstrooiing in een medium met andere verstrooiers.
De zelfconsistente $t-$matrix
\BE t=\frac{u}{1-u\Lambda-\frac{i}{4\pi}u \sqrt{k_0^2+nt}} \label{tzelf} \EE
leidt
tot een kloof\index{toestandsdichtheid!kloof} (`gap')\footnote{Kloven
zijn interessant, omdat bij kloven gemakkelijk gelocaliseerde
toestanden kunnen optreden.} in de toestandsdichtheid zoals uitgerekend door
Polishchuk {\it et al} \cite{poli}. Dit komt omdat $t$ re\"eel negatief kan 
zijn, op zo'n manier dat het argument van de wortel negatief is.
Door hogere orde correcties is er 
vermoedelijk geen echte gap. Toch blijft de toestandsdichtheid klein,
waardoor Anderson localisatie\index{Anderson localisatie} kan optreden. Dit
wordt in het volgend hoofdstuk toegelicht.

De $t$--matrix (\ref{tzelf}) is fysisch interessant omdat zij nog steeds
een resonantie heeft. Als er geen absorptie is moet het optisch
theorema gelden. Dit is equivalent met de eis: $a=1$.  Het
albedo\index{albedo!zelfconsistente $t-$matrix} $a$ was gedefinieerd als:
\beq
\label{aaext=}
a\equiv\frac{\sigma_{sc}}{\sigma_{ex}}=
\frac{k\bar{t}t}{4\pi {\rm Im} t}
\eeq
\BO Laat zien dat $a=1$  voor de $t$--matrix gegeven door:
\beq
t=\frac{1}{\frac{1}{u}-\Lambda-\frac{i}{4\pi}\sqrt{k_0^2+nt}}
\eeq
{\bf Uitwerking:}
Schrijf: $\sqrt{k_0^2+nt}=k+\frac{i}{2\ell} $ dan
\beqa
k^2 & = & \frac{1}{2}(k_0^2+nt_r+\{(k_0^2+nt_r)^2+n^2t_i^2\}^\frac{1}{2})
\nonumber\\
\frac{1}{4\ell^2} & = & \frac{1}{2}(-(k_0^2+nt_r)+
\{(k_0^2+nt_r)^2+n^2t_i^2\}^\frac{1}{2}).
\nonumber
\eeqa
Waarin $t_r$ het reele en $t_i$ het imaginaire deel van t is.  Invullen
geeft \beqa
t & = & \frac{ (1/u-\Lambda+1/4\pi 2\ell )+ (k/4\pi)i }
{ (1/u-\Lambda+1/4\pi 2 \ell)^2+k^2/16\pi^2}\nonumber\\
t\bar{t} & = &
\frac{1}{ (1/u-\Lambda+1/4\pi 2\ell)^2+k^2/16\pi^2 } \nonumber\\
& = & \frac{4\pi}{k} {\rm Im} t = \frac{4\pi}{k} \times 
\frac{ k/4\pi }{ (1/u-\Lambda+1/4\pi 2\ell )^2+k^2/16\pi^2}
\eeqa\EO
Dus de zelfconsistente t-matrix voldoet aan het optisch
theorema\index{optisch theorema!zelfconsistente $t-$matrix} voor alle ordes
in de dichtheid. Let wel dat $k$, het re\"ele deel van het 
zelf-consistente golfgetal $K$, 
en niet $k_0$, in het optisch theorema opduikt.

\subsection{Opmerkingen}
\begin{itemize}
\item{Voor het uitrekenen van gemiddelde Greense funkties in
wanordelijke systemen bestaan
veel benaderingstechnieken, bijvoorbeeld de `coherente potentiaal
benadering' ofwel CPA-methode (zie Economou, \cite{ecou})}
\item{Tot nu toe is alleen de Greense funktie $g$ en zijn gemiddelde $G$
behandeld. Dit zijn zogenaamde amplitude-eigenschappen. In het Lloyd model
zagen we
al snelle afval over \'e\'en vrije weglengte van de Greense functie. We weten
echter
dat diffuus voortgeplante golven veel verder kunnen propageren. Voor dit
soort effecten zullen we de intensiteit $\overline{|g^2|}$ moeten
uitrekenen, in plaats van $|G|^2$. Amplitudes en intensiteiten
gedragen zich vaak tamelijk onafhankelijk (zie verdere hoofdstukken).
Dit komt omdat je bij de amplitude eerst middelt over wanorde, waardoor er
veel cancellaties optreden, terwijl je voor de intensiteit eerst 
de absolute waarde kwadrateert, waarna er weinig cancellaties optreden.
}\end{itemize}

\renewcommand{\thesection}{\arabic{section}}
\section{Intermezzo: verstrooiing van fononen\index{fononen}}
\setcounter{equation}{0}\setcounter{figure}{0}

\label{intermezzo}
{\it Naar een lezing van dr. A.L. Burin op 6 mei 1993, vdWZL, Amsterdam.}\\

Dit hoofdstuk bespreekt het verwante probleem van akoestische golven in
een wanordelijk kristal. Het behandelt de stof van ref.\cite{PolBurMaks},
in een soms afwijkende notatie. Met name de Greense functie verschilt met een factor $-m{\rm v}^2$.

\subsection{Inleiding}
We bestuderen een kubisch kristal met daarin een fractie $1-x$ van atomen met
massa $m$ en een kleine fractie $x$ \impurs met een grotere massa $M$.
We nemen aan dat $x \ll 1$ en $ M/m \gg1 $.
Als $\omega_D$ een karakteristieke frekwentie
(Debye frequentie)\index{Debye frequentie} van het pure systeem (met alleen
maar massa's $m$) is, dan geeft de aanwezigheid van \'e\'en enkele \impur
aanleiding tot een mode met frekwentie
\bea
\omega_0 &\approx& \omega_D \sqrt{\frac{m}{M}} \, \ll \, \omega_D
\claus
We willen nu zien wat er gebeurt bij een kleine maar eindige
dichtheid van zware massa's.
\index{$m$, massa}\index{$M$, zware massa}\index{$$x0@$\omega_0$,
resonantiefrequentie} \index{$$xD@$\omega_D$ Debije frequentie}
\index{$x$, fractie zware massa's}
\index{$c_i$, stochast}
\index{$A$, koppelingsmatrix}

\subsection{De Hamiltoniaan van het model}
We beschouwen dit effect nader in 
een \'e\'endimensionaal `kolom' model
van het kristal. We defini\"eren 
de stochasten\index{stochasten} $c_i$ met als waarde
\bea
c_i&\equiv&\left\{
\begin{array}{ll}
0 & \mbox{indien massa $m$ op plaats $i$} 
\\
1 & \mbox{indien massa $M$ op plaats $i$}
\end{array} \right.
\\ \label{verwachtingswaarden}
\langle c_i\rangle &=&x\; ; \; \langle c_i^2\rangle =x \; ; \; \langle c_ic_j\rangle _{j\neq i}=x^2
\claus
Hiermee kunnen we de Hamiltoniaan
opschrijven, opgesplitst in een
kinetisch deel $K$ en een potenti\"eel
deel $U$
\bea
H&=& K+U  \label{mcw3}
\\
K&=&\sum_i \frac{p_i^2}{2m}(1-c_i)+ 
\sum_i \frac{p_i^2}{2M}c_i
\\
U&=&\vrd\sum_{i \neq j} A_{ij}(u_i-u_j)^2=\frac{1}{2} \sum_{i < j} A_{ij} (u_i-u_j)^2
\claus
Quantummechanisch is deze vergelijking
een operatorvergelijking, $u$ en $p$
worden operatoren die aan kanonieke
commutatierelaties moeten voldoen.
We willen de Greense functie $G$ van 
dit systeem berekenen. Om de berekeningen
te vergemakkelijken kiezen we eenheden
waarin $\hbar =1$. De 
kanonieke commutatie relaties 
in dit systeem worden dan
\bea
\label{kanoniek}
[u_i,p_j]&=&\rmi \delta_{ij}
\\ \nolabel
[u_i,u_j]&=&[p_i,p_j]=0
\claus
We noteren  het gemiddelde over alle configuraties van de
\impurs met op elke roosterplaats kans $x$ op een \impur als $\langle ..\rangle _x$.
De quantummechanische verwachtingswaarde in de grondtoestand van een operator
noteren we als $\langle ..\rangle _{T=0}$$=$$\langle T=0|..|T=0\rangle $.
Hierbij is $T$ de temperatuur.
De gemiddelde Greense functie noemen we $G$, die voor middeling $g$.
Verder voeren we de Laplace getransformeerde Greense functie in:
\bea
G_{i0}(\omega)&\equiv&\langle g_{i0}(\omega)\rangle _x
\\ \nolabel
g_{i0}(\omega)&=&-\rmi \int_0^\infty \dint t \eexp{\rmi(\omega+\rmi \delta)t}
g_{i0}(t)
\\ \nolabel
&=&-\rmi \int_0^\infty \dint t \eexp{\rmi(\omega+\rmi \delta)t}
\langle [u_i(t),u_0(0)]\rangle _{T=0}
\claus
We maken nu gebruik van een expansie in evolutieoperatoren om een
fundamenteel `Poisson-haakje' af te leiden:
\bea
\langle \Psi|u_i(t)|\Psi\rangle &=&\langle \Psi|\eexp{\rmi Ht}u_i(0)\eexp{-\rmi Ht}|\Psi\rangle 
\\ \nolabel
\frac{\dpa}{\dpa t}u_i&=&\rmi[H,u_i]
\claus
Met deze relatie en de kanonieke commutatierelaties (\ref{kanoniek}) kunnen
we rekenen aan de tijdsafgeleide van de Greense functie $g$:
\bea
\frac{\dpa}{\dpa t}g_{0i}(t) &=&-\rmi\langle [\frac{\dpa}{\dpa t}u_i(t),u_0(0)]\rangle _{T=0}
\\ \nonumber
&=& \langle [[H,u_i(t)],u_0(0)] \rangle _{T=0}
\\ \nonumber
&=& -\rmi(\frac{1}{m}+c_i(\frac{1}{M}-\frac{1}{m}))\langle [p_i(t),u_0(0)]\rangle _{T=0}
\claus
Dit lossen we met behulp van een Laplacetransformatie
\bea
-\rmi \omega g_{i0}(\omega)&=&
-\rmi(\frac{1}{m}+c_i(\frac{1}{M}-\frac{1}{m}))
\int_0^\infty \dint t \eexp{\rmi \omega t}\langle [p_i(t),u_0(0)]\rangle _{T=0}
\claus
De stokterm $g_{i0}(t=0)$ die optreedt bij de transformatie
van $g'(t)$ is nul voor alle $i$.
Om het rechterlid te kunnen transformeren beschouwen we de afgeleide:
\bea
-\rmi\langle [\frac{\dpa}{\dpa t}p_i(t),u_0(0)]\rangle _{T=0}&=&
\langle [[H,p_i(t)],u_0(0)]\rangle _{T=0}
\\ \nolabel
&=&\langle [-\sum_j A_{ij}(u_i(t)-u_j(t)),u_0(0)]\rangle _{T=0}
\\ \nolabel
&=&-\sum_j A_{ij}(g_{i0}-g_{j0})
\claus
Bij het Laplacetransformeren hiervan vinden we wel een stokterm $\delta_{i0}$.
Enig reorganiseren van termen geeft ons dan
\bea
\label{laplaceg}
-m\omega^2g_{i0}(1-c_i)-c_iM\omega^2g_{i0}+\delta_{i0}&=&
-\sum_j A_{ij} (g_{i0}-g_{j0})
\claus
Deze vergelijking bevat geen operatoren meer.
\BO
Toon aan dat je precies dezelfde vergelijkingen krijgt als je vergelijking
(\ref{mcw3}) opvat voor klassiek oscilatoren. Schrijf hiertoe eerst de
klassieke bewegingsvergelijkingen voor $u_i$ op. \EO

\subsection{De kale
propagator \index{propagator!kale}}
We beschouwen nu het geval $x=0$, \impurs zijn geheel afwezig. De propagator
is nu eenvoudig uit te rekenen door middel van een Fourierreeks:
\bea
\label{ppropagator}
g_\vcp &=& \sum_i \eexp{\rmi \vcp \cdot (\vcr_i-\vcr_0)}g_{i0}
\claus
We vullen deze reeks in, in de bewegingsvergelijking (\ref{laplaceg})
voor $x=0$
\bea
-m\omega^2g^0_{i0}+\delta_{i0}&=&-\sum_j A_{ij} (g^0_{i0}-g^0_{j0})
\\ \nolabel
-m\omega^2g^0_\vcp+1&=&-(A_0-A_\vcp)g^0_\vcp
\\ \nolabel
g_\vcp &=& \frac{1}{m(\omega^2-(A_0-A_\vcp)/m)}
\\ \nolabel
&\equiv& \frac{1}{m(\omega^2-\omega_\vcp^2)}
\claus
Nu maken we een aanname voor $A_{ij}$: we nemen aan dat
het een naaste buur wisselwerking is
\bea
A_{ij}&=&\left\{
   \begin{array}{ll}
   A&\mbox{indien $i,j$ naaste buren} \\
   0&\mbox{in alle andere gevallen}
   \end{array}
 \right.
\\ \nolabel
A_0 &=& \sum_j A_{ij} = 6A \; \; \mbox{(kubisch kristal)}
\\ \nolabel
A_\vcp &=& 2A(\cos p_x a + \cos p_y a + \cos p_z a)
\claus
Voor niet al te hoge frekwenties kunnen we de cosinussen in een Taylor
reeks ontwikkelen ($\cos p_x a $$\,\approx\,$$ 1-(p_x a)^2/2$) en vinden
we een eenvoudige uitdrukking voor de kale Greense functie $g^0$
\index{$g^0$, kale Greense functie}
\bea
A_\vcp&=& 2A(3-\frac{\vcp^2 a^2}{2})
\\ \nolabel
\omega_\vcp &=& A\sqrt{\frac{a}{m}} \; p \equiv {\rm v}p
\\ \nolabel
g^0_\vcp &=& \frac{1}{m(\omega^2-{\rm v}^2 p^2)}
\claus
waarbij v de geluidssnelheid is.
\index{${\rm v}$, geluidssnelheid}

\subsection{De aangeklede propagator\index{propagator!aangeklede} }
Voor het geval $0<x\ll 1$ worden de vergelijkingen een stuk ingewikkelder.
We defini\"eren weer een Fourier reeks voor de propagator, maar nu is
er geen translatie-symmetrie zodat we bij de Fourier reeks moeten
aangeven ten opzichte van welk punt we getransformeerd hebben.
\bea
g(\vcp;0;\omega)&=&\sum_i \eexp{\rmi \vcp \cdot (\vcr_i-\vcr_0)}g_{i0}
\\ \nolabel
g_{i0}&=& \frac{1}{N} \sum_\vcp \eexp{-\rmi \vcp \cdot (\vcr_i-\vcr_0)}
g(\vcp;0;\omega)
\claus
We vullen deze Fourier-expansie in in (\ref{laplaceg}) en vinden
\bea
(\omega^2-\omega_\vcp^2)g(\vcp;0;\omega)&=&
\frac{1}{m}-
\sum_i \eexp{\rmi \vcp \cdot (\vcr_i-\vcr_0)}c_i(\frac{M}{m}-1)\omega^2g_{i0}
\claus
Dit kunnen we ontwikkelen in ordes van de $c_i$'s, zie
(\ref{verwachtingswaarden}) \bea
g^0&=&\frac{1}{m(\omega^2-\omega_\vcp^2)}
\\ \nolabel
g^1&=&-g^0\sum
_i\eexp{\rmi \vcp\cdot(\vcr_i-\vcr_0)}c_i(\frac{M}{m}-1)\omega^2
\frac{1}{N}\sum_{\vc{q}} \eexp{-\rmi\vc{q}\cdot(\vcr_i-\vcr_0)}g^0_\vc{q}
\\  \nolabel
g^2&=&\sum_{ij\vc{q}\vc{s}} \eexp{\rmi(\vcp-\vc{q})\cdot\vcr_i}c_i
\eexp{\rmi(\vc{q}-\vc{s})\cdot \vcr_j}c_j
(\frac{M}{m} -1)^2\omega^4 g^0_\vcp g^0_\vc{q} g^0_\vc{s}
\claus
Dit wordt diagrammatisch weergegeven in  figuur \ref{burinfig1}
\begin{figure}[htbp]
\caption{}
\centerline{\includegraphics[width=10cm]{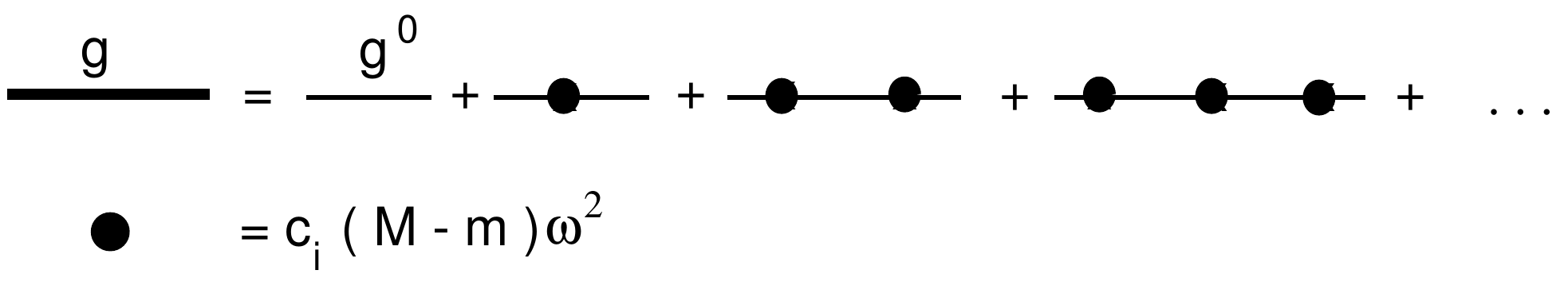}}
\label{burinfig1}
\end{figure}

\subsection{De gemiddelde propagator\index{propagator!gemiddelde}}
Nu nemen we het gemiddelde door gebruik te maken van de relaties
(zie \ref{verwachtingswaarden})
\bea
\langle c_i\rangle &=&x
\\ \nolabel
\langle c_i c_j\rangle &=&\left\{
\begin{array}{ll}
x&i=j      \\
x^2&i\neq j
\end{array}
\right.
\claus
Dit betekent dat de diagrammen waarin twee keer dezelfde \impur voorkomt
een orde lager in $x$ zijn. We geven dat aan met boogjes in figuur
\ref{burinfig2}. In de tweede regel van dit diagram is
de Dyson vergelijking opgeschreven: \begin{figure}[htbp]
\caption{}
\centerline{\includegraphics[width=10cm]{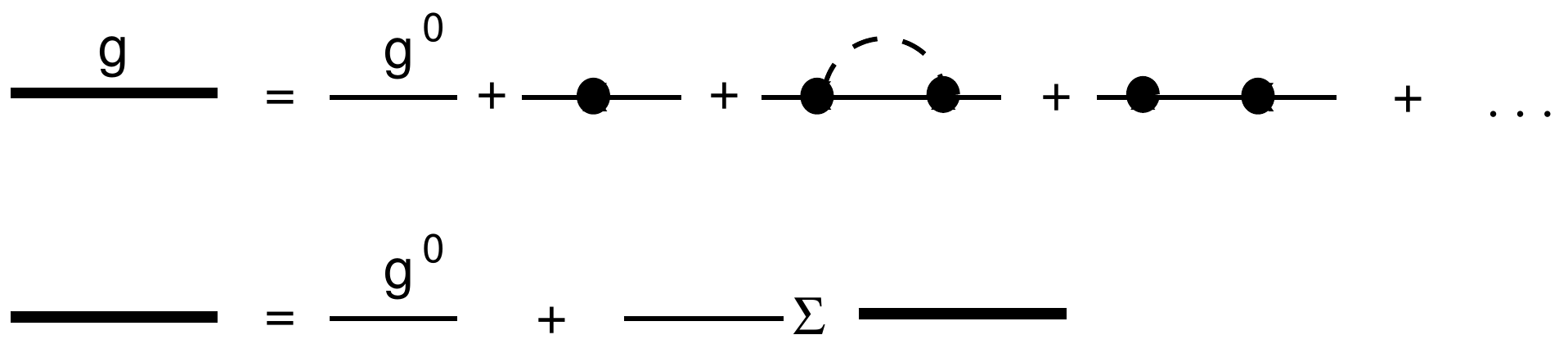}}

\label{burinfig2}
\end{figure}

In geval van een lage dichtheid van \impurs kunnen we de zelfenergie
$\Sigma$ benaderen door de laagste orde in $x$ te nemen, deze diagrammen
vormen de $t$-matrix (zie hoofdstuk \ref{microscopie} en \ref{greense}).
Gebruiken
we in deze diagrammen vervolgens de aangeklede propagator, dan krijgen we de
zelfconsistente $t-$matrix als benadering voor de zelfenergie (figuur
\ref{burinfig3}):
\begin{figure}[htbp]
\caption{De zelfenergie}
\centerline{\includegraphics[width=10cm]{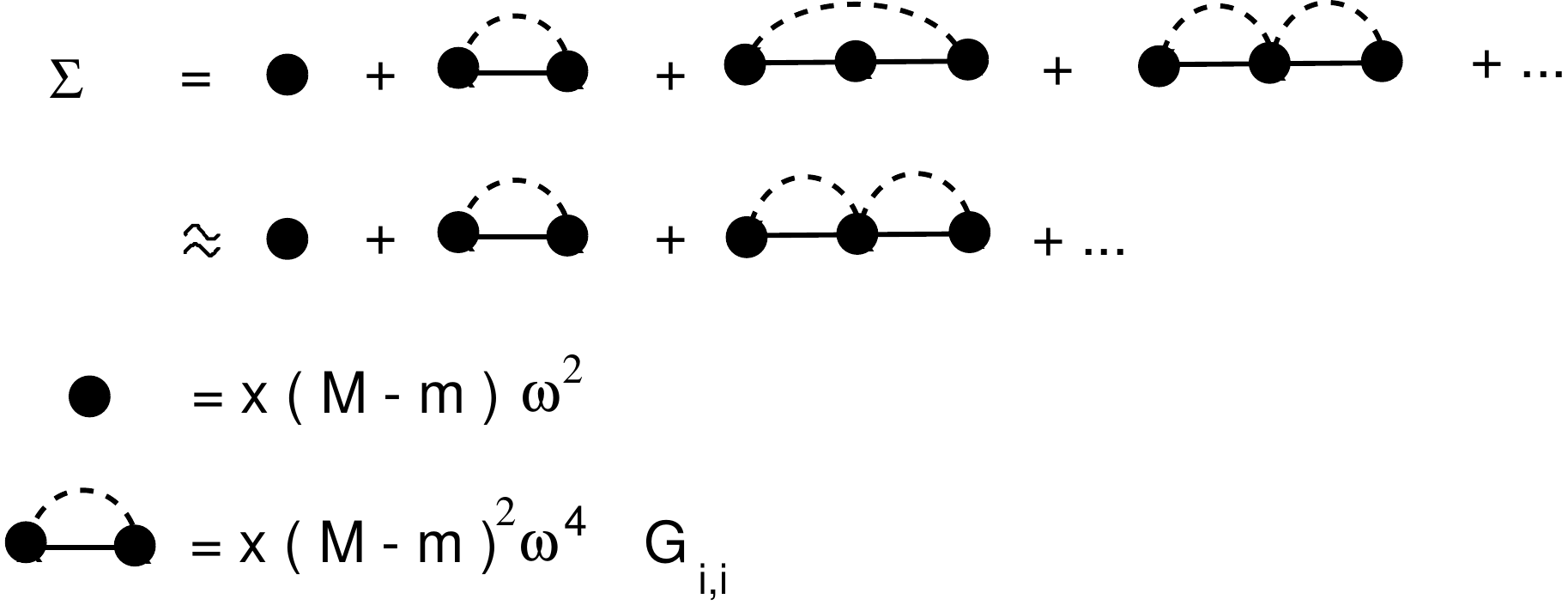}}
\label{burinfig3}
\end{figure}

Met behulp van de Fourier terugtransformatie (inverse Fourier transformatie)
\bea
\label{gterug}
G_{ii}&=&a^3\int \frac{\dint^3 \vcp}{(2\pi)^3} G_\vcp
\claus
wordt de zelfenergie gelijk aan
\bea
\Sigma&=&\frac{x\omega^2(\frac{M}{m}-1)}
{1-\frac{\omega^2 (M-m)}{(2\pi)^3 m}a^3\int \dint^3 \vcp G_\vcp(\omega)}
\\ \nolabel
&=& xt
\claus

Deze vergelijking is zelfconsistent aangezien $\Sigma$ weer nodig is om
$G$ te berekenen, volgens de Dyson vergelijking
\bea
G&=& \frac{1}{\omega^2-{\rm v}^2p^2-\Sigma}
\claus
We gaan nu de $p$-integraal van de Greense functie, die in de zelfenergie
voorkomt, berekenen. We splitsen hem op in zijn re\"ele en imaginaire deel
\bea
\int \dint^3\vcp\, G&=& \int \frac{\dint^3 \vcp}{\omega^2-{\rm v}^2p^2-\Sigma}
\\ \nolabel
&=& \int \frac{4\pi p^2 \dint p (\omega^2-{\rre \Sigma} -{\rm v}^2p^2)}
{(\omega^2-{\rre \Sigma} -{\rm v}^2p^2)^2 + (\rim \Sigma)^2}
\\ \nolabel
&&+ {\rmi \; {\rim \Sigma}} \int \frac{4\pi p^2 \dint p}
{(\omega^2-{\rre \Sigma} -{\rm v}^2p^2)^2 + (\rim \Sigma)^2}
\claus
Het re\"ele deel van deze integraal is linear divergent: teller en noemer zijn
van orde 4 in $p$. Kennelijk is de continu\"um-benadering die we gebruikt
hebben in de terugtransformatie  (\ref{gterug}), met de \impur met 
afmetingen 0, niet goed. We kappen daarom
de impuls-integraal voor het re\"ele deel af bij $\xi/a$ waarbij
$a$ de roosterconstante en $\xi$ een parameter van orde 1.
\index{$$nxi@$\xi$, constante}
Dit geeft ingevuld in de uitdrukking voor $\Sigma$
\bea
\Sigma&=&\frac{
               \frac{x}{m}(M-m)\omega^2}{
               1-\omega^2\frac{(M-m)\xi}{m\omega_D^2}
               -\rmi\omega^2(\frac{M}{m}-1)
               \frac{a^3 \rre \sqrt{\omega^2-\Sigma}}{4\pi v^3}}
\\ \nolabel
\omega_D &\approx& {\rm v}/a
\claus
We kunnen de zelfenergie nu schrijven op een manier die sterk doet
denken aan die bij verstrooiing van licht:
\bea
\Sigma&\sim&\frac{1}{\omega_0^2 -\omega^2 -2 \omega_0\rmi \Gamma}
\claus
Het verschil met lichtverstrooiing is dat in het geval van licht
$\Gamma$ (bij benadering) constant is in $\omega$, terwijl bij
fononen $\Gamma \approx \rim \Sigma/\omega$.

\subsection{De pseudo-bandkloof\index{toestandsdichtheid!pseudo-bandkloof}
en Anderson localisatie\index{Anderson localisatie}}
In het regime van sterke verstrooiing, $x>(m/M)^{3/2}$ ontstaat er een
bandkloof: een gedeelte van het spectrum waarin de toestandsdichtheid nul
is. (In feite is dat een gevolg van de gemaakte benadering, in een betere
benadering blijkt er een kleine, eindige toestandsdichtheid te zijn).
In de `pseudo-bandkloof' geldt:
\bea
\rre \Sigma &\gg& \omega^2
 \nolabel   \\
\rim \Sigma   &\approx& 0
\claus
Van Tiggelen, Lagendijk en Tip \cite{tiggelen2} hebben een criterium
geformuleerd voor Anderson localisatie :
\bea
(\omega^2- \rre \Sigma) &<& 0.7\, |\rim \Sigma|
\claus
In de bandkloof is aan dit criterium voldaan en dus zijn de toestanden
gelocaliseerd. Een verstoring kan zich dan niet door het kristal voortplanten.
Dit is wat er overblijft van een resonantie onder invloed 
van veelvoudige verstrooiing.

\renewcommand{\thesection}{\arabic{section}}
\section{ Isotrope verstrooiing: transport in oneindige media }
\setcounter{equation}{0}\setcounter{figure}{0}

\label{transport}\index{isotrope verstrooiing}
Zoals de amplitude Greense functie uit de Dyson vergelijking
\index{Dyson vergelijking} opgelost moet worden, zo volgt de intensiteit uit
de zog. Bethe-Salpeter vergelijking\index{Bethe-Salpeter vergelijking}. In
dit hoofdstuk zullen we ons beperken tot een vereenvoudigde vorm van deze
vergelijking. Dit zijn dit ladderdiagrammen\index{diagrammen!ladder}, die
leiden tot een transportvergelijking\index{transportvergelijking} equivalent
aan de Boltzmann vergelijking\index{Boltzmann vergelijking} uit de 
mesoscopische benadering, zie hoofdstuk \ref{mesoscopie}.

Transport in een systeem waar je niet op de randen hoeft te letten is
relatief eenvoudig. In de praktijk is dit van toepassing zo gauw je enkele
vrije weglengtes van de wand verwijderd bent. We beschouwen daarom eerst deze
situatie. We leiden de diffusievergelijking af en vinden de diffusieconstante
en de transportsnelheid.

\subsection{De ladderbenadering in de Bethe-Salpeter vergelijking}

Transport is een intensiteitseigenschap. We moeten dus de amplitude Greense
functie kwadrateren en dan pas middelen over de wanorde. De diagrammatische
ontwikkeling
moeten we vermenigvuldigen met z'n complex geconjugeerde, ofwel de geavanceerde 
Greense functie. Dit geven we aan door
dezelfde diagrammen er onder te schrijven, zie figuur \ref{diapeter1}.
\begin{figure}[htbp]
\caption{Bovenste regel: Greense functie. Onder: enkele
bijdragen tot intensiteitstransport, voor middeling.}
\vspace{3mm}
\centerline{\includegraphics[width=13cm]{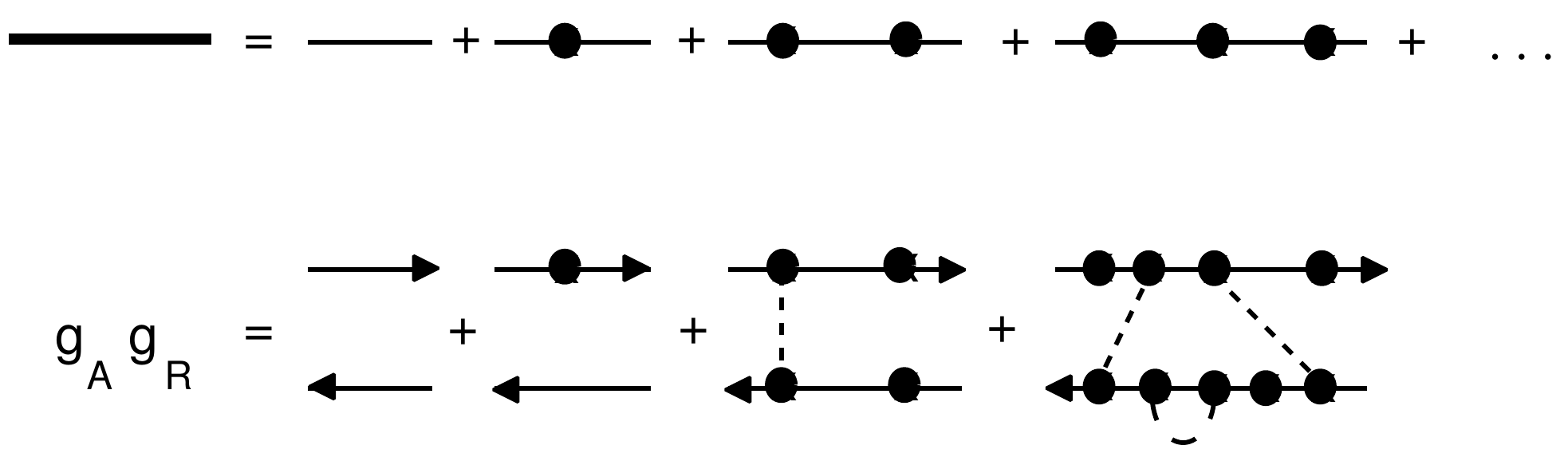}}
\label{diapeter1}
\end{figure}

Boven en onder kunnen dezelfde verstrooiers voorkomen, dit is aangegeven met
stippellijntjes. In het bijzonder zijn die diagrammen van belang waarbij de
propagatoren de gemeenschappelijke verstrooiers in dezelfde volgorde
tegenkomen. Dit is een nieuwe bijdrage die nog niet meegenomen is in de
amplitude Greense functie\index{Greense functie!amplitude} en die je ook niet
krijgt wanneer je de gemiddelde Greense functie kwadrateert. Het kwadraat van
de gemiddelde Greense functie valt exponentieel af met de afstand, terwijl de
gemiddelde intensiteit omgekeerd evenredig met de afstand gaat. 
Zo'n diagram
waarbij $G_0$\index{$G_0$, kale Greense functie} en ${\overline G}_0
\equiv G_0^*$ dezelfde verstrooier
bezoeken stelt verstrooiing van de
intensiteit voor. \BO Teken het resulterende diagram als je eerst middelt en
dan kwadrateert. \EO

Tussen de twee stippellijntjes komen, als gevolg van middelling over de
wanorde, alle mogelijke configuraties van $x, x-x, x-x-x.$ etc.  voor.
Dit zijn de zelfenergie bijdragen uit
hoofdstuk~\ref{greense}. Door de kale amplitude Greense functie $G_0$ te
vervangen door de Greense functie van het gemiddelde medium $G$ brengen we al
deze verschillende bijdragen in \'e\'en keer in rekening. De aangeklede Greense
functie wordt weer als een dikke streep genoteerd. Het resultaat staat
in figuur \ref{diapeter2}. We defini\"eren $\Phi=\langle g_A g_R \rangle $,
dat is $ \sum_{n=0}^\infty [n$ {\it keer verstrooide intensiteit in uitgaande
richting}].\index{$$UXF@$\Phi$, uitgaande intensiteit} Dit is de som van
ladderdiagrammen (`laddersom', `diffuson', `particle-hole channel').
Men kan de uitgaande lijnen weglaten door te defini\"eren:
\BE \Phi=G_R G_A +G_R G_A \L=G_R G_A +G_R G_A \frac{4\pi}{\ell_{sc}}I.
\EE De functie $\L$\index{$L@$\L$, ladderpropagator} noemen we de {\it
ladderpropagator},
zie figuur \ref{diapeter2}. Zij is evenredig met de {\it diffuse
intensiteit}
$I$: \BEQ \L(\r)=\frac{4\pi}{\ell_{sc}}I(\r)=n\overline{t} t I(\r).\EEQ
Het diagram uit figuur \ref{diapeter2} kunnen we ook uitschrijven als:
\BEA\label{laddereqn}
I(\r)&= |\Psi_{in} (\r)|^2& + n\bar{t}t \int d^3 \r' G(\r-\r') G^*(\r-\r')
I(\r')\nonumber\\
\L(\r)&=n\bar{t}t|\Psi_{in} (\r)|^2& + 
n\bar{t}t \int d^3 \r' G(\r-\r') G^*(\r-\r')\L(\r')
\label{mcw4} \EEA

\begin{figure}[htbp]
\caption{De ladderdiagrammen. Bovenste regel: in de vergelijking voor
$\Phi$  kleden we de Greense functies aan. Onderste regel: uitdrukking voor
de ladderpropagator $\L$. De eerste term beschrijft eerste orde verstrooiing
van de inkomende bundel;
hogere orde verstrooiingen\index{hogere orde verstrooing} worden door de
tweede term gegenereerd.}
\vspace{3mm}
\centerline{\includegraphics[width=10cm]{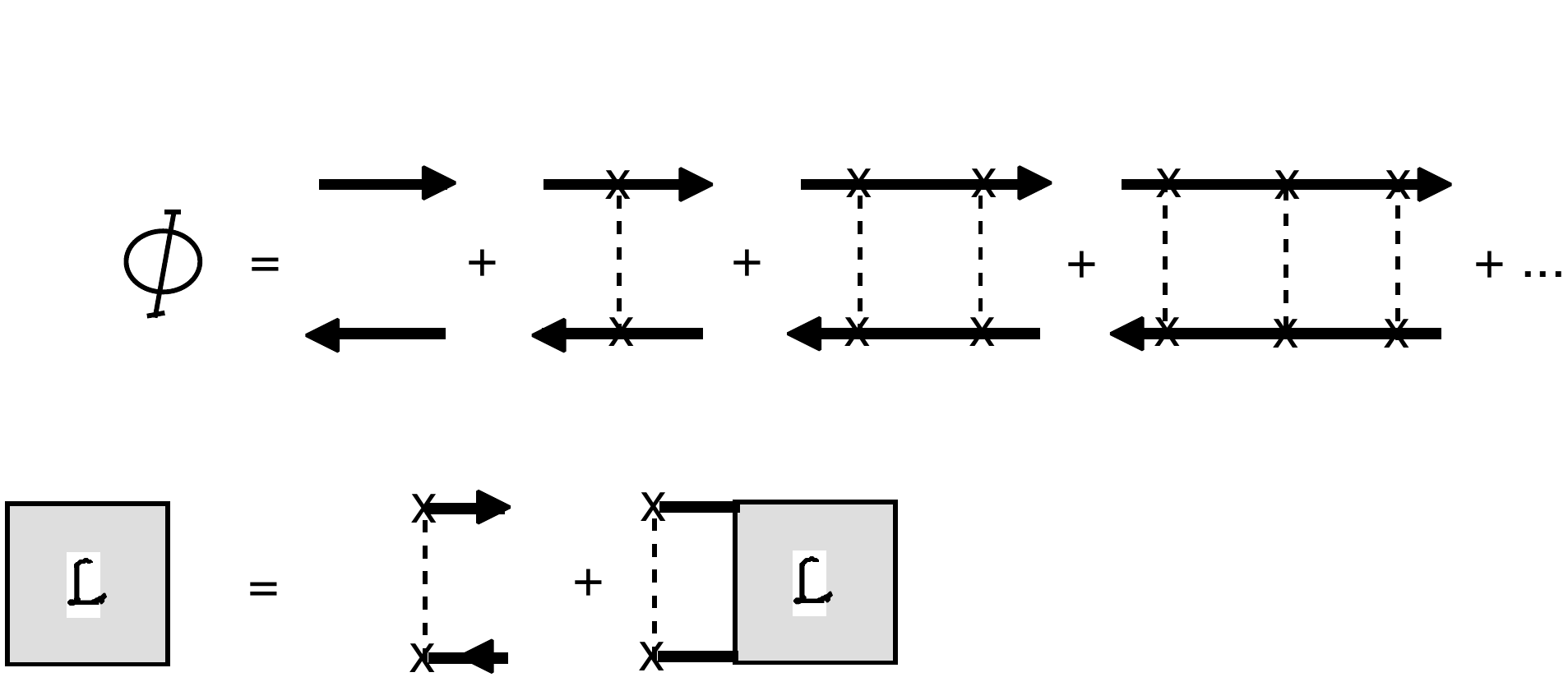}}
\label{diapeter2}
\end{figure}


\subsection{Diffusie uit de stationaire laddervergelijking}
De laddervergelijking gaat in de bulk over in
\beq
\L(\r)= n\bar{t}t\int\mid G(\r-\r')\mid^2 \L(\r')d^3\r'. \label{blad}
\eeq
Neem nu aan dat $\L(\r)$ in de bulk een langzaam vari\"{e}rende functie
is. We kunnen dan $\L(\r')$ ontwikkelen rond $\r$:
\BE
\L(\r')=\L(\r+(\r'-\r))=\L(\r)+(\r-\r')\cdot\nabla \L(\r)+
\frac{1}{2}(\r-\r')(\r-\r'):\nabla \nabla \L(\r). \label{mvr}
\EE
Invullen in (\ref{blad}) geeft drie bijdragen. De makkelijkste uit te rekenen
bijdrage is de tweede term waarbij de integrand lineair is in
$(\r-\r')$. Deze term is nul wegens symmetrie.
 Voor de eerste bijdrage volgt:
\BEA
n\bar{t}t\L(\r)\int d^3\r' \frac{\eexp{-r'/\ell}}{(4\pi r')^2}&=& n\bar{t}t
\L(\r)4\pi\int_0^\infty \frac{r'^2 dr' \eexp{-r'/\ell}}{16\pi^2 r'^2} 
= n\bar{t}t\frac{\ell}{4\pi}\L(\r)\nonumber\\&=&
\frac{\bar{t}t/4\pi}{{\rm  Im}(t)/k}\L(\r) \equiv a\L(\r)
\EEA
In de voorfactor herkennen we het albedo $a$  gedefinieerd
als $a\equiv\sigma_{sc}/\sigma_{ext}$, zie hoofdstuk \ref{microscopie}.
 Uitwerken van de contractie in de derde term geeft: 
\BA
&&\int d^3\r' G^*(\r-\r') G(\r-\r') (\r-\r')(\r-\r'):\nabla\nabla \L(\r)
\nonumber \\ &&= \int d^3\r'
 G^*(\r') G(\r')r_i' r_j' \partial_i\partial_j  \L(\r) \nonumber \\
&&= \frac{1}{3} \int d^3\r' G^*(\r') G(\r')
r'^2 \nabla^2 \L(\r) \EA
We vinden voor de derde term in (\ref{mvr}) derhalve:
\beq
 \frac{1}{3} nt\bar{t}\int 4\pi r'^2 dr' \frac{r'^2\eexp{-r'/\ell}}{(4\pi
r')^2} \nabla^2 \L(\r)
   = nt\bar{t}\frac{1}{2}\,\times\frac{1}{3}\times 2\ell^2\nabla^2 \L(\r).
\eeq
Invullen levert de stationaire
diffusievergelijking\index{diffusievergelijking!stationaire} \beqa
\L(\r) & = & a\L(\r)+\frac{1}{3}\ell^2\nabla^2 \L(\r)\nonumber\\
\Longrightarrow \nabla^2 \L(\r) & = & \frac{3(1-a)} {\ell^2}
\L(\r)\equiv \frac{1}{L_{abs}^2}\L( r) \label{difff} \eeqa 
waarbij de absorptielengte\index{absorptielengte} gegeven is door
\BE L_{abs} \equiv \frac{\ell}{\sqrt{3(1-a)}}, \EE
in overeenstemming met hoofdstuk \ref{mesoscopie}.
We zien dat we voor $a=1$ pure diffusie krijgen. 
Een oplossing is dan $\L(\r)= A + Bz$. Voor $a<1$
is er absorptie\index{albedo}, een oplossing is dan
$\L(\r)=A\exp(-z/L_{abs})+B\exp(z/L_{abs})$.
\subsubsection{De Fourier-getransformeerde bulkkern}\index{bulkkern}

In een stationaire  situatie met isotrope verstrooiing
kan de kern uit vgl. (\ref{laddereqn}) in de bulk geschreven worden als
\BEQ M(\r,\r';\Omega=0)=
n{\bar t}t\vert G(\r-\r')\vert^2=\frac{\eexp{-|\r-\r'|/\ell}}
{4\pi\ell |\r-\r'|^2}\EEQ
De Fouriergetransformeerde kan exact bepaald worden.
\BO Toon aan dat\index{$M@$M$, Milne-kern}
\BEQ M(\q,\Omega=0)=a\frac{\arctan{q\ell}}{q\ell}\EEQ \EO
Voor kleine $\q$ en $(1-a)$ geeft dit
\BEQ M(\q,\Omega=0)=a-\frac{1}{3}q^2\ell^2 \EEQ
Dit houdt in dat in de plaatsruimte geldt
\BEQ 1-M=1-a-\frac{\ell^2}{3}\nabla^2,\EEQ
hetgeen wederom tot vergelijking (\ref{difff}) leidt.

\subsection{De diffusieconstante en de
transportsnelheid}\index{diffusievergelijking!niet stationair}
Hier geven we een eenvoudige doch leerzame\index{eenvoudig}
afleiding van de niet-stationaire
diffusievergelijking. Dit laat direct zien dat de transportsnelheid
voor licht verkleind wordt door resonantie effecten. 
We zullen een bijdrage tegenkomen die in het geval van
electronen met puntverstrooiers afwezig is.

De interessante grootheid is de intensiteit\index{intensiteit!definitie}
\BEQ I(\r,t)=\Psi^*(\r,t)\Psi(\r,t) \EEQ
Zijn Fourier-Laplace-getransformeerde is
\BEA I(\q,\Omega)&=&\int d^3\r\, \eexp{-i\q\cdot\r}
\int_0^\infty dt\,\eexp{i\Omega t}\Psi^*(\r,t)\Psi(\r,t) \nonumber \\ 
&=& \int      d^3\r             \eexp{-i\q\cdot\r} 
    \int\frac{d^3\p_1}{(2\pi)^3}\eexp{-i\p_1\cdot\r}
    \int\frac{d^3\p_2}{(2\pi)^3}\eexp{ i\p_2\cdot\r} 
 \int_0^\infty dt_1 \int_0^\infty dt_2\,\nonumber\\
&& \int_{-\infty}^\infty\frac{d\omega}{2\pi}
\eexp{i\omega(t_2-t_1)} \Psi^*(\p_1,t_1)\Psi(\p_2,t_2) 
\eexp{\frac{1}{2}i\Omega(t_1+t_2)}
\nonumber\\
&=&\int \frac{d^3\p}{(2\pi)^3} \int_{-\infty}^\infty\frac{d\omega}{2\pi}
\Psi^*(\p-\frac{1}{2}\q,
\omega-\frac{1}{2}\Omega)\Psi(\p+\frac{1}{2}\q,\omega+\frac{1}{2}\Omega)\EEA
Deze grootheid is gerelateerd aan 
\BEQ \Phi_{\p}(\q,\Omega)=\langle g^R(\p_+,\omega_+)g^A(\p_-,\omega_-)
\rangle=\langle g(\p_+,\omega_+)g^*(\p_-,\omega_-)\rangle \EEQ
We hebben ingevoerd
\BEQ \p_\pm=\p\pm\frac{1}{2}\q;\qquad \omega_\pm=\omega\pm\frac{1}{2}\Omega.
 \EEQ
De parameters $\p$ en $\omega$ zijn `interne' of `snelle' variabelen, die
te maken hebben met slechts \'e\'en periode van de golf.
 $\q$ en $\Omega$ zijn macroscopische (langzaam vari\"erende)
parameters, golfgetal en frequentie. 
Zij beschrijven variaties over afstanden van
vele golflengtes en over tijden veel langer dan de oscillatieperiode.

We beschouwen isotrope verstrooiers, met de $t$-matrix uit hoofdstuk
\ref{microscopie}. In een oneindig systeem is het handig de
Fourier-Laplace getransformeerde
laddervergelijking\index{laddervergelijking} te beschouwen. Hij heeft de vorm
\BEA 
\label{LqOm}
\L(\q,\Omega)&=&S(\q,\Omega)+n\bar t(\omega_-)t(\omega_+)
\int\frac{d^3\p}{(2\pi)^3}
G(p_+,\omega_+)G^*(p_-,\omega_-)\L(\q,\Omega)\nonumber\\
&\equiv&S(\q,\Omega)+M(\q,\Omega)\L(\q,\Omega)\nonumber\\
&=&\frac{S(\q,\Omega)}{1-M(\q,\Omega)} \label{mcw15} \EEA
waarbij $S$ de bronterm is; zijn precieze vorm is nu niet van belang.
Zoals we in hoofdstuk \ref{macroscopie} gezien hebben, willen we het
gedrag weten voor kleine $\q$ en $\Omega$.
We ontwikkelen daarom tot op orde $\Omega$ en $q^2$. Er geldt,
met $G\equiv G(\p,\omega)=1/(p^2-\omega^2c^{-2}-nt)$ uit (\ref{Gpknt}),
\BEA 
\label{Gexpand}
G(\p+\frac{1}{2}\q,\omega+\frac{1}{2}\Omega)&\approx&
\left(p^2+\p\cdot\q+\frac{1}{4}q^2-
\frac{\omega^2}{c^2}-\Omega\frac{\omega}{c^2}-nt-\Omega\frac{n}{2}\,
\frac{dt}{d\omega}\right)^{-1} 
 \\
&\approx& G +\left[ -(\p\cdot\q)-\frac{1}{4}q^2+
\Omega(\frac{\omega}{c^2}+\frac{n}{2}\,\frac{dt}{d\omega} ) \right] G^2
+(\p\cdot\q)^2G^3 \nonumber \EEA
Dit impliceert
\BEA M(\q,\Omega)&=&n{\bar t}t\left\{1+\frac{1}{2}\Omega(\frac{d\log t}
{d\omega}-\frac{d\log\bar t}{d\omega}) \right\}\times  \\
&& \left\{ I_{11}
+\Omega(\frac{\omega}{c^2}+\frac{nt'     }{2})I_{21}
-\Omega(\frac{\omega}{c^2}+\frac{n\bar t'}{2})I_{12} \right.  \nonumber \\
&&
\left.
-\frac{q^2}{4}(I_{12}+I_{21})+\frac{k_0^2q^2}{3}(I_{31}+I_{13}-I_{22})
\right\}\nonumber\EEA
Hierbij hebben we ingevoerd de integralen
\BEQ I_{kl}=\int\frac{d^3\p}{(2\pi)^3} G^k(\p)G^{*\,l}(\p)\EEQ
Deze worden in de appendix uitgewerkt.
Invullen van hun waarden levert
\BEQ M(\q,\Omega)= a+i\Omega\tau-\frac{1}{3}q^2\ell^2\EEQ
Met $a$ het albedo, $\ell$ de vrije weglengte. $\tau$ heeft precies de
decompositie besproken in hoofdstuk~\ref{inleiding}:
\BEQ \tau=\frac{\ell}{c}+\tau_{dw}\equiv\tau_{sc}+\tau_{dw}.\EEQ
We zien hier als eerste de `reistijd' $\tau_{sc}=\ell/c$.
De uitdrukking voor de verblijftijd\index{verblijftijd} (dweiltijd) $\tau_{dw}$ volgt nu
expliciet \BEQ \tau_{dw}= {\rm Im}\frac{d\log t}{d\omega}
+\frac{2\pi}{k_0\bar tt}{\rm Re}\frac{dt}{d\omega}\EEQ
In hoofdstuk \ref{microscopie} hebben we besproken dat een puntverstrooier
de $t$-matrix heeft
\BEQ t=\left( \frac{1}{U_{\rm eff}}-\frac{ik_0}{4\pi} \right)^{-1} \EEQ
Dit levert
\BEA \tau_{dw}&=&{\bar tt} \left( \frac{1}{4\pi cU_{\rm eff}}
+\frac{k_0U_{\rm eff}'} {4\pi U_{\rm eff}^2}\right)
+\frac{2\pi({\bar tt})^2}{k_0{\bar tt}}
\left(\frac{ U_{\rm eff}'} {U_{\rm eff}^4}-\frac{ k_0^2U_{\rm eff}'}
{16\pi^2 U_{\rm eff}^2}
-\frac{k_0}{8\pi^2 cU_{\rm eff}}\right) \nonumber\\
&=&\frac{k_0\bar tt}{8\pi U_{\rm eff}^2} \left( 1+\left[\frac{4\pi}{k_0U_{\rm eff}}
\right]^2 \right) \frac{dU_{\rm eff}}{d\omega}\EEA
We zien dat de termen zonder $U_{\rm eff}'$ precies tegen elkaar wegvallen.
Voor electronen, waar $U_{\rm eff}$ een constante is, is $\tau_{dw}$
dus gelijk
aan nul: er zijn helemaal geen opladingseffecten door de `stijfheid' van de
verstrooiers.
~\footnote{ Deze cancellatie volgt meer algemeen uit een Ward-identiteit,
zie ref. \cite{vATLT2}.} 

Voor akoestische golven en lichtgolven is de situatie interessanter:
omdat $U_{\rm eff}$ expliciet van de frequentie afhangt, is $U_{\rm eff}'$
niet nul en deze termen vallen niet weg tegen elkaar. Er volgt een eindige
opladingstijd $\tau_{dw}$.
\BO Toon aan dat de expliciete vorm 
$U_{\rm eff}=\alfa k^2/(1-k^2/k_*^2)$, levert dat op
resonantie\index{resonantie} geldt: \BEQ \tau_{dw}=\frac{2\pi}{\omega}\,\cdot
\frac{1}{\alfa k^3}=\frac{{\rm periode}}
{\rm koppeling}\qquad{\rm op\quad resonantie}
\EEQ \EO
Deze `oplaadtijd' wordt langer naarmate de koppeling met de omgeving, dat
wil zeggen de genormeerde verstrooiingsterkte $k^3\alfa=4\pi k^3a^3
(m^2-1)/3$, zwakker is! Dientengevolge is dit een experimenteel belangrijk
effect. Het leidt er toe dat de transportsnelheid
\BEQ \label{vspeed}
v=c\frac{\tau_{sc}}{\tau_{sc}+\tau_{dw}} \EEQ
aanmerkelijk kleiner kan zijn dan de lichtsnelheid voor realistische
waarden van de dichtheid $n$.

Dit is de meest direkte afleiding van het resultaat besproken in hoofdstuk
\ref{inleiding}.\index{levenswijsheid} Het enige wat we gedaan hebben is
alle correcties van orde $\Omega$ meenemen
\footnote{Onder het motto: `Je moet doen wat je doen moet'.}. 
Voor de algemene situatie en voor meer details, zie \cite{vATLT2}.
Hier merken we nog op dat door dit effect de diffusieconstante $D=v\ell/3$
sterk gereduceerd wordt, zoals experimenteel ook waargenomen is.

Met dit resultaat vinden we voor de ladderpropagator (\ref{mcw15})
\BA \L(\q,\Omega)&=&\frac{S( {\bf 0},0 )}{\tau}\frac{1}{Dq^2+D\kappa^2
-i\Omega} \nonumber \\
&=& \frac{3S({\bf 0},0)}{\ell^2} \frac{1}{q^2+\kappa^2-i\tOmega}.
\EA
met de gereduceerde externe frequentie 
\BE\label{tOmega}
 \tOmega\equiv \frac{\Omega}{D}
 =\frac{3\Omega}{c\ell}(1+\frac{\tau_{dw}}{\tau_{sc}})
 .\EE
\index{$$Xt@$\tOmega$, gereduceerde externe frequentie}
\index{externe frequentie} Dit is precies de propagator uit
hoofdstuk
\ref{macroscopie} met absorptielengte\index{absorptielengte}
$L_{abs}\equiv 1/\kappa\equiv\ell/\sqrt{3(1-a)}$.

Vaak beschouwt  men de oplossing van de Schwarzschild-Milne vergelijking
die als bron heeft $n\bar t t\delta(\r)=4\pi\ell^{-1}\delta(\r)$.
In de diffusiebenadering wordt dit
\BE\label{difladder} \L(\q,\Omega)=\frac{12
\pi}{\ell^3} \frac{1}{q^2+\kappa^2-i\tOmega} \label{mcw16} \EE

\renewcommand{\thesection}{\arabic{section}}
\section{Transport in een halfoneindig medium\index{halfoneindig medium}}
\setcounter{equation}{0}\setcounter{figure}{0}

In dit hoofdstuk kijken we heel gedetailleerd naar de transport
vergelijking voor een realistische situatie: een heel dikke plak.
We volgen hier het artikel `Skin layer of diffusive media' [NL].
We bekijken ook het geval dat er een verschil in brekingsindices is, en het
geval dat er een lucht-glas-medium rand is. Er wordt aangetoond dat er 
behoud van flux geldt.

\subsection{Vlakke golf invallend op een halfoneindig medium}
We bepalen de vorm van de inkomende golf in een verstrooiend medium.
Daartoe vervangen de werking van de verstrooiers weer door
de zelfenergie. De gemiddelde golfvergelijkingen voor een vlakke
golf invallend op een
halfoneindig wanordelijk medium zijn, (wanorde voor $z>0$; geen verschil in
brekingsindex)
\beqa
-\nabla^2\Psi-(k_0^2+nt)\Psi=0, \mbox{\ \ } z>0\nonumber\\
-\nabla^2\Psi-k_0^2\Psi=0, \mbox{\ \ } z<0 \label{golfslab}
\eeqa
Fourier transformatie van de transversale vector
$\kt=(k_x,k_y)$\index{$kl@$\kt$, loodrechte component inkomende golfvector}
geeft voor (\ref{golfslab}) \beqa
\frac{d^2}{dz^2}\Psi(z)+ P^2 \Psi(z) & =& 0, \mbox{\ \ } z>0 \\
\frac{d^2}{dz^2}\Psi(z)+ p^2 \Psi(z) & =& 0, \mbox{\ \ } z<0 \\
 P^2= k_0^2 - \kt^2 +nt & \mbox{en} & p^2=k_0^2 - \kt^2 \label{Pp}
\index{$p$, $z$-component inkomende golfvector}
\index{$P$, $z$-component golfvector in medium}\eeqa
Indien de reflectieterm verwaarloosd kan worden ($n$ klein), is de oplossing
\beqa
\Psi & =  \exp (iPz+i\kt\cdot\rhoo  ), &\qquad z>0\label{ain}\\
\Psi & =  \exp (ipz+i\kt\cdot\rhoo ), &\qquad z<0
\eeqa
 In laagste orde in de dichtheid
vinden we voor het re\"{e}le en het imaginaire deel van $P\equiv P_1+iP_2$
\BE
P_1=k\cos \theta, \qquad  P_2=\frac{1}{2\ell\cos \theta}
\EE
wanneer de inkomende golf een hoek $\theta$ met de z-as maakt.
De vrije weglengte is gedefinieerd door
$\ell^{-1}\equiv\frac{n}{2k} \mbox{Im}(t)$.

Het kwadraat van $\Psi_{in}$ is exponentieel gedempt,
\beq
\mid\Psi_{in}\mid^2=\exp(-2P_2z)=\exp(-\frac{z}{\ell\cos\theta})
\eeq
dus al op enige vrije weglengtes van de rand is deze bijdrage aan de
intensiteit
volledig te verwaarlozen. Dit is wederom de wet van
Lambert-Beer\index{Lambert-Beer wet}, nu `echt' afgeleid.

\subsection{Berekening van de hoekafhankelijkheid van de reflectie}
\subsubsection{Zonder brekingsindexcontrast}
\label{zobr}
In de berekening voor de hoekafhankelijkheid van de reflectie zullen we in eerste instantie
het brekingsindex contrast niet meenemen.  Stel de oplossing van vergelijking
(\ref{mcw4}) is $I(z,\rhoo)$, dan kunnen we de gereflecteerde
intensiteit\index{intensiteit!gereflecteerde} in het punt $(z',\rhoo')$
schrijven als: \beq
I_R(z',\rhoo')=\int dz d^2\rhoo \: \frac{4\pi}{\ell} I(z,\rhoo)\mid
G(z,\rhoo;z',\rhoo')\mid^2 \label{Ir}
\eeq
De defini\"erende vergelijking voor
de gemiddelde Greense functie in het medium is:
\beq
\{\frac{d^2}{dz^2}-\ql^2+(k_0^2+\frac{i}{2\ell})^2\}G(z,z',\ql)=-\delta(z-z');
\eeq
De oplossing voor $z,z'>0$ is:
\beq
G(z,z';\ql)=\frac{\exp(iP\mid z-z'\mid)}{-2iP}.
\eeq
Volgens (\ref{Pp}) kunnen we voor lage dichtheden ($n$ klein) in de noemer $P$ door $p$
vervangen. Verder geldt dat buiten de plak $P\rightarrow p$, dus pikken we
hier alleen een fase-factor, $\exp(-ipz')$, op. De Greense functie voor
$z'<0$, $z>0$ wordt dus (we hebben nu een opgelegde impuls $\kl$):
\beq
G(z,z',\kl)= \frac{i\exp(iPz-ipz')}{2p} \Longrightarrow
\mid G(z,z',\kl)\mid^2 =\frac{\exp(-z/\ell\cos\theta_b)}{4\mid p\mid^2}.
\eeq
Het is handig om eerst een eindig oppervlak $A$ te bekijken en dit periodiek
voort te zetten. We kunnen dan de $\rhoo$-afhankelijkheid uitintegreren;
later
nemen we de limiet $A\rightarrow\infty$. Gebruikmakend van de periodiciteit
van het oppervlak, is de Greense functie te schrijven als:\index{$A$, oppervlak}
\beq
G(z,\rhoo;z',\rhoo')=\frac{1}{A}\sum_q G(z;z';q)\eexp{i\ql\cdot(\rhoo-\rhoo')}
\label{mcw10}
\eeq
Omdat we een oneindige vlakke inkomende golf hebben, hangt de diffuse
intensiteit $I$ niet van $\rhoo$ af.
 De integratie over $\rhoo$ wordt eenvoudig:
\BA
\int d^2\rhoo \mid G(z,\rhoo;z',\rhoo')\mid^2&=&\int d^2\rhoo\frac{1}{A^2}
\sum_{\qt,\qt'}G(z,z',\qt)G^*(z,z',\qt')
\times \nonumber \\ &&\eexp{i\qt\cdot(\rhoo-\rhoo')-i\qt'\cdot (\rhoo-\rhoo')}.
\nonumber \\ &=&\frac{1}{A^2}\sum_{\qt,\qt'}GG^*A\delta_{\qt,\qt'}
=\frac{1}{A}\sum_\qt GG^*\nonumber\\
 &\approx& \int \frac{d^2\qt}{4\pi^2} \mid
G(z,z',\qt)\mid^2 \EA
De laatste stap geldt in de limiet $A\rightarrow\infty$. Als we de
uitgaande $\qt$ met ${\bf q}_\perp^b$ aangeven kunnen we $d^2\qt$
uitwerken tot:
\BE
d^2{\bf q}_\perp^b=q_\perp^b dq_\perp^b d\phi_b
=k_0^2\sin\theta_b\cos\theta_bd\theta_bd\phi_b=
k_0^2\cos\theta_bd\Omega_b
\index{$dO@$d\Omega$, infinitesimale ruimtehoek}\EE
Invullen  in (\ref{Ir}) geeft een resultaat dat niet van het observatiepunt
afhangt
\beq
I_R(z',\rhoo')=\int dz\int \frac{k_0^2\cos\theta_b}{4\pi^2}d\Omega_b
\frac{1}{4\mid p\mid^2}\L(z)\eexp{-z/\ell\cos\theta_b}
\eeq
Tenslotte vinden we voor de hoekafhankelijkheid van de gereflecteerde diffuse
intensiteit\index{intensiteit!hoekafhankelijkheid diffuse} (zie ook figuur
\ref{jmlar}): \BA A_R\equiv \frac{dR(a\rightarrow
b)}{d \Omega_b}=\frac{d I_R}{d\Omega_b}&=& \frac{k_0^2}{4\pi^2}
\cos\theta_b \frac{1}{4k_0^2\cos^2\theta_b}\int dz
\L(z)\eexp{-z/\ell\cos\theta_b}\nonumber \\
&=& \frac{1}{16 \pi^2 \cos \theta_b} \int_0^\infty dz
 \L(z)\eexp{-z/\ell\cos\theta_b}
\EA

\subsubsection{Met brekingsindexcontrast\index{brekingsindexcontrast}}

We zullen nu weer de hoekafhankelijkheid van het intensiteitsprofiel berekenen, maar nu
voor de situatie dat de ree\"{e}le delen van de brekingsindices verschillend
zijn. Dit beschrijft
de experimentele situatie van een monster met een lage dichtheid
van verstrooiers achter glas of lucht.  Als gevolg van het indexcontrast
zal het licht
gedeeltelijk gerefracteerd worden wanneer het de grenslaag passeert.
 Ook zullen er speculaire reflecties
aan de grenslaag optreden \footnote{De vorm van de reflectie is
onder meer afhankelijk van de oppervlakte ruwheid. We
beschouwen hier alleen speculaire reflectie aan een glad oppervlak.}.
Bekijk eerst een half-oneindig medium. Voor $z>0$
is er een verstrooiend medium met brekingsindex $n_0$. Voor $z<0$ is
er een di\"{e}lectricum\index{di\"electricum} met
brekingsindex $n_1\equiv n_0/m$. De notaties zijn in figuur \ref{skin.eps}
aangegeven. Voor een van $z<0$ invallende golf krijgen
we een speculair gereflecteerde golf voor $z<0$ en een gerefracteerde golf
in $z>0$: \beqa
\Psi_{in} & = &
\eexp{i
{\bf k}^a_\perp\cdot\rhoo+ip_az}-\frac{P_a-p_a}{P_a+p_a}\eexp{i
{\bf k}^a_\perp \cdot\rhoo -ip_az} \mbox{\ \ \ } (z<0) \\
\Psi_{in} & = & \frac{2p_a}{P_a+p_a}\eexp{i {\bf k}^a_\perp\cdot\rhoo+iP_az}
\mbox{\ \ \ } (z>0) \eeqa
met
\BE
P_a=\sqrt{k_0^2+nt-k^{a\,2}_\perp}, \mbox{\ \ \ \ \ \
}p_a=\sqrt{k_1^2-k^{a\,2}_\perp} \EE
De tweede term van $\Psi_{in}$ voor $z<0$ is natuurlijk de direct
gereflecteerde golf. De voorfactoren
 $(P_a-p_a)/(P_a+p_a)$ en $2p_a/(P_a+p_a)$ volgen uit de eis dat
$\Psi$ overal continu differentieerbaar moet zijn.  De bronterm in het medium,
de binnenkomende intensiteit, kunnen we schrijven als:
\beq
I_{in}=\left|
\Psi_{in}\right|^2=\mid\frac{2p_a}{P_a+p_a}\mid^2\eexp{-z/\ell\cos\theta_a}
= \frac{p_a}{P_a}T_a\eexp{-z/\ell\mu_a}
\eeq
met, tot op leidende orde in $1/k\ell$~\footnote{Deze benadering leidt er
toe dat de vrije weglengte $\ell$ alleen in exponenten voorkomt. Na herschaling
van $z$ naar $\tau=z/\ell$ komt de vrije weglengte nergens meer expliciet voor.
Later zal blijken dat dit precies tot behoud van intensiteit leidt voor de
ladderdiagrammen.}, 
\index{$T@$T(\mu)$, transmissieco\"efficient} \beq
T_a=\frac{4p_aP_a}{P_a+p_a}=1-R(\mu_a), \mbox{\ \ \ }
R(\mu)=\left|\frac{|\mu|-\sqrt{\mu^2-1+1/m^2}}
                  {|\mu|+\sqrt{\mu^2-1+1/m^2}}\right|^2,
\mbox{\ \ \ } \mu_a=\cos\theta'_a.  \label{mcw22}
\eeq
$R(\mu)$ is de hoekafhankelijke reflectieco\"{e}fficient voor scalaire
golven.
Dit komt overeen met reflectie van s-gepolariseerd licht. In deze uitdrukking
voor $R(\mu)$ wordt het imaginaire deel van $P_a$ verwaarloosd. Merk op
dat $R=1$ voor totale reflectie, $m>1/\sqrt{1-\mu^2}$. Definieer
$\Gamma (\tau)$ volgens
\beq
I(z)=\frac{\ell}{4\pi}\L(z)=\frac{p_a}{P_a}T_a\Gamma (\tau). \eeq
$\Gamma$\index{$$C@$\Gamma$, genormeerde diffuse intensiteit} is de
intensiteit per eenheid van in het medium binnenkomende
intensiteit. Hiermee wordt de laddervergelijking (\ref{mcw4}) dimensieloos:
\beq
\Gamma(\tau)=\eexp{-\tau/\mu_a}+\int_0^\infty d\tau'M(\tau,\tau')\Gamma(\tau')
\label{dlL} \eeq
$M(\tau,\tau')$ volgt uit het kwadraat van de Greense functie: $|G|^2$. $G$
bestaat nu in principe uit twee termen; net als bij de inkomende golf hebben
we nu ook een bijdrage als gevolg van reflectie aan de wand.
De Greense functie wordt gegeven door
\beq
G(z,z',\qt)=\frac{i}{2P}\{\eexp{iP\mid z-z'\mid}+
\frac{P-p}{P+p}\eexp{iP(z+z')}\}
\mbox{; \ \ \ vergelijk met (\ref{mcw10})}
\eeq
De bijdragen van de kruistermen in $\mid G\mid^2$ kunnen verwaarloosd worden
omdat ze snel oscilleren. We hebben dus een volume (`bulk') term
$M_B$ en de randlaag (`layer') term $M_L$, $M=M_B+M_L$.\index{$Ml@$M_L$,
randlaagkern}\index{$Mb@$M_B$, bulkkern} Voor de bulk term vinden we
\BE
M_B(z,z')= 4\pi \int d^2\rhoo\mid G(z,\rhoo;z',\rhoo')\mid^2.
\EE
Neem even $z'=0$ dan
\BE
M_B(z,0)=\int dx dy \frac{1}{4\pi}
\frac{\eexp{-\sqrt{x^2+y^2+z^2}/\ell}}{x^2 +y^2+z^2}
\EE
Er geldt \BE
r^2=z^2+\rho^2=z^2/\mu^2\Rightarrow \rho^2=z^2(1/\mu^2-1)\Rightarrow
\rho d\rho=-z^2 d\mu/\mu^3.
\EE
dus
\BA\label{Mbulk}
M_B(z,0) & = &4\pi \int dxdy \frac{1}{(4\pi r)^2} \eexp{-z/\ell\mu} =
\int_0^\infty \frac{\rho d\rho}{2r^2}\eexp{-z/\ell\mu} \nonumber \\
&=& \int_0^1 \frac{d\mu}{2\mu} \eexp{-z/\ell\mu}
\nonumber\\ M_B(\tau,\tau') & = &
\int_0^1\frac{d\mu}{2\mu}\eexp{-|\tau-\tau'|/\mu}
=\frac{1}{2}E_1(|\tau-\tau'|) \EA
waarbij $\tau \equiv z/\ell$ de optische diepte is.
Evenzo vinden we voor de tweede term
\BA\label{Mlayer}
M_L(\tau,\tau')&=&4\pi \int dx dy\left| \frac{i}{2P} \frac{P-p}{P+p}e^
{iP(z+z')}\right|^2=
\int_0^1\frac{d\mu}{2\mu}\eexp{-(z+z')/\ell\mu}R(\mu) \nonumber \\
&=&\int_0^1\frac{d\mu}{2\mu}\eexp{-(\tau+\tau')/\mu}R(\mu)
\EA
We kunnen de speciale oplossing, $G_S(\tau,\tau')$,\index{$Gs@$G_S$,
speciale oplossing} defini\"eren door de vergelijking: \beq
G_S(\tau,\tau')=\delta(\tau-\tau')+\int_0^\infty d\tau''
\{M_B(\tau,\tau'')+M_L(\tau,\tau'')\}G_S(\tau'',\tau')
\eeq
Met als randvoorwaarde
$G_S(\infty,\tau') < \infty $. Dan wordt de speciale oplossing voor
(\ref{dlL}) gegeven door \beq
\Gamma_S(\tau)=\int_0^\infty d\tau' G_S(\tau,\tau')\eexp{-\tau'/\mu_a}
\eeq
Wanneer we de intensiteit in het punt $(z',\rho')$ buiten de plak (\ref{Ir})
weer op dezelfde manier uitwerken als in de vorige paragraaf (\ref{zobr})
vinden we: \beqa
I_R(z',\rho') & = &
\int\frac{k_1^2}{(2\pi)^2}\cos\theta_bd\Omega_b
\frac{1}{P_b+p_b}\int_0^\infty \L(z)\eexp{-k_0z/P_bl}dz
\nonumber\\ & = &
\int\frac{k_1^2}{(2\pi)^2}\cos\theta_bd\Omega_b
\frac{1}{P_b+p_b} 4\pi \frac{p_aT_a}{P_a}
\int_0^\infty\Gamma(\tau)\eexp{-\tau/\mu_b}d\tau\nonumber\\
\eeqa
De voorfactor van de integraal  kan met formule (\ref{mcw22}) uitgedrukt
worden als \beqa
\frac{T_b}{4P_b p_b}
\frac{k_1^2}{(2\pi)^2}\cos\theta_b4\pi\frac{p_aT_a}{P_a}
& = & \frac{k^2_1p_a}{4\pi p_b}\cos\theta_b \frac{T_aT_b}{P_aP_b}\nonumber
\\ &=& \frac{k^2_1}{4\pi}
\frac{\cos\theta_a}{\cos\theta_b}\cos\theta_b\frac{T_aT_b}{k_0\mu_ak_0\mu_b}
\nonumber\\
& = &\frac{\cos{\theta_a}}{4\pi m^2}\frac{T_aT_b}{\mu_a\mu_b}
\eeqa
Voor de hoekafhankelijkheid van de reflectie\index{reflectie!hoekafhankelijkheid} van een
half-oneindig medium vinden we dus: \beq
A_R(\theta_a,\theta_b)=\frac{\cos{\theta_a}}{4\pi m^2}\frac{T_aT_b}{\mu_a\mu_b}
\gamma(\mu_a,\mu_b)
\label{mcw5}
\eeq
met
\beq
\gamma(\mu_a,\mu_b)=\int_0^\infty d\tau\Gamma_S(\mu_a,\tau)\eexp{-\tau/\mu_b}=
\int_0^\infty d\tau d\tau' G_S(\tau,\tau')\eexp{-\tau/\mu_a}\eexp{-\tau'/\mu_b}
\label{mcw6}
\index{$$cm@$\gamma(\mu_a,\mu_b)$, genormeerde bistatische \\
co\"efficient}\index{bistatische co\"efficient}\eeq
Voor het geval van loodrechte inval $(\theta_a=0)$, is $A_R$ numeriek
bepaald voor verschillende $m$. De resultaten staan in figuur \ref{jmlar}.
\begin{figure}
\caption{De hoekafhankelijkheid van de reflectieco\"efficient voor loodrechte inval op
een half oneindig medium voor verschillende waarden van $m$.}
\centerline{\includegraphics[width=8cm]{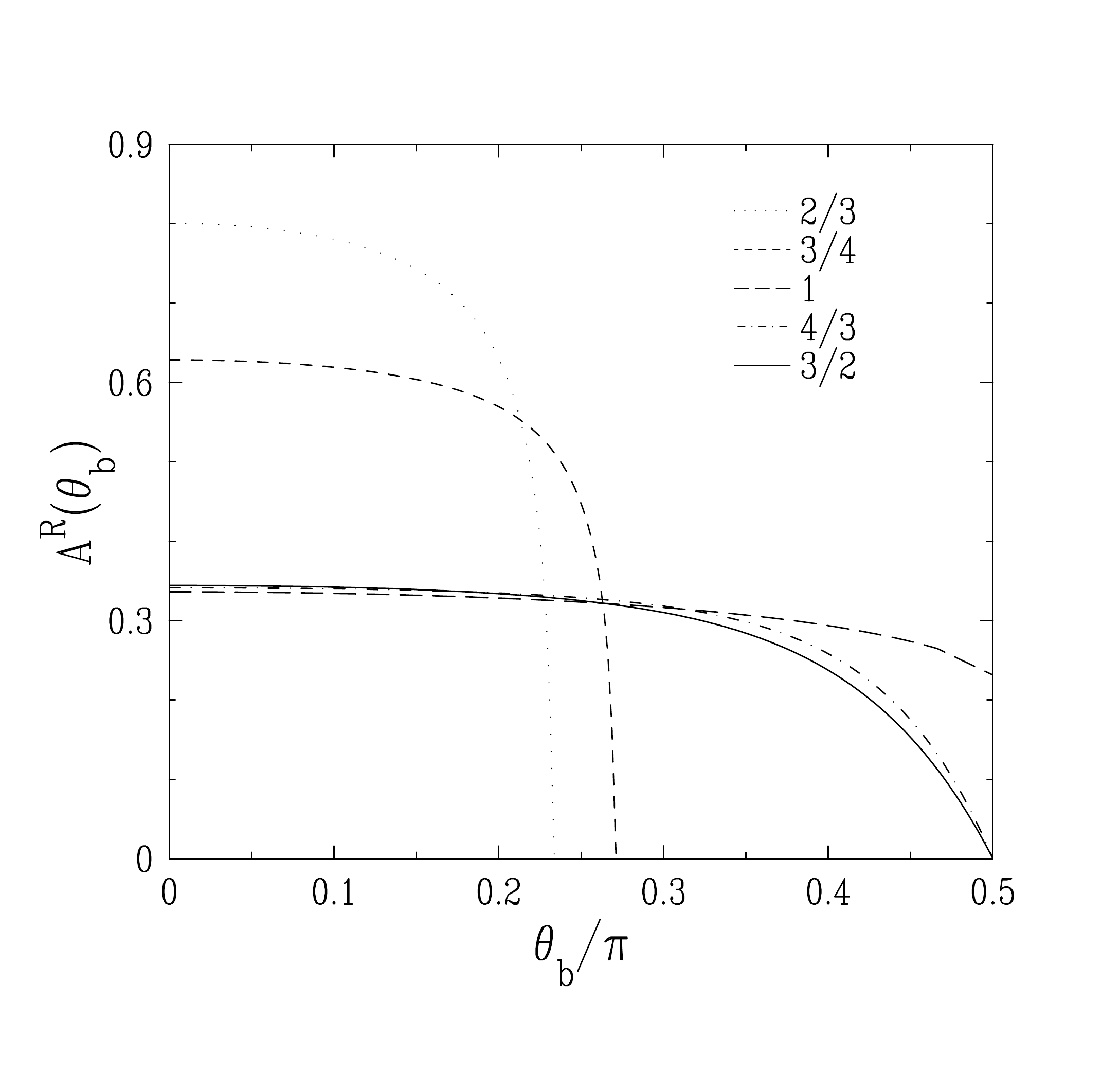}}
\label{jmlar}
\end{figure}

\subsection{De lucht-glas-medium overgang\index{lucht-glas-medium overgang}}
We beschouwen nu een half-oneindig verstrooiend medium, gescheiden van de
lucht door een
glasplaat met dikte d. Voor $z<-{\rm d}$ hebben we lucht, voor $-{\rm
d}<z<0$ glas en voor
$z>0$ het verstrooiende medium. We laten van links een op \'{e}\'{e}n
genormeerde vlakke golf invallen. De golfvektor loodrecht op de
voortplantingsrichting ($\kl$) is behouden. De golfvektor in de
voortplantingsrichting in medium $i$ noemen we $p_{i}$, waarbij $i=0$ correspondeert
met het verstrooiende medium, $i=1$ met lucht en $i=2$ met glas.
De golfgetallen in de drie media zijn $k_0=\omega/c_0$, $k_1=\omega/c_1$
en $k_2=\omega/c_2$ met $c_i$ de voortplantingssnelheden in de drie media.
De binnenkomende, afgebogen en speculair gereflecteerde golven in de drie
media worden gegeven door \beqa
\Psi({\bf r}) = \left \{ \begin{array}{ll}
{\rm e}^{i\kl\cdot\rhoo+ip_{1}z}+r{\rm e}^{i\kl\cdot\rhoo-ip_{1}z} &
(z<-{\rm d}) \nonumber \\
t_{1}{\rm e}^{i\kl\cdot\rhoo+ip_{2}z} + r_1{\rm e}^{i\kl\cdot\rhoo
-ip_{2}z}  &(-{\rm d}<z<0)  \nonumber \\
t{\rm e}^{i\kl\cdot\rhoo+ip_{0}z}    & (z>0)
\end{array}
\right.
\index{$t$, transmissieamplitude}\index{$r$, reflectieamplitude}\eeqa
Hierin is $r$ de reflectieamplitude van het systeem, $t$ de
transmissieamplitude en $p_i=\sqrt{k_i^2-\kl^2}$.
 Door op te leggen dat de golffunktie en de eerste afgeleide
continu moeten zijn in $z=-{\rm d}$ en $z=0$ kunnen $r$ en $t$ bepaald
worden. 
\BO
Bereken $t,t_1,r$ en $r_1$. Laat zien dat:
\beqa
r &=& \frac{(p_{0}+p_{2})(p_{1}-p_{2})-(p_{0}-p_{2})(p_{1}+p_{2}){\rm
e}^{2ip_{2}{\rm d}}}
{(p_{0}+p_{2})(p_{1}+p_{2})-(p_{0}-p_{2})(p_{1}-p_{2}){\rm e}^{2ip_{2}{\rm
d}}} \nonumber \\
t &=& \frac{4p_{1}p_{2}{\rm e}^{i(p_{2}-p_{1}){\rm d}}}
{(p_{0}+p_{2})(p_{1}+p_{2})-(p_{0}-p_{2})(p_{1}-p_{2}){\rm e}^{2ip_{2}{\rm
d}}}
\eeqa
\EO
De reflectie- en transmissieco\"{e}ffici\"{e}nt van het systeem zijn
\beq |r|^{2}\;\;\; ; \;\;\; 
\frac{p_{0}}{p_1}|t|^{2}\eeq
\index{reflectieco\"efficient}\index{transmissieco\"efficient}
Omdat de dikte van de glasplaat niet overal binnen \'{e}\'{e}n golflengte
bepaald zal zijn, moeten deze uitdrukkingen over de spreiding in de dikte
gemiddeld worden. Dat komt op hetzelfde neer als middelen over de fase
$2p_{2}{\rm d} \equiv \varphi$. Het leidt tot de gemiddelde transmissie-
en reflectieco\"eficienten 
\beqa
 T_{3}  & =&  1-R_{3} =
\frac{p_{0}}{p_{1}}\int^{\pi}_{-\pi}\frac{d\varphi}{2\pi} |t|^{2}
\nonumber \\
\EA
\BEN =
\frac{p_{0}}{p_{1}}\int^{\pi}_{-\pi}\frac{d\varphi}{2\pi}
\frac{16p_{1}^{2}p_{2}^{2}} {(p_{0}+\!p_{2})^{2}(p_{1}+\!p_{2})^{2}+(p_{0
}-\!p_{2})^{2}(p_{1}-\!p_{2})^{2}-\!(p_{0}^{2}
-\!p_{2}^{2})(p_{1}^{2}-\!p_{2}^{2})2\cos\varphi}
\EEN
\BA
 & \stackrel{(\ref{ident2})}{=} &
\frac{p_{0}}{p_{1}}16p_{1}^{2}p_{2}^{2}
\frac{1}{(p_{0}+p_{2})^{2}(p_{1}+p_{2})^{2}-(p_{0}-p_{2})^{2}
(p_{1}-p_{2})^{2}}\EEA
waarbij gebruik is gemaakt van formule (\ref{ident2}).
Het eindresultaat is
\BEQ T_{3}=1-R_3=
\frac{4p_{0}p_{1}p_{2}}{(p_{0}+p_{1})(p_{0}p_{1}+p_{2}^{2})}
\EEQ
Invullen van $p_{2}=p_{1}=p$ en $p_{0}=P$ geeft het vroegere
resultaat voor twee media.

 Speculaire reflecties in het glas zijn nu
meegenomen. Dit is goed voor dunne glasplaatjes. (Voor dikker glas
kan het licht tot ver van de invallende bundel heen en weer spiegelen, of
zelfs buiten het medium terecht komen).
We kunnen dit systeem dus beschrijven door in de voorgaande vergelijkingen
$T(\mu)$ door $T_3$ te vervangen en in de
Milne kern\index{Milne kern} de
reflectieco\"efficient $R(\mu)$ door $R_3$.  
$\mu\equiv p_0/k_0$ geeft nog immer de cosinus van de hoek van de straling 
in het medium en de $z$-as aan.

\subsection{Oplossingen van de Schwarzschild-Milne
vergelijking\index{Schwarzschild-Milne vergelijking}} \label{oplossing}

We volgen ref. [NL] en bekijken
de eigenschappen van de transportvergelijking in een
half-oneindige ruimte. De diffuse intensiteit kan geschreven worden als
\beq
I(z) = \frac{p_{a} T_{a}}{P_{a}}
\Gamma(\tau), \mbox{ met } \tau = z/\ell,
\index{$$t@$\tau$, optische diepte}\eeq
Dit leidt tot een dimensieloze transportvergelijking voor de genormeerde
diffuse intensiteit $\Gamma$, \beq
\Gamma(\tau)=\eexp{-\tau/\mu_a}+\int_0^{\infty} M(\tau,\tau')
\Gamma(\tau')\d\tau',
\label{milne}
\eeq
Deze vergelijking zagen we al in het vorige hoofdstuk.
De bronterm $\exp(-\tau/\mu_a)$ beschrijft de afval van de invallende
bundel ten gevolge van verstrooiing.
$\Gamma(\tau)$ representeert de genormeerde diffuse intensiteit
die verstrooid is op een diepte $z=\ell\tau$, ontstaan uit een
genormeerde vlakke golf die invalt onder een hoek $\theta_a$.
Vergelijking (\ref{milne}) heet de Schwarzschild-Milne of kortweg
Milne vergelijking\index{Milne vergelijking} van het probleem.
De Milne -kern\index{Milne kern} $M(\tau,\tau')$ heeft een bulk bijdrage
$M_B$, en een oppervlaktebijdrage $M_L$, gegeven door
\beq
M(\tau,\tau')=M_B(\tau,\tau')+M_L(\tau,\tau'), 
\eeq
\BA \mbox{ waarin \hspace{1em}}
M_B(\tau,\tau')&=&\int_0^1{\d\mu\over 2\mu}
\eexp{-\vert\tau-\tau'\vert/\mu}, \\
M_L(\tau,\tau')&=&\int_0^1{\d\mu\over 2\mu} R(\mu) \eexp{-(\tau+\tau')/\mu}.
\EA
De Milne vergelijking vergelijking (\ref{milne}) heeft een speciale oplossing
$\Gamma_S(\mu_a;\tau)$, de ge\"associeerde homogene vergelijking
(zonder bronterm) heeft een oplossing $\Gamma_H(\tau)$. We zijn
ge\"interesseerd in het asymptotisch gedrag van $\Gamma_S(\mu_a;\tau)$ en
$\Gamma_H(\tau)$. Diep in de bulk $(\tau\gg 1)$ kun je $\Gamma(\tau)$
in een Taylor expansie schrijven (je verwacht dat diep in de bulk
$\Gamma(\tau)$ langzaam varieert over \'e\'en vrije weglengte
$\vert\tau-\tau'\vert\approx 1$, zodat hogere
orde afgeleiden snel afvallen),
\beq
\Gamma(\tau') = \Gamma(\tau) + (\tau'-\tau)\Gamma'(\tau) +
\frac{1}{2}(\tau'-\tau)^2\Gamma''(\tau) + \cdots.
\eeq
Vul nu deze expansie voor $\Gamma(\tau)$ in vergelijking ~(\ref{milne}) in
waarbij de bronterm verwaarloosd wordt, en neem alleen de bulkbijdrage van de kern mee. Voor $\tau\gg 1$ vind je dan
\beq
\Gamma(\tau) \approx \Gamma(\tau) + \frac{1}{3} \Gamma''(\tau) +
O(\Gamma^{(4)}).
\label{gamtau}
\eeq
Hierbij is gebruik gemaakt van de integraal
\beq
\frac{1}{2} \int_0^1\frac{\d\mu}{2\mu}\int_0^\infty\d\tau'
\eexp{-\vert\tau-\tau'\vert / \mu}(\tau-\tau')^{2} \approx \frac{1}{3}
\mbox{  als $\tau\gg 1$.} \label{eendrieint}
\eeq
$O(\Gamma')$ en $O(\Gamma''')$ zijn nul want $(\tau'-\tau)^{2n+1}$
ingevuld in vergelijking (\ref{milne}) geeft een oneven integraal voor
$\tau\gg 1$.
Uit vergelijking (\ref{gamtau}) volgt logischerwijs dat $\Gamma''(\tau)
\approx 0$ en dat de
asymptotische oplossing voor $\Gamma(\tau)$ hoogstens lineair is in $\tau$.

We beschouwen de homogene oplossing $\Gamma_H$ en de speciale oplossing
 $\Gamma_S$  waarvoor geldt\footnote{Correcties op dit asymptotische gedrag vallen exponentieel in
$\tau$ af. Bij het numeriek oplossen kan men dan de vergelijkingen voor
$\delta\Gamma_S(\tau)=\Gamma_S(\tau)-\tau_1(\mu_a)$ en $\delta\Gamma_H(\tau)=
\Gamma_H(\tau)-\tau-\tau_0$ itereren vanuit $\tau=\infty$. De eis dat zij
 inderdaad naar nul gaan legt dan vanzelf $\tau_0$ en $\tau_1$ vast. }:
\beqa
\left\{ \begin{array}{l} \Gamma_H(\tau)\approx\tau+\tau_0 \\
\Gamma_S(\tau;\mu_a)\approx \tau_1(\mu_a) \end{array}  
\qquad(\tau\to\infty). \right. \label{gammaasymp}
\index{$$t0@$\tau_0$, genormeerde injectiediepte} \index{$$t1@$\tau_1$,
genormeerde limiet intensiteit} \index{$$Ch@$\Gamma_H$, homogene
oplossing}\index{$$Cs@$\Gamma_S$, speciale oplossing}
\index{injectiediepte}\eeqa De constante $\tau_0$ hangt alleen van de
verhouding $m$ van
de brekingsindices af, omdat de homogene vergelijking geen bronterm heeft.
 $\tau_1(\mu_a)$ hangt ook van de richting van de invallende bundel af.
We kunnen de speciale Greense functie $G_S(\tau,\tau')$ voor de Milne
vergelijking opschrijven. Deze is gedefinieerd als de oplossing van
de volgende vergelijking:
\beq
G_S(\tau,\tau')=\delta(\tau-\tau')+\int_0^\infty M(\tau,\tau'')
G_S(\tau'',\tau')\d\tau'', \label{greenmil}
\eeq
onder de conditie dat $G_S(\tau,\tau')$ eindig blijft als $\tau$ of 
$\tau'$ naar oneindig gaan. Zij heeft de volgende symmetrie eigenschappen,
\beq
G_S(\tau,\tau')=G_S(\tau',\tau),
\eeq
en de limiet
\beq
\lim_{\tau'\to\infty}G_S(\tau,\tau')={1\over D}\Gamma_H(\tau).
\label{greenlim}
\eeq
Deze laatste vergelijking kan bewezen worden door in vergelijking
 (\ref{greenmil}) de limiet $\tau'\to\infty$ te nemen. De delta-functie
valt weg, en de vergelijking is gelijk aan de die voor de
homogene vgl. voor $\Gamma_H(\tau)$. De Greense functie
$G_S(\tau,\tau'\to\infty)$ is dus, op een constante factor na, gelijk aan
$\Gamma_H(\tau)$. De factor kan bepaald worden door
$G_S(\tau,\tau')$ te ontwikkelen in een Taylorreeks rond $\tau$ en de
Taylorexpansie in het rechterlid van vergelijking (\ref{greenmil}) in te
vullen. Voor $\tau,\tau'\gg 1$ en als hogere dan tweede orde afgeleiden
van $G_S(\tau,\tau')$ worden verwaarloosd (diffusie benadering), vind
je de volgende vergelijking, waarbij ook weer gebruik is gemaakt van
vergelijking (\ref{eendrieint}),
\beq
0 = \delta(\tau-\tau') + \frac{1}{3} \frac{\d^2}{\d\tau^2} G_S(\tau,\tau').
\eeq
De oplossing wordt als volgt gevonden
\beq
\frac{\d^2}{\d\tau^2} {\rm min}(\tau,\tau') = \frac{\d^2}{\d\tau^2}\left\{
\begin{array}{c}\tau' \hspace{1em} \tau>\tau'\\ \tau \hspace{1em}
\tau<\tau'\end{array} \right\} =
\frac{\d}{\d\tau}\left\{
\begin{array}{c} 0\hspace{1em} \tau>\tau'\\ 1\hspace{1em}
\tau<\tau'\end{array} \right\} = -\delta(\tau-\tau').
\eeq
We vinden dus $G_S(\tau,\tau')=3 \, {\rm min}(\tau,\tau')$ in het regime
$(\tau$, $\tau'$, $\vert\tau-\tau'\vert\gg 1)$.  De diffusieco\"efficient $D$
in vergelijking (\ref{greenlim}) is $1/3$ in gereduceerde
eenheden, ofwel $D=v\ell/3$ in fysische eenheden.
Uit vergelijking (\ref{greenmil}) volgt
\beq
\Gamma_S(\tau;\mu)=\int_0^\infty G_S(\tau,\tau')\eexp{-\tau'/\mu}\d\tau',
\label{gammas}\eeq
en in het bijzonder, gebruik makend van vergelijkingen (\ref{gammaasymp}) en
(\ref{greenlim}),
\beq
\tau_1(\mu)=\lim_{\tau\to\infty} \Gamma_S(\tau;\mu) = 
{1\over D}\int_0^\infty\Gamma_H(\tau)\eexp{-\tau/\mu}\d\tau. \label{taueen}
\eeq
De fysische interpretatie van $\tau_1(\mu)$ is de
limietintensiteit in het halfoneindige medium.

Numerieke waarden van de injectiediepte $\tau_0$, de genormeerde
limietintensiteit $\tau_1(1)$ en de genormeerde bistatische co\"efficient
$\gamma(1,1)$ voor verschillende waarden van de brekingsindexverhouding
$m$ zijn te vinden in de tabel 
aan het eind van hoofdstuk \ref{terugstrooikegel}.

\subsection{Fluxbehoud\index{fluxbehoud}}

Het is belangrijk dat ondanks de gemaakte benaderingen, behoudswetten
blijven gelden. We tonen hier het behoud van flux aan voor het geval er
 geen absorptie is.
De diffuse reflectie van het licht in ruimtehoek b dat onder richting a is
ingevallen op het halfoneindig medium wordt gegeven door (zie hoofdstuk
\ref{transport}) \beq
\frac{\d R_{a b}}{\d\Omega_b} = A^R(\theta_a, \theta_b) =
\frac{\cos(\theta_a) T_a
T_b}
{4\pi m^2 \mu_a \mu_b}\gamma(\mu_a, \mu_b),
\eeq
met 
\beq
\gamma(\mu_a,\mu_b) = \int_0^\infty\d\tau\Gamma_S(\tau;\mu_a)
\eexp{-\tau/\mu_b}. \label{gammaab}
\eeq
De vraag is nu of de hoeveelheid licht die binnendringt in het
halfoneindig medium gelijk is aan de hoeveelheid licht die er uit
komt. Fysisch wel, maar mathematisch ook na al onze benaderingen?
De vraag is:
\beq
\int_0^{4\pi} A^R(\theta_a,\theta_b) \cos(\theta_b)\d\Omega_b \stackrel{?}
{=} \cos \theta_a -R_a \cos\theta_a =T_a\cos(\theta_a), \label{conservatie}
\eeq
met de inkomende flux $\cos(\theta_a)$ en de speculaire gereflecteerde flux
$R_a \cos(\theta_a)$.
Het linker lid is het diffuus gereflecteerde licht, het rechterlid
is het licht dat het halfoneindig medium is binnengedrongen. Na middeling
over azimutale richtingen \beq
\int_{\phi_b} \frac{\cos(\theta_b)\d\Omega_b}{4\pi m^2 \mu_b} =
\frac{\cos(\theta_b)\sin(\theta_b)}{4\pi m^2 \mu_b}
\d\theta_b \int_0^{2\pi} d\phi_b =- \frac{1}{2} \d\mu_b,
\eeq
(in het laatste gelijk teken is gebruik gemaakt van
$\cos\theta\sin\theta\d\theta = -m^2\mu\d\mu$)
reduceert vergelijking (\ref{conservatie}) tot
\beq
\int_0^1\frac{\d\mu_b}{2} T(\mu_b)\gamma(\mu_a,\mu_b) \stackrel{?}{=}
\mu_a. \label{tussencon}
\eeq
Dit gaan we nu bewijzen. Hiertoe hebben we de volgende gelijkheid nodig:
\beq
\int_0^1 \frac{\d \mu_b}{2} T(\mu_b) \eexp{-\tau'/\mu_b} =
1-\int_0^\infty M(\tau', \tau'')\d\tau'' \mbox{ met $M = M_B + M_L$},
\eeq
waarvan het bewijs nu eerst volgt. Omdat
\beq
M_L = \int_0^1 \eexp{-(\tau+\tau')/\mu} \frac{1-T(\mu)}{2\mu} \d\mu.
\eeq
volgt
\BA
&&\int_0^\infty M(\tau',\tau'')\d\tau'' \nonumber \\ &&=
\int_0^1\frac{\d\mu}{2\mu}
\int_0^\infty \d\tau'' \left[ \eexp{-\vert\tau'-\tau''\vert/\mu}  + 
\eexp{-(\tau'+\tau'')/\mu} - T(\mu)
\eexp{-(\tau'+\tau'')/\mu} \right]\nonumber  \EA
\BA
&&\int_0^\infty M(\tau',\tau'')\d\tau'' \nonumber \\ &&= \underbrace{
\int_0^1\frac{\d\mu}{2\mu} \int_{-\infty}^\infty
\eexp{-\vert\tau'-\tau''\vert/\mu} \d\tau''}_{=1} - \int_0^1\frac{\d\mu}{2\mu} 
\int_0^\infty T(\mu)\eexp{-(\tau'+\tau'')/\mu} \d\tau'' \nonumber
\EA
Dus
\beq
1-\int_0^\infty M(\tau',\tau'')\d\tau'' = \int_0^1\frac{\d\mu}{2\mu}
T(\mu) \int_0^\infty \eexp{-(\tau'+\tau'')/\mu} \d\tau'' =
\int_0^1\frac{\d\mu}{2} T(\mu) \eexp{-\tau'/\mu}. \label{tussenres}
\eeq
Schrijf nu het linkerlid van vergelijking (\ref{tussencon}) uit met behulp van
vergelijkingen (\ref{gammas}) en (\ref{gammaab}).
\beq
\int_0^1 \frac{\d\mu_b}{2} T(\mu_b) \gamma(\mu_a, \mu_b) =
\int_0^\infty \d\tau \int_0^\infty \d\tau' \underline{\int_0^1
\frac{\d\mu_b}{2} T(\mu_b) \eexp{-\tau'/\mu_b}}
\eexp{-\tau/\mu_a} G_S(\tau, \tau').
\eeq
De onderstreepte term geeft precies het rechterlid van vergelijking (\ref{tussenres}).
Vul je het linkerlid van deze vergelijking in, dan krijg je:
\beq
\int_0^1 \frac{\d\mu_b}{2} T(\mu_b) \gamma(\mu_a, \mu_b) =
\int_0^\infty\! \d\tau \int_0^\infty \!\d\tau' \eexp{-\tau/\mu_a}
G_S(\tau,\tau') \left[1-\int_0^\infty \d\tau'' M(\tau', \tau'') \right]
\eeq
Gebruik vergelijking (\ref{greenmil}) om dit te herschrijven in
\BA
\int_0^1 \frac{\d\mu_b}{2} T(\mu_b) \gamma(\mu_a, \mu_b)& = &
\int_0^\infty \d\tau \eexp{-\tau/\mu_a} \left[\int_0^\infty \d\tau'
G_S(\tau, \tau')\! \nonumber \right. \\ && \left. -\!\int_0^\infty
\d\tau''
\left[G_S(\tau, \tau'')\! -\! \delta(\tau\!-\!\tau'') \right] \right]
\EA
Er volgt
\beq
\int_0^1 \frac{\d\mu_b}{2} T(\mu_b) \gamma(\mu_a, \mu_b) = \mu_a
\mbox{\hspace{3em}} \Box,
\eeq
en hiermee is bewezen dat ook mathematisch voldaan wordt aan fluxbehoud
voor een half-oneindig medium.

\renewcommand{\thesection}{\arabic{section}}
\section{Transport door een plak}\label{plak}\index{plak}
\setcounter{equation}{0}\setcounter{figure}{0}

We bestuderen nu de transmissie\-eigenschappen van een optisch dikke plak
met isotrope verstrooiers. We vinden het `Ohmse' gedrag $T\sim \ell/L$ uit
hoofdstuk \ref{macroscopie}, alsmede
de volledige hoekafhankelijkheid. Dit resultaat wordt dan gebruikt voor
de berekening van de weerstand van een geidealiseerde geleider.

\subsection{De diffuse transmissie\index{transmissie!diffuse}}

We bekijken een wanordelijk medium met een eindige dikte, $L$. Het
medium wordt wel optisch dik geacht: $ b=L/\ell\gg 1$.
We willen de exponentieel afvallende effecten van de rand verwaarlozen.
Daarom nemen aan dat we niet te dicht bij de rand
zitten (`10' vrije weglengtes is een goede maat). De oplossing voor
$\Gamma(\tau)$ is een lineaire combinatie van de speciale en de
homogene oplossing,
\beq\label{beginSM}
\Gamma(\tau) = \Gamma_S(\tau) - \alpha \Gamma_H(\tau) \mbox{  voor  $0\leq
\tau
\leq 10$;    } \Gamma(\tau) = \tau_1(\mu) - \alpha(\tau +\tau_0) \mbox{
voor $\tau>10$.}
\index{$$a@$\alpha$, constante}\eeq
\beq\label{eindSM}
\Gamma(\tau) = \alpha' \Gamma_H(b-\tau) \mbox{  voor  $0\leq b-\tau \leq
10$;    }
\Gamma(\tau) = \alpha'(b-\tau+\tau_0) \mbox{  voor $b-\tau > 10.$}
\eeq
Als beide
asymptotische oplossingen aan elkaar worden geknoopt in het midden van de
plak, en continuiteit en continu-differentieerbaarheid wordt opgelegd, dan
vind je
\beq \label{al=al=}
\alpha=\alpha'= \frac{\tau_1(\mu)}{b+2 \tau_0}.
\eeq
We hebben nu een oplossing voor de intensiteit in de slab, uitgedrukt in
$\Gamma_S$
en $\Gamma_H$, en we hebben een expliciete uitdrukking voor $\Gamma(\tau)$
in het
regime $10
\leq \tau \leq b-10$. Om de hoekafhankelijke
transmissie\index{transmissie!hoekafhankelijk}
uit te rekenen volgen we de afleiding voor de diffuse reflectie.
 Combineren van  vergelijking (\ref{mcw5}) en
(\ref{mcw6}) levert \beq
\frac{dR(a\rightarrow b)}{d\Omega_b} = \frac{\cos(\theta_a) T_a T_b}{4\pi
m^2 \mu_a \mu_b}
\int_0^\infty \d \tau \Gamma(\tau) \eexp{-\tau/\mu_b}.
\eeq
De uitdrukking voor de transmissieco\"effcient per eenheid van ruimtehoek
$d\Omega_b$ voor een bundel invallend onder hoek $\theta_a$
 wordt op analoge wijze gegeven door
\beq \label{dTabdO=}
\frac{dT(a\rightarrow b)}{d\Omega_b}
 = \frac{\cos(\theta_a) T_a T_b}{4\pi m^2 \mu_a
\mu_b}
\int_0^b \d \tau \Gamma(\tau) \eexp{-(b-\tau)/\mu_b}.
\eeq
De integraal kan omgeschreven worden, gebruik makend van de oplossing voor
$\Gamma(\tau)$,
\beqa
\int_0^b \d \tau \Gamma(\tau) \eexp{-(b-\tau)/\mu_b} &\approx& \alpha
\int_{-\infty}^b
\d \tau \Gamma_H(b-\tau) \eexp{-(b-\tau)/\mu_b} \nonumber \\ &=& \alpha
\int_0^\infty \d \tau
\Gamma_H(\tau) \eexp{-\tau/\mu_b} \nonumber \\ & = &\frac{\tau_1(\mu_a)}{b+2
\tau_0} \tau_1(\mu_b)
D,
\eeqa
waarbij gebruik is gemaakt van vergelijking (\ref{taueen}). We hebben nu de
hoekafhankelijke differentiele
transmissieco\"efficient gevonden voor een laag met dikte $b$,
\beq\label{Tabiso}
\frac{dT(a\rightarrow b)}{d\Omega_b}\equiv \frac{A^T(\theta_a,
\theta_b)}{b+2\tau_0} \equiv  \frac{\cos(\theta_a) T_a T_b}{12\pi m^2 \mu_a
\mu_b (b+2 \tau_0)} \tau_1(\mu_a) \tau_1(\mu_b)
\eeq
Aangezien de intensiteit aan de intreekant van de laag  $\Gamma(\tau) =
\Gamma_S(\tau) -
\alpha \Gamma_H(\tau)$ is, zien we direct dat de transmissieterm
 $\alpha \Gamma_H(b-\tau)$ ten koste van de reflectie gaat. Er is dus ook
in het geval $1\ll b< \infty$ aan fluxbehoud voldaan.

In figuur \ref{jmlat} is $A^T$\index{$At@$A^T$, genormeerde hoekopgeloste
transmissie} weergegeven
voor loodrechte inval $(\theta_a=0)$ voor diverse waarden van $m$.
\begin{figure}
\caption{De amplitude van de hoekopgeloste transmissieco\"efficient voor
een bundel loodrecht invallend op een dikke plak. De verhouding van
brekingindices, $m$, is aangegeven.}
\centerline{\includegraphics[width=10cm]{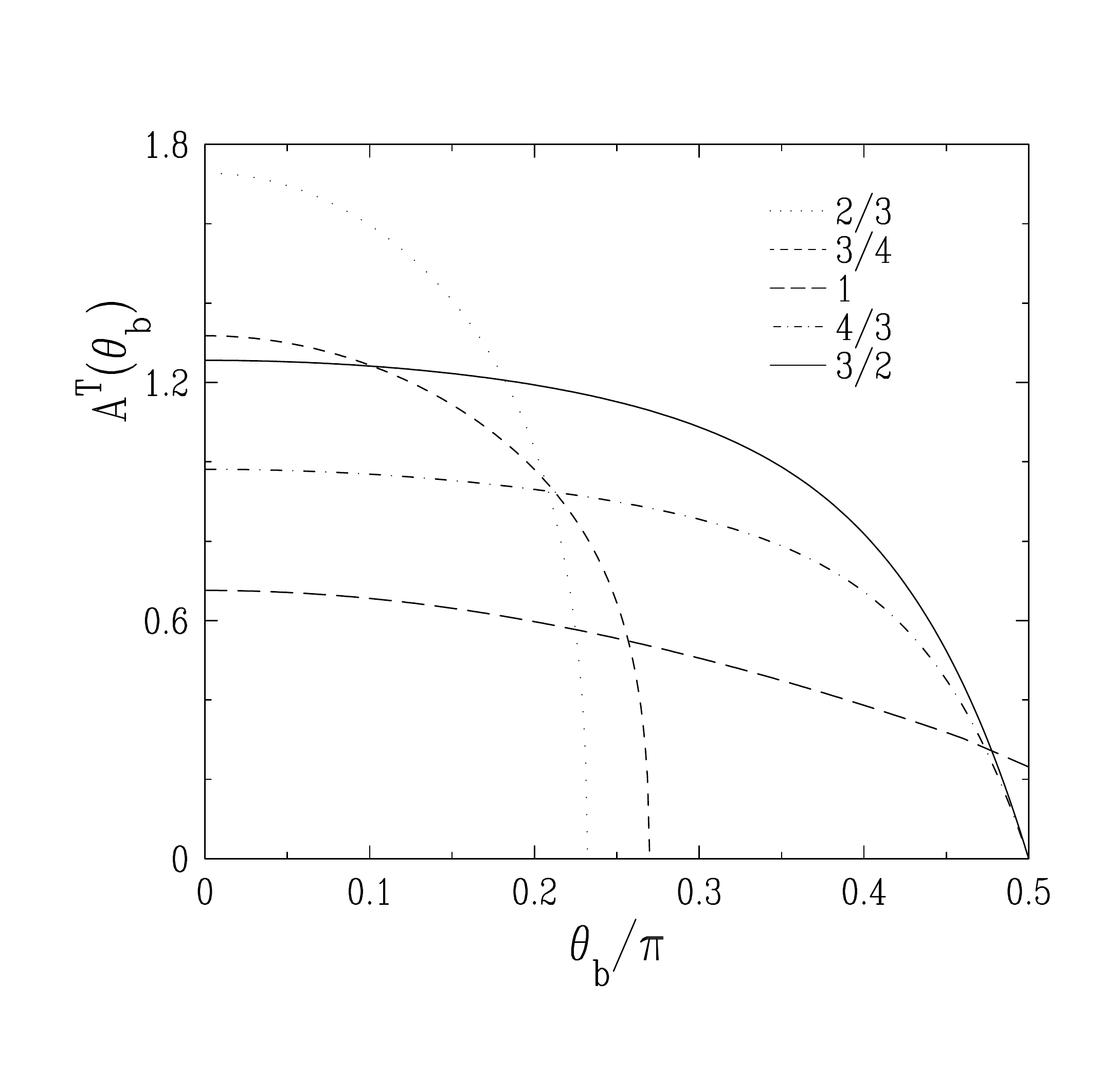}}
\label{jmlat}
\end{figure}

\subsection{Electrische geleidbaarheid\index{geleidbaarheid} en de
contactweerstand\index{contactweerstand}}
Voor metallische geleiders in het mesoscopische regime $(\lambda\ll \ell
\ll L \ll L_\phi)$, kan men de
geleidbaarheid bestuderen ($L_\phi$\index{$Lf@$L_\phi$,
fasecoherentielengte}
\index{fasecoherentielengte} is de fasecoherentielengte). Zij
wordt gegeven door de Landauer formule\index{Landauer formule}, \beq
G = \frac{2e^2}{h} \sum_{a,b} T_{ab}^{flux},
\eeq
waarbij $h/e^2 \approx 25k\Omega$. Deze formule telt simpelweg het
aantal kanalen dat aan de transmissie bijdraagt. In de bovenstaande
beschrijving kan de golfvector $k_0$ vervangen worden door de Fermi
\index{$k_F$, Fermi golfvector}\index{Fermi golfvector}
golfvector $k_F$. Het analogon van het brekingsindexcontrast is hier
het potentiaalverschil tussen de geleider en de contacten ($V_1 \neq V_0)$.
In ons formalisme wordt de geleidbaarheid gegeven door
\beq
G = \frac{2e^2}{h} \sum_{a,b} T_{ab}^{flux} = \frac{2e^2}{h} \sum_{a,b}
T_{ab}
\frac{\cos(\theta_b)}{\cos(\theta_a)} = \frac{2e^2}{h}\,\frac{k_F^2 A}{3 \pi
(b+2\tau_0)}, \eeq
met $T_{ab} =A^T(\theta_a,\theta_b)/(b+2\tau_0)$. Hiermee wordt de
geleidbaarheid, \beq
G= \frac{A \sigma_B}{L+2z_0} \mbox{ , } \sigma_B = \frac{2 e^2
k_F^2 \ell}{3 \pi h},
\eeq
waarbij $z_0=\tau_0 \ell$.
Het bulkgeleidingsvermogen\index{bulkgeleidingsvermogen}
\index{$$sb@$\sigma_B$, bulkgeleidingsvermogen} is gedefinieerd als
$\sigma_B$.
We kunnen een bulkweerstand en een contactweerstand $(R_c)$ onderscheiden,
\beq
R=\frac{1}{G} = \frac{L}{A \sigma_B} +2 R_c \mbox{ , met }
R_c=\frac{3 \pi \hbar}{2 e^2 A k_F^2} \tau_0.
\eeq
Het aantal kanalen (of modes) wordt gegeven door\index{$N@$N$, aantal modes}
$N \approx A k_F^2$. Bovenstaande uitdrukking laat zien dat $R_c$
evenredig is met  de dimensieloze dikte $\tau_0$ en omgekeerd
evenredig met het totaal aantal kanalen. Alles wat niet triviaal is zit in
$\tau_0$. $\tau_0$ hangt af van het potentiaalverschil maar niet van de 
dichtheid van verstrooiers. Aangezien
\beq
k_F^2 - V_1 = k_1^2 \mbox{  ,  } k_F^2 - V_0 = k_0^2
\eeq
kunnen we de analogie met lichtverstrooiing maken door $k_0^2=m^2 k_1^2 $
te stellen: \beq
m^2 \Longrightarrow  \frac{k_F^2-V_0}{k_F^2-V_1}
\index{$V_0\,(V_1)$ potentiaal binnen (buiten) geleider}\eeq

\subsection{De Ward identiteit voor algemene situaties}
\index{Ward identiteit}
In systemen waar geen absorptie optreedt, is er een behoudswet.
\index{behoudswet}
Voor electronen zal de waarschijnlijkheid (het aantal electronen)
behouden zijn; voor klassieke golven de totale energie.
Het is van groot belang dat dit in de theoretische beschrijving ingebouwd
 zit. Immers, schending van de behoudswet zal leiden tot schijnbare
absorptie of, nog erger, tot een onfysische bron van intensiteit.

In de Boltzmann beschrijving is het niet zo moeilijk aan de behoudswet te
voldoen. In de microscopische beschrijving moet je echter goed opletten. Het
blijkt dat je aan twee criteria moet voldoen. Je berekent eerst, in een of
andere benadering, de zelfenergie {\it zelfconsistent}. Dit levert de
benadering voor de amplitude Greense functie $G=\langle g\rangle$. Vervolgens moet de
irreducibele vertex van de (benadering voor de) Bethe-Salpeter vergelijking
aan een {\it Ward identiteit} voldoen. Met andere woorden: er is vrijheid om
de zelfenergie te benaderen, maar de gemaakte keuze legt wel de vertex vast.
\index{Bethe-Salpeter vergelijking}

We beschouwen nu een stationaire situatie. De {\it Ward identiteit} voor de
ladderdiagrammen in een systeem met puntverstrooiers kan alsvolgt worden
weergegeven \index{ladderdiagrammen} \index{puntverstrooier} \BEQ
\Sigma-{\overline \Sigma}=n\bar tt\left\{G(\r,\r)- {\overline
G}(\r,\r)\right\}. \EEQ Laten we kijken naar een systeem met kleine, maar
{\it inhomogene} dichtheid $n(\r)$ van verstrooiers.
\index{dichtheid!inhomogeen} Zo'n beschrijving is vooral van belang voor
een medium met niet
vlakke randen, bijv. hobbelige laagjes verf. Maar zelfs bij een scherpe rand
is er een inhomogeniteit, want er is een sprong in de verstrooiersdichtheid.
Zoals uitgelegd in hoofdstuk \ref{microscopie} zal ook de terugkeer Greense
functie, en dus de $t$-matrix, plaatsafhankelijk zijn. Bovenstaande
vergelijking heeft dan de vorm \BEQ
n(\r)t(\r)-n(\r)\bar{t}(\r)=n(r)\bar{t}(\r)t(\r)\left\{ G(\r,\r)-{\overline
G}(\r,\r)\right\}. \EEQ

We leiden hier af dat deze Ward identiteit precies leidt tot behoud
van stationaire flux {\it voor een willekeurige geometrie}\footnote{Dit is
een niet-gepubliceerd resultaat van Th.M. Nieuwenhuizen (1991).}.
\index{geometrie, willekeurige}
Laten we kijken naar een situatie waar we de bron van intensiteit aangeven
met $\psi_s(\r_a)$. De inkomende bundel is dan
\BEQ \Psi_{in}(\r)=\int d^3\r_a G(\r,\r_a)\psi_s(\r_a) \EEQ
\BO Toon aan dat een inkomende vlakke golf verkregen kan worden uit een
puntbron in $z=-\infty$ en ook uit een vlak met puntbronnen in $z=-\infty$.
\EO
De ladderbenadering van de Bethe-Salpetervergelijking heeft de vorm
\BA \label{BSeq}
\Phi(\r)&=&\int d^3\r_a d^3\r_b G(\r,\r_a){\overline G}(\r,\r_b)
\psi_s(\r_a){\overline \psi}_s(\r_b)\nonumber \\
&+&\int d^3\r'G(\r,\r'){\overline G}(\r,\r')n(\r')\bar
t(\r')t(\r')\Phi(\r') \EA
De eerste term geeft de onverstrooide, maar door verstrooiing gedeeltelijk
 uitgedoofde, intensiteit weer. De tweede term is de diffuse bundel,
 genereerd door verstrooiing.

We  vermenigvuldigen deze vergelijking met
$\Sigma-\overline{\Sigma}$ en integreren over het volume $V$ waar de
verstrooiers zitten. We gebruiken dat $G$ symmetrisch is, dwz.
$G(\r,\r')=G(\r',\r)$.  Er geldt:
\BEA &&\int_V d^3\r \, G(\r_a,\r)\left\{\Sigma(\r)-
\overline{\Sigma}(\r)\right\}{\overline G}(\r,\r_b)\nonumber \\
&=& \int_V d^3\r \,G(\r_a,\r)\left\{
{\stackrel{\rightarrow}{\nabla}}^2+ k_0^2+\Sigma(\r)-
{\stackrel{\rightarrow}{\nabla}}^2-k_0^2-{\overline \Sigma}(\r) \right\}
{\overline G}(\r,\r_b)\nonumber\\
&=& \int_V d^3\r \, G(\r_a,\r)\left\{
{\stackrel{\leftarrow}{\nabla}}^2+ k_0^2+\Sigma(\r)-
{\stackrel{\rightarrow}{\nabla}}^2-k_0^2-{\overline \Sigma}(\r) \right\}
{\overline G}(\r,\r_b)\nonumber\\
&+&\int_{\delta V}d^2{\bf S}\cdot G(\r_a,\r)
\{ \stackrel{\rightarrow}{\nabla}
  -\stackrel{\leftarrow} {\nabla} \} 
{\overline G}(\r,\r_b)\nonumber\\
&=&\int_V d^3\r\left\{ G(\r_a,\r)\delta(\r-\r_b)-\delta(\r_a-\r)
{\overline G}(\r,\r_b)\right\}\nonumber\\
&+&\int_{\delta V}d^2{\bf S}\cdot G(\r_a,\r)
\{  {\stackrel{\rightarrow}{\nabla}} 
   -{\stackrel{\leftarrow} {\nabla}} \}
{\overline G}(\r,\r_b)
\label{lange} \EEA
Deze vergelijking gebruik je nu twee keer. Voor de bronterm van (\ref{BSeq})
geldt dat de posities van de bron, $\r_a$, {\it buiten} het verstrooiersvolume
$V$ liggen; dientengevolge geeft de eerste term uit de laatste gelijkheid
van (\ref{lange}) geen
bijdrage. Voor de verstrooiingsterm van (\ref{BSeq}) gebruik je ook vgl.
(\ref{lange}) met $\r_a,\r_b\to\r'$. Je krijgt dan
\BEA
&&\int_Vd^3\r \left\{\Sigma(\r)-{\overline \Sigma}(\r)\right\}\Phi(\r)=
\nonumber\\
&&\int d^3\r_ad^3\r_b\int_{\delta V}d^2{\bf S}\cdot \psi_s(\r_a)G(\r_a,\r)
\{{\stackrel{ \rightarrow}{\nabla}}
-{\stackrel{\leftarrow}{\nabla}} \}{\overline G}(\r,\r_b)
{\overline\psi}_s(\r_b)+\nonumber\\
&&\int_Vd^3\r'\left\{\Sigma(\r')-{\overline \Sigma}(\r')\right\}\Phi(\r')+
\nonumber\\
&&\int_Vd^3\r'\int_{\delta V}d^2{\bf S}\cdot G(\r',\r)
\{{\stackrel{ \rightarrow}{\nabla}}-{\stackrel{\leftarrow}{\nabla}}\}
{\overline G}(\r,\r')n(\r')\bar t(\r')t(\r')\Phi(\r')
\EEA
De volumetermen vallen weg tegen elkaar. De oppervlaktetermen kunnen
geschreven worden als
\BEQ -\int_{\delta V}d^2{\bf S}\cdot \left<\Psi(\r)\right>
\{{\stackrel{ \rightarrow}{\nabla}}-{\stackrel{\leftarrow}{\nabla}}\}
\left<{\overline \Psi}(\r)\right>=
\int_{\delta V}d^2 {\bf S}\cdot \left<\Psi(\r)
\{{\stackrel{ \rightarrow}{\nabla}}-{\stackrel{\leftarrow}{\nabla}} \}
{\overline \Psi}(\r)\right>_{ladder} \EEQ
Dit beschrijft dat de netto inkomende flux (dwz. de binnenkomende minus
de onverstrooid uitgaande flux) precies gelijk is aan de uitgaande diffuse
flux, ongeacht de precieze geometrie of de precieze locatie van de bronnen!

Voor onze plak hebben we de zelfenergie onafhankelijk van de positie gekozen.
Een preciezere beschouwing levert dat de zelfenergie positieafhankelijk wordt
voor afstanden van de orde van \'e\'en golflengte van de rand. Dit leidt tot
een veel moeilijker probleem. In het zwakke-localisatieregiem $1/k\ell\ll 1$
kan dit effect verwaarloosd worden. Zoals boven gevonden, dient men dan wel de
transmissie- en reflectieco\"efficienten in {\it nulde} orde in $1/k\ell$ te
nemen.

De hier behandelde behoudswet heeft alleen betrekking op de ladderdiagrammen.
Waar de intensiteit van de terugstrooikegel vandaan komt, is een nog steeds
niet beantwoorde vraag.\footnote{Meint van Albada verwacht dat de achtergrond
iets kleiner wordt ten gunste van de kegel.\index{Meint} (M.P. van Albada,
priv\'ediscussie, 1989-1993)} \footnote{Men kan proberen deze vraag te
beantwoorden door de {\it minimale} set van zelfenergiediagrammen te
beschouwen (de `hendel-diagrammen') die de {\it maximaal} gekruiste
diagrammen in de vertex genereren.  Zie figuur \ref{hendeldiag} (Th.M. N., niet-gepubliceerd 1991).}

\begin{figure}
\caption{Hendeldiagrammen, de maximaal gekruiste bijdragen tot de zelfenergie.}
\centerline{\includegraphics[width=10cm]{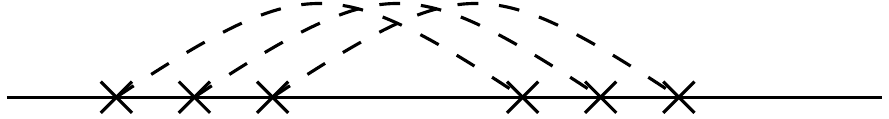}}
\label{hendeldiag}
\end{figure}

\renewcommand{\thesection}{\arabic{section}}
\section{De terugstrooikegel\index{terugstrooikegel}}
\setcounter{equation}{0}\setcounter{figure}{0}
\label{terugstrooikegel}

Zoals besproken in de inleiding is de terugstrooikegel een in het oog
springend zwakke-localisatie-effect. Hij wordt veroorzaakt door de 
interferentietermen tussen paden en hun tijdsomgekeerde equivalenten. Als
je precies in de terugstrooirichting kijkt is er een {\it extra factor twee} 
voor zulke paden. Naast de terugstrooirichting is er gedeeltelijk destructieve
interferentie; dat uit zich in het massief worden van propagatoren en het
verdwijnen van dit effect.
We volgen hier de discussie van ref. [NL].
\subsection{De Milne kern\index{Milne kern!terugstrooi}}

We stralen licht in een halfoneindig verstrooiend medium. In de
terugstrooirichting, de
richting van waaruit we het licht instralen, zit een piek in de intensiteit
van het uit het medium tredende licht. Dit komt doordat in
de terugstrooirichting alle paden in fase zijn met hun
tijdsomgekeerde\index{tijdsomgekeerde paden} pad, wat
een extra bijdrage in de intensiteit geeft. Deze extra bijdrage 
is de verzameling van maximaal gekruiste
ladderdiagrammen\index{diagrammen!maximaal gekruiste}, waarin de golffunktie
het tijdsomgekeerde pad doorloopt van de complex toegevoegde golffunktie.
Bij de eerste verstrooier in het medium geldt
\index{$kin@${\bf k}_{in}$, inkomende golfvector}
\index{$kuit@${\bf k}_{uit}$, uitgaande golfvector}
\beq  {\bf k}_{\rm in} = k_1(\sin\theta_{a}\cos\phi_{a},
\sin\theta_{a}\sin\phi_{a},
\cos\theta_{a})\equiv ({\bf k}_{\perp}^a, p_a 
) \;,
\eeq \beq  {\bf k}_{\rm uit} = k_1(\sin\theta_{b}\cos\phi_{b},
\sin\theta_{b}\sin\phi_{b},
\cos\theta_{b}) \equiv ({\bf k}_{\perp}^{b},p_b 
)\;.\eeq 
Dit geeft voor de intensiteit
\beq  \Psi^{\ast}\Psi =\frac{p_aT_a}{P_a}
 {\rm e}^{-i{\bf k}_{b}^{\ast}\cdot{\bf r}}\cdot
{\rm e}^{i{\bf k}_{a}\cdot{\bf r}} = \frac{p_aT_a}{P_a}{\rm e}^{i({\bf k}^{a}_{\perp}-{\bf k}^{b}
_{\perp})\cdot\roo + i(p_a-p_a^*)z 
}\;,\eeq
met $\roo\equiv (x,y)$.
We defini\"{e}ren  \({\bf Q}\index{$Qv@${\bf Q}$, $Q$,
 genormeerde transversale impuls}
\equiv \ell ({\bf k}^{a}_{\perp}-{\bf
k}^{ b}_{\perp})\), $Q$ is de genormeerde transversale impuls.
 We beschouwen loodrecht inval
($\theta_a=0$; ${\bf k}^{a}_{\perp}={\bf 0}$)
en kijken in het gebied rond loodrecht terugstrooiing (\(\theta_b\simeq0\)).
Dan geldt
\beq \frac{|{\bf Q}|}{\ell}\simeq k_{1}\theta_{b}\eeq
en
\beq  \Psi^{\ast}\Psi =\frac{T(1)}{m}
\eexp{i{\bf Q}\cdot\roo/\ell} \eexp{-z/\ell}\;.\eeq
We kijken nu naar de diffuse intensiteit $I$ in de
terugstrooipiek. Deze wordt gegeven door de volgende vergelijking
\beq I({\bf r}) =\frac{T(1)}{m}
\eexp{i{\bf Q}\cdot\roo/\ell} \eexp{-z/\ell} +
\frac{4\pi}{\ell}\int 
d{\bf r}\, '|G({\bf r}-{\bf r}\, ')|^{2} I({\bf r}\, ')\;.\eeq
Merk op dat de Q-afhankelijkheid alleen in de bronterm van de 
integraalvergelijking zit. Door te itereren komt zij echter in de
kern terecht. Dit ziet men als volgt: Invullen
van $I({\bf r})={\rm e}^{i{\bf Q}\cdot\roo/\ell}I(z,Q)$ geeft de
Milne vergelijking voor $I(z,Q)$
\beq I(z,Q) =\frac{T(1)}{m}{\rm e}^{-z/\ell}
 + \int \frac{dz'}{\ell}M_C(z,z',Q) I(z',Q);,\eeq
met de Q-afhankelijke Milne kern\index{$Mc@$M_C$, Milne kern kegel}
\beq M_C(z,z',Q) = 4\pi\int d^{2}\rho '{\rm e}^{i{\bf Q}{\bf\cdot}(\roo\,
'-\roo)/\ell}|G({\bf r},{\bf r}\,')|^{2}\;.\eeq
De genormeerde intensiteit van de maximaal gekruiste diagrammen voldoet
aan\BA \Gamma_C(\tau,Q)&=& \frac{m I(z,Q)}{T(1)} \nonumber \\
\Gamma_C(\tau,Q)&=& \eexp{-\tau} +\int d\tau' M_C(\tau,\tau',Q)
\Gamma_C(\tau',Q) \EA
De Milne kern \BEQ M_C(\tau,\tau',Q)=M_B(\tau,\tau',Q)+M_L(\tau,\tau',Q)
\EEQ heeft een bulk bijdrage $M_{B}$ en een layer bijdrage $M_{L}$. De
`bulk'-bijdrage is afkomstig van de `ladings'-term van de Greense funktie, de
`laag'-bijdrage van de `spiegelladings'-term.
Invullen van \[|G({\bf r})|^{2}=\frac{{\rm e}^{-r/\ell}}{(4\pi
r)^{2}}\] en \[ \roo-\roo' \rightarrow \roo\] geeft
\beq M_{\rm B}(z,z',Q) = 4\pi\int\rho d\rho d\phi 
\frac{1}{(4\pi)^{2}((z-z')^{2}+\rho^{2})}{\rm e}^{i Q \rho\cos\phi/\ell}
{\rm e}^{-\sqrt{(z-z')^{2}+\rho^{2}}/\ell}
\label{mbulk}\;.\eeq
Ga over op de variabelen \( \tau = z/\ell \) en \( \mu = \cos\theta'
\) , met $\theta'$ de hoek van ${\bf \rho}$ t.o.v. de z-as.
Dan geldt 
\[ \frac{\sqrt{(z-z')^{2}+\rho^{2}}}{\ell} = \frac{|\tau-\tau'|}{\mu}\;
\mbox{  ,   }
\frac{\rho d\rho}{\ell^{2}}= -\frac{|\tau-\tau'|^{2}}{\mu^{3}}d\mu\;. \]
Dit invullen in (\ref{mbulk}) geeft
\beqa
M_{\rm B}(|\tau - \tau'|,Q)=\int^{1}_{0}\frac{d\mu}{2\mu}\int^{\pi}_{-\pi}d\phi\; 
{\rm e}^{iQ\cos\phi|\tau-\tau'|\sqrt{\mu^{-2}-1}}{\rm
e}^{-|\tau-\tau'|/\mu} \nonumber \\
=
\int^{1}_{0}\frac{d\mu}{2\mu}
{\rm J}_0\left(Q|\tau-\tau'|\sqrt{\mu^{-2}-1}\right)
\; {\rm e}^{-|\tau-\tau'|/\mu} \;,
\label{mb}
\eeqa
met J$_0$ de nulde orde Besselfunktie.

Een gelijksoortige berekening geeft voor de oppervlakte term
\beq
M_{\rm L}(\tau + \tau',Q) =  
\int_{0}^{1}\frac{d\mu}{2\mu}R(\mu){\rm J}_{0}\left(Q(\tau+\tau')
\sqrt{\mu^{-2}-1}\right){\rm e}^{-(\tau+\tau')/\mu} \;.
\label{ml} \eeq
                         
\subsection{Vorm van de terugstrooikegel}
De intensiteit in de terugstrooirichting is opgebouwd uit twee bijdragen : een 
achtergrondbijdrage $A^{\rm R}$ van de diffuse
 gereflecteerde intensiteit en een bijdrage $A^{\rm C}$ van de
  maximaal gekruiste diagrammen.
Bij loodrechte inval hebben we voor de achtergrondbijdrage (\ref{mcw5})
\beq
A^{\rm R}(0,0) = \frac{T(1)^{2}\gamma(1,1)}{4\pi m^{2}}\;,
\index{$Ar@$A^R$, diffuse reflectie}\eeq
waarbij de amplitude $\gamma$ gegeven wordt door
\beqa
\gamma(\mu_{a},\mu_b) &  = & \int^{\infty}_{0}d\tau\,{\rm
e}^{-\tau/\mu_{b}} \Gamma_{\rm C}(\tau,\mu_{a}) \label{GrGamma}  \\
  &=& \int^{\infty}_{0}d\tau \,d\tau'\,{\rm e}^{-\tau/\mu_{b}\: - \:\tau'/\mu_{a}}G_{\rm S}(\tau,\tau')\;,
\label{gamma}   
\eeqa
met $G_{\rm S}(\tau,\tau')$ de oplossing van de Milne vergelijking met 
$\delta(\tau-\tau)$ als bronterm.
De bijdrage van de maximaal gekruiste diagrammen, genormeerd op de
diffuse achtergrond, wordt gegeven door \beq
A^{\rm C}(Q) = \frac{\gamma_{\rm C}(Q)- \gamma_{\rm TR}}{\gamma(1,1)}
\;,
\index{$Ac@$A^C$, bijdrage maximaal gekruisten}\eeq
waarin $\gamma_{\rm TR}$ de amplitude is van de paden die hun eigen 
tijdsomgekeerde (`time-reversed') zijn en die dus geen extra 
bijdrage in de terugstrooiing geven. Bij lage dichtheid is voor
 $\gamma_{\rm TR}$ alleen de direkte terugstrooiing aan \'{e}\'{e}n 
verstrooier van belang.\index{$$ctr@$\gamma_{TR}$, direkte
terugstrooi} \index{$$cc@$\gamma_C$, terugstrooi} (\ref{gamma}) geeft
\beq
\gamma_{\rm TR} = \int^{\infty}_{0}d\tau\,d\tau'\,{\rm 
e}^{-\tau-\tau'}\delta(\tau-\tau') = \frac{1}{2}\;,
\eeq
\beq
\gamma_{\rm C}(Q) =\int_0^\infty d\tau\eexp{-\tau}\Gamma_C(\tau,Q)=
 \int^{\infty}_{0}d\tau\,d\tau'\,{\rm 
e}^{-\tau-\tau'}G_{\rm C}(\tau,\tau',Q)\;,
\label{gammac}
\eeq
waarbij $G_C$ naar nul gaat voor $\tau\to\infty$ en voldoet aan
\beq 
G_{\rm C}(\tau,\tau';Q)= \delta(\tau-\tau') + \int^{\infty}_{0}d\tau''\,M_{\rm 
C}(\tau,\tau'';Q)G_{\rm C}(\tau'',\tau';Q)\;.
\label{Milne}
\index{$Gc@$G_C$, terugstrooi Greense functie}\eeq

De Milne kern\index{Milne kern!terugstrooi} voor de maximaal gekruiste
diagrammen $M_{\rm C}$ wordt
gegeven door $M_{\rm B} + M_{\rm L}$ uit de vorige paragraaf.
In figuur \ref{jmlac} wordt het numerieke resultaat voor $A^C(Q)$
weergegeven bij verschillende $m$.
\begin{figure}
\caption{De terugstrooikegel genormeerd op de achtergrond, bij loodrechte
inval op een halfoneindig medium, voor verschillende
brekingsindexverhoudingen.} 
\centerline{\includegraphics[width=10cm]{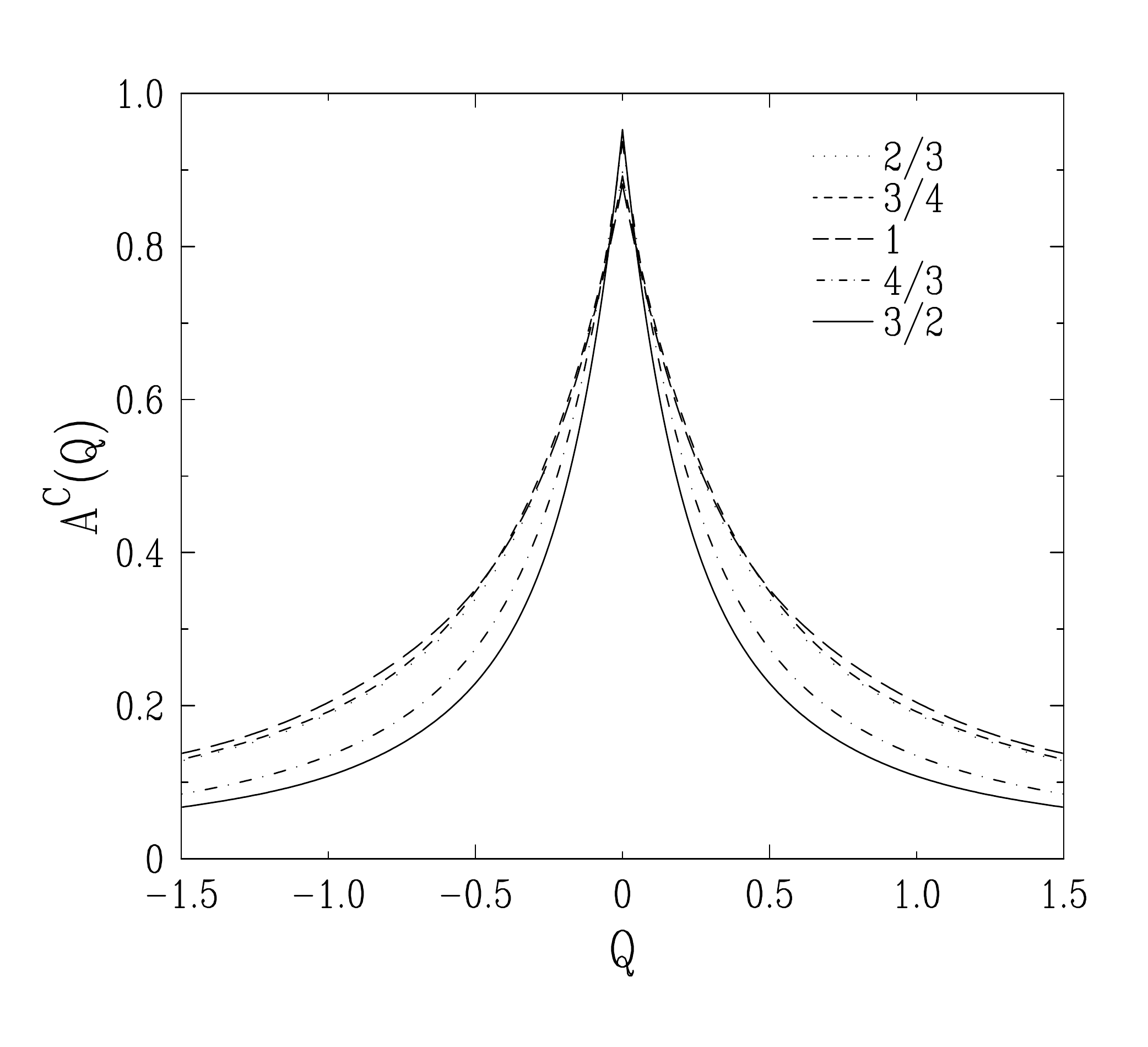}}

\label{jmlac} \end{figure}

\subsection{Gedrag voor grote hoeken}
Uit (\ref{mb}) en (\ref{ml}) volgt dat $M_{\rm B}$ en $M_{\rm L}$
voor grote $Q$ snel
afvallen als funktie van $\tau$. We kunnen daardoor (\ref{gammac})
benaderen
tot lage orde verstrooiing. Benaderen tot eerste orde verstrooiing geeft 
$\gamma_{\rm C}=\gamma_{\rm TR} = \frac{1}{2}$.
 We benaderen tot tweede orde door $G_{\rm C}(\tau,\tau';Q) = 
 \delta(\tau-\tau') + M_{\rm C}(\tau,\tau';Q)$ te nemen.
Dit geeft
\beqa
\gamma_{\rm C}(Q)= \frac{1}{2} + \int^{\infty}_{0}d\tau\,d\tau'\,{\rm 
e}^{-\tau-\tau'}(M_{\rm B}(|\tau-\tau'|;Q)+M_{\rm L}(\tau+\tau';Q))\;.
\label{gambml} \eeqa
We substitueren in de volumeterm $\sigma = \tau-\tau'$ en voeren de
integratie over $\tau'$ uit.
\beqa
\int^{\infty}_{0}d\tau\,d\tau'\,{\rm 
e}^{-\tau-\tau'}M_{\rm B}(|\tau-\tau'|;Q) \!&=& \!
\int^{\infty}_{0}d\sigma\,{\rm e}^{-\sigma}M_{\rm B}(\sigma,Q) \nonumber \\
\!&\stackrel{(\ref{mb})}{=}&  \!
\int_0^1 \frac{d\mu}{2\mu} \int_{-\pi}^\pi \frac{d\phi}{1+1/\mu-iQ
\cos \phi\sqrt{1/\mu^2-1}} \nonumber \\
\!&=& \!
\int^{1}_{0}\frac{d\mu}{2}\frac{1}{\sqrt{(\mu+1)^{2}+Q^{2}(1-\mu^{2})}}
\nonumber  \\
 &\stackrel{Q groot}{\simeq}& \int^{1}_{0}\frac{d\mu}{2}\frac{1}{Q\sqrt{1-\mu^{2}}} \nonumber \\
\!& =&\! \frac{\pi}{4Q}\;.
\eeqa
Waarbij gebruik gemaakt is van de identiteit: 
\beqa
\int^{\pi}_{-\pi}\frac{d\phi}{2\pi}\frac{1}{A+B\cos\phi} = 
\frac{1}{\sqrt{A^{2}-B^{2}}}
\label{ident2}
\eeqa
Invullen van (\ref{ml}) geeft voor de randlaagterm in (\ref{gambml})
\beqa
\int^{\infty}_{0}d\tau\,d\tau'\,{\rm e}^{-\tau-\tau'}M_{\rm L}(\tau+\tau';Q)
\nonumber \\ = \int^{1}_{0}\frac{d\mu}{2}\int^{\pi}_{-\pi}
\frac{d\phi}{2\pi}R(\mu)\frac{\mu}{(\mu+1-iQ\cos\phi\sqrt{1-\mu^{2}})^{2}}\;.
\eeqa
\BO
Leid uit (\ref{ident2}) af:
\beq
\int^{\pi}_{-\pi}\frac{d\phi}{2\pi}\frac{1}{(A+B\cos\phi)^{2}} = 
\frac{A}{(A^{2}-B^{2})^{3/2}}.
\label{ident.2} \eeq
\EO
Dit geeft
\beq
\frac{1}{2}\int^{1}_{0}d\mu\,\mu 
R(\mu)\frac{\mu+1}{[(\mu+1)^{2}+Q^{2}(1-\mu^{2})]^{3/2}}
\eeq
De integrand gaat snel naar nul voor $Q^{2}(1-\mu^{2})$ groot. We ontwikkelen 
$\mu$ rond $\mu=1$, we stellen  $\mu=1-\frac{x}{Q^{2}}$.
Hiermee wordt de randlaagterm
\beqa
\frac{R(1)}{Q^{2}}\int^{\infty}_{0}\frac{dx}{(4+2x)^{3/2}} = 
\frac{R(1)}{2Q^{2}} \;,
\eeqa
waarbij gebruikt is $\int^{Q}_{0} \simeq \int^{\infty}_{0} $.
Dit geeft uiteindelijk
\beq
A^{\rm C}(Q) \simeq \frac{T(1)^{2}}{4\pi m^{2}}\left(\frac{\pi}{4Q} + 
\frac{R(1)}{2Q^{2}}\right)
\eeq
voor Q groot. De randlaagterm is in essentie het kwadraat van de bulkterm.
Dit komt doordat de betreffende paden in essentie twee keer zo lang zijn,
zie ook figuur \ref{back2}. \BO
Bereken de $1/Q^2$ correctie ten gevolge van $M_B$ en alle $1/Q^3$
correcties. Toon
aan dat het precieze gedrag is:
\BEA
A^{\rm C}(Q)&=&\frac{T(1)^2}{4\pi m^2} \left[
\frac{\pi}{4Q}+\frac{1}{Q^2} \left(
\frac{(m-1)^2}{2(m+1)^2} -1 \right) \right. \nonumber\\
&& \left. +\frac{1}{Q^3}\left(
\frac{\pi}{8}+\frac{1}{2}+\frac{(m-1)^2}{(m+1)^2}
(2m-\frac{\pi}{4}-\frac{1}{2}) \right) \right] \EA
\EO
\begin{figure}[htb]
\caption{De terugstrooikegel voor grote hoeken wordt voornamelijk door
tweede orde verstrooiing bepaald. De term met interne reflecties heeft een
langer pad en valt daarom sneller af.}
\centerline{\includegraphics[width=9cm]{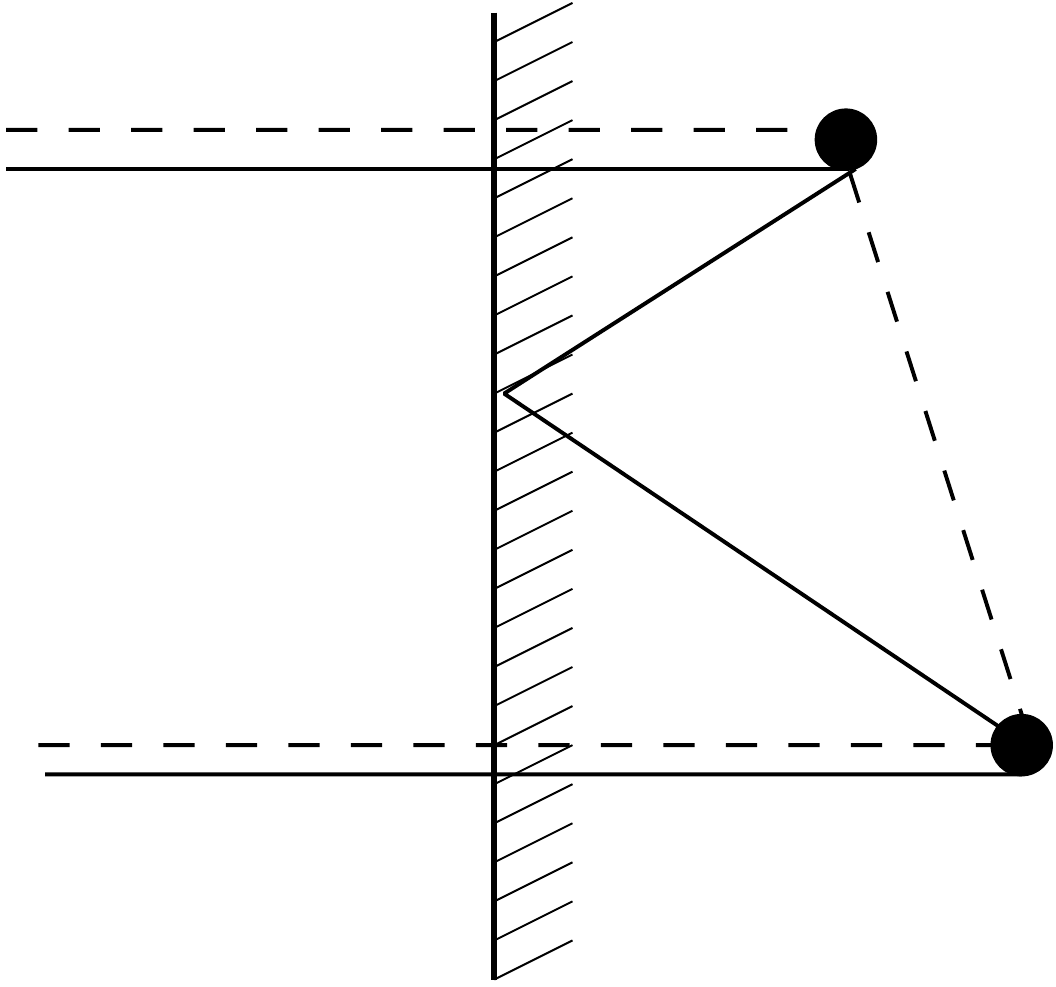}}
\label{back2} \end{figure}

\subsection{Gedrag voor kleine hoeken}
We ontwikkelen $\Gamma_{\rm C}(\tau;Q)\equiv m I(z;Q)/T(1)$ rond $Q = 0$
onder de aanname dat dit lineair gaat \beq
\Gamma_{\rm C}(\tau;Q) \simeq \Gamma_{\rm C}(\tau;Q=0)-Q\tilde{\Gamma}_{\rm 
C}(\tau;Q)
\label{rondQ=0}
\eeq
Voor $\Gamma_{\rm C}$ geldt (\ref{milne})
\beq
\Gamma_{\rm C}(\tau;Q) = {\rm e}^{-\tau} + \int^{\infty}_{0}d\tau'\,M_{\rm 
C}(\tau,\tau';Q)\Gamma_{\rm C}(\tau';Q)
\eeq
Hierin (\ref{rondQ=0}) invullen en differenti\"{e}ren naar $Q$ geeft de
vergelijking voor $\tilde{\Gamma}_{\rm C}$
\beq
\tilde{\Gamma}_{\rm C}(\tau;Q) = \int^{\infty}_{0}d\tau'\,M_{\rm C}(\tau,
\tau';Q)\tilde{\Gamma}_{\rm C}(\tau';Q) \;.
\eeq
Er geldt dus
\beq
\tilde{\Gamma}_{\rm C}(\tau;Q) \sim \Gamma_{\rm H}(\tau;Q)\;.
\eeq
(\ref{rondQ=0}) wordt hiermee
\beqa
\Gamma_{\rm C}(\tau;Q) & =& \Gamma_{\rm S}(\tau) - \alpha Q
\Gamma_{\rm H}(\tau) 
\nonumber \\
& \stackrel{(4.8)}{=}& \tau_{1}- \alpha Q(\tau + \tau_{0}) \nonumber \\
& \stackrel{Q klein}{\simeq}& \tau_{1} - \alpha Q\tau\;. \label{alfa}
\eeqa      
Diep in het verstrooiende medium ($\tau \gg 1$) moet aan de diffusiebenadering 
voldaan zijn. Dus
\beqa
 &\Gamma_{\rm C}''(\tau;Q) = Q^{2}\Gamma_{\rm C}(\tau;Q)  \nonumber \\
 \Rightarrow &  \Gamma_{\rm C}(\tau;Q) = \tilde{\alpha}{\rm e}^{-Q\tau}\;,
\eeqa
voor $\tau \gg 1$.
Vergelijken we dit met (\ref{alfa})  dan volgt
\beqa
\tilde{\alpha} = \alpha = \tau_{1} \;.
\eeqa
Met (\ref{GrGamma}) geeft dit uiteindelijk
\beqa
\gamma_{\rm C}(Q) & = \gamma_{\rm C}(0) - 3Q\tau_{1}^{2}(1) \nonumber \\
& = \gamma_{\rm C}(0)\left(1-\frac{Q}{\Delta Q}\right)\;,
\eeqa
waarbij gebruikt is (\ref{taueen}) met $D= \frac{1}{3}$. De genormeerde
openingshoek $\Delta Q$ wordt gegeven  door \beq
\Delta Q = 3\frac{\gamma_{\rm C}(0)}{\tau_{1}^{2}(1)}
\index{$$Dq@$\Delta Q$, genormeerde openingshoek}\eeq

Numerieke berekeningen geven de volgende resultaten:
\vspace{.5cm}

\begin{tabular}{|c|c|c|c|c|}
\hline $m$ & $\tau_0$ &$\tau_1(1)$ & $\gamma(1,1)$ &$\Delta Q$ \\ \hline
\hline 2 &6.08 &21.7 &21.5 &0.136 \\ \hline
3/2 &2.50 &10.8 &10.6 &0.269 \\ \hline
4/3 &1.69 &8.34 &7.94 &0.343 \\ \hline
1 &0.710446 &5.03648 &4.22768 &1/2 \\ \hline
3/4 &0.815 &5.39 &4.63 &0.479 \\ \hline
2/3 &0.881 &5.60 &4.85 &0.465 \\ \hline
1/2 &1.09 &6.25 &5.55 &0.427 \\ \hline
\end{tabular}


\subsection{Toepassing in de astronomie}

E\'en van de toepassingen van de versterkte terugstrooikegel betreft het oppositie\"effect van de maan, Mars en kometen:
die blijken helderder wanneer de zon precies achter ons staat (`opposition surge', oppositie-aanzwelling).

Voor de uitwerking betreffende de manen van Jupiter, zie \cite{MLN}.

\renewcommand{\thesection}{\arabic{section}}
\section{Groot verschil in brekingsindices\index{brekingsindexcontrast!groot}} 
\setcounter{equation}{0}\setcounter{figure}{0}

\label{groot}

Aan de rand van een diffuus verstrooiend medium treden reflecties op als
de brekingsindices $n_1$ buiten het medium en $n_0$ binnen het medium
verschillen. In het algemeen leidt dat er toe dat we de \smv
niet meer analytisch kunnen oplossen. Voor grote verschillen in
brekingsindex,
$m \equiv n_0/n_1 \to \infty$ of $ m \to 0 $ kunnen we dat echter wel. Dit
komt doordat in
zulke situaties de wanden goede spiegels zijn. Hierdoor wordt de intensiteit
bijna constant.

\subsection{De \smv}
In navolging van ref. [NL]
 beschouwen we nu de
Schwarzschild-Milne\index{Schwarzschild-Milne vergelijking} vergelijking in
het
regime van grote verschillen in brekingsindices ($m \to 0 , m \to \infty$).
Er geldt hier voor de
reflectieco\"effici\"ent $R(\mu)=1-T(\mu)$ waarbij de 
transmissieco\"effici\"ent
$T(\mu)$ klein is. We ontwikkelen de oplossing in $T$. We bekijken eerst
de Greense functie van de \smv en de bijbehorende Milne
kern\index{Milne kern} : \begin{eqnarray}
G_S(\tau,\tau')&=& \delta(\tau-\tau') + \int_0^\infty \dint \tau''
 M(\tau,\tau'')G_S(\tau'',\tau')\\
M&=& M_{\rm B}+M_{\rm L}\\ 
M_{\rm B} &=& \int_0^1 \frac{\dint \mu}{2\mu} \eexp{-\frac{|\tau -\tau'|}{\mu}}
\nonumber\\
M_{\rm L} &=& \int_0^1 \frac{\dint \mu}{2\mu} \eexp{-\frac{\tau +\tau'}{\mu}}
(1-T(\mu))\equiv M_{\rm B}(\tau+\tau'')-N(\tau+\tau'')\nonumber\\ \label{definN}
N&=&\int_0^1 \frac{\dint \mu}{2\mu} \eexp{-\frac{\tau+\tau''}{\mu}}T(\mu)
\index{$N$, verschilkern}\EA
$N$ is van de orde van $T$, dus een kleine constante, geschikt
om in te ontwikkelen. Hiertoe herschrijf je de integraalvergelijking als
\begin{eqnarray}
\label{gsn}
G_S(\tau,\tau')&=& \delta(\tau-\tau') + \int_0^\infty \dint \tau''
 \{M_{\rm B}(\tau-\tau'')+M_{\rm B}(\tau + \tau'')\}G_S(\tau'',\tau')
 \nonumber \\
&& -\int_0^\infty \dint \tau'' N(\tau+\tau'') G_S(\tau'',\tau')
\end{eqnarray}
Stel dat $N=0$ (dit gebeurt als we de limiet $m \to 0$ of $m \to \infty$ nemen),
dan heeft het rechterlid van vergelijking (\ref{gsn}) een eigenwaarde $a=1$
voor $G_S = const$,
d.w.z. bij een oplossing van (\ref{gsn}) kan een willekeurige constante
opgeteld worden en dit levert weer een oplossing. Dit geeft het feit weer
dat in een ruimte met perfect spiegelende wanden de intensiteit niet
afhankelijk is van randcondities { \em buiten} deze ruimte.
Voor kleine $N$ zal de waarde van $N$ grote invloed hebben op deze constante.
We schrijven, voor $N \ll 1$, de Greense functie als
\begin{eqnarray}
\label{schrijfwijze}
G_S(\tau,\tau')&=&C_S+G_0(\tau,\tau')+G_1(\tau,\tau')
\\ \nonumber
&&C_S \gg 1 ,\;G_0 \approx 1 ,\; G_1 \ll 1
\end{eqnarray}\index{$C_S$, constante term} \index{$G_0$, eerste correctie
op $C_S$} \index{$G_1$, tweede correctie op $C_S$} We vullen dit in in
(\ref{gsn}) en vergelijken termen van dezelfde orde : \begin{eqnarray}
C_S+G_0+G_1&=& \delta(\tau-\tau')+C_S\nonumber \\ &&+\int_0^\infty\dint
\tau''
\{M_{\rm B}(\tau -\tau'') + M_{\rm B}(\tau+\tau'')\}(G_0+G_1) \nonumber \\
&& -\int_0^\infty \dint \tau'' N(\tau+\tau'')(C_S+G_0+G_1)
\end{eqnarray}
Vergelijken van termen van dezelfde orde geeft
\begin{eqnarray}
C_S&=& C_S
\\ \nonumber
G_0(\tau,\tau')&=&\delta(\tau-\tau')+\int_0^\infty \dint \tau''
\{M_{\rm B}(\tau -\tau'') + M_{\rm B}(\tau+\tau'')\}G_0(\tau'',\tau')
\nonumber \\ &&-C_S\int_0^\infty \dint \tau''N(\tau+\tau'')
\\ \nonumber
G_1(\tau,\tau')&=&\int_0^\infty \dint \tau''
\{M_{\rm B}(\tau -\tau'') + M_{\rm B}(\tau+\tau'')\}G_1(\tau'',\tau')
\nonumber \\
&&-\int_0^\infty \dint \tau''N(\tau+\tau'')G_0(\tau'',\tau')
\end{eqnarray}
Merk op dat we de term van orde $N G_1\sim {\cal T}^2 $ verwaarloosd
hebben. Integreren van deze resultaten over $\tau$ geeft:
\begin{eqnarray}
\int_0^\infty \dint \tau G_0(\tau,\tau') &=& 1 + \int_0^\infty \dint \tau
G_0(\tau,\tau')  -C_S\int_0^\infty \int_0^\infty \dint \tau
\dint \tau'' N(\tau +\tau'') \nonumber \\
\int_0^\infty \dint \tau G_1(\tau,\tau') &=& \int_0^\infty \dint \tau G_1(\tau,\tau')
-\int_0^\infty \dint \tau \dint \tau'' N(\tau +\tau'') G_0(\tau'',\tau')\,\,\,
\end{eqnarray}
Dit levert de uitdrukking voor $C_S$ op en een conditie op $G_0$ om de
consistentie van (\ref{schrijfwijze}) te garanderen:
\begin{eqnarray}
C_S=\left\{\int_0^\infty \int_0^\infty \dint \tau \dint \tau'' N(\tau +\tau'') \right\}
^{-1} \\ \nolabel
\int_0^\infty \dint \tau \dint \tau'' N(\tau +\tau'') G_0(\tau'',\tau')=0
\; \mbox{     voor alle } \tau' \end{eqnarray}
We kunnen de uitdrukking voor $C_S$ vereenvoudigen door invoering van
de hoekgemiddelde flux\-\transco
$\tcrl$\index{$T$, transmissieco\"efficient}
\index{transmissie!hoekgemiddelde}: \begin{eqnarray}
C_S&=& \left\{ \int_0^\infty \int_0^\infty \dint \tau \dint \tau''
\int_0^1 \frac{\dint \mu}{2\mu} \eexp{-\frac{\tau+\tau''}{\mu}}T(\mu)
\right\} ^{-1}
\\ \nonumber
&=& 2 \left\{
\int_0^1 \mu \dint \mu T(\mu)
\right\} ^{-1} \equiv \frac{4}{\tcrl}
 \nonumber
\EA
waarbij
\BA
\tcrl&\equiv &2\int_0^1 \mu \dint \mu T(\mu)
= 2\int_0^{\pi/2} \cos \theta \sin \theta\, \dint \theta\, T(\cos
\theta) \end{eqnarray}
Merk op dat de factor $\mu=\cos \theta$ hier staat voor de
fluxprojectiefactor
in de $z-$richting. De integraal voor $\tcrl$ kunnen we uitwerken met behulp
van de uitdrukking voor de \transco $T(\mu)$,
\beatrix
T(\mu)&=&\frac{4\mu \sqrt{\mu^2-1+m^{-2}}}{(\mu+\sqrt{\mu^2-1+m^{-2}})^2}
\qquad (\mu>0)\claus
waarbij we de gevallen $m>1$ en $m<1$ apart behandelen. Voor $m<1$, dus de
interne reflectie van een medium met lagere brekingsindex dan de
buitenwereld, introduceren we $\epsilon^2 = m^{-2} -1$ als hulpconstante:
\beatrix
\tcrl&=&2\int_0^1 \mu \dint \mu
\frac{4\mu \sqrt{\mu^2+\epsilon^2}}{(\mu+\sqrt{\mu^2+\epsilon^2})^2}
\label{mvr2} \claus
Dit wordt met de substitutie $\phi$$=$$\sinh^{-1}({\mu}/{\epsilon})$
\beatrix
\\ \nolabel
\tcrl&=&8\epsilon^2\int_0^{\phi_+}\dint \phi
\frac{\sinh^2 \phi \cosh^2\phi}{\eexp{2\phi}}
\\ \nolabel
&=& 2\epsilon^2 \int_0^{\phi_+} \dint \phi \sinh^2 2\phi \eexp{-2\phi}
=\frac{\epsilon^2}{2} \int_0^{\phi_+} \dint \phi
\left[ \eexp{2\phi}-2\eexp{-2\phi}+\eexp{-6\phi} \right] \\ \nolabel
&=&\frac{\epsilon^2}{2}\left\{ \frac{\eexp{2\phi_+}}{2}+\eexp{-2\phi_+}-\frac{\eexp{-6\phi_+}}{6}-\hlf-1+\zsd \right\}
\claus
Gebruik makend van
\BE
\sinh\phi_+=\frac{1}{\epsilon}=\frac{m}{\sqrt{1-m^2}} \mbox{  en }
\cosh\phi_+=\frac{1}{\sqrt{1-m^2}}
\EE
vinden we $\exp{(2\phi_+)}=(1+m)/(1-m)$. Dit levert
\bea
\tcrl&=&\frac{1-m^2}{2m^2} \left\{ \hlf \frac{1+m}{1-m}+\frac{1-m}{1+m}-
\zsd \frac{(1-m)^3}{(1+m)^3}-\frac{4}{3} \right\}
\\ \nolabel
&=&\frac{4m(m+2)}{3(m+1)^2}\qquad \mbox{ voor } \,\,m<1
\claus
Het gedrag voor $m \to 0$ is sneller te vinden door in formule
(\ref{mvr2}) de teller en de noemer tot op leidende orde te nemen :
\beatrix
\tcrl \approx 2\int_0^1 \mu d\mu \frac{4\mu\epsilon}{\epsilon^2}
= \frac{8}{3}m\qquad \mbox{ voor } \,\,m\rightarrow 0 \claus

Nu doen we hetzelfde voor $m>1$. Hierbij moeten we letten op de grenshoek:
voor $\mu < (1 - m^{-2})^{\hlf}$ is er totale reflectie, de \transco is hier
nul. Dit doen we door de hulpconstante $\epsilon^2 = 1-m^{-2}$ te kiezen
en nu de integratie uit te voeren van $\epsilon$ tot $1$:
\beatrix
\tcrl&=&8\int_\epsilon^1 \mu^2 \dint \mu \frac{\sqrt{\mu^2-\epsilon^2}}
{(\mu+\sqrt{\mu^2-\epsilon^2})^2}
\\ \nolabel
&=&\frac{4(2m+1)}{3m^2(1+m)^2}\qquad \mbox{      voor   } m>1
\claus
Het gedrag voor zeer grote $m$ kan weer gevonden worden door alleen de
termen van leidende orde te behouden:
\beatrix
\tcrl \approx \frac{8}{3m^3} \qquad \mbox{ voor} \; m\rightarrow \infty.
\claus
Merk op dat $\T=1$ als $m=1$, zoals het hoort.

\subsection{De diffuus gereflecteerde intensiteit}\index{reflectie!diffuse}
Met behulp van de gevonden waarde voor $\tcrl$ kunnen we eenvoudig het
leidende gedrag van de genormeerde bistatische \cof
$\gamma(\mu_a,\mu_b)$ vinden: \beatrix
\gamma(\mu_a,\mu_b)&=&\int_0^\infty \dint \tau \dint \tau'
G_S(\tau,\tau') \eexp{-\tau/\mu_b} \eexp{-\tau'/\mu_a}
\\ \nolabel
&=& \int_0^\infty \dint \tau \dint \tau'
C_S \eexp{-\tau/\mu_b} \eexp{-\tau'/\mu_a} + {\cal O}(\T^0)
\\ \nolabel
&=& \frac{4 \mu_a\mu_b}{\tcrl} + {\cal O}(\T^0)
\claus
We kunnen ook een benadering vinden voor de gereflecteerde intensi\-teit
$A^{\rm R} (\theta_a,\theta_b)$, in het geval $m\gg1$.
\beatrix
A^{\rm R}&=&\frac{\cos \theta_a}{4\pi m^2} \frac{T_aT_b}{\mu_a\mu_b}
\gamma(\mu_a,\mu_b)
\claus
Voor $m\gg1$ geldt $T(\mu) \approx 4 \mu/m$ (want de hoek $\theta$ kan
niet groter zijn dan de grenshoek, die klein wordt voor grote $m$).
\beatrix
A^{\rm R}&=&\frac{\cos \theta_a}{4\pi m^2} \frac{4 \cos \theta_a}{m}
\frac{4 \cos \theta_b}{m} \frac{4}{\tcrl}
\\ \nolabel
&=& \frac{16 \cos^2 \theta_a \cos \theta_b}{m^4\pi\tcrl}
\\ \nolabel
&=& \frac{16 \cos^2 \theta_a \cos \theta_b}{m^4\pi \frac{8}{3}m^{-3}}
=\frac{6}{\pi m} \cos^2 \theta_a \cos \theta_b
\claus

\subsection{De limietintensiteit\index{limietintensiteit} en de
injectiediepte\index{injectiediepte}}
De constante $\tau_1$, de limiet van de speciale oplossing van de \smv
(zie vergelijking (\ref{gammaasymp})) is ook te berekenen. Uit
hoofdstuk \ref{oplossing} hebben we nog: \beatrix
\tau_1(\mu)&=&\int_0^\infty \dint \tau' G_S(\infty,\tau')\eexp{-\tau'/\mu}
\\ \nolabel
&=&\frac{1}{D} \int_0^\infty \dint \tau' \Gamma_H(\tau')\eexp{-\tau'/\mu}
\\ \nolabel
&=&\lim_{\mu_b \to \infty} \frac{\gamma(\mu,\mu_b)}{\mu_b}
\\ \nolabel
&=& \frac{4\mu}{\tcrl}
\claus
Ook de indringdiepte $z_0=\tau_0 \ell$ kan berekend worden.
Nemen we nu de (diffusie) benadering $\Gamma_H=\tau_0+\tau\approx\tau_0$
 dan vinden we door invullen in bovenstaande vergelijking
\beatrix
\tau_1(\mu)=\frac{\tau_0}{D}\mu&=&\frac{4\mu}{\tcrl}
\\ \label{mcw7}
\tau_0&\approx&\frac{4}{3\tcrl}
\claus
We kunnen nu, net zoals bij de berekening van $G_S$, de homogene oplossing
$\Gamma_H$ schrijven als de som van drie delen van verschillende orde:
\beatrix
\Gamma_H(\tau)&=&\frac{4}{3\tcrl} + \Gamma_0(\tau)+\Gamma_1(\tau)
\\ \nolabel
&& \Gamma_0(\tau)={\cal O}(1),\qquad \Gamma_1(\tau) =\olandau(\tcrl)
\claus
Verder eisen we dat $\Gamma_0$ zich gedraagt als $\tau
+\tau_{00}$\index{$$t00@$\tau_{00}$, constante} voor
$\tau \to \infty$. De constante $\tau_{00}$ is een $\olandau(1)$-correctie
op $\tau_0$
in formule (\ref{mcw7}). Voor $m \to \infty$ kunnen we gebruik maken van de
definitie (\ref{definN})
en van het feit dat $\mu \approx 1$ omdat de grenshoek klein is:
\beatrix
N(\tau + \tau')&=&\int_0^1 \frac{\dint \mu}{2\mu} \eexp{-\frac{\tau+\tau''}{\mu}}T(\mu)
\\ \nolabel
&=&\int_0^1 {2\mu}T(\mu){\dint \mu}\frac{1}{4\mu^2}
\eexp{-\frac{\tau+\tau''}{\mu}}
\\ \nolabel
&\approx& \frac{\tcrl}{4} \eexp{-(\tau+\tau')}
\\ \nolabel
\Gamma_H(\tau)&=&\int_0^\infty \dint \tau' M(\tau,\tau') \Gamma_H(\tau')
\\ \nolabel
&=&\int_0^\infty \dint \tau' [M_{\rm B}(\tau-\tau')+M_{\rm B}(\tau+\tau')]
\Gamma_H(\tau')
\\ \nolabel
&=&-\frac{\tcrl}{4}\eexp{-\tau} \int_0^\infty \dint \tau' \eexp{-\tau'}
\Gamma_H(\tau')
\claus
Nu vullen we de expansie voor $\Gamma_H$ in en vergelijken termen van
gelijke orde, waarbij we de orde $\tcrl^2$ weglaten:
\beatrix
\frac{4}{3\tcrl}+\Gamma_0+\Gamma_1&=&\frac{4}{3\tcrl}+
\int_0^\infty \dint \tau' [M_{\rm B}(\tau-\tau')+M_{\rm B}(\tau+\tau')]
(\Gamma_0+\Gamma_1) \nonumber
\\ \nolabel
&&-\frac{\tcrl}{4}\eexp{-\tau}
[\frac{4}{3\tcrl}+\int_0^\infty \dint \tau'\eexp{-\tau'}\Gamma_0(\tau')]
\\ \nolabel
\Gamma_0&=-&\drd \eexp{-\tau}+\int_0^\infty \dint \tau' 
[M_{\rm B}(\tau-\tau')+M_{\rm B}(\tau+\tau')]
\Gamma_0
\\ \nolabel
\Gamma_1&=&\int_0^\infty \dint \tau' [M_{\rm B}(\tau-\tau')+M_{\rm B}(\tau+\tau')]
\Gamma_1\nonumber \\ &&- \frac{\tcrl}{4}\eexp{-\tau}
\int_0^\infty \dint \tau'\eexp{-\tau'}\Gamma_0(\tau')
\claus
Door de vergelijking voor $\Gamma_1$ te integreren kan men zien dat de laatste
term $0$ moet zijn, dit vormt een conditie op de oplossing voor $\Gamma_0$:
\BE \int_0^\infty d\tau \eexp{-\tau} \Gamma_0(\tau)=0 \label{mcw9}\EE
Deze vergelijkingen kunnen we niet analytisch oplossen. De numerieke
oplossing geeft [NL]
\beatrix
\tau_{00}&=&-1.0357
\\ \label{mcw8}
\tau_0^{\rm exact}&=&\frac{4}{3\tcrl} - 1.0357 \qquad (m\to\infty)
\claus

In de limiet $m\to 0$ is de transmissieco\"efficient klein, $T(\mu)
\approx 3{\cal T}\mu/2$. Daardoor wordt de laatste term in vgl.
(\ref{gsn}) gelijk aan
\BA -\frac{3}{2}{\cal T}\int_0^\infty d\tau''\int_0^1\frac{d\mu}{2\mu}
\eexp{-(\tau+\tau'')/ \mu} \mu\frac{4}{{\cal T}} &=&-3\int_0^1 \mu\, d\mu
\eexp{-\tau/\mu} \label{mcw12} \nonumber \\
\int_0^1 \d\mu' \mu' \int_0^\infty d\tau \eexp{-\tau/\mu'} {\rm J}_0
(\tau,\mu')&=& 0 \nonumber \\
\Rightarrow \int_0^1 \d\mu \mu \int_0^\infty d\tau \eexp{-\tau/\mu}
\Gamma_0(\tau) \EA
\BO
Toon aan dat de diffusieve oplossing $\tau+\tau_{00}$ de exacte oplossing is
van de besproken vergelijking.
Laat zien dat dit leidt tot:
\beatrix
\tau_0^{\rm exact}&=&\frac{4}{3\tcrl}-\frac{3}{4}+{\cal O}(\T)  
\,\;\; \,(m\to 0)
\label{mcw13}
\claus
\EO
In figuur \ref{jmltau} zijn numerieke waarden van $\tau_1(1)$, $\gamma(1,1)$
en $ 3\tau_0$ als functie van $m$ getekend. De doorgetrokken lijn is $4/\T$,
de gestreepte lijn is (\ref{mcw8})+(\ref{mcw13}). Merk op dat deze
asymptotische resultaten goed overeenstemmen, zelfs voor $m$ dichtbij
\'e\'en.

\subsection{Vergelijking met de verbeterde diffusie\-benadering
\index{diffusiebenadering!verbeterde}}
We vergelijken de verkregen exacte resultaten voor $\tau_0$ met die
van de hierboven besproken verbeterde diffusie benadering.
Voor $m\to\infty$ kunnen we de \cofn $C_1,\,C_2$ (zie formule \ref{C1=C2=})
benaderen:
\BE
C_1=\hlf(1-\tcrl) \; ,\; C_2\approx\drd-\int_0^1\mu\dint\mu T(\mu) = \drd-\hlf \tcrl
\EE
Dit geeft met vergelijking (\ref{z0===})
\BE
\tau_0^{\rm diff}=\frac{4}{3\tcrl}-1
\EE
Dit is erg dicht bij de exacte waarde van formule (\ref{mcw8}). Merk op dat
hetzelfde resultaat volgt door de diffusiebenadering 
$\Gamma_0(\tau)=\tau+\tau_{00}$
(voor alle $\tau$) in vergelijking (\ref{mcw9}) in te vullen.

Voor $m=1$, dus geen interne reflecties, is de exacte waarde numeriek
berekend. Wij vonden in hoofdstuk \ref{mesoscopie} uit de
diffusie-benadering: \beatrix
\tau_0^{\rm diff} & = &\frac{2}{3}=0.666666
\\ \nolabel
\tau_0^{\rm exact}&=&0.710449
\claus
\begin{figure}
\caption{Hoe goed is de verbeterde diffusiebenadering? Getekend zijn de
numerieke waarden van $ \tau_1(1) $, $ \gamma(1,1)$
en $ 3\tau_0 $. De doorgetrokken lijn is $ 4/\T $,
de gestreepte lijn is (11.33), (11.35)
Zelfs voor $m$ rond
\'e\'en stemmen asymptotische resultaten goed overeen met de numerieke
waarden. }
\vspace{3mm}
\centerline{\includegraphics[width=10cm]{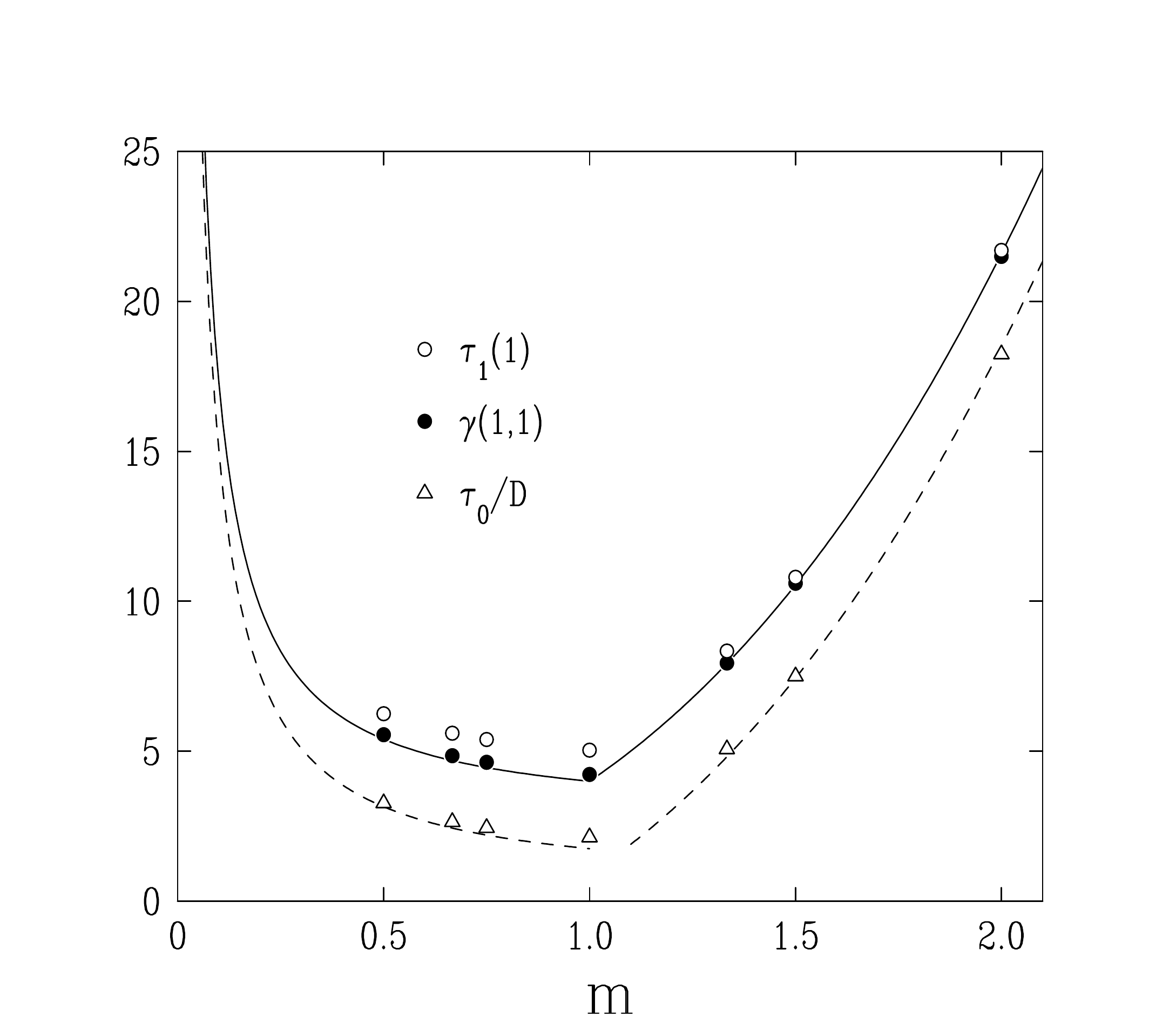}}
 \label{jmltau}
\end{figure}
\typeout{"pas op absolute ref"}

Voor $m \to 0$ vinden we de exacte waarde ook in de diffusie-benadering,
verschillen treden pas op in hogere ordes:
\beatrix
\tau_0^{\rm diff}=\tau_0^{\rm exact}=\frac{4}{3\tcrl}-\frac{3}{4}+
{\cal O}({\cal T})\qquad (m\rightarrow 0) \claus

\subsection{De terugstrooikegel\index{terugstrooikegel!groot
brekingsindexconstrast}}
Nu rekenen we de vorm van de terugstrooikegel in aanwezigheid van
interne reflecties uit, in het regime $m \gg 1$.
\beatrix
\label{back1}
\Gamma_C(Q,\tau)&=&\eexp{-\tau}+
\int_0^\infty \dint \tau' [M_{\rm B}(Q,\tau-\tau')+M_{\rm B}(Q,\tau+\tau')]\Gamma_C(Q,\tau')
\nonumber \\
&&-\int_0^\infty \dint \tau' N(Q,\tau+\tau')\Gamma_C(Q,\tau')
\claus
De oplossing zal voor kleine $Q$ de vorm hebben:
\beatrix
\Gamma_C(Q,\tau)\approx \gamma_C(Q)\eexp{-Q\tau}
\claus
We integreren (\ref{back1}) over $\tau$ en maken gebruik van de het feit dat
de bulk-kern $M_{\rm B}$ een eigenwaarde $1-\drd Q^2$ heeft, en van het feit
dat $N$ nauwelijks van $Q$ afhangt voor kleine $Q$
\beatrix
\nolabel
0&=&1+\int_0^\infty \dint \tau \dint \tau'
[M_{\rm B}(Q,\tau-\tau')-M_{\rm B}(0,\tau-\tau')+    \nonumber \\
&&M_{\rm B}(Q,\tau+\tau')
-M_{\rm B}(0,\tau+\tau')]\gamma_C(Q)\eexp{-Q\tau'}
\\ \nolabel
&&-\int_0^\infty \dint d\tau\tau' N(0,\tau+\tau')\gamma_C(Q)\eexp{-Q\tau}
\\ \nolabel
1&=&\int_{-\infty}^\infty \dint\tau\int_0^\infty \dint \tau'[
M_{\rm B}(0,\tau-\tau')-M_{\rm B}(Q,\tau-\tau') ]\gamma_C(Q)\eexp{-Q\tau'}
+\frac{\tcrl}{4}\gamma_C(Q)
\\ \nolabel
1&=&\int_0^\infty \dint \tau [1-1+\drd Q^2]\gamma_C(Q)\eexp{-Q\tau}
+\frac{\tcrl}{4}\gamma_C(Q)
\\ \nolabel
1&=&\gamma_C(Q)[\drd Q + \frac{1}{4} \tcrl]
\\ \nolabel
\claus
Het resultaat is dus
\BE\label{gamaCQ}
\gamma_C(Q)=\frac{4}{\tcrl+\frac{4}{3}Q} =\frac{4}{{\cal T}}\frac{1}{1+Q\tau_0}
=\frac{3}{Q+1/\tau_0} \label{mvr3} \EE
Hiermee hebben we de driehoekige vorm van de terugstrooikegel gevonden.
Het is interessant de parameter $1/\gamma_C$ te plotten als functie van $Q$.
Doe dit voor steeds hetzelfde medium, maar met verschillende containers,
waardoor $n_1$ en dus $m$ vari\"eert.
De bewering die uit formule (\ref{mvr3}) volgt, is dat de vorm van de
lijnen een rechte is, met asafsnede bepaald door $m$.

Zeer recentelijk is een experimentele bepaling gedaan van de breedte van
de terugstrooikegel van een melkachtige vloeistof~\cite{dOLag}. Door met
verschillende glazen containers te werken kon de brekingsindexverhouding $m$
gevari\"eerd worden. De metingen zijn vergeleken met de oorspronkelijke 
theorie van Lagendijk, Vreeker en de Vries~\cite{LVVries}, met die van Zhu,
Pine en Weitz ~\cite{ZPWeitz} en die van van Nieuwenhuizen en Luck [NL]. De 
resultaten van de eerste twee werken zijn al besproken in hoofdstuk \ref{vbdb}.
Het bleek dat de experimenten zeer goed beschreven worden door de
theorie\"en van Zhu, Pine en Weitz alsmede met die van Nieuwenhuizen en Luck.

Op het eerste gezicht is het vreemd dat een experiment met vectorgolven 
die anisotroop verstrooid worden, beschreven kan worden door een theorie voor
isotrope verstrooiing van scalaire golven.
 In [NL] werd echter aangetoond dat de theorie voor
grote indexverhouding al voor realistische verhoudingen een goede beschrijving
geeft. Dit komt doordat goede spiegels er voor zorgen dat ook dicht bij 
de rand de straling 
isotroop verdeeld is. {\it In zo'n geval is noch de anisotropie van de
verstrooiing, noch het vectorkarakter van belang}. Dat verklaart waarom 
de overeenstemming met het experiment van ref.~\cite{dOLag} zo goed is.

\renewcommand{\thesection}{\arabic{section}}
\section{Anisotrope verstrooiing: verschil in brekingsindices\index{anisotrope verstrooiing}}
\setcounter{equation}{0}\setcounter{figure}{0}

\label{anisotrope}

Tot dusver hebben we ons hoofdzakelijk beperkt tot isotrope verstrooiing.
In dit hoofdstuk bespreken we de resultaten van [NL] voor anisotrope 
verstrooiing. Vervolgens breiden we ze uit met nieuwe resultaten
\cite{ThMNunp93}.

In hoofdstuk \ref{mesoscopie} hebben we de stralingstransportvergelijking
 besproken. In een plak-geometrie heeft hij de vorm:
\beq
\mu'\frac{d}{d\tau}\I(\tau,\mu') = \int^{1}_{-1}\frac{d\mu''}{2}
p_{0}(\mu',\mu'')\I(\tau,\mu'') - \I(\tau,\mu')
\label{r.t.e.}
\eeq
We bekijken weer een ingaande bundel die rotatie-symmetrisch is
rond de $z$-as, dwz. de inkomende bundel is een kegel met openingshoek
$\theta_a$. In het medium is de gerefracteerde bundel dan beschreven door
hoek $\theta_a'$ en $\mu_a=\cos\theta_a'$, $I_{in}(z, \theta',\phi')=
(p_aT_a/P_a)\delta(\mu-\mu_a)\exp(-z/\ell\mu_a)/2\pi$

Als we interne reflecties aan de oppervlakken willen meenemen, moeten voor
$\mu'>0$ de volgende randcondities worden gesteld op $\tau=0^{+}$
(linkerrand) en $\tau=b^{-}$ (rechterrand) 
\BE
\I(0^{+},\mu') =\frac{p_aT_a}{P_a} \delta(\mu'-\mu_{a}) 
+ R(\mu')\int_{0}^{b}d\tau'\,\eexp{-\tau'/\mu'}
\int^{1}_{-1}\frac{d\mu''}{2}p_{0}(-\mu',\mu'')\I(\tau',\mu'') \label{I0}
\EE
\BE
\I(b^{-},-\mu') = R(-\mu')\int^{b}_{0}d\tau'{\rm
e}^{-(b-\tau')/\mu'}\int^{1}_{-1}\frac{d\mu''}{2}p_{0}(\mu',
\mu'')\I(\tau',\mu'')  \label{Ib}
\EE
De eerste term in het rechterlid van (\ref{I0}) geeft de rechtstreeks invallende intensiteit
in richting $\mu'$. De tweede term geeft de intensiteit die in richting
$\mu'$ terug in het medium wordt gereflecteerd aan de linkerrand.

De bronfunctie $\J(\tau,\mu)$ is de intensiteit die op optische diepte 
$\tau$ in richting $\mu$ wordt verstrooid. Er geldt volgens (\ref{bronfie})
\beq \label{angav}
\J(\tau,\mu) = \int^{1}_{-1}\frac{d\mu}{2}p_{0}(\mu,\mu')\I(\tau,\mu')
\label{g=pidm}
\eeq
We proberen met behulp van de bovenstaande
vergelijkingen de Schwarzschild-Milne vergelijking voor $\J$ af te leiden.
Invullen van (\ref{g=pidm}) in (\ref{r.t.e.}) geeft
\beq
\left(\mu'\frac{d}{d\tau} + 1 \right)\I(\tau,\mu') = \J(\tau,\mu')
\eeq
Voor $\mu'>0$ geeft dit
\beqa
 & \frac{d}{d\tau} \left({\rm e}^{\tau/\mu'}\I(\tau,\mu')\right) =
\frac{1}{\mu'}{\rm e}^{\tau/\mu'}\J(\tau,\mu')  \nonumber \\
\Rightarrow & \I(\tau,\mu')=\I(0^{+},\mu'){\rm e}^{-\tau/\mu'} +\frac{1}{\mu'}
\int_{0}^{\tau}d\tau' \,{\rm e}^{-|\tau-\tau'|/\mu'}\J(\tau',\mu')
\eeqa
Voor $\mu'<0$ vind je op soortgelijke wijze
\beq
 \I(\tau,\mu')=\I(b^{-},\mu'){\rm e}^{(b-\tau)/\mu'} +\frac{1}{|\mu'|}
\int_{\tau}^b d\tau' \,{\rm e}^{-|\tau-\tau'|/|\mu'|}\J(\tau',\mu')
\eeq

Vul dit in (\ref{g=pidm}) in en maak gebruik van de rand\-condities
(\ref{I0}) en (\ref{Ib}). Defini\"eren we
\BEQ \J(\tau,\mu;\mu_a)=\frac{p_aT_a}{P_a} \Gamma(\tau,\mu;\mu_a)\EEQ
dan geeft dit een Schwarz\-schild-Milne vergelijking voor 
de genormeerde bronfunctie
$\Gamma(\tau,\mu)\equiv \Gamma(\tau,\mu;\mu_a)$. 
\index{$C@$\Gamma(\tau,\mu)$, genormeerde bronfunctie}
\index{bronfunctie, genormeerd}
Voor een half-oneindig
medium wordt de vergelijking \beqa \label{anisSM}
\Gamma(\tau,\mu) &=& p_{0}(\mu;\mu_a ){\rm e}^{-\tau/\mu_{a}} \\
 &+& \int^{\tau}_{0}d\tau'\int^{1}_{0}\frac{d\mu'}{2\mu'}p_{0}(\mu;\mu')
 {\rm e}^{-|\tau-\tau'|/\mu'}\Gamma(\tau',\mu') \nonumber \\
 &+& \int^{\infty}_{\tau}d\tau'\int^{1}_{0}\frac{d\mu'}{2\mu'}p_{0}(\mu;-\mu')
 {\rm e}^{-|\tau-\tau'|/\mu'}\Gamma(\tau',-\mu') \nonumber \\
 &+& \int^{\infty}_{0}d\tau'\int^{1}_{0}\frac{d\mu'}{2\mu'}R(\mu')
p_{0}(\mu;\mu') {\rm e}^{-(\tau+\tau')/\mu'}\Gamma(\tau',-\mu')\nonumber
\eeqa
\BO Toon aan dat voor eindige $b$
\beqa \label{anisSMb}
\Gamma(\tau,\mu) &=& p_{0}(\mu;\mu_a ){\rm e}^{-\tau/\mu_{a}} \\
 &+& \int^{\tau}_{0}d\tau'\int^{1}_{0}\frac{d\mu'}{2\mu'}p_{0}(\mu;\mu')
 {\rm e}^{-|\tau-\tau'|/\mu'}\Gamma(\tau',\mu')\nonumber \\
 &+& \int^{b}_{\tau}d\tau'\int^{1}_{0}\frac{d\mu'}{2\mu'}p_{0}(\mu;-\mu')
 {\rm e}^{-|\tau-\tau'|/\mu'}\Gamma(\tau',-\mu') \nonumber \\
 &+& \int^{b}_{0}d\tau'\int^{1}_{0}\frac{d\mu'}{2\mu'}R(\mu')p_{0}(\mu;\mu')
 {\rm e}^{-(\tau+\tau')/\mu'}\Gamma(\tau',-\mu') \nonumber \\
 &+& \int^{b}_{0}d\tau'\int^{1}_{0}\frac{d\mu'}{2\mu'}R(-\mu')p_{0}(\mu;-\mu')
 {\rm e}^{-(2b-\tau-\tau')/\mu'}\Gamma(\tau',\mu') \nonumber
\eeqa \EO
De eerste term geeft de intensiteit die voor de eerste keer wordt verstrooid
(op diepte $\tau$ in richting $\mu$). De tweede respectivelijk derde term
geven de
verstrooide intensiteit die van links ($0< \tau'<\tau$) resp. van rechts
($\tau<\tau'<b$) komt. De laatste termen geven de bijdrage van interne
reflecties. Ze geven de intensiteit die op een willekeurige diepte is
uitgezonden, gereflecteerd aan \'e\'en van de randen van het systeem en
vervolgens verstrooid op diepte $\tau$ in richting $\mu$.

Tot zover de resultaten van [NL]. We beschouwen nu enkele 
consequenties.\cite{ThMNunp93}

In hoofdstuk \ref{mesoscopie}, boven vgl. (\ref{avcos}), hebben we
de relatie $\int d\n'p(\n,\n')\n'/4\pi=\langle\cos\Theta\rangle\, \n$ 
besproken.\BO $\bullet$  Toon aan dat de $z$-component hiervan leidt tot
\BEQ \label{avp0} \int_{-1}^1\frac{d\mu'}{2}p_0(\mu,\mu')\mu'=
\langle\cos\Theta\rangle\,\mu \EEQ 
$\bullet$ Gebruik dit om het asymptotisch gedrag van de speciale en homogene
oplossingen af te leiden:
\BEA \label{asympt}
     &\Gamma_S(\tau,\mu;\mu_a)\to \tau_1(\mu_a)\qquad
     &\Gamma_H(\tau,\mu)      \to \tau-\mu(\tau_{tr}-1)+\tau_0\tau_{tr}
\nonumber\\
     &\I_S(\tau,\mu;\mu_a)    \to \frac{p_aT_a}{P_a}\tau_1(\mu_a)\qquad
     &\I_H(\tau,\mu)          \to \tau-\mu\tau_{tr}+\tau_0\tau_{tr} \EEA
Hierin zijn $\tau_1(\mu_a)$ en $\tau_0$ nader te bepalen. $\tau_{tr}$ is
de genormeerde transport-vrije-weglengte
\index{transport-vrije weglengte!genormeerd}
\index{$$tautr@ $\tau_{tr}$, transport-vrije weglengte!genormeerd}
\BEQ\label{tautr} \tau_{tr}\equiv\frac{\ell_{tr}}{\ell_{sc}}=
\frac{1}{1-\langle\cos\Theta\rangle} \EEQ
\EO

\subsection{De hoekafhankelijkheid van de reflectieco\"efficient}

In de uitdrukking (\ref{mcw5}) voor de hoek-opgeloste reflectie\-co\"efficient
voor een half-oneindig medium moet men nu voor de genormeerde 
bistatische co\"efficient invullen:
\BEQ \gamma(\mu_a,\mu_b)=\int_0^\infty d\tau\Gamma_S(\tau,-\mu_b;\mu_a)
\eexp{-\tau/\mu_b},\EEQ
want $\Gamma_S(\tau,-\mu_b;\mu_a)$ is de straling die op diepte
$\tau$ uitgezonden wordt in de terugwaardse richting $\pi-\theta_b'$.
Voor isotrope verstrooiing is hij onafhankelijk van $\mu_b$.

\subsection{De hoekafhankelijkheid van de transmissieco\"efficient}

In analogie met de uitdrukking (\ref{dTabdO=}) voor isotrope verstrooiing, 
hebben we nu
\beq\label{ttaabb}
\frac{dT(a\rightarrow b)}{d\Omega_b}
 = \frac{\cos(\theta_a) T_a T_b}{4\pi m^2 \mu_a \mu_b}
\int_0^b \d \tau \Gamma(\tau,\mu_b;\mu_a) \eexp{-(b-\tau)/\mu_b}.
\eeq
In vgl. (\ref{beginSM}), (\ref{eindSM}) en (\ref{al=al=})
hebben we, voor isotrope verstrooiing, de integraal uitgedrukt in de speciale
en de homogene oplossing. Dit doen we nu weer.
\BO Toon aan dat het asymptotisch gedrag (\ref{asympt}) 
inhoudt dat voor $b\gg 1$
\BEA \Gamma(\tau,\mu;\mu_a)=&\Gamma_S(\tau,\mu;\mu_a)-\frac{\tau_1(\mu_a)}
{b+2\tau_0\tau_{tr}}\Gamma_H(\tau,\mu) &0<\tau<b-10\nonumber\\
=&\frac{\tau_1(\mu_a)}{b+2\tau_0\tau_{tr}}\Gamma_H(b-\tau,-\mu)
\qquad &\tau>10
\EEA \EO
De integraal in (\ref{ttaabb}) kan daarom weer omgeschreven worden tot
\beqa
\int_0^b d \tau \Gamma(\tau,\mu_b;\mu_a) \eexp{-(b-\tau)/\mu_b} 
&\approx& \frac{\tau_1(\mu_a)}{b+2\tau_0\tau_{tr}}
\int_{-\infty}^b
d \tau \Gamma_H(b-\tau,-\mu_b) \eexp{-(b-\tau)/\mu_b} \nonumber \\ 
&=& \frac{\tau_1(\mu_a)}{b+2\tau_0\tau_{tr}}\int_0^\infty \d \tau
            \Gamma_H(\tau,-\mu_b) \eexp{-\tau/\mu_b} 
\eeqa
We tonen verderop aan dat de laatste integraal gelijk is aan 
\BEQ \int_0^\infty \d \tau \Gamma_H(\tau,-\mu_b) \eexp{-\tau/\mu_b} 
=D\tau_1(\mu_b)\label{hulp1}\EEQ
met gereduceerde diffusieconstante $D=\tau_{tr}/3$, dwz. $D=v\ell_{tr}/3$
in fysische eenheden. We vinden dus
\beq
\frac{dT(a\rightarrow b)}{d\Omega_b}\equiv \frac{A^T(\theta_a,
\theta_b)}{b+2\tau_0\tau_{tr}} \equiv  \frac{\cos{\theta_a}}{12\pi m^2}
\frac{  T_a\tau_1(\mu_a)}{\mu_a}\,
\frac{  T_b\tau_1(\mu_b)}{\mu_b}\,
\frac{\tau_{tr}} {b+2 \tau_0\tau_{tr}}.
\eeq
Merk op dat de factor $\tau_{tr}/(b+2 \tau_0\tau_{tr})$ gelijk is aan 
$\ell_{tr}/(L+2\tau_0\ell_{tr})$. Dit toont het welbekende feit aan dat
 de {\it transport-vrije-weglengte} de relevante vrije weglengte is
voor transmissie. We hebben dat nu door {\it exacte} manipulaties afgeleid.

\subsubsection{Een eenvoudige doch leerzame identiteit}
\index{eenvoudig!doch leerzaam}
We leiden nu, met Uw hulp, vgl. (\ref{hulp1}) af. De speciale
oplossing van de Schwarz\-schild-Milne vergelijking (\ref{anisSM}) kan 
opgesplitst worden in zijn limiet waarde en de rest 
\BEQ
\Gamma_S(\tau,\mu;\mu_a)\equiv \tau_1(\mu_a)+F_S(\tau,\mu;\mu_a) 
\EEQ\index{$F_S$, deel van speciale oplossing}
De  Schwarzschild-Milne vergelijking voor de speciale
oplossing kan hiermee uit gedrukt worden als
\BEQ p_0(\mu,\mu_a)\eexp{-\tau/\mu_a}=F_S(\tau,\mu)-M*F_S
+\frac{\tau_1(\mu_a)}{2}\int_0^1d\mu' p_0(\mu,\mu')T(\mu')\eexp{-\tau/\mu'},
\label{leebel} \EEQ
De kern $M=M_B+M_L$ uit (\ref{Mbulk}), (\ref{Mlayer}) kan in matrixnotatie
worden geschreven als
\BEQ M_{\tau\tau'}^{\mu\mu'}=\theta(\mu'\tau-\mu'\tau')
\frac{p_0(\mu,\mu')}{2|\mu'|}\eexp{-\frac{|\tau-\tau'|}{|\mu'|}}
+\theta(-\mu')\frac{p_0(\mu,-\mu')}{2|\mu'|}
R(\mu') \eexp{-\frac{\tau+\tau'}{|\mu'|}}.\EEQ 
\BO We weten dat $\Gamma_H(\tau,\mu)$ de rechtereigenfunctie is van $M$ met
eigenwaarde \'e\'en. 

$\bullet$ Toon aan dat $\I_H(\tau,-\mu)$ de 
linkereigenfunctie is. 

Dit is een gevolg van tijdsomkeerinvariantie, wat ook al
impliceert dat \BEQ p_0(\mu,\mu')=p_0(-\mu',-\mu).\EEQ

$\bullet$ Toon ook aan dat $\I_H$ en $\Gamma_H$ alsvolgt samenhangen
\BEQ \Gamma_H(\tau,\mu)=
\int^{1}_{-1}\frac{d\mu}{2}p_{0}(\mu',\mu)\I_H(\tau,\mu').
\label{aabbcc} \EEQ \EO

Vermenigvuldig je nu vgl. (\ref{leebel}) met $\I_H(\tau,-\mu)$ en integreer
je over $\tau$ van $0$ tot $\infty$, en over $\mu$ van $-1$ tot $1$,
dan vallen de termen met $F_S$ weg tegen elkaar. Gebruik je de symmetrie van
$p_0$, dan hou je vgl. (\ref{hulp1}) over, mits 
\BEQ \frac{\tau_{tr}}{3}=
\int_0^1\frac{d\mu}{2}\int_0^\infty d\tau \Gamma_H(\tau,-\mu)T(\mu)
 \eexp{-\tau/\mu}.\label{D=tau3}\EEQ
Om dit aan te tonen schrijven we $\Gamma_H$ als
\BEQ \Gamma_H(\tau,\mu) 
=\tau-\mu(\tau_{tr}-1)+\tau_{tr} F_H(\tau,\mu).\label{GHFH} \EEQ
\index{$F_H$, deel van homogene oplossing}
De functie $ F_H$ voldoet aan een vergelijking waar 
$\tau_{tr}$ niet in voorkomt
\BEQ \label{FHvgl} F_H(\tau,\mu)=M*F_H +\int_0^1\mu'd\mu'
(1-\frac{1}{2}T(\mu'))p_0(\mu,\mu')\eexp{-\tau/\mu'} \EEQ
en gaat naar een constante, $\tau_0$, in oneindig.
\BO $\bullet$ Toon aan dat
\BEA M*F_H(\tau,\mu)&=&\int_{-1}^1\frac{d\mu'}{2}p_0(\mu,\mu')F_H(\tau,\mu')\\
&-&\frac{d}{d\tau}\int_0^\tau d\tau'\int_0^1\frac{d\mu'}{2}
p_0(\mu,\mu')F_H(\tau',\mu')\eexp{-\frac{|\tau-\tau'|}{\mu'}}
\nonumber\\
&+&\frac{d}{d\tau}\int_\tau^\infty d\tau'\int_0^1\frac{d\mu'}{2}
p_0(\mu,-\mu')F_H(\tau',-\mu')\eexp{-\frac{|\tau-\tau'|}{\mu'}}
\nonumber\\
&-&\frac{d}{d\tau}\int_0^\infty d\tau'\int_0^1\frac{d\mu'}{2}
p_0(\mu,\mu')R(\mu') F_H(\tau',-\mu')\eexp{-\frac{\tau+\tau'}{\mu'}}
\nonumber\EEA
$\bullet$ Integreer deze uitdrukking over $\tau$ van $0$ tot $\infty$ en
over $\mu$ van $-1$ tot $1$. Toon aan dit leidt tot
\BEQ \int_0^\infty d\tau \int_0^1\frac{d\mu}{2} F_H(\tau,-\mu)T(\mu)
\eexp{-\frac{\tau}{\mu}}=\int_0^1\mu^2d\mu
(1-\frac{1}{2}T(\mu))\label{constraintFH} \EEQ
$\bullet$ Toon aan dat dit met vgl. (\ref{GHFH}) precies
tot vgl. (\ref{D=tau3}) leidt. $\Box$ \EO

\subsection{Rayleigh verstrooiing\index{Rayleigh verstrooiing}}

Zoals in hoofdstuk \ref{mesoscopie} aangetoond werd, geldt voor
Rayleigh verstrooiing: \BE
p_0(\mu,\mu')=\frac{3}{8} (3-\mu^2-\mu'^2+3 \mu^2\mu'^2) \EE
De sferische symmetrie van de verstrooiier impliceert ook spiegelsymmetrie.
Dit betekent dat $\Gamma(\tau, \mu)=\Gamma (\tau, -\mu) $.
Voor $b=\infty$ hebben we daarom:
\BEA
\Gamma(\tau,\mu)=&&p_0(\mu,\mu_a)\eexp{-\tau/\mu_a}+\\&&
 \int_0^\infty d\tau' \int_0^1
\frac{d \mu'}{2 \mu'} p_0 (\mu,\mu') \left[ \eexp{-\frac{|\tau-\tau'|}{\mu'}}
+R(\mu') \eexp{-\frac{\tau+\tau'} {\mu'}} \right] 
\Gamma({\tau',\mu'})\nonumber
\EEA
In formule (\ref{GammaDelta})
 hebben we de functies $\Gamma(\tau)$ en $\Delta(\tau)$
ingevoerd voor Rayleigh verstrooiers.
\BO Toon aan dat de bronfunctie gegeven wordt door:
\BE \frac{P_a}{p_aT_a}\J(\tau,\mu)= 
\frac{9}{16} \Gamma(\tau)-\frac{3}{16} \Delta(\tau) +\mu^2
\left[ \frac{9}{16} \Delta(\tau) -\frac{3}{16} \Gamma(\tau) \right] \EE \EO
Het blijkt ook in de aanwezigheid van interne reflecties mogelijk een
gesloten set vergelijkingen voor $\Gamma$ en $\Delta$ op te schrijven.
\BO
Leid af dat vergelijkingen analoog aan (\ref{GammaDeltavgl}) geldig blijven,
\BEA \Gamma(\tau)&=&\eexp{-\tau/\mu_a}     
+(\frac{9}{16}\tilde E_1-\frac{3}{16}\tilde E_3)*\Gamma+
(\frac{9}{16}\tilde E_3-\frac{3}{16}\tilde E_1)*\Delta  
          \label{GammaDeltavgl2}\\
\Delta(\tau)&=&\mu_a^2\eexp{-\tau/\mu_a}
+(\frac{9}{16}\tilde E_3-\frac{3}{16}\tilde E_5)*\Gamma+
(\frac{9}{16}\tilde E_5-\frac{3}{16}\tilde E_3)*\Delta \nonumber
\EEA
waarin we nu de door interne reflecties gemodificeerde exponenti\"ele 
integralen hebben
\BE \tilde{E}_k(\tau,\tau')=\int_0^1 \frac{d\mu}{\mu}\mu^{k-1} \left[
\eexp{-|\tau-\tau'|/\mu}+ R(\mu) \eexp{-(\tau+\tau')/\mu}
\right] \EE
terwijl het product is gedefinieerd door:
\BE
\left(\tilde{E}_k*f\right) (\tau)= \int_0^\infty d\tau' \tilde{E}_k
(\tau,\tau')f(\tau') \index{$Ekt@$\tilde{E}_k$, gemodificeerde
exponenti\"ele integraal}  \EE
\EO

\subsection{Groot verschil in brekingsindices}

We hebben in het vorige hoofdstuk gezien dat er vereenvoudigingen optreden
in de limieten $m\to 0$ en $m\to \infty$. We willen dat nu bekijken voor
niet-isotrope verstrooiing. We hebben al gezien dat anisotrope verstrooiing
er vaak toe leidt dat de verstrooiings-vrije weglengte vervangen wordt
door de transport-vrije weglengte. Een verdere vraag is bijvoorbeeld:
 wat voor invloed heeft de vorm 
van de fasefunctie op het asymptotisch gedrag van $\tau_0$ ? (Zie
(\ref{mcw8}) en (\ref{mcw13}) voor isotrope verstrooiing).


\subsubsection{De injectiediepte}
\label{injectiediepte!anisotropie}
We beschouwen de limiet $m\to 0$ en volgen dezelfde 
stappen als in vorig hoofdstuk. We 
ontbinden de oplossing van vgl. (\ref{FHvgl})
 volgens de {\it Aanzet}
\BEQ F_H(\tau,\mu)=C_H+F_0(\tau,\mu)+\cdots\label{ontb}\EEQ
in termen van orde $\T^{-1}$, $\T^0$, etc. Er geldt weer
$T(\mu)\approx \mu T(1)\approx 3\T\mu/2$. 
\BO $\bullet$ Laat zien dat vgl. (\ref{FHvgl}) met (\ref{constraintFH})
 leiden tot
\BEA &&C_H=\frac{4}{3\T};\\ 
&&F_0=M(m=0)*F_0\nonumber\\
&&\int_0^1d\mu \mu \int_0^\infty d\tau\eexp{-\tau/\mu}
F_0(\tau,-\mu)=-\frac{1}{4}.\nonumber \EEA

$\bullet$ Toon aan dat de oplossing gegeven is door

\BEQ F_0(\tau,\mu)=-\frac{3}{4}. \EEQ \EO


Dit bewijst dat vgl. (\ref{mcw13}), $\tau_0\approx 4/3\T-3/4$,
 geldig blijft voor anisotrope verstrooiing. De anisotropie leidt er
alleen toe dat de transport-vrije-weglengte naar voren komt. Verdere
eigenschappen van de fasefunctie spelen pas een rol in de
correctieterm van orde $\T$.


We beschouwen hetzelfde argument voor de limiet $m\to\infty$. We zoeken 
dus het analogon van  vgl. (\ref{mcw8}) voor anisotrope verstrooiing.
We maken weer de {\it Aanzet} (\ref{ontb}) en defini\"eren
\BEQ F_0(\tau,\mu)=\tilde F_0(\tau,\mu)-1.\EEQ
\BO Toon aan dat dit leidt tot de vergelijkingen
\BEA
&&\tilde F_0(\tau,\mu)=M(m=\infty)*\tilde F_0-
\frac{1}{3}p_0(\mu,1)\eexp{-\tau}
+\int_0^1d\mu'\mu'p_0(\mu,\mu')\eexp{-\tau/\mu'}\nonumber\\
&&\int_0^\infty d\tau \eexp{-\tau}\tilde F_0(\tau,-1)=0\nonumber\\
&&\lim_{\tau\to\infty}\tilde\Gamma(\tau,\mu)=\delta\tau_{00}<\infty. \EEA\EO
Dit resulteert in het asymptotische gedrag
\BEQ \Gamma_H(\tau,\mu)\to \tau-\mu(\tau_{tr}-1)+\tau_0\tau_{tr} \EEQ
met dimensieloze injectiediepte
~\footnote{Omdat $\tau_0$ zelf ook nog van de fasefunctie afhangt, is de
opsplitsing $\tau_0\tau_{tr}$ enigzins arbitrair.}
\BEQ \tau_0\equiv \frac{z_0}{\ell_{tr}}=\frac{4}{3\T}-1+\delta\tau_{00}. \EEQ
Voor isotrope verstrooiing weten we uit (\ref{mcw8}) 
dat $\delta\tau_{00}=-0.0357$ een kleine correctie geeft. Voor niet-isotrope
verstrooiing zal $\delta\tau_{00}$ veranderen, maar waarschijnlijk toch
klein blijven. In tegenstelling tot de limiet $m\to 0$ vinden we hier dus
dat anisotropie, behalve tot het invoeren van de transport-vrije-weglengte, 
ook tot een ${\cal O}(\T^0)$  effect leidt op de injectiediepte. 

\subsubsection{De limietintensiteit}

We beschouwen nu weer een half-oneindig medium en generaliseren de behandeling
van de speciale oplossing van de Schwarz\-schild-Milne vergelijking uit
hoofdstuk~\ref{groot}. We maken dus de weer de {\it Aanzet}
\BEQ \Gamma_H(\tau,\mu)=C_H+\Gamma_0(\tau,\mu)+\cdots\EEQ
en vullen dit in in vgl. (\ref{anisSM}).
\BO Toon aan door deze vergelijking over $\tau$ en $\mu$ te integreren 
dat $C_H=4\mu_a/\T$. \EO
Dit levert dus een limietintensiteit
\BEQ \tau_1(\mu_a)\equiv \Gamma_S(\infty,\mu;\mu_a)
\approx \frac{4\mu_a}{\T} \EEQ
met dezelfde $\T$ als  in hoofdstuk  \ref{groot} voor isotrope 
verstrooiing. 

We zien dus dat er, in leidende orde in $\T$, geen enkele invloed is van
de anisotropie van de verstrooiing op de limietintensiteit in een heel
dikke plak. 

\subsubsection{De terugstrooikegel}

We beschouwen weer de limiet $m\to\infty$ en volgen de afleiding voor
het analoge probleem in vorig hoofdstuk. Uit de diffusievergelijking
voor anisotrope verstrooiing verwachten we in de bulk 
\BEQ \Gamma_C(\tau,\mu;\mu_a;Q)\approx [\tau_1(\mu_a)+\beta\mu]
\eexp{-Q\tau}\EEQ
met constante factor $\beta$. Net als in vorig hoofdstuk zal
voor $\tau>10$ en $Q$ klein de ontbinding in homogene en speciale oplossing
gelden
\BEA \Gamma_C(\tau,\mu;\mu_a;Q)&=&\Gamma_S(\tau,\mu;\mu_a;0)-
\alpha Q \Gamma_H(\tau,\mu;0)\nonumber\\
&\approx&\tau_1(\mu_a)-\alpha Q[\tau-\mu(\tau_{tr}-1)+\tau_0\tau_{tr}]\EEA
Deze uitdrukkingen zijn  consistent indien
\BEQ \Gamma_C(\tau,\mu;\mu_a;Q)=\frac{\tau_1(\mu_a)}{1+Q\tau_0\tau_{tr}}
[1+\mu Q(\tau_{tr}-1)]\eexp{-Q\tau} \EEQ
Voor loodrechte inval volgt hieruit de kegel
\BEQ \gamma_C(Q)=\int_0^\infty d\tau \Gamma_C(\tau,-1;Q;1)\eexp{-\tau}
=\tau_1(1)\frac{1+Q-Q\tau_{tr}}{(1+Q)(1+Q\tau_0\tau_{tr})} \EEQ
Binnen de gemaakte benaderingen is dit gelijk aan
\footnote{Dit suggereert dat de uitdrukking 
$ \frac{3-\gamma_C/\tau_0}{3+\gamma_C} $
over een redelijk groot interval van hoeken constant is. Om dit 
verband te testen
hoef je niet de transport-vrije-weglengte te kennen, doch slechts 
de brekings\-index\-verhouding $m$.}
\BEQ \gamma_C(Q)=\tau_1(1)\frac{1-\tilde Q}{1+\tilde Q\tau_0}
=3\frac{1-\tilde Q}{\tilde Q+1/\tau_0} \EEQ
met op de {\it transport-vrije-weglengte} genormeerde hoek
\BEQ \tilde Q=Q\tau_{tr}=\theta_b k_1\ell_{tr}\EEQ

Experimenteel is het wel bekend dat ook voor de terugstrooikegel
de transport-vrije-weglengte de relevante mesoscopische afstandsschaal is.

De toepasbaarheid van bovenstaande theorie
geverifi\"eerd voor de terugstrooikegel in experimenten met licht, 
zie de discussie aan het eind van hoofdstuk ~\ref{groot}.

We concluderen dat de resultaten uit de verbeterde diffusie\-benadering 
in leidende orde correct blijven voor anisotrope verstrooiing, mits
de verstrooiings-vrije weglengte wordt vervangen door de 
transport-vrije-weglengte. 
Zoals opgemerkt aan het eind van hoofdstuk \ref{mesoscopie}, is de fysische
verklaring hiervoor dat de aanwezigheid van goede spiegels zorgt voor een
nagenoeg isotrope intensiteits\-verdeling. Hierdoor speelt de vorm van de 
fasefunctie en het eventuele vectorkarakter van de
straling een ondergeschikte rol.

\renewcommand{\thesection}{\arabic{section}}
\section{De invloed van \'e\'en enkele verstrooier op diffusie}
\setcounter{equation}{0}\setcounter{figure}{0}

\index{localisatie!van objecten} \label{invloed}

In hoofdstuk \ref{macroscopie}  hebben we het effect bekeken van een
macroscopisch object (potlood, glasfiber) op de transmissie door een
melkachtige vloeistof. Als een toepassing van de microscopische theorie
bekijken we nu, in navolging van ref. \cite{berkovits2}, de invloed van het
kleinst mogelijke object, \'e\'en enkele verstrooier. We zullen vinden dat de
microscopische theorie inderdaad hetzelfde resultaat als de macroscopische
theorie levert, voor het geval van een vloeistof met een extra bolletje.
We volgen referentie \cite{hik0}.
 
\subsection{Verstrooing aan \'e\'en extra verstrooier\index{extra
verstrooier}}
De invloed van een enkele verstrooier is van praktisch belang, aangezien men
objecten probeert te localiseren in diffuse media, zie hoofdstuk
\ref{macroscopie} en ref.\cite{berkovits2}. Dit kan gebeuren door met gevoelige
apparatuur naar de transmissie te kijken \cite{denouter}. 

We behandelen dit probleem diagrammatisch. Wetende dat er een aankomende
en een uitgaande ladderpropagator bij de extra verstrooier zullen zijn,
is de vraag: hoe moeten deze twee ladderpropagatoren aan elkaar moeten
 worden geknoopt? In figuur \ref{tnbox}
zijn alle relevante bijdragen aan de verstrooiing getekend. De eerste
bijdrage in de onderste regel ligt voor de hand, het is de
bouwsteen van ladderdiagrammen. De twee diagrammen rechtsonder
zijn zelfenergie diagrammen welke nog niet in de ladderpropagator $\L$ zitten,
want die heeft betrekking op het ongestoorde medium.
De bijdrage van deze diagrammen is heel belangrijk.
 Ze vallen namelijk in leidende orde weg tegen het
eerste diagram, rechts van het gelijkteken in de figuur.

\begin{figure}[htbp]
\centerline{\includegraphics[width=10cm]{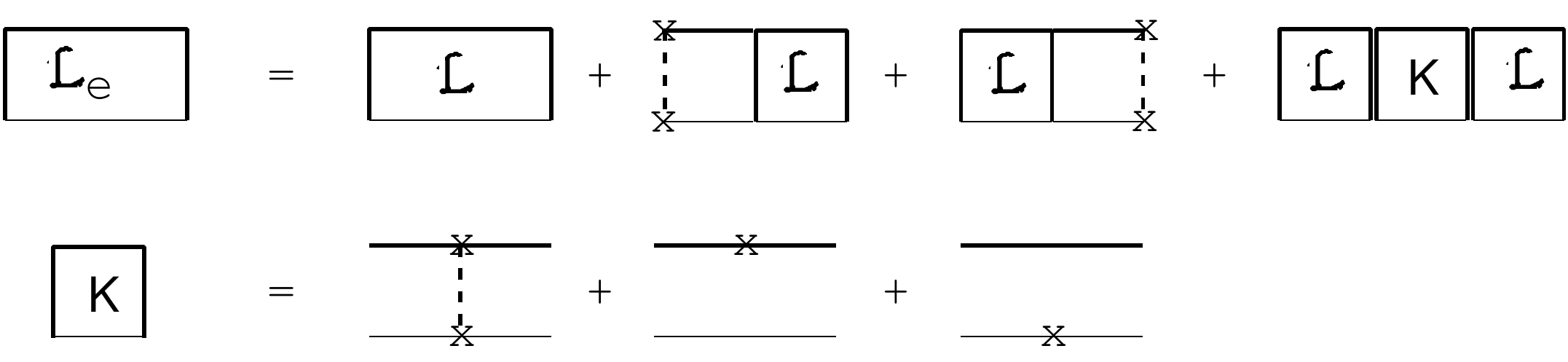}}

\caption{Bijdragen van de extra verstrooier tot de
ladder intensiteit. $\L$
is de ladderpropagator zonder de extra verstrooier, $\L_e$ de
ladderpropagator  met bijdragen van de extra verstrooier. Dikke lijnen
geven geretardeerde aan, dunne lijnen geven geavanceerde Greense functies
aan. De $t_0$--matrix van de extra verstrooier is weergegeven met een {\sc x}.
Haar totale effect is met $K$ aangegeven.}\label{tnbox} \end{figure}

Laten we de $t-$matrix van de extra verstrooier met $t_0$\index{$t_0$, $t-$
matrix extra verstrooier} aangeven, zijn albedo met $a_0$.\index{$a_0$,
albedo extra verstrooier}
We plaatsen de verstrooier op $\r_0=(0,0,L-Z_0)$, $Z_0$ is de afstand tot
het uitgaande vlak.
Stel dat de ladder propagatoren die aan de extra verstrooier vastzitten eencimpuls $\q_1$ and $\q_2$ hebben 
en frequentie parameters $\Omega_1$ en $\Omega_2$. Beide impulsen wijzen naar
de extra verstrooier. In laagste orde van deze impulsen, berekenen we
de onderste diagrammen van figuur \ref{tnbox}.
Analoog met (\ref{LqOm}) en met gebruik van (\ref{Gpknt}), (\ref{Gexpand}) en de notatie uit hoofdstuk \ref{hikami}, vinden we
voor de eerste term: \BA
K_1 &=&t_0\overline{t}_0  \int d^3\p
G(\p+\frac{1}{2}\q_1,\omega+\frac{\Omega_1}{2})G^*(\p-\frac{1}{2}\q_1,\omega-\frac{\Omega_1}{2})
 \int d^3
\p'  G(\p'-\frac{1}{2}\q_2) G^*(\p'+\frac{1}{2}\q_2)\nonumber \\ &\approx
&t_0 \overline{t}_0 [ I_{1,1}
-\frac{1}{3} k^2 q_1^2 I_{2,2}+ \frac{2}{3} k^2 q_1^2 I_{1,3} 
+\frac{1}{3} k\ell \tOmega I_{2,1}- \frac{1}{3}k\ell \tOmega I_{1,2}]\times
\left[ \q_1 \rightarrow \q_2 \right] \nonumber \\
&=&t_0\overline {t}_0 \frac{\ell^2}{16\pi^2} \left[
1-\frac{1}{3}(q_1^2+q_2^2)\ell^2+\frac{2}{3}
i\tOmega\ell^2 -2(1-a) \right] \nonumber \EA
\index{$K$, TN-doos}\index{$K_{1,2,3}$ diagrammen uit TN-doos}
De integralen $I_{k,l}$ zijn gedefini\"eerd in de Appendix.
Voor de tweede term
vinden we \BA
K_2 &=& t_0 \int d^3\p G^*(\p)G(\p+\q_1)G(\p-\q_2)
\nonumber\\ &\approx& t_0 \left[ I_{2,1}+ \frac{4}{3}
(q_1^2+q_2^2-\q_1 \cdot \q_2)k^2 I_{4,1}-\frac{2}{3} k\ell
\tOmega (-I_{2,2}+2I_{1,3}) \right] \nonumber \\
&=& \nonumber -\frac{it_0 \ell^2}{8\pi k} \left[
-1+\frac{1}{3}(q_1^2+q_2^2-\q_1\cdot \q_2)
\ell^2 -\frac{2}{3} i \tOmega \ell^2 +2(1-a)
 \right]   \EA Er geldt: $K_3=K_2^*$ en uit (\ref{aaext=}) ${\rm Im}\, t_0\approx\bar t_0t_0k(1+(1-a_0))/4\pi$.
De som
van de drie diagrammen wordt hiermee
\BE K=- \frac{t_0 \overline{t}_0 \ell^2}{48 \pi^2} [\, \q_1 \! \cdot \!
\q_2 \, \ell^2 +3(1-a_0) ] \; .\EE
Volgens de eerste regel van figuur \ref{tnbox} is de uitdrukking voor
de ladder met de extra verstrooier in ruimtelijke
coordinaten\index{$Le@$\L_e$, ladderpropagator met extra verstrooier} :\BA
\L_e(\r,\r')&=&\L(\r,\r')+ \frac{t_0\overline{t}_0}{nt \overline{t}}
[\delta(\r'-\r_0)+\delta(\r-\r_0)]\L(\r,\r')
\nonumber \\ &&\frac{t_0 \overline{t}_0 \ell^2}{48\pi^2}
\int d\r_1 d\r_2 [\ell^2 \nabla_1 \! \cdot \nabla_2-3(1-a_0)]\L(\r,\r_1)
\times \nonumber \\
&&\L(\r_2,\r')\delta(\r_1-\r_0) \delta(\r_2-\r_0) \label{ltn} \; .\EA
We nemen aan dat de extra
verstrooiing zwak is ten opzichte van de totale verstrooiing
$n_0t_0\overline{t}_0
\ll nt\overline{t}$, met $n_0=N_0/V$, $N_0=1$ en $V$ is het volume van de
plak.\index{$n_0$, dichtheid extra verstrooiers}

\subsection{Middeling over de positie van de extra verstrooier}
Eerst middelen we de positie van de extra verstrooier over het hele
volume. We hopen nu de difusie propagator terug te
vinden, maar ditmaal voor een dichtheid van $ n+n_0 $ verstrooiers.
Inderdaad vind je
zo'n propagator terug met een gereduceerde vrije weglengte
 $\ell_e=4\pi/(nt \overline{t}+n_0 t_0 \overline{t}_0)$.\index{$le@$\ell_e$,
vrije weglengte met extra verstrooier} Uitgaande van vergelijking
(\ref{ltn}) en na substitutie van (\ref{difladder}), 
de ladderpropagator zonder de extra
verstrooier $\L(\q)=12\pi\ell^{-3}(q^2+\kappa^2-i\tOmega) ^{-1}$, 
vindt men in leidende orde van $n_0$ 
\BA \L_e(\q)&=&\L(\q)+2 \frac{n_0 t_0 \overline{t}_0} {nt\overline{t}}
\L(\q)+[q^2\ell^2-3(1-a_0)]
\frac{n_0 t_0 \overline{t}_0 \ell^2}{48 \pi^2} \L^2(\q)  \nonumber \\
&\approx&\frac{12\pi(1+3(\ell/\ell_e-1))}
              {\ell^3(q^2+\kappa^2-i\tOmega)}  
\left(1-(\ell/\ell_e-1)\frac{-3\ell^{-2}(1-a_0)-\kappa^2+i\tOmega}
{q^2+\kappa^2-i\tOmega}\right)^{-1} \nonumber \\
&\approx& \frac{12\pi}{\ell_e^3 (q^2+\kappa_e^2-i\tOmega_e)} \; ,
\label{Ie} \EA
met $\tOmega_e \equiv \tOmega\ell/\ell_e $\index{$$Xt@$\tOmega_e$,
genormeerde
frequentie met extra verstrooier}. Dit is consistent met onze definitie:
$\tOmega=3\Omega/v\ell$, waar de vrije weglengte in voorkomt. De
combinatie $\kappa_e^2 \ell_e \equiv 3[nt\overline{t}(1-a)+
n_0 t_0\overline{t}_0(1-a_0)] /4\pi$ is bij nader inzien precies een gewogen som van
albedo's.\index{$$ka@$\kappa_e$, inverse absorptie lengte met extra
verstrooier} \BO
Controleer de afleiding van formule (\ref{Ie})  \EO

\subsection{Vastgeprikte verstrooier}

In plaats van te middelen over de positie kunnen we
de verstrooier ook vastzetten. Dit is veel interessanter.
Beschouw de transmissie van een vlakke golf door een niet
absorberende laag met dikte $L$.\index{$I_e$, diffuse intensiteit met extra
verstrooier} De transmissie is bij benadering $T=-\ell \frac{dI_e(z)}{dz}
|_{z=L}$. $I_e=\L_e\ell/4\pi$
kan worden berekend met de beeldladingsmethode\index{beeldladingen}
\cite{denouter}. Deze geeft: \BA I_e(\r)&=& I_0
\frac{L-z}{L}\nonumber \\&& +q\sum_{n=-\infty}^\infty
\left\{ \frac{1}{|\r-\r_0+2nL
\hat{z}|} -\frac{1}{|\r+\r_0+2nL \hat{z}|} \right\} \nonumber \\
&&-p\sum_{n=-\infty}^\infty
\frac{z-Z_0+2nL}{[x^2+y^2+(z-Z_0+2nL)^2]^{3/2}} \nonumber \\
&&-p\sum_{n=-\infty}^\infty
\frac{z+Z_0+2nL}{[x^2+y^2+(z+Z_0+2nL)^2]^{3/2}}
\EA
met \BA \label{negen}
q&=&-\frac{Z_0}{L}\frac{3t_0\overline{t}_0}{16 \pi^2\ell} (1-a_0) \; ,
\nonumber \\
p&=&\frac{\ell}{L}\frac{t_0\overline{t}_0}{16 \pi^2} \; .
\EA

(We hebben het
tweede en derde van
de bovenste diagrammen verwaarloosd). De transmissie in het nabije veld
wordt\BA
T({\bf \rho})&=&\frac{\ell}{L} +2 q\ell
\sum_{n=-\infty}^{\infty}\frac{Z_0+2nL}{(\rho^2+(Z_0+2nL)^2)^{3/2}}
\nonumber \\
&&+2p\ell
\sum_{n=-\infty}^{\infty}\frac{\rho^2-2(Z_0+2nL)^2}
{(\rho^2+(Z_0+2nL)^2)^{5/2}} \; ,
\EA
waarbij $\rhoo=(x,y)$
\BO Controleer dit. \EO

Twee gevallen zijn van belang: Als de extra
verstrooier niet absorbeert is er alleen
de $p$--term. Dit is analoog aan een dipool\index{dipool} in een statisch
electrisch veld tussen twee condensator platen. In de doorgelaten intensiteit
zie je een kronkel als $Z_0 < L/2$. Als $Z_0>L/2$, zie je een dip in
de transmissie. Als de extra verstrooier wel absorbeert, is de $q$--term
dominant en het object werkt als een afvoer waar intensiteit in verdwijnt.
Het resultaat is een dip in de transmissie. Dit is equivalent aan een 
lading in de electrostatica.

\subsection{Vergelijking met de macroscopie}

Het is leerzaam om onze resultaten te vergelijken met
resultaten van de diffusie benadering door den Outer
\etal~\cite{denouter}. Voor cylinders is dit beschreven in
hoofdstuk \ref{macroscopie}, nu beschouwen we een bolletje.
Het omringende medium heeft een diffusie constante $D_1$. Het sferische
object is veel groter dan een paar vrije weglengtes. We kunnen het
beschrijven door het een extra dichtheid $n_0$ van extra verstrooiers te
geven ten opzichte van de vloeistof.
De macroscopische resultaten van den Outer \etal blijken consistent te zijn
met formule (\ref{negen}).

Stel namelijk dat het object een straal $R$, een inverse absorptielengte
 $\kappa_2$, en diffusie constante $D_2\neq D_1$ heeft. We nemen de
limiet van zwakke absorptie
en vrijwel gelijke diffusie constanten ($0<D_1-D_2\ll
D_1$).
Het resultaat van den Outer {\it et al.} reduceert in dit geval tot:
$q_D=-\kappa_2^2
R^3 Z_0/3L$ en $p_D=R^3(D_1-D_2)/3LD $. Dit komt overeen met
vergelijking
(\ref{negen}). Om dit aan te tonen nemen we aan dat
het object  $N_0$  extra verstrooiers bevat, met een dichtheid
$n_0=3N_0/ 4\pi R^3$. Net als in de electrostatica
is het veld buiten het object gelijk aan dat van \'e\'en puntlading met een
sterkte $N_0q$ en een dipool met sterkte $N_0p$.
Anderzijds hebben we de diffusie constanten
$D_1=\frac{1}{3}v\ell$ en
$D_2=\frac{1}{3} v \ell_e$, en de absorptie
 $\kappa_2=(3n_0 t_0{\overline t}_0(1-a_0)/4\pi\ell)^{1/2}$.
\BO
Bereken hiermee $q_D$ en $p_D$. Laat zien dat zij
hetzelfde zijn als $N_0q$ en $N_0p$.
\EO

Hoe belangrijk de zelfenergie diagrammen in figuur \ref{tnbox} zijn, blijkt
uit vergelijking met de resultaten van ref. \cite{berkovits2}.
Deze auteurs hebben alleen het eerste diagram voor $K$ beschouwd. Zij
vonden
het interessante maar onfysische resultaat dat een extra verstrooier werkt
als een bron van intensiteit (`een lampje'). De behoudswet wordt door het
eerste diagram geschonden, maar niet door de combinatie van diagrammen.

\renewcommand{\thesection}{\arabic{section}}
\section{De doos van Hikami\index{Hikami doos}}
\setcounter{equation}{0}\setcounter{figure}{0}

\label{hikami}
Uitrekenen van de verstrooiing beter dan in de tweede orde Born benadering
resulteerde (via de andere $t-$matrix) in een simpele herdefinitie van de
vrije weglengte in fysisch observeerbare grootheden (hoofdstuk
\ref{microscopie}).

Soms zijn de effecten van deze hogere ordes subtieler, bijvoorbeeld in
correlatiefuncties. Twee intensiteitspropagatoren kun wisselwerken door
\'e\'en van hun amplitudepropagator uit te wisselen (`partnerruil'). 
In de tweede orde Born benadering blijkt dat deze interactie beschreven
wordt door een set van drie diagrammen, de `Hikami box'. (Hikami merkte de
rol van dit type diagrammen op voor wanordelijke electronsystemen 
\cite{hikami1}). Bij de langedrachts-correlatiefunctie $C_2$ van
veelvoudig verstrooid licht treed dezelfde interactie op.

De diagrammen
worden meestal met ruit-vormige blokken weergeven. Aan de armpjes en beentjes
van de diagrammen moeten ladderpropagatoren worden geknoopt; de inkomende
aan de verticale poten en de uitgaande aan de horizontale. We zullen hier
uitrekenen wat er gebeurt als je de hele Born reeks\index{Born-reeks}
meeneemt. We volgen hierbij referentie \cite{hik2}.

\subsection{Tweede-orde Born-benadering}\index{Born-benadering!tweede
orde} 
Zoals opgemerkt door Hikami \cite{hikami1}, moeten de eerste drie 
diagrammen van figuur~\ref{hikbox} uitgerekend worden. 
Dit zijn de relevante "partnerruil"-diagrammen in de tweede-orde
 Born-benadering. Twee ladderpropagatoren in de bulk
van het medium verwisselen een amplitude met elkaar.
 De basis voor de diagrammen is de volgende integraal:
\beq
\int_{-\infty}^\infty\!\! \d \r_1 \cdots \d\r_4 \,\L_1(\r_1) \L_2(\r_2)
\L_3(\r_3) \L_4(\r_4) \, G(\r_1-\r_2) G^\ast(\r_3-\r_2)
G(\r_3-\r_4) G^\ast(\r_1-\r_4) \label{hikrspace}
\eeq
In bovenstaande integraal zijn $\L_1(\r_1)$ en $\L_3(\r_3)$ de 
ladderpropagatoren
die de intensiteit {\it naar} $\r_{1,3}$ transporteren, en $\L_2(\r_2)$ en
$\L_4(\r_4)$ de ladderpropagatoren
 die de intensiteit {\it van} $\r_{2,4}$ {\it weg}
transporteren. Het product van de vier amplitude propagatoren $(G G
G^\ast G^\ast)$
vormt het lege Hikami diagram, de eerste uit de reeks van
acht in Figuur~\ref{hikbox}.
Het gemakkelijkst kunnen de diagrammen in de $\q$-ruimte uitgerekend
worden. De Fourrier-getransformeerde van vergelijking (\ref{hikrspace})
wordt gegeven door \beqa
&&\int \L_1(\q_1) \L_2(\q_2) \L_3(\q_3) \L_4(\q_4)
\eexp{i\q_1\r_1 + i\q_2\r_2 + i\q_3\r_3 +i\q_4\r_4}\nonumber\\
&&G(\p_1)G^\ast(\p_2) G(\p_3) G^\ast(\p_4) \nonumber \\
&&\eexp{i\p_1\cdot(\r_1-\r_2)
+i\p_2\cdot(\r_2-\r_3) +i\p_3\cdot(\r_3-\r_4)
+i\p_4\cdot(\r_4-\r_1)}\nonumber\\ && \frac{\d^3\q_1
\cdots \d^3\q_4 \d^3\r_1 \cdots \d^3\r_4 \d^3\p_1 \cdots
\d^3\p_4}{(2\pi)^{3 \times 8}} \label{hikqspace}
\eeqa
Integratie over $\r_{1,2,3,4}$ levert een aantal delta-functies op:
\beqa
\int \d^3\r_1 \rightarrow \delta(\q_1+\p_1-\p_4)(2\pi)^3 \hspace{2em} &
\p_4 \rightarrow \p & \p_1 = \p - \q_1 \nonumber \\
\int \d^3\r_2 \rightarrow \delta(\q_2+\p_2-\p_1)(2\pi)^3 \hspace{2em} &
 & \p_2 = \p - \q_1 -\q_2 \nonumber \\
\int \d^3\r_3 \rightarrow \delta(\q_3+\p_3-\p_2)(2\pi)^3 \hspace{2em} &
 & \p_3 = \p - \q_1-\q_2 -\q_3 \nonumber \\
\int \d^3\r_4 \rightarrow \delta(\q_4+\p_4-\p_3)(2\pi)^3 \hspace{2em} &
 & \p_4 = \p - \q_1 -\q_2 -\q_3 -\q_4 \nonumber \\
 && \Rightarrow \q_1 +\q_2+\q_3+\q_4 = 0  \nonumber
\eeqa
Dit houdt in dat $\p$ (voorheen $\p_4$) een vrije impuls is, terwijl dat de
andere impulsen ($\p_{1,2,3}$) zijn vastgelegd.\index{$H$, doos van Hikami}
De integratie over het $\p$-afhankelijke deel van
vergelijking (\ref{hikqspace}) wordt
\BEA     \label{ipbeh}
\int \cdots \frac{\d^3\p_1\d^3\p_2\d^3\p_3\d^3\p}{(2\pi)^{12}}\nonumber\\
\Rightarrow (2\pi)^3 \delta(\q_1 +\q_2 +\q_3 +\q_4)H_1 \EEA

Bovenstaande term is het eerste diagram van de acht in Figuur~\ref{hikbox}
We nummeren ze als $H_1 \cdots H_8$. $H_1$ wordt gegeven door
\beq
H_1 = \int \frac{\d^3\p}{(2\pi)^3} G(\p-\q_1) G^\ast(\p+\q_3+\q_4)
G(\p+\q_4) G^\ast(\p)
\eeq

We berekenen $H_1$ in de limiet $\q_i\rightarrow 0$. Dan is hij precies 
\'e\'en van de $I$-integralen gedefinieerd in de
Appendix:\index{$H_{1\ldots8}$, Hikami doos-diagrammen} \beq
H_1(\q_i \rightarrow {\bf 0}) = I_{2,2} + O(\q_i^2) = \frac{\ell^3}{8\pi
k_0^2} + O(\q_i^2)
\eeq
Nu de diagrammen met \'e\'en extra verbindingslijn, $H_2$ en $H_3$.
\beq
H_2 = n t^2\int\frac{\d^3\p}{(2\pi)^3}G(\p-\q_1)G^\ast(\p)G(\p+\q_4)
\int\frac{\d^3\p'}{(2\pi)^3} G(\p'-\q_3)G^\ast(\p')G(\p'+\q_2)
\eeq
\beq
H_3 = n \bar{t}^2\int\frac{\d^3\p}{(2\pi)^3}G^\ast (\p-\q_4)G(\p)G^\ast
(\p+\q_3)
\int\frac{\d^3\p'}{(2\pi)^3} G^\ast (\p'-\q_2)G(\p')G^\ast (\p'+\q_1)
\eeq
Je ziet dat beide diagrammen in twee\"en gesplitst kunnen worden met in elk
deel een vrije impuls; dit zie je ook direct aan de diagrammen. Immers,
ieder gestreept lijntje stelt een factor \'e\'en voor, maar er kan een
willekeurige impuls doorlopen!!
We berekenen ook $H_2$ en $H_3$ in de limiet $\q_i \rightarrow {\bf 0}$.
\beq
H_2(\q_i = {\bf 0}) = n t^2 I_{2,1}^2 \approx n t^2 \frac{-\ell^4}{64
\pi^2 k_0^2} \approx \frac{t}{\bar{t}} \frac{-\ell^3}{16 k_0^2 \pi}
\label{h2qis0}
\eeq
\beq
H_3(\q_i = {\bf 0}) = n \bar{t}^2 I_{1,2}^2 \approx n \bar{t}^2
\frac{-\ell^4}{64
\pi^2 k_0^2} \approx \frac{\bar{t}}{t} \frac{-\ell^3}{16 k_0^2 \pi}
\label{h3qis0}
\eeq
In de tweede-orde Born benadering geldt:
\beq
t \rightarrow u+iu^2 \frac{k}{4\pi}, \qquad \ell = \frac{4 \pi}{n u^2}
\qquad \frac{\overline{t}}{t}=1 \qquad \mbox{ ($2^e$ orde Born benadering)}
\label{bornapprox} \eeq
en hiermee krijg je
\beq
H_2({\bf 0}) = \frac{-\ell^3}{16 \pi k_0^2} \mbox{;} \qquad H_3({\bf 0}) =
\frac{-\ell^3}{16 \pi k_0^2} \label{h1h2qis0}
\eeq
Er blijkt:
\beq H_1({\bf 0}) + H_2({\bf 0}) + H_3({\bf 0}) = 0 + O(\q^2) \label{sumh1h2h3qis0}
\eeq
Je ziet dat de leidende term in het eerste diagram ($H_1$) weg valt
tegen de leidende termen van het tweede en het derde diagram! Dit hangt
sterk samen met energiebehoud (zie de opmerking aan het eind van hoofdstuk
\ref{invloed}). De diagrammen
moeten dus tot op orde $\q^2$ uitgerekend worden. Hiertoe
gaan we $G(\p-\q)$ benaderen tot op orde $\q^2$ met een Taylorreeks rond
$\q = {\bf 0}$.
\[
G(\p-\q) = \frac{1}{(\p-\q)^2 +\mu^2} = \frac{1}{\p^2 +\mu^2
-2\p\cdot\q + \q^2} \approx
\]
\beq
G(\p) + (2\p\cdot\q - \q^2)G^2 (\p)
+4(\p\cdot\q)^2 G^3(\p) + O(\q^3) \label{gpminqexp}
\eeq
Vervolgens gebruiken de volgende benadering
 voor de derde term in het rechterlid
van vergelijking (\ref{gpminqexp}),
\beq
\int \d^3\p (\p\cdot\q)^2 \, G^k(\p)G^{\ast\,l}(\p) \approx
\frac{1}{3}  \q^2 k_0^2 \int \d^3\p G^k(\p)G^{\ast\,l}(\p)= \frac{1}{3}
\q^2 k_0^2 I_{k,l}. \label{hoekapprox}
\eeq
De factor 1/3 komt van de middeling over de hoek in het inproduct,
$\p^2 \rightarrow k_0^2$ omdat daar de pool ligt die de
belangrijkste bijdrage levert aan de integraal.
Er kan nog een tweede vereenvoudiging gemaakt worden, die niet voor de
hand ligt. De tweede factor $-\q^2 G^2(\p)$ in de tweede term van
het rechterlid van
vergelijking (\ref{gpminqexp}) kan verwaarloosd worden omdat het een
bijdrage geeft in lagere orde van de vrije weglengte dan de andere termen.
De eerste factor $2(\p \cdot \q) G^2(\p)$ van de tweede term is $0$
na integratie over $\p$ omdat de integrand asymmetrisch is in $\p$.
Wat niet wegvalt zijn producten van de tweede term: $(2\p \cdot
\q_i)(2\p \cdot \q_j)$. Deze bijdragen kunnen we op dezelfde manier
benaderen als gedaan is in vergelijking (\ref{hoekapprox}), waarbij termen
asymmetrisch in $\p_i$ weer $0$ zijn. Je houdt over:
\beq
\int\d^3\p \, (\p\cdot\q_i)(\p \cdot \q_j) \, G^k(\p)G^{\ast\,l}(\p) \approx
\frac{1}{3} k_0^2 \,(\q_i \cdot \q_j) \, \int \d^3\p G^k(\p)G^{\ast\,l}(\p).
\label{hoekapproxtwee}
\eeq

Met behulp van de $I_{k,l}$ integralen uit de Appendix kunnen we de
diagrammen $H_1(\q_i)$, $H_2(\q_i)$ en $H_3(\q_i)$ tot op orde $\q^2$
opschrijven,\BA
H_1(\q_i)& =& \frac{\ell^3}{8 \pi k_0^2} - \frac{\ell^5}{24 \pi k_0^2}
(\q_1^2 + \q_2^2 + \q_3^2 + \q_4^2 + \q_1\cdot\q_3 + \q_2\cdot\q_4)
 \\
H_2(\q_i)& =& \frac{t}{\bar{t}} \left\{ \frac{- \ell^3}{16 \pi k_0^2} +
\frac{\ell^5}{48 \pi k_0^2} (\q_1^2 + \q_2^2 + \q_3^2 + \q_4^2 -
\q_1\cdot\q_4 - \q_2\cdot\q_3) \right\} \nonumber \\
H_3(\q_i)&=&\frac{\bar{t}}{t} \left\{ \frac{- \ell^3}{16 \pi k_0^2} +
\frac{\ell^5}{48 \pi k_0^2} (\q_1^2 + \q_2^2 + \q_3^2 + \q_4^2 -
\q_1\cdot\q_2 - \q_3\cdot\q_4) \right\} \nonumber
\EA
Omdat in tweede orde Born geldt $\overline{t}/t\approx1$, kunnen we
nu het eerste tot en met het derde
diagram optellen en vergelijking (\ref{bornapprox}) gebruiken. Je vindt dan
\BA \label{him2born}
\sum_{j=1}^3 H_j(\q_i)& =& \frac{\ell^5}{96\pi k_0^2} (-2\q_1\cdot \q_3
-\!2\q_2\cdot \q_4 -\! \q_1\cdot\q_4 -\!\q_1\cdot\q_2 -\!\q_2\cdot\q_3-\!
\q_3\cdot\q_4) \nonumber \\ &=& \frac{-\ell^5}{96 \pi k_0^2} (-2\q_1\cdot
\q_3 -2\q_2\cdot\q_4 +\q_1^2 +\q_2^2 +\q_3^2
+\q_4^2\nonumber \\ && -(\q_1+\q_2+\q_3+\q_4)^2)
\nonumber \\&\stackrel{\ref{ipbeh}}{=}& \frac{\ell^5}{96 \pi k_0^2}
(2\q_1\cdot
\q_3 + 2\q_2\cdot \q_4 -\q_1^2 -\q_2^2 -\q_3^2 -\q_4^2) \label{hikborn}
\EA
Deze uitdrukking, geldig voor een stationaire situatie zonder diffusie,
kan nog vereenvoudigd worden. Hij komt namelijk in praktijk altijd voor
vermenigvuldigd
 met \linebreak $\L_1(\q_1)\L_2(\q_2)\L_3(\q_3)\L_4(\q_4)$. 
 Volgens vgl. (\ref{difladder}) brengt ieder van
deze  propagatoren een factor $\L(\q)=12\pi/(\ell^3\q^2)$. Daarom eten de
 factoren $\q_i^2$ in (\ref{him2born}) \'e\'en van de diffusiepolen op
 ~\footnote{In de plaatsruimte geeft een bijdrage $\q_1^2\to -\nabla_1^2$ 
 werkend op een ladderpropagator: $\nabla_1^2\L_1(\r_1)=S_1(\r_1)$. Je raakt
 dus $\L_1$ kwijt ten gunste van de onschuldige bronterm $S_1$, die 
 meestal alleen aan de rand van belang is.}.
 Dus zorgen ze voor een niet-leidende bijdrage, die weggelaten wordt.
 Het resultaat is:
\beq
\sum_{j=1}^3 H_j(\q_i) = \frac{ - \ell^5}{48 \pi k_0^2} (\q_1 \cdot\q_3 +
\q_2 \cdot \q_4 ) \label{hikborndef}
\eeq
Dit is een symmetrische vorm, maar voor berekeningen is een andere vorm soms
handiger. Gebruik impulsbehoud voor $\q_2$ en $\q_4$ in de $\q_2 \cdot
\q_4$ in formule (\ref{hikborn}), dan vindt je:
\BE \sum_{j=1}^3 H_j(\q_i) = \frac{ - \ell^5}{48 \pi k_0^2} (2 \q_1
\cdot\q_3 - \q_2^2 - \q_4^2 ) \EE

\subsection{Hogere-orde Born-benadering}\index{Born-benadering}

We kunnen de doos van Hikami ook in hogere orde verstrooiing uitrekenen. 
Zoals opgemerkt in ref. \cite{hik2}, moeten dan 
wel meer dan de eerste drie diagrammen worden mee\-ge\-no\-men
($H_1$ tot en met $H_8$). In de
hogere dan tweede-orde Born-benadering kan namelijk de term $O(u)$
in $t/\bar{t} = 1 + O(u)$ groot zijn. In vergelijkingen (\ref{h1h2qis0}) en
(\ref{hikborn}) werd
aangenomen dat $O(u)$ klein is.
Voor de volledigheid geven we de uitdrukking voor $H_4$ ($H_5$,
$H_6$ en $H_7$ zijn hiermee equivalent) en voor $H_8$.
\BA
H_4 &=& n t^2 \bar{t} \int\frac{\d^3\p}{(2\pi)^3} G(\p-\q_1)G^\ast(\p)
G(\p+\q_4) \times \nonumber \\
&&\int \frac{\d^3\p'}{(2\pi)^3} G(\p')G^\ast(\p'+\q_3) \int
\frac{\d^3\p''}{(2\pi)^3} G^\ast(\p'')G(\p''+\q_2)
\EA
\BA
H_8 &=& n t^2 \bar{t}^2 \int \frac{\d^3\p}{(2\pi)^3} G(\p)G^\ast(\p+\q_1)
\int \frac{\d^3\p}{(2\pi)^3} G^\ast(\p)G(\p+\q_2) \times
\nonumber \\ &&
\int \frac{\d^3\p}{(2\pi)^3} G(\p)G^\ast(\p+\q_3)
\int \frac{\d^3\p}{(2\pi)^3} G^\ast(\p)G(\p+\q_4)
\EA
\BO
Toon aan dat in de limiet $\q \rightarrow {\bf 0}$ geldt:
\BE \begin{array}{lll}
H_1 =& \frac{\ell^3}{8 \pi k_0^2};  
&H_2 = \frac{t}{\bar{t}} \,\frac{- \ell^3}{16 \pi k_0^2}; \\
H_3 =& \frac{\bar{t}}{t} \,\frac{- \ell^3}{16 \pi k_0^2};
&H_4 = \frac{t\,Im\, t}{t\bar{t}}\, \frac{i \ell^3}{8 \pi
k_0^2}; \\
H_5 =& \frac{\bar{t}\, Im \, t}{t \bar{t}} \, \frac{-i \ell^3}{8 \pi
k_0^2};
&H_6 = \frac{\bar{t}\, Im \, t }{t \bar{t}} \, \frac{-i
\ell^3}{8 \pi k_0^2};\\
H_7 =& \frac{t\, Im \, t}{t \bar{t}} \, \frac{i \ell^3}{8 \pi
k_0^2};
&H_8 = \frac{ (Im \,t)^2}{t \bar{t}} \, \frac{\ell^3}{4 \pi k_0^2}
\end{array}\EE
Toon ook aan dat voor $\q_i = 0$ deze termen elkaar opheffen, dwz
\BEQ
H({\bf 0}) = \sum_{i=0}^8 H_i({\bf 0}) = 0
\EEQ
Je mag nu geen gebruik maken van vergelijking (\ref{bornapprox}),
alleen van het optisch theorema, zie ook vergelijking (\ref{opttheorema}).
\EO
Er geldt dus $H({\bf 0}) = 0$, ook in de hogere dan
$2^e$-orde Born-benadering, als maar de juiste diagrammen meegenomen
worden! Wat is nu het resultaat tot op orde $\q^2$? Er blijkt weer
\cite{hik2} \beq
H(\q_i) = \sum_{j=1}^8 H_j(\q_i) = \frac{\ell^5}{96 \pi k_0^2} \{-2\q_1\cdot
\q_3 - 2 \q_2 \cdot \q_4 + \q_1^2+ \q_2^2+ \q_3^2 + \q_4^2 \}
\eeq
Als de $t$-matrix niet unitair\index{$t$-matrix!unitariteit} is (in het
geval van absorptie) dan komt er een tweede term bij:
\[
\frac{ \ell^5 \kappa^2}{24 \pi k_0^2} \mbox{, met } \kappa^2 =
\frac{3(1-a)}{\ell^2} \; \mbox{, $a$ is het albedo.}
\]
Als de inkomende
bundels verschillende frequenties hebben, komt er nog een derde term
bij:
\[
-\frac{i \ell^5}{96 \pi k_0^2}(\Omegat_1+\Omegat_2+\Omegat_3+\Omegat_4) \;
\mbox{, met } \Omegat = \Omega/D \]
en $D=v \ell/3$ is de diffusie constante. De som van al deze termen geeft
\beq
H(\q) = \frac{\ell^5}{96 \pi k_0^2} \{-2\q_1\cdot
\q_3 - 2 \q_2 \cdot \q_4 + \q_1^2+ \q_2^2+ \q_3^2 + \q_4^2
+ 4 \kappa^2 - i (\Omegat_1+\Omegat_2+\Omegat_3+\Omegat_4) \}
\eeq
In deze algemenere situatie geldt volgens vgl. (\ref{difladder})
$\L(\q,\Omega)=12\pi\ell^{-3}/(\q^2+\kappa^2-i\Omegat)$.
Dus als we de ladderpropagatoren aan de doos van Hikami vastmaken,
vallen er weer diffusiepolen weg. We houden over
\beq
H(\q_1, \q_2, \q_3, \q_4) = \frac{- \ell^5}{48 \pi k_0^2} (\q_1 \cdot
\q_3 + \q_2 \cdot \q_4) = \frac{ - \ell^5}{48 \pi k_0^2} (2 \q_1
\cdot\q_3 - \q_2^2 - \q_4^2 ) \label{mcw17}
\eeq

Om de uitdrukking in de re\"ele ruimte te krijgen moeten we
bovenstaande uitdrukking invullen in vergelijking (\ref{hikrspace}) voor het
product van de vier Greense functies en het resultaat Fourier transformeren.
\BEA
\int \d^3\r_1 \cdots \d^3\r_4 \int \d^3\r \frac{\ell^5}{48 \pi k_0^2}
(\nabla_1 \nabla_3 + \nabla_2 \nabla_4)\nonumber\\
 \delta(\r_1 - \r) \delta(\r_2 - \r) \delta(\r_3 - \r) 
 \delta(\r_4 - \r) L(\r_1) L(\r_2) L(\r_3) L(\r_4)
\EEA
Dit is de vertex voor het uitwisselen van partners.
\BO Ga na
dat ook geldt:
$ \tOmega_1-\tOmega_2+ \tOmega_3-\tOmega_4 =0 $.\EO

\begin{figure}
\centerline{\includegraphics[width=10cm]{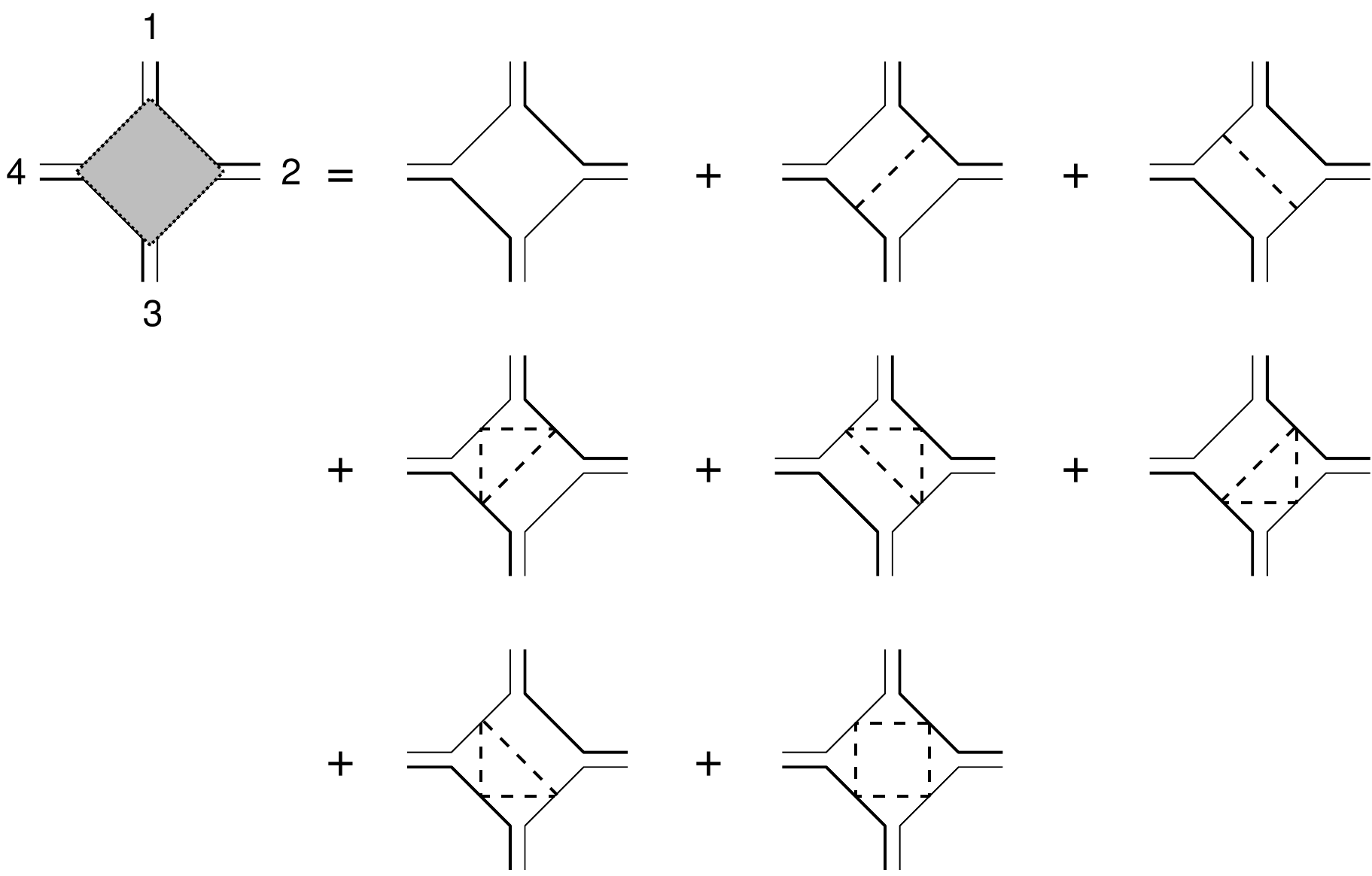}}

\caption{De volledige doos van Hikami. In de tweede orde Born zijn
alleen de bovenste drie
diagrammen belangrijk, zonder deze benadering zijn alle acht van
belang.
Dikke lijnen geven geretardeerde propagatoren weer, dunne lijnen geven
geavanceerde propagatoren weer. Gestreepte lijnen geven verstrooiing aan
\'e\'en verstrooier aan.} \label{hikbox} \end{figure}
De gevonden uitdrukking (\ref{mcw17}) is weer dezelfde als die
van Hikami \cite{hikami1}
en bijvoorbeeld Stephen en Cwilich \cite{stephen0} in de tweede orde
Born, waar $\ell=4\pi/(nu^2)$.
{\it (Pas op: er staan nogal wat foute uitdrukkingen voor de Hikami doos in
de literatuur.)}
In onze benadering is wel de definitie van de vrije weglengte anders,
namelijk $\ell=4\pi/(n\overline{t}t)$.
We concluderen: als we hogere orde Born termen meenemen in
\'e\'en deel
van de theorie (de $t-$matrix), moeten we dit overal doen.

\renewcommand{\thesection}{\arabic{section}}
\section{Correlatiefuncties van spikkels\index{spikkels}}
\setcounter{equation}{0}\setcounter{figure}{0}

 \label{correlatie}\index{correlatiefuncties}

\subsection{Correlaties met korte, lange en oneindige dracht}
Als een monochromatische vlakke golf op een wanordelijk medium schijnt,
ontstaan grote intensiteits fluctuaties (spikkels of `speckles') in
transmissie. Dit
effect is het gevolg van interferentie tussen de verschillende paden.
Wij zijn ge\"interesseerd in de correlaties tussen de
spikkelpatronen van
twee verschillende bundels. De bundels kunnen verschillende inkomende hoeken,
verschillende frequentie of verschillende posities hebben. Misschien
zou je verwachten dat alle grootheden volstrekt willekeurig zijn, dat
blijkt niet zo te zijn. Ondanks de wanorde
van het medium, zijn er correlaties met korte, lange en `oneindige' dracht
\cite{feng}. Correlatie functies in wanordelijke media werden voor het eerst
bestudeerd in electronsystemen. Meestal wordt daar voor berekeningen een
diagrammatische techniek toegepast.
Stephen en Cwilich gebruikten een soortgelijke techniek om
correlaties van veelvoudig verstrooid licht uit te rekenen \cite{stephen1}.
Zyuzin en Spivak introduceerden een Langevin
benadering\index{Langevin benadering} om de berekening van
correlatie functies te vereenvoudigen \cite{zyuzin1}. Pnini en Shapiro
gebruikten deze methode om de correlatie functies van licht dat ofwel
reflecteerde aan of doorgelaten werd door medium \cite{pnini2}.

Correlatie functies met korte en lange dracht zijn gemeten in
verschillende experimenten, bijvoorbeeld \cite{albada1}, \cite{genack1} en
\cite{deboer}. Effecten van absorptie \cite{pnini1} en randlagen
\cite{lisyansky} zijn berekend. Overigens verschillen de theoretische
voorspellingen nogal eens in voorfactoren (bijvoorbeeld \cite{stephenboek},
\cite{berkovits} en \cite{pnini2}). Het was niet helemaal duidelijk of de
Langevin benadering volledig correct is. Die voorfactor is belangrijk,
zij kan immers experimenteel bepaald worden \cite{deboer}.

De invallende bundel is een vlakke golf met oppervlakte $A$ ($A \gg
L^2$)\footnote{In echte experimenten moet men rekening houden met de
eindige grootte van de bundel \cite{deboer}}.
Stel dat $a,a'$ de hoeken en frequenties van de inkomende bundels aangeven
en $b,b'$ evenzo voor de uitgaande bundels. De relatieve correlatie functie
is gedefinieerd als: \BE C_{aba'b'} \equiv \frac{\expect{T_{ab}
T_{a'b'}}-\expect{T_{ab}}\expect{T_{a'b'}} } {\expect{T_{ab}}
\expect{T_{a'b'}} } \label{defcor} \; .   \EE

Een belangrijk verschil met electronische systemen is dat in een optisch
experiment
de impuls van de golven precies gecontroleerd kan worden. Daarom kan men de
korte, lange en oneindige dracht bijdragen tot de correlatiefunctie
onderscheiden: $C=C_1+C_2+C_3$ \cite{feng}.\index{$C$, $C_1$, $C_2$, $C_3$,
correlatiefuncties}
 Schematisch staan deze
bijdragen in figuur \ref{c123}.
 \begin{figure}
\caption{De bijdragen tot de correlatie functie. De `zwarte doos' geeft
een alle bijdragen aan $<T T>$ aan, waarvan $<T><T>$  verreweg het
belangrijkste is, maar ook correlaties zijn aanwezig. De
pijlen geven aan of we met een geretardeerde of geavanceerde propagator te
maken hebben. Propagatoren met dezelfde massa hebben
hetzelfde lijntype (doorgetrokken of gestreept).
De sporten van de ladders zijn
weggelaten, dicht
bij elkaar lopende lijnen zijn ladderpropagatoren.
 Voor $C_2$ en $C_3$ zijn er ook
diagrammen met gepermuteerde lijnen}
\vspace{5mm}
\centerline{\includegraphics[width=10cm]{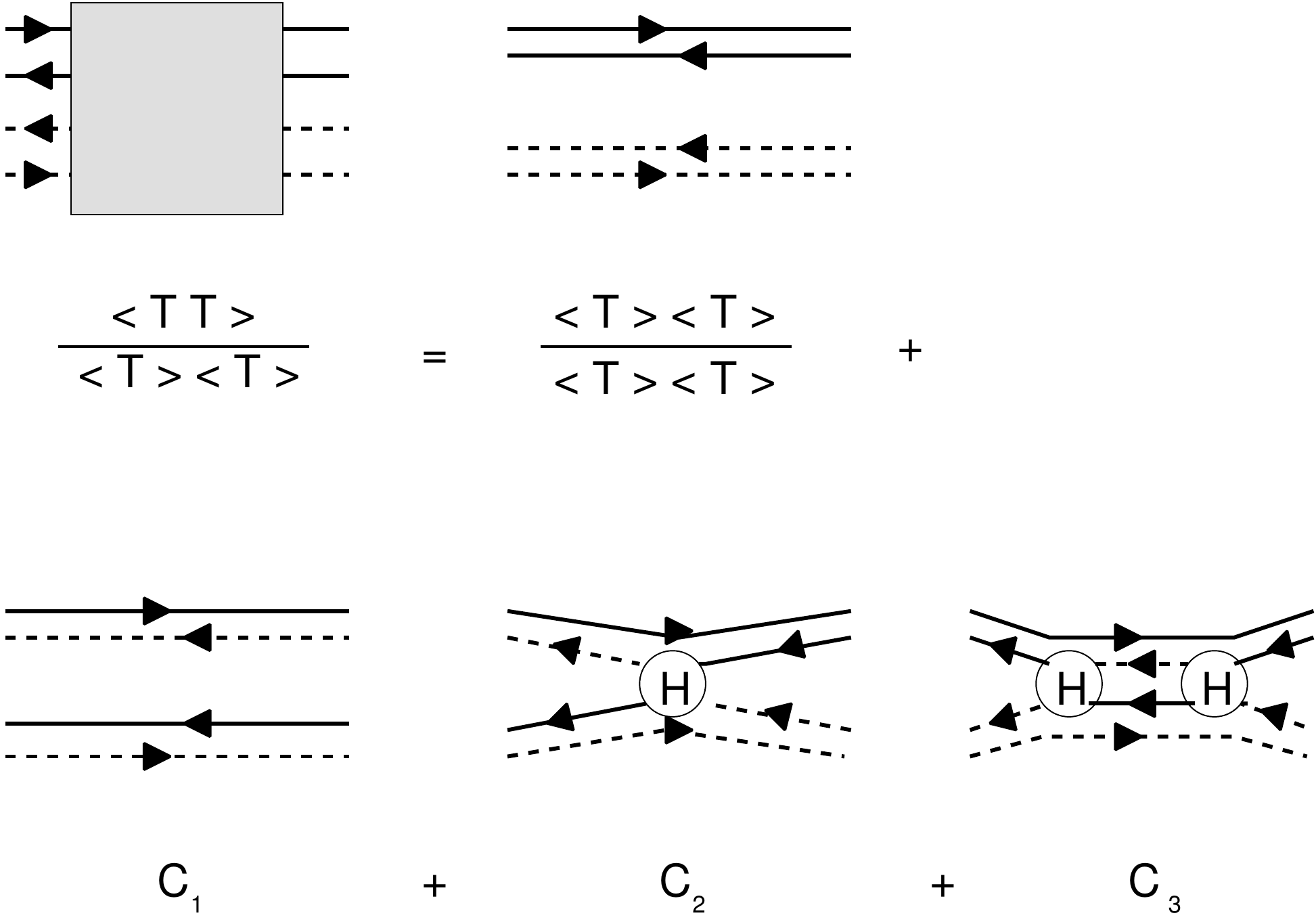}}
\label{c123}
\end{figure}
Deze bijdragen kunnen
afzonderlijk gemeten worden. De scheiding kan uitgelegd worden door de
analogie met een golfpijp te bekijken. In een golfpijp corresponderen de
verschillende bijdragen met verschillende machten van de dimensieloze
geleidbaarheid $g$, waarbij $g \sim N
\ell/L$. ($N$ is het aantal golfpijp 'modes', evenredig met $k_0^2 A$).
$C_1, C_2 , C_3$ zijn evenredig met $g^0,g^{-1},g^{-2}$, respectievelijk.
\index{$g$, dimensieloze geleidbaarheid} De
factor $1/g$ is ruwweg de kans dat twee diffusie-paden elkaar ontmoeten in
een punt in het medium. Beschouw de diffusiepaden als buisjes met een
oppervlakte $1/k_0^2$, door diffusie is hun lengte $\ell N_{stap}
\sim \ell (L/\ell)^2 =L^2/\ell$. Het volume van het buisje ten
opzichte van het totale volume is de trefkans en wordt: $[L^2/(\ell
k_0^2)]/[AL]=g^{-1}$. Ofschoon $g$ normaliter erg groot is (100-1000),
kunnen $C_2$ en $C_3$ bijdragen toch afzonderlijk gemeten worden. Dit komt
doordat de relatieve sterkte van de termen afhangt van het type experiment
dat men uitvoert.

Beschouw een \'e\'en-kanaal-in, \'e\'en-kanaal-uit experiment. Als de
frequentie van de inkomende bundel verandert, vervormen de spikkels van de
uitgaande bundel. Uiteindelijk is er niets meer van de spikkel over en de
correlatie vervalt exponentieel. Dit is de kortedrachtsfunctie $C_1$. We
kunnen ook de hoek van de inkomende bundel iets veranderen. De uitgaande bundel
verandert nu om {\it twee} redenen. Ten eerste volgt de spikkel de inkomende
bundel
(`memory effect'). Ten tweede vervormt de spikkel weer willekeurig door de
decorrelatie. $C_1$ is een scherp gepiekte functie. Zij is alleen
ongelijk nul als de inkomende en uitgaande hoek evenveel veranderen.

Correlaties met lange dracht worden door de $C_2$ beschreven. Beschouw
opnieuw
een \'e\'en-kanaal-in, \'e\'en-kanaal-uit meting. Echter ditmaal meten we
het
kruis-correlaat van twee ver uiteengelegen spikkels \cite{garcia}. Omdat er
nu een hoekverschil is bij de uitgaande maar niet bij de inkomende bundels,
geeft de $C_1$ geen bijdrage. In plaats daarvan ziet men een veel
zwakkere correlatie.
Deze correlatie vervalt algebraisch, als we de frequentie veranderen. Dit is
nu $C_2$, hij beschrijft correlaties tussen spikkels die ver van elkaar
liggen. In een \'e\'en-kanaal-in, \'e\'en-kanaal-uit meting, kan men de
zwakke effecten van deze hogere orde correlaties slechts zien in zeer sterk
verstrooiende media \cite{genack2}. $C_2$ kan eenvoudiger gemeten worden
in een opstelling met \'e\'en-kanaal-in en integrerend over alle uitgaande
kanalen, zie van Albada {\it et al.} \cite{deboer}. De scherp gepiekte
$C_1-$correlatiefunctie wordt dan overstemd door  $C_2$.

De spikkelpatronen veranderen als de positie van de verstrooiers verandert,
of de frequent van het licht. De verdeling van de $C_2$ is voorspeld \cite{NvRC2}
en experimenteel geverifieerd voor verstrooiing aan metalen kogels \cite{StGenack1}.
Men heeft zelfs een universaliteit waargenomen: dezelfde $C_2$ verdeling 
treedt ook op rond een nulpunt in het spikkelpatroon, en alsmede indien er absorptie of localisatie is \cite{SZGenack2}
en dynamisch.

Ten slotte is er de $C_3$, deze is verantwoordelijk voor de correlaties met
'oneindige' dracht. In een alle-kanalen-in, alle-kanalen-uit meting, is
$C_3$ dominant. In zo'n meting middelt men over een veelvoud van
spikkelpatronen, het gemiddelde bestaat uit een gemiddelde waarde en een fluctuatie
daarop. De $C_3$ beschrijft de frequentie correlaties van de fluctuaties in
de totale transmissie. Dit deel van de correlatie functie is de oorzaak van
zogenaamde `universele geleidbaarheids fluctuaties'\index{universele
geleidbaarheids fluctuaties} in electron systemen, zie bijv. \cite{vRALN}. Voor licht is zij nog
niet waargenomen.

\subsection{Invloed van randlagen\index{correlatiefuncties!invloed van
randlagen} op spikkelcorrelaties}
\subsubsection{De ladderpropagator incluis randlagen, absorptie,
frequentie-
en hoekverschil}
Bij het uitrekenen van correlatiefuncties rekent men gewoonlijk in de
diffusie benadering met dezelfde brekingsindices binnen en buiten het medium.
De diffuse intensiteit wordt nul verondersteld iets buiten het medium op
\'e\'en extrapolatie lengte, $z_0$. Randeffecten kunnen belangrijk
worden ten gevolge van interne reflecties. Zij veranderen immers de
vorm van de diffuse intensiteit in het medium. Deze interne reflecties worden
bepaald door de verhouding van de brekingsindices binnen en buiten het
medium, $m \equiv n_0/n_1$.

Hoewel de diffusie benadering niet meer goed is bij de randen, hebben we
gezien dat
er correcties op de diffusiebenadering zijn gemaakt om randeffecten mee te
nemen (hoofdstuk \ref{mesoscopie}), \cite{lagendijk}. De invloed van
randlagen op correlatie functies is
bestudeerd met zo'n verbeterde diffusie benadering, en men dacht te zien dat
het verval van de lange-drachtscorrelatie functie door de randlagen
verandert \cite{lisyansky}.

In het voorgaande en door Nieuwenhuizen en Luck is aangetoond dat de effecten
van randlagen beter op het niveau van ladderdiagrammen behandeld kan worden
[NL]. In tegenstelling tot in de diffusie benadering, zijn er dan
geen vrije parameters. Zij hebben er ook op gewezen 
dat de verbeterde diffusiebenadering exact wordt in de limiet van grote
brekingsindexverhouding. Dat is fysisch te begrijpen, immers diffusietheorie
is geldig op lengteschalen die veel groter zijn dan de gemiddelde vrije
weglengte. Bij grote brekingsindexverhoudingen zijn de randlagen inderdaad
dik ten op zichte van de vrije weglengte. Zelfs bij realistische, vrij kleine
waarden ($m \approx 1.5$ $ \mbox{\`a}$ $2$) geeft de verbeterde 
diffusiebenadering een goede beschrijving van de randlagen [NL].

We gaan uit van de Schwarz\-schild-Milne vergelijking
voor de genormaliseerde diffuse intensiteit $\Gamma(z)$ in het medium.
\BE \Gamma(z)=\eexp{-z/\ell
\cos\theta_a} +\int_0^L M(z,z') \Gamma(z') dz' \; . \label{sm}
\EE
Na Fourier transformatie in de $z$-richting wordt deze vergelijking
op grote afstand: \BE (-\partial_z^2+M^2)\Gamma(z,M) \equiv (-\partial_z^2+Q^2
+\kappa^2-i\Omegat)\Gamma(z,M) =0 \label{ladderverg} \; . \EE $M$ speelt dezelfde
rol als een massa (inverse afvallengte), 
hij beschrijft het exponenti\"ele verval van de ladderpropagator
in de $z-$richting. We nemen aan  dat $M\ell \ll 1$. De loodrechte
externe impuls ${\Q=\qt}$, hangt samen met de afgebogen inkomende hoeken van de
bundels: \BE 
{\bf Q}={\qt}= k_0\sin\vartheta'_a (\cos \phi_a , \sin \phi_a )-
k_0\sin\vartheta'_{a'} (\cos \phi_{a'}, \sin \phi_{a'})  \;  \EE
zijn absolute waarde is
\index{$ql@$\qt$, transversale impuls}\index{$Q@$\Q$, $Q$, transversale impuls}
\BEQ Q=\left| \qt \right|. \EEQ
$\tOmega$ is evenredig met
het frequentieverschil van de propagatoren, $\tOmega=\Omega/D$.

Voor optisch dikke media ($L\gg \ell$) beschouwen we weer
de homogene en speciale oplossingen van vergelijking (\ref{sm}) in de
limiet van oneindige $L$. Deze zien er ditmaal als volgt uit:\BA
\Gamma_H(z,M)&=&\frac{\sinh(M
z)}{M \ell} +\frac{z_0(M;m)\eexp{-M z}}{\ell} \; , \nonumber \\ \Gamma_S(z,M)&=&
\frac{z_1(M;m) \eexp{-M z}}{\ell} \; .
\label{homspec} \EA
\BO
Ga  na dat als $M\rightarrow 0$, je het bekende gedrag terugvindt: $\Gamma_H
\sim \tau+\tau_0 \, ;\, \Gamma_S\sim \tau_1$. \EO
Als we de parameters $z_0(M;m)$ en $z_1(M;m)$ (numeriek) bepaald hebben ligt
de diffuse intensiteit vast. Een extra moeilijkheid duikt op omdat $z_0$ en
$z_1$ van $M$ afhankelijk zijn, en we willen juist $M$ vari\"eren. De
afhankelijkheid kan op een soortgelijke manier worden behandeld als de
behandeling van grote brekingsindex verhouding in referentie [NL] en
hoofdstuk \ref{groot}. Men kan aantonen dat voor grote
brekingsindexverhouding
$z_{0,1}\rightarrow 1/M$, als $m \rightarrow \infty$. Een bruikbare
interpolatie tussen de regimes is \BE z_{0,1}(M;m)=\frac{z_{0,1}(m)}{1+M
z_{0,1}(m)} \label{zm} \; .\EE Omdat  $m$ per experiment constant is,
schrijven we $z_{0,1}$ voor  $z_{0,1}(m)$.
De parameters $z_0$ en $z_1$ kunnen nu numeriek berekend worden [NL].
$z_1$ hangt ook nog van de hoek van de inkomende bundel af. Maar bij de
beschouwing van hoekcorrelaties zijn de typische hoekveranderingen zo klein
dat $z_1$ constant mag worden genomen.

We zijn nog steeds op zoek naar de diffuse intensiteit. De
intensiteitspropagator  kan in het gebied $0<z<L-10 \ell $ geschreven
worden als \BE
\Gamma(z,M)=\alpha \Gamma_H(z,M)+\Gamma_S(z,M) \; ,  \\ \EE en de propagator in het gebied
$10 \ell <z<L$ \BE \Gamma(z,M)=\beta \Gamma_H (L-z,M)  \; . \EE
\BO
Knoop beide oplossingen aan elkaar in het midden van het medium. Bepaal
daaruit $\alpha$
en $\beta$. Laat zien dat dit de ladderintensiteit geeft
\BE
\Gamma(z,M) =\frac{z_1}{\ell} \; \frac{\sinh(M L-M z)+M z_0
\cosh(M L-M z)}{(1+M^2 z_0^2) \sinh(M L)+2M z_0\cosh(M L)} \;.
\label{rtladder} \EE
Dit is de formule die we in het vervolg nodig
hebben.
\EO
De formule lijkt sterk op die van de verbeterde diffusie benadering
\cite{lisyansky}. Echter hier zijn de parameters $z_0$ en $z_1$ precies
voorgeschreven. We zijn nu in staat de correlatie van bundels met
verschillende frequentie en verschillende hoeken en met absorptie te
berekenen. (De ruimtelijke correlaties kunnen berekend worden met de Fourier
getransformeerde van de hoek\-correlatiefunctie).

\subsection{Correlaties met korte dracht}
De $C_1$ is het product van twee onafhankelijke ladders met uitgewisselde
partners. Beide ladders bestaan uit een propagator van de ene bundel en een
propagator van de andere bundel. Het loodrechte impulsverschil van de twee
inkomende bundels, $Q$, is gelijk aan het impulsverschil van de uitgaande
bundels. We berekenen het genormeerde product van een ladder met
gekwadrateerde `massa' $M^2=Q^2+\kappa^{2}-i\Omegat$ en zijn complex
geconjugeerde.

We moeten nu nog (net als bij de transmissie) de ladderpropagator
 naar buiten integreren. Voor kleine massa ($ M
\ell \ll 1$ ) krijgen we dezelfde integralen in teller en noemer.
Het resultaat wordt dan eenvoudig gegeven door de waarde op $z=L$. Dit
geeft: \BA C_1 &=&\left| \frac{\Gamma(z=L,M)}{\Gamma(z=L, \kappa)} \right|^2 \nonumber
\\ &=&\frac{|M |^2}{\kappa^2} \left| \frac{(1+\kappa^2 z_0^2)\sinh \kappa
L+2\kappa z_0\cosh \kappa L} {(1+M^2 z_0^2)\sinh M
L+2M z_0 \cosh ML} \right|^2 \; .
\EA
\begin{figure}
\centerline{\includegraphics[width=8cm]{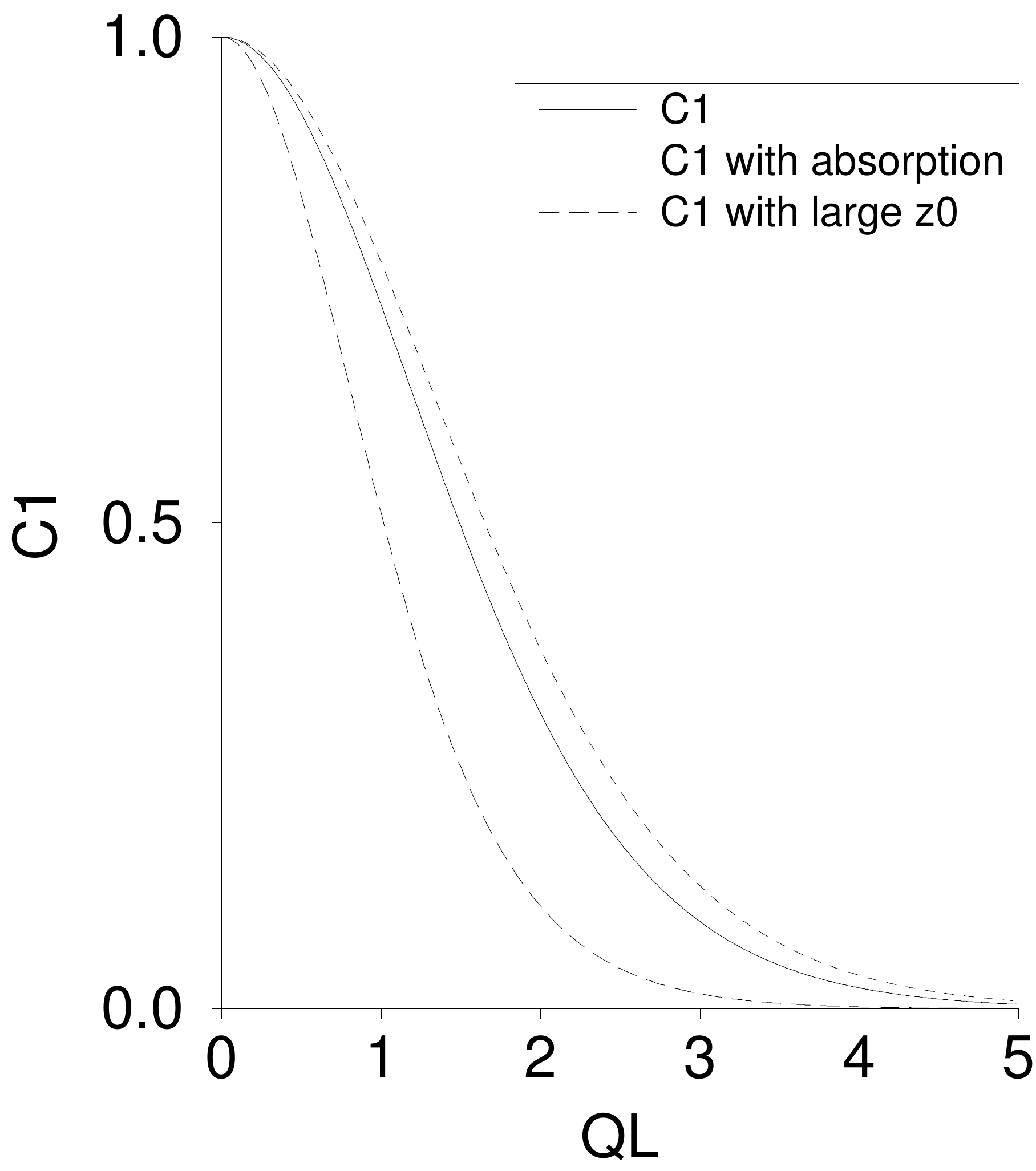}}

\caption{De $C_1$ hoekcorrelatie functie versus het geschaalde
loodrechte impulsverschil. Doorgetrokken lijn: kleine $z_0$,
geen absorptie. Lijn met korte strepen: absorptie ($\kappa=2/L$), kleine $z_0$.
Lijn met lange strepen: dikke randlagen ($z_0=L/3$), geen absorptie.}
\label{figc1}
\end{figure}

In figuur \ref{figc1} is $C_1$ voor verscheidene gevallen getekend.
In een dikke plak in de limiet $M z_0 \ll 1$ en geen absorptie 
vinden we het resultaat uit de 
diffusie benadering \cite{feng} \BE
C_1 = \left| \frac{ML}{ \sinh M L} \right|^2 \; . \EE 
In dit geval spelen de randlagen geen rol. Voor grote hoek of
frequentieverschillen ($ML \gg 1$), valt $C_1$ exponentieel af.

\subsection{Correlaties met lange dracht}
De correlatiefunctie met lange dracht, $C_2$, wordt gegeven door:
\BA
C_2 & =&\frac{1}{\langle T \rangle \langle T \rangle } \int_0^L dz dz_1
dz_2 dz_3 dz_4 \L(z_1,M_1) \L(z_3,M_3) H(z_1,z_2,z_3,z_4)\\
 \times
&& \delta(z-z_1) \delta(z-z_2)\delta(z-z_3)\delta(z-z_4) \L(L-z_2,M_2)
\L(L-z_4,M_4)\nonumber\label{hikc2}
\EA
(Voor het gemak hebben de indices maar weggelaten.)
We hier gebruikt dat ook $C_2$ kan worden benaderd door waarden op rand.
$C_2$ komt van de interactie van de diffusie paden die `ergens in het
verborgene partnertje ruilen'. Deze interactie wordt gegeven door de
Hikami-doos
\cite{hikami2}. In de $z-$coordinaat representatie is de Hikami-doos
\BE
H(z_1,z_2,z_3,z_4)= \frac{\ell^5}{48\pi k_0^2} (\partial_{z_1}
\partial_{z_3} +\partial_{z_2} \partial_{z_4}+Q^2 )
 \; .   \label{hik}
\EE
Hierbij is ook gebruikt dat $\q_{1\perp}=\q_{3\perp}=0 \;
,\q_{2\perp}=-\q_{4\perp}\equiv {\bf Q}$.
 We berekenen  $C_2$ door in te vullen
$ M_1=M_3=\kappa
,\, M^2_{2,4}=Q^2+\kappa^{2} \pm i
\tOmega$. Dit geeft \'e\'en bijdrage, een andere bijdrage volgt uit
 het verwisselen
van ingaande en uitkomende bundels (verwisselen van $M_{1,3}$ met $M_{2,4}$). 
Maar in een $C_2$-experiment wordt vaak over de uitgaande kanalen
ge\"integreerd. Die tweede term middelt
 dan en laat alleen een constante bijdrage achter.
Merk op dat $C_2$ van de oppervlakte van de bundel afhangt.
In het algemeen is ook het intensiteitsprofiel van de bundel van belang
\cite{deboer}.
De uitdrukking die we voor $C_2$ vinden lijkt sterk op het resultaat van
de {\it Langevin-methode}. Dit kun je nagaan door formule (39) van
referentie
\cite{pnini2} met formules (\ref{hik}) en (\ref{hikc2}) hier te vergelijken. 
Dat is niet zo vreemd, in de
Langevin-benadering neem je aan dat er een macroscopisch intensiteit is, die
de gemiddelde diffusie beschrijft, en daar bovenop een ongecorreleerde
stochastische ruis stroom. In onze benadering hebben we een langzaam
vari\"erende diffuse intensiteit, de ongecorreleerde random puntinteracties
komen van de Hikami-box: de twee benaderingen zijn, per definitie, equivalent.

De uiteindelijke uitdrukking voor de $C_2$ is lang en staat hieronder.
We zullen hier ons tot enkele speciale gevallen beperken.
Neem eerst het geval van hoekcorrelaties zonder absorptie
maar met randlaageffecten:
\BE
C_2(Q)= \frac{3\pi L}{k_0^2 l A} F_2(Q) \label{f2c2} \; ,
\EE waarbij $F_2$ is gedefinieerd als:
\BA
F_2(Q)&=& [ \sinh
2Q L-2Q L+Qz_0 \; (6\sinh^2QL-2Q^2L^2) \nonumber \\
&&+4 Q^2z_0^2 \; (\sinh 2QL-QL) +6Q^3 z_0^3 \; \sinh^2 QL
+Q^4z_0^4 \; \sinh 2QL ] \times  \nonumber \\
&& [2QL \left\{(1+Q^2 z_0^2)\sinh QL+2Q z_0 \cosh QL\right\}^{2}]^{-1}
 \; .\index{$F_2$, genormeerde $C_2$}\EA
dit valt als $1/Q$ af voor grote Q (grote hoeken).
Als we de randeffecten verwaarlozen, vinden we het resultaat van de diffusie
benadering terug \cite{pnini1}:
\BE
F_2(Q)=\frac{\sinh(2QL)-2QL}{2QL \sinh^2
(QL)} \label{c2dq} \; . \EE
In figuur \ref{figc2} is de $F_2$ voor verschillende gevallen getekend.
\begin{figure}
\centerline{\includegraphics[width=8cm]{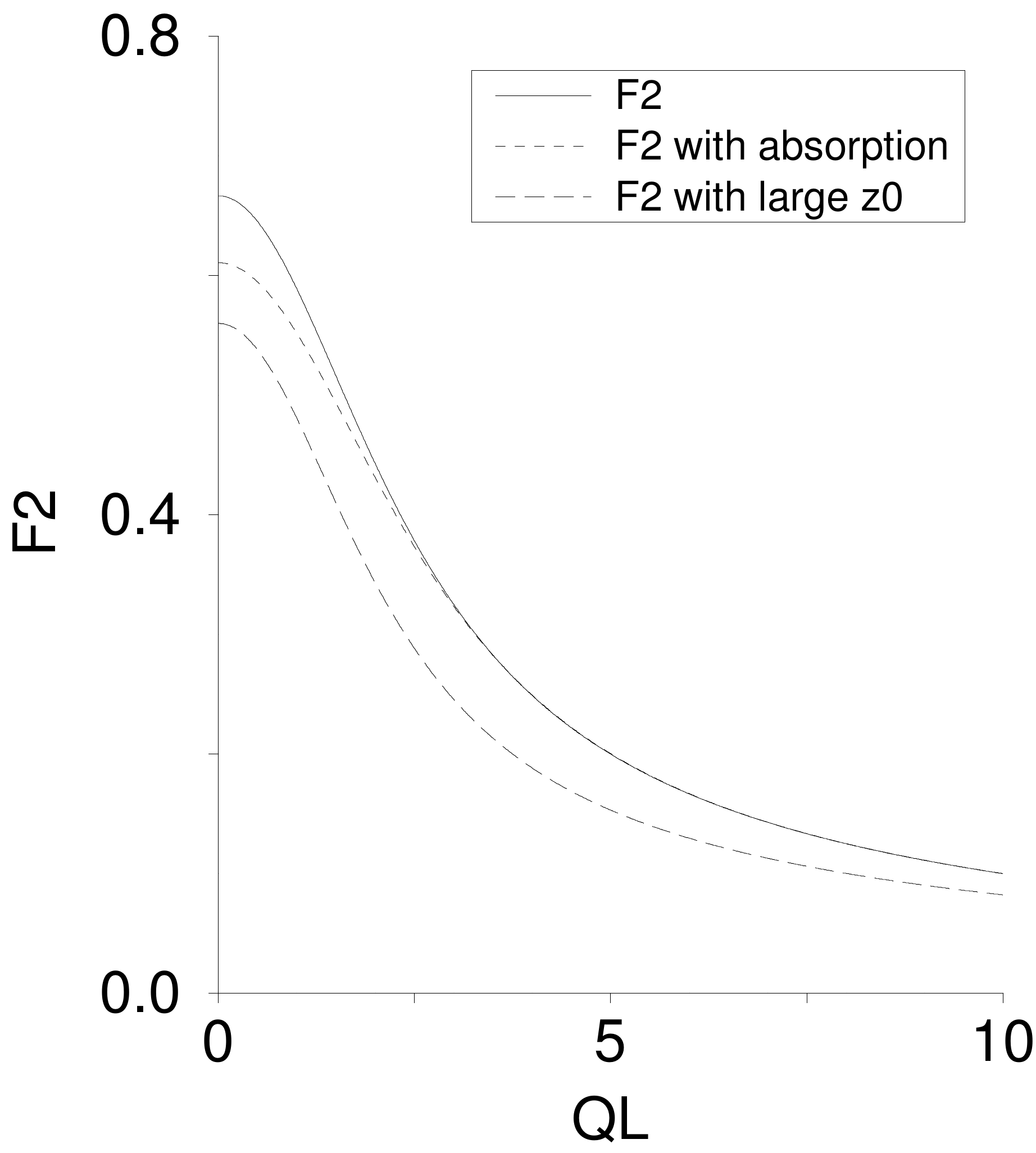}}

\caption{$F_2$ hoekcorrelatie functie versus geschaalde
loodrechte impulsverschil. Doorgetrokken lijn: kleine $z_0$,
geen absorptie. Lijn met korte strepen: absorptie ($\kappa=2/L$), kleine $z_0$.
Lijn met lange strepen: dikke randlagen ($z_0=L/3$), geen absorptie.}
\label{figc2}
\end{figure}

We kunnen ook frequentiecorrelaties bekijken;
zonder absorptie en met verwaarlozing van de randen vinden we \cite{deboer}:
\BA F_2 (\Omega) &=& \frac{2}{\sqrt{2\Omega} \, L}
\left( \frac{\sinh(\sqrt{2\Omega}\,L)-\sin(\sqrt{2\Omega}\,L)}
{\cosh(\sqrt{2\Omega}\,L)-\cos(\sqrt{2\Omega}\,L)} \right)  \label{c2dw}
\; . \EA
Zoals je hieronder kunt zien, verandert de vorm als we randlagen meenemen.
Zonder interne reflecties ($m=1$) vervalt $C_2$ als $1/|M|$ voor
grote $|M|$. Lisyanski en
Livdan beweren dat dit verandert bij sterke
interne reflectie \cite{lisyansky}. Hier vinden we dat echter niet;
bij grote $z_0$ blijft het verval $1/|M|$.

Deze gebruikte technieken kunnen ook voor de $C_3$ gebruikt worden. Ook
re\-flec\-tie\-cor\-relaties zijn interessant, zij zullen immers gevoeliger
zijn voor randlagen.

\subsection{De algemene uitdrukking voor de $C_2$-correlatie}
Hier is dan de algemene uitdrukking voor de correlatie met lange dracht,
deze wordt gegeven door formule (\ref{hikc2}) met de ladderpropagatoren van
formule (\ref{rtladder}).
De uitdrukking is een vierde-orde polynoom in $z_0$.
Het is handzaam om $M_2$ en $M_4$ te splitsen in hun re\"ele en imaginaire
deel, $M_{2,4}=\sqrt{Q^2+\kappa^2\pm i\Omega}\equiv a \pm i b$.
Men vindt\BE
C_2=\frac{3\pi L}{k_0^2 \ell A} \frac{N_a+N_b+N_{\kappa}}{BL} \; ,
\EE
waarbij:
\BA
B&=&[1+2z_0^2(a^2-b^2)+z_0^4(a^2+b^2)](\cosh 2aL-\cos
2bL) \nonumber \\ && +2z_0[1+z_0^2(a^2+b^2)](a\sinh
2aL+b\sin 2bL)
\nonumber \\ && +4z_0^2(a^2+b^2)(\cosh 2aL +\cos 2bL)  \nonumber \\
%
N_a&=&\frac{1}{2a(a^2-\kappa^2)} \times
\{[ (2a^2-\kappa^2)+z_0^2 (8a^4+2a^2
b^2-b^2\kappa^2-3a^2\kappa^2+\kappa^4)+ \nonumber \\ &&
z_0^4(a^2+b^2)(2a^4-2a^2\kappa^2+\kappa^4)] \sinh2aL+\nonumber \\ &&
[ 4z_0a(3a^2-\kappa^2)+4z_0^3a(3a^4+a^2
 b^2-2a^2\kappa^2+\kappa^4) ]\sinh^2 aL \} \nonumber \\
N_b&=&\frac{1}{2b(b^2+\kappa^2)}\times
\{[-(2 b^2+\kappa^2)+z_0^2 (2a^2
b^2+8b^4+a^2\kappa^2+3b^2\kappa^2+\kappa^4)- \nonumber \\ &&
z_0^4(a^2+b^2)(2b^4+2b^2\kappa^2+\kappa^4)] \sin 2bL+   \nonumber \\&&
[ 4z_0b(3b^2+\kappa^2)-4z_0^3b(3b^4+a^2 b^2+
2b^2\kappa^2+\kappa^4)] \sin^2 bL \} \nonumber \\
N_{\kappa}&=&\frac{a^2+b^2}{2\kappa(a^2-\kappa^2)(b^2+\kappa^2)} \times
\{ [-\kappa^2-z_0^2 (6a^2 b^2+5a^2 \kappa^2-5b^2 \kappa^2+\kappa^4)+ \nonumber \\
&& z_0^4 \kappa^2
(-2a^2 b^2-a^2\kappa^2+b^2\kappa^2) ]  \sinh 2\kappa L+
\nonumber \\&&
[4z_0 (-a^2 b^2-a^2\kappa^2+b^2\kappa^2-\kappa^4)+ \nonumber \\ &&
4z_0^3  (-3a^2
b^2-2a^2\kappa^2+2b^2 \kappa^2) ] \sinh^2 \kappa L \}
\EA
Zonder absorptie wordt de frequentiecorrelatie
($Q=\kappa=0$, $a=b=\sqrt{\Omega/2}$): 
\BA
C_2&=& \frac{3\pi}{k_0^2 \ell A a}
\{ (\sinh2aL- \sin2aL)+
z_0a(-4a^2 L^2 + 6\sinh^2 aL + 6\sin^2 aL)+ \nonumber \\
&&z_0^2a^2(-12a L + 5\sinh2aL +5\sin2aL)+
8z_0^3a^3(\sinh^2 aL - \sin^2 aL)+ \nonumber \\
&&2z_0^4a^4(\sinh2aL- \sin 2aL) \} \times \nonumber \\
&&\{ (1+2a^2z_0^4)(\cosh 2aL-\cos
2aL) +2az_0(1+2a^2z_0^2)(\sinh
2aL+\sin 2aL)
\nonumber \\ && +8a^2z_0^2(\cosh 2aL +\cos 2aL) \}^{-1}
\EA
\index{$N_a$, $N_b$, $N_\kappa$, co\"efficienten van $C_2$}


 \renewcommand{\thesection}{\alph{section}}
  \renewcommand{\thesubsection}{\thesection\arabic{subsection}}
   \renewcommand{\theequation}{\thesection\arabic{equation}}

\setcounter{section}{0}
\setcounter{equation}{0}

\renewcommand{\thesection}{\Alph{section}.}
\section{Appendix}
\label{AppendixA}

\label{iklint}
We berekenen enkele integralen, die herhaald voorkomen.
\[I_{k,l} = \int \frac{\d^3\p}{(2\pi)^3} G^k(\p)G^{\ast\,l}(\p)
\]
met\index{$I_{k,l}$, integralen}
\beq G(\p) = \frac{1}{\p^2 - k^2_0 - nt} = \frac{1}{\p^2 + \mu^2} \mbox{
 ;}\qquad G^\ast(\p) = \frac{1}{\p^2 + \bar{\mu}^2}
\eeq
en $\mu^2 = -k^2_0-nt, \, \mu = -ik_0 \sqrt{1+nt/k^2_0} \approx
-ik-int/2k_0$. De keuze van
het teken is arbitrair, het teken bij $\bar{\mu}$ wordt wel vastgelegd
door de keuze bij $\mu$. Dus $\bar{\mu} \approx
ik_0 + in\bar{t}/2k_0$. De keuze van het teken bepaalt waar de polen
komen te liggen in het complexe vlak, maar be\"invloedt de waarde van de
contourintegraal niet. Het hulpmiddel is dat als je $I_{1,1}$ kent,
je $I_{k,l}$ kunt uitrekenen volgens
\beq
I_{k+1,l}= \frac{-1}{2k\mu}\frac{\d}{\d\mu}I_{k,l} \mbox{   ;} \qquad
I_{k,l+1}= \frac{-1}{2l\bar\mu}\frac{\d}{\d\bar \mu}I_{k,l}
\eeq
We hebben $I_{1,1}$ nodig.
\beq
I_{1,1}=\int\frac{\d^3\p}{ (2\pi)^3 } G(\p)G^\ast(\p) =
\int_{-\infty}^\infty \frac{ \d p }{ (2\pi)^2 } \,\frac{p^2}{(p^2
+\mu^2)(p^2 +\bar{\mu}^2)}
\eeq
Deze integraal doe je met behulp van contourintegratie. Neem aan dat
Im$t>0$.
De som van de residuen geeft je $1/4\pi(\mu+\bar{\mu})$. We hebben nog
de volgende gelijkheid nodig,
\beq
\mu + \bar{\mu} = \frac{-in(t-\bar{t})}{2k_0} = \frac{n\,{\rm Im}\,t}{k_0}
= \frac{n\,t\,\bar{t}}{4\pi a} = \frac{1}{a\ell}. \label{opttheorema}
\eeq
Bij het voorlaatste gelijkteken is het optisch theorema gebruikt, $\ell$
is de vrije weglengte. Met
behulp van bovenstaande gelijkheid vinden we de waarden van de
volgende integralen,
\BEN \begin{array}{llll}
I_{1,1}  = &\frac{1}{4\pi(\mu+\bar{\mu})} = \frac{a\ell}{4\pi}
  & I_{2,2} = &
\frac{1}{8\pi\mu\bar{\mu}(\mu+\bar{\mu})^3} \approx \frac{a^3\ell^3}{8\pi
k_0^2}   \\
I_{1,2} =&\frac{1}{8\pi\bar{\mu}(\mu+\bar{\mu})^2} \approx \frac{-i
a^2\ell^2}{8\pi
k_0} &I_{2,1} =& \frac{1}{8\pi\mu(\mu+\bar{\mu})^2}
\approx \frac{i a^2\ell^2}{8\pi k_0} \\
I_{1,3} =& \frac{1}{16\pi \bar{\mu}^2 (\mu+\bar{\mu})^3} \approx
\frac{- a^3\ell^3}{16 \pi k_0^2} ;
&I_{3,1} =& \frac{1}{16\pi \mu^2 (\mu+\bar{\mu})^3} \approx
\frac{- a^3\ell^3}{16 \pi k_0^2}  \\
I_{2,3}= &\frac{3}{32 \pi \mu^1 \bar{\mu}^2 (\mu+\bar{\mu})^4}
\approx \frac{-3 i a^4\ell^4}{32 \pi k_0^3} ;
&I_{3,2} =& \frac{3}{32 \pi \mu^2 \bar{\mu}^1 (\mu+\bar{\mu})^4}
\approx \frac{3 i a^4\ell^4}{32 \pi k_0^3}   \\
I_{1,4} =& \frac{1}{32 \pi \bar{\mu}^3 (\mu+ \bar{\mu})^4} \approx
\frac{i a^4\ell^4}{32 \pi k_0^3} ;
&I_{4,1} =& \frac{1}{32 \pi \mu^3 (\mu+ \bar{\mu})^4} \approx
\frac{-i a^4\ell^4}{32 \pi k_0^3}  \\
I_{2,4} =& \frac{1}{16 \pi \mu \bar{\mu}^3 (\mu+ \bar{\mu})^5} \approx
\frac{- a^5\ell^5}{16 \pi k_0^4} ;
&I_{4,2} =& \frac{1}{16 \pi \mu^3 \bar{\mu} (\mu+ \bar{\mu})^5} \approx
\frac{- a^5\ell^5}{16 \pi k_0^4} \\
I_{3,3}= &\frac{3}{32 \pi \mu^2 \bar{\mu}^2 (\mu + \bar{\mu})^5}
\approx \frac{3 a^5\ell^5}{32 \pi k_0^4} &&.
\end{array} \EEN

 \newcommand{\myskip}[1]{}
\myskip{

\beq
\mu + \bar{\mu} = \frac{-in(t-\bar{t})}{2k_0} = \frac{n\,{\rm Im}\,t}{k_0}
= \frac{n\,t\,\bar{t}}{4\pi} = \frac{1}{\ell}. \label{opttheorema}
\eeq
Bij het laatste gelijkteken is het optisch theorema gebruikt, $\ell$
is de vrije weglengte. Met
behulp van bovenstaande gelijkheid vinden we de waarden van de
volgende integralen,
\BEN \begin{array}{llll}
I_{1,1}  = &\frac{1}{4\pi(\mu+\bar{\mu})} = \frac{\ell}{4\pi}
  & I_{2,2} = &
\frac{1}{8\pi\mu\bar{\mu}(\mu+\bar{\mu})^3} \approx \frac{\ell^3}{8\pi
k_0^2}   \\
I_{1,2} =&\frac{1}{8\pi\bar{\mu}(\mu+\bar{\mu})^2} \approx \frac{-i
\ell^2}{8\pi
k_0} &I_{2,1} =& \frac{1}{8\pi\mu(\mu+\bar{\mu})^2}
\approx \frac{i \ell^2}{8\pi k_0} \\
I_{1,3} =& \frac{1}{16\pi \bar{\mu}^2 (\mu+\bar{\mu})^3} \approx
\frac{- \ell^3}{16 \pi k_0^2} ;
&I_{3,1} =& \frac{1}{16\pi \mu^2 (\mu+\bar{\mu})^3} \approx
\frac{- \ell^3}{16 \pi k_0^2}  \\
I_{2,3}= &\frac{3}{32 \pi \mu^1 \bar{\mu}^2 (\mu+\bar{\mu})^4}
\approx \frac{-3 i \ell^4}{32 \pi k_0^3} ;
&I_{3,2} =& \frac{3}{32 \pi \mu^2 \bar{\mu}^1 (\mu+\bar{\mu})^4}
\approx \frac{3 i \ell^4}{32 \pi k_0^3}   \\
I_{1,4} =& \frac{1}{32 \pi \bar{\mu}^3 (\mu+ \bar{\mu})^4} \approx
\frac{i \ell^4}{32 \pi k_0^3} ;
&I_{4,1} =& \frac{1}{32 \pi \mu^3 (\mu+ \bar{\mu})^4} \approx
\frac{-i \ell^4}{32 \pi k_0^3}  \\
I_{2,4} =& \frac{1}{16 \pi \mu \bar{\mu}^3 (\mu+ \bar{\mu})^5} \approx
\frac{- \ell^5}{16 \pi k_0^4} ;
&I_{4,2} =& \frac{1}{16 \pi \mu^3 \bar{\mu} (\mu+ \bar{\mu})^5} \approx
\frac{- \ell^5}{16 \pi k_0^4} \\
I_{3,3}= &\frac{3}{32 \pi \mu^2 \bar{\mu}^2 (\mu + \bar{\mu})^5}
\approx \frac{3 \ell^5}{32 \pi k_0^4} &&.
\end{array} \EEN
}
\section{Tabellen met conventies}
\label{tabel}

\indent Verschillende grootheden binnen en buiten het medium:
\vspace{.5cm}

\begin{tabular}{|l|c|c|}
\hline
  &{\bf buiten medium}&{\bf in het medium} \\
\hline
brekingsindex & $n_1$ & $n_0=mn_1$ \\ \hline
golfgetal & $k_1=n_1\omega/c$ & $k_0=n_0\omega/c=mk_1$ \\ \hline
hoek van inval & $\theta$ & $\theta'$ \\ \hline
$z-$component golfvector
& $p=k_1\cos\theta$ & $P=k_0\cos\theta'$ \\
& $=k_0\sqrt{\mu^2-1+1/m^2}$ &
$=k_0\mu$  \\ \hline
totale reflectie conditie &
$m<1$ and $\sin\theta>m$   &
$m>1$ and $\sin\theta'>1/m$ \\
 &i.e. $p$
imaginair  &
 i.e.  $P$
imaginair \\  \hline
\end{tabular}

\vspace{1.cm}
De transversale golfvector en reflectie- en transmissieco\"efficienten:
\vspace{.5cm}

\begin{tabular}{|c|c|}
\hline transversale golfvector &
$|\q_\perp|=q_\perp=k_1\sin\theta=k_0\sin\theta'=k_0\sqrt{1-\mu^2}$ \\
\hline \begin{tabular}{c} reflectie- en  \\  transmissie- \\
co\"efficienten
\end{tabular}
&
$\left\{ \begin{array}{ccc}
R
&=&\left(\frac{P-p}{P+p}\right)^2=
\left(\frac{|\mu|-\sqrt{\mu^2-1+1/m^2}}{|\mu|+\sqrt{\mu^2-1+1/m^2}}\right)^2
\\ T&=&\frac{4Pp}{(P+p)^2}=
\frac{4|\mu|\sqrt{\mu^2-1+1/m^2}}{\left(|\mu|+\sqrt{\mu^2-1+1/m^2}\right)^2}
\end{array} \right. $
\\ \hline
\end{tabular}\index{$T@$T(\mu)$, transmissieco\"efficient}
\index{$R@$R(\mu)$, reflectieco\"efficient} \vspace{1.cm}

Fourier-transformaties defini\"eren we als:
\vspace{.5cm}

\begin{tabular}{|c|c|}
\hline continuum & $f(\r)= \int\frac{d^d\p}{(2\pi)^d}\, \eexp{i \p\cdot \r}
f(\p)$ \\
	& $f(\p)= \int d^d\r \, \eexp{-i \p\cdot \r}
f(\r)$ \\ \hline
       discreet & $f_{\r}= \int\frac{d^d\p}{(2\pi)^d} \, \eexp{i \p\cdot \r}
f(\p)$ \\
	& $f(\p)= a^d \sum_{\r}\, \eexp{-i \p\cdot \r}
f_{\r}$ \\ \hline
\end{tabular}

\vspace{.5cm}

Laplace-transformaties defini\"eren we als:
\vspace{.5cm}

\begin{tabular}{|c|}
\hline  $f(\Omega)= \int_0^\infty dt \, \eexp{i \Omega
t} f(t)$ \\
	$f(t)= \int_{-\infty}^\infty \frac{d\Omega}{2\pi} \, \eexp{-i \Omega
t} f(\Omega)$ \\ \hline
\end{tabular}

\addcontentsline{toc}{section}{Index}
\printindex
\addtocontents{toc}{\protect\vspace{2cm}}
\addcontentsline{toc}{section}{Dit werk is opgedragen aan Jurriaan}
\index{Jurriaan}
\pagestyle{empty}
\section*{Voor Jurriaan}\index{Jurriaan}
\mbox{}
\clearpage
\section*{Voor Jurriaan}
\mbox{}


\begin{thebibliography}{9}
\bibitem{NL11} Th.M. Nieuwenhuizen and J.M. Luck, {\it Skin layer of diffusive media},
Phys. Rev. E {\bf 48} (1993) 569-588
\end{thebibliography}

\begin{thebibliography}{9}


\bibitem{vRN}
M. C. W. van Rossum en Th. M. Nieuwenhuizen, {\it  Multiple scattering of classical waves: microscopy, mesoscopy, and diffusion},
Rev. Mod. Phys. {\bf 71} (1999) 313-371.


\bibitem{NvR}
Th. M. Nieuwenhuizen en M. C. W. van Rossum. {\it Intensity distributions of waves transmitted through a multiple scattering medium},
Phys. Rev. Lett. {\bf 74} (1995) 2674-2677. 

\bibitem{ALN1}
E. Amic, J. M.  Luck en Th. M. Nieuwenhuizen, {\it Anisotropic multiple scattering in diffusive media},
J. Phys. A Math. Gen. {\bf 29} (1996) 4915-4955.


\bibitem{vTMN} B. A.  van Tiggelen, R. Maynard en Th.  M. Nieuwenhuizen, 
{\it Theory for multiple light scattering from Rayleigh scatterers in magnetic fields},
Phys. Rev. E {\bf 53} (1996) 2881-2908.



\bibitem{ALN2}
E. Amic, J. M.  Luck en Th. M. Nieuwenhuizen, {\it Multiple Rayleigh scattering of electromagnetic waves}, 
J. de Physique {\bf 7} (1997) 445-483.

\end{thebibliography}

\begin{thebibliography}{99}
\bibitem{Chandra} S. Chandrasekhar, {\it Radiative Transfer} 
(Dover Publications, New York, 1960)
\bibitem{vdHulst} H.C. van de Hulst, {\it Light Scattering by Small
Particles}, (Dover, New York, 1957/1981; {\it Multiple Light Scattering},
Vol. 1 and 2 (Academic, New York, 1980)
\bibitem{Ishim} A. Ishimaru {\it Wave Propagation and Scattering in
 Random Media}, Vols. 1 and 2 (Academic, New York, 1978)   
\bibitem{AAS} B.L. Altshuler, A.G. Aronov and B.Z. Spivak, JETP Lett.
{\bf 33} (1981) 94
\bibitem{SharvinSharvin} Y.D. Sharvin and Y.V. Sharvin, JETP Lett.
{\bf 34} (1981) 272
\bibitem{FisLee} D.S. Fisher and P.A. Lee, Phys. Rev. B{\bf 23} (1981) 6851
\bibitem{Barbara} Yu.N. Barbaranenkov, {\it Wave Corrections to the Transfer
 Equation for Backscattering}, Izv. Vysch. Uch. Zav.-Radiofiz. {\bf 16}
 (1973) 88  
\bibitem{KugaIsh} Y. Kuga and A. Ishimaru, J. Opt. Soc. Am. {\bf A1} (1984) 831
\bibitem{vALag} M. van Albada en A. Lagendijk, Phys. Rev. Lett. {\bf 55}
(1985) 2692
\bibitem{WMM} P.E. Wolf and G. Maret, Phys. Rev. Lett. {\bf 55} (1985) 2696
\bibitem{vHL} W. van Haeringen and D. Lenstra, {\it Analogies in Optics and
Micro Electronics} (Kluwer Academic, Haarlem, 1990)
\bibitem{AWL} B.L. Altshuler, P.A. Lee, and R.A. Webb, 
{\it Mesoscopic Phenomena in Solids} (Modern Problems in Condensed 
Matter Sciences, Volume 30;
 V.M. Agranovich and A.A. Maradudin, algemene editeuren;
North Holland, Amsterdam, 1991)
\bibitem{vATLT} M.P. van Albada, B.A. van Tiggelen, A. Lagendijk,
and A. Tip, Phys. Rev. Lett. {\bf 66} (1991) 3112; B.A. van Tiggelen,
A. Lagendijk, M.P. van Albada, and A. Tip, Phys. Rev. {\bf B 45}
(1992) 12233 
\bibitem{PWA} P.W. Anderson, Phys. Rev. {\bf 109} (1958) 1493
\bibitem{VW} D. Vollhardt and P. W\"olfle Phys. Rev. Lett. {\bf 45} (1980) 482;
Phys. Rev. {\bf B22} (1980) 4666; zie ook hun overzichtsartikel
`Self-consistent Theory of Anderson Localization' in
{\it  Electronic Phase Transitions}, W. Hanke and Yu.V. Kopaev, editeuren.
(Elsevier Science Publishers B. V., Amsterdam, 1992)
\bibitem{Wegner} F.J.  Wegner, Z. Phys. B {\bf 35} (1979) 207
\bibitem{Hikami2} S. Hikami and A. Fujita, (1991)
\end{thebibliography}

\begin{thebibliography}{99}

\bibitem{dOuter} P.N. den Outer, Th.M. Nieuwenhuizen, and A. Lagendijk,
{\it Location of objects in multiple-scattering media},
J. Opt. Soc. Am. {\bf A 10} (1993) 1209-1218


\bibitem{LanN1998}
D. Lancaster en Th.M. Nieuwenhuizen, {\it Scattering from objects immersed in a diffusive medium},
Physica A {\bf 256} (1998) 417-438

\bibitem{LN1999}
J.M. Luck en Th.M. Nieuwenhuizen, 
{\it Light scattering from mesoscopic objects in diffusive media},
Eur. Phys. J. B {\bf 7} (1999) 483-500



\end{thebibliography}

\begin{thebibliography}{9}
\bibitem{Chandra2} S. Chandrasekhar, {\it Radiative Transfer}
(Dover Publications, New York, 1960)

\bibitem{Ishimaru} A. Ishimaru, {\it Wave Propagation and Scattering in
 Random Media}, Vols. 1 and 2 (Academic, New York, 1978)   

\bibitem{LVdV} A. Lagendijk, R. Vreeker and P. de Vries,
Phys. Lett. A{\bf 136} (1989) 81
\bibitem{ZPW} J.X. Zhu, D.J. Pine, and D.A. Weitz, Phys. Rev 
A {\bf 44} (1991) 3948
\end{thebibliography}

\begin{thebibliography}{99}
\bibitem{Paepe} P.J. de Paepe, {\em Calculus II voor natuurkundigen},
syllabus verkrijgbaar bij de portier van de diamantslijperij op het Roeterseilandcomplex.
\bibitem{rps} Th.M. Nieuwenhuizen, A. Lagendijk and B.A. van Tiggelen,
Phys. Lett. A {\bf 169}, 191 (1992)
\bibitem{merz} E. Merzbacher, Quantum Mechanics, (J. Wiley \& Sons, New
York, 1970)
\bibitem{vdHSP} H.C. van de Hulst, {\it Light Scattering by Small Particles}
(Dover, New York, 1957; 1981)
\bibitem{Ishi} A. Ishimaru, {\it Wave propagation and scattering in
 random media}, Vols. 1 and 2 (Academic, New York, 1978) 
\end{thebibliography}

\begin{thebibliography}{9}
\bibitem{thmnaio}{Th.M. Nieuwenhuizen, Lezing AIO-cursus Amsterdam (1993)}
\bibitem{ziman}{J.M. Ziman, {\it Models of Disorder}, 
(Cambridge, 1978), mooi overzicht over
allerlei wanordelijke systemen.}
\bibitem{ecou}{E.N. Economou, {\it Green's functions in Quantum Physics},
 Springer
(1990), goed leesbaar boek over Greense funkties, toepassingen en
benaderingsstrategie\"en.}
\bibitem{halp} B.I. Halperin, Phys. Rev. A~{\bf 139}  (1965) 104.

\bibitem{thmn6} Th.M. Nieuwenhuizen, Physica A {\bf 120} (1983) 197.
\bibitem{thmn7} Th.M. Nieuwenhuizen, Physica A {\bf 125} (1984) 125.
\bibitem{poli}{I.Ya. Polishchuk, A.L. Burin en L.A. Maksimov, JETP Lett. {\bf 51}
(1990) 731. }
\end{thebibliography}

\begin{thebibliography}{1}
\bibitem{PolBurMaks}{I.Ya. Polishchuk, 
A.L. Burin en L.A. Maksimov, {\it Anderson localization in crystals with heavy isotopic impurities}, JETP Lett. 51 (1990) 730}
\bibitem{tiggelen2}B.A. van Tiggelen, A. Lagendijk and A.
Tip, J. Phys. CM {\bf 2} (1990), 7653
\end{thebibliography}

\begin{thebibliography}{9}
\bibitem{vATLT2} M.P. van Albada, B.A. van Tiggelen, A. Lagendijk,
and A. Tip, Phys. Rev. Lett. {\bf 66} (1991) 3112; \\ 
B.A. van Tiggelen,
A. Lagendijk, M.P. van Albada, and A. Tip, Phys. Rev. {\bf B 45}
(1992) 12233 
\end{thebibliography}

\begin{thebibliography}{9}
\bibitem{MLN}
M. Mishchenko, J. M. Luck en T.M. Nieuwenhuizen, {\it 
Full angular profile of the coherent polarization opposition effect}, J. Opt. Soc. Am. A {\bf 17} (2000) 888-891.
\end{thebibliography}

\begin{thebibliography}{9}

\bibitem{dOLag} P.N. den Outer en A. Lagendijk, {\it Influence of the
refractive index contrast on coherent backscattering}, 
Optics Communications {\bf 103} (1993)  169-173
\bibitem{LVVries} A. Lagendijk, R. Vreeker and P. de Vries,
Phys. Lett. A{\bf 136} (1989) 81
\bibitem{ZPWeitz} J.X. Zhu, D.J. Pine, and D.A. Weitz, Phys. Rev 
A {\bf 44} (1991) 3948
\end{thebibliography}

\begin{thebibliography} {9}
\bibitem {ThMNunp93} Th.M. Nieuwenhuizen, {\it Veelvoudige Verstrooiing van 
Golven}, (NvR Producties, Amsterdam, Noord-Holland, 1993)
\end{thebibliography}

\begin{thebibliography}{9}
\bibitem{berkovits2} R.~Berkovits and S.~Feng, Phys. Rev. Lett., {\bf 65},
(1990)   3120.
\bibitem{hik0} Th. M. Nieuwenhuizen and M. C. W. van
Rossum, Phys. Let. A, {\bf 177} (1993) 102
\bibitem{denouter}
P.N. den Outer, Th.M. Nieuwenhuizen, and A.~Lagendijk, J.
  Opt. Soc. Am. A, {\bf 10}, (1993), 1209.
\end{thebibliography}

\newpage

\begin{thebibliography}{9}

\bibitem{hikami1}
S.~Hikami,  \,  Phys. Rev. B  {\bf 24} (1981)    2671.
\bibitem{hik2} Th.M. Nieuwenhuizen and M.C.W. van Rossum,
 Phys. Let. A {\bf 177} (1993) 102.
\bibitem{stephen0}
M.J. Stephen and G.~Cwilich,  \,  Phys. Rev. Lett.  {\bf 59} (1987)    285.
\end{thebibliography}

\begin{thebibliography}{99}
\bibitem{feng}
S.~Feng, C.~Kane, P.~Lee, and A.~D. Stone,  \,  Phys. Rev. Lett.  {\bf 61}
  (1988)    834.
\bibitem{stephen1}
M.J. Stephen and G.~Cwilich,  \,  Phys. Rev. Lett.  {\bf 59} (1987)    285.
\bibitem{zyuzin1}
A.Yu. Zyuzin and B.Z. Spivak,  \,  Sov. Phys. JETP  {\bf 66} (1987).
\bibitem{pnini2}
R.Pnini and B.Shapiro,  \,  Phys. Rev. B  {\bf 39} (1989)
  6986.
\bibitem{albada1}
M.P. van Albada, J.F. de~Boer, and A.~Lagendijk,  \,  Phys. Rev. Lett.  {\bf
  64} (1990)    2787.
\bibitem{genack1}
A.Z. Genack, N.~Garcia, and W.~Polkosnik,  \,  Phys. Rev. Lett.  {\bf 65}
  (1990)    2129.
\bibitem{deboer}
J.F. de~Boer, M.P. van Albada, and A.~Lagendijk,  \,  Phys. Rev. B  {\bf 45}
  (1992)    658.
\bibitem{pnini1}
R.~Pnini and B.~Shapiro,  \,  Phys. Lett. A  {\bf 157} (1991)
  265.
\bibitem{lisyansky}
A.A. Lisyansky and D.~Livdan,  \,  Phys. Lett. A  {\bf 170} (1992)   ~53.
\bibitem{stephenboek}
M.J. Stephen,
\newblock In {\em
  Mesoscopic Phenomena in Solids}  volume~30, B.L. Altshuler, P.A. Lee, and
R.A. Webb, editors, ~81. North-Holland (1991).
\bibitem{berkovits}
R.~Berkovits and S.~Feng,  Phys. Rep. {\bf 283} (1994) 135.
\bibitem{garcia}
N.~Garcia, A.Z. Genack, R.~Pnini, and B.~Shapiro,  \,  Phys. Let. A {\bf 176}   (1993) 458.
\bibitem{genack2}
A.Z. Genack and N.~Garcia,  \,  Europhys. Lett. {\bf 21}   (1993)  Pages: 753.

\bibitem{NvRC2}
T.M. Nieuwenhuizen en M.C.W. van Rossum,
Phys. Rev. Lett. {\bf 74} (1995) 2674. 


\bibitem{StGenack1}
M.  Stoytchev and A. Z.  Genack, {\it Measurement of the probability distribution of total transmission in random waveguides},
Phys. Rev. Lett. {\bf 79} (1997) 309-312.

\bibitem{SZGenack2}
Sheng Zhang en Ariel Z. Genack,
Phys. Rev. Lett. {\bf 99} (2007) 203901.

\bibitem{vRALN}
M.C.W. van Rossum, I.V. Lerner, B.L.  Altshuler and T.M. Nieuwenhuizen, 
 {\it Deviations from the Gaussian distribution of mesoscopic conductance fluctuations,}
Phys. Rev. B {\bf 55} (1997) 4710-4716.



\bibitem{lagendijk}
A.~Lagendijk, R.~Vreeker, and P.~de~Vries,  \,  Phys. Lett. A  {\bf 136} (1989)   ~81.
\bibitem{hikami2}
S.~Hikami,  \,  Phys. Rev. B  {\bf 24} (1981)    2671.
\bibitem{skin}
M.C.W. van Rossum and Th.M. Nieuwenhuizen, Phys. Let. A {\bf 177}, 452
(1993).
\end{thebibliography}
\end{document}